\documentclass{article}
\usepackage[english]{babel}
\usepackage{microtype}

\usepackage[nonatbib,final]{neurips_2025}

\usepackage[utf8]{inputenc} % allow utf-8 input
\usepackage[T1]{fontenc}    % use 8-bit T1 fonts
\usepackage[hyphens]{url}
\usepackage[colorlinks=true,allcolors=blue,pdfborder={0 0 0}]{hyperref}       
\usepackage{booktabs}       % professional-quality tables
\usepackage{amsfonts}       % blackboard math symbols
\usepackage{nicefrac}       % compact symbols for 1/2, etc.
\usepackage[table]{xcolor}
\usepackage{amsmath,amsfonts,bm}

\def\eqref#1{equation~\ref{#1}}
\def\ceil#1{\lceil #1 \rceil}
\def\floor#1{\lfloor #1 \rfloor}
\def\1{\bm{1}}

\def\vx{{\bm{x}}}

\def\vz{{\bm{z}}}

\DeclareMathAlphabet{\mathsfit}{\encodingdefault}{\sfdefault}{m}{sl}
\SetMathAlphabet{\mathsfit}{bold}{\encodingdefault}{\sfdefault}{bx}{n}

\def\sB{{\mathbb{B}}}

\def\sD{{\mathbb{D}}}
\def\sS{{\mathbb{S}}}

\def\sZ{{\mathbb{Z}}}

\newcommand{\E}{\mathbb{E}}

\newcommand{\R}{\mathbb{R}}

\newcommand{\Var}{\mathrm{Var}}

\newcommand{\Cov}{\mathrm{Cov}}
\usepackage{graphicx}
\usepackage{amsmath}
\usepackage{amssymb}
\usepackage{mathtools}
\usepackage{cleveref}
\usepackage{enumitem}
\usepackage{multirow}
\usepackage{bbm}
\usepackage{subcaption}
\usepackage{csquotes}
\usepackage{float}
\usepackage{placeins}
\usepackage{wrapfig}
\usepackage[numbers,sort,compress]{natbib}
\usepackage{tcolorbox}

\newcommand*\samethanks[1][\value{footnote}]{\footnotemark[#1]}

\newif\ifshowcomments
\showcommentstrue  % <<< --- SET TO TRUE TO SHOW COMMENTS
\ifshowcomments
    \newcommand{\jamie}[1]{%
      \par % Ensure we start on a new line
      \noindent % Prevent paragraph indentation
      \begin{minipage}{\linewidth}% Create a box spanning the line width
        \color{red}% Switch color for the content of the minipage
        \textbf{Jamie:} % Add the prefix
        #1% The actual content provided by the user
      \end{minipage}%
      \par % Ensure space after the minipage
    }
    \newcommand{\jamieinline}[1]{\textcolor{red}{Jamie: #1}}
    
    \newcommand{\matthew}[1]{\textcolor{orange}{#1}}
    \newcommand{\cc}[1]{\textcolor{orange}{#1}}
    
    \newcommand{\ilia}[1]{\textcolor{red}{ilia: #1}}
    \newcommand{\cooper}[1]{\textcolor{blue}{cooper: #1}}
    \newcommand{\yva}[1]{\textcolor{green}{yva: #1}}
    \newcommand{\nlfr}[1]{\textcolor{violet}{nlfr: #1}}
    \newcommand{\sundar}[1]{\textcolor{magenta}{sundar: #1}}
    \newcommand{\matthieu}[1]{\textcolor{orange}{matthieu: #1}}
    \newcommand{\igor}[1]{\textcolor{cyan}{igor: #1}}
    \newcommand{\adam}[1]{\textcolor{blue}{adam: #1}}
    \newcommand{\franzi}[1]{\textcolor{gray}{franzi: #1}}
    \newcommand{\todo}[1]{\textcolor{red}{TODO: #1}}

\else
  \newcommand{\jamie}[1]{}
  \newcommand{\jamieinline}[1]{}
  \newcommand{\matthew}[1]{}
  \newcommand{\cc}[1]{}
  \newcommand{\ilia}[1]{}
  \newcommand{\cooper}[1]{}
  \newcommand{\yva}[1]{}
  \newcommand{\nlfr}[1]{}
  \newcommand{\sundar}[1]{}
  \newcommand{\adam}[1]{}
  \newcommand{\franzi}[1]{}
  \newcommand{\matthieu}[1]{}
  \newcommand{\todo}[1]{}
  \newcommand{\igor}[1]{}
\fi

\newcommand{\custompar}[1]{\vspace{.1cm} \noindent{\bf #1.}\:}

\title{Exploring the limits of strong membership \\ inference attacks on large language models}

\author{\textbf{Jamie Hayes$^{1}$\thanks{Equal contribution; corresponding authors: \texttt{jamhay@google.com}, \texttt{afedercooper@gmail.com}} \quad Ilia Shumailov$^{1}$ \quad Christopher A. Choquette-Choo$^{1}$}\\ 
\textbf{Matthew Jagielski$^{1}$ \quad Georgios Kaissis$^{1}$ \quad Milad Nasr$^{1}$ \quad Sahra Ghalebikesabi$^{1}$}\\
\textbf{Meenatchi Sundaram Muthu Selva Annamalai$^{2}$ \quad  Niloofar Mireshghallah$^{3}$ \quad Igor Shilov$^{4}$}\\
\textbf{Matthieu Meeus$^{4}$ \quad Yves-Alexandre de Montjoye}$^{4}$ \quad \textbf{Katherine Lee$^{1}$}\\
\textbf{Franziska Boenisch$^{5}$ \quad Adam Dziedzic$^{5}$ \quad A. Feder Cooper$^{6,7}$\samethanks}\vspace{.2cm}\\
$^{1}$Google DeepMind \quad $^{2}$University College London \quad $^{3}$University of Washington\\
$^{4}$Imperial College London \quad $^{5}$CISPA Helmholtz Center for Information Security\\
$^{6}$Stanford University \quad
$^{7}$Microsoft Research
}

\usepackage{xspace}
\newcommand{\shadow}{reference\@\xspace}

\newcommand{\roc}{\mathrm{ROC}}
\newcommand{\auc}{\mathrm{AUC}}
\newcommand{\rocauc}{\mathrm{ROC{\text{-}}AUC}}

\newcommand{\tpr}{\mathrm{TPR}}
\newcommand{\fpr}{\mathrm{FPR}}

\begin{document}

\maketitle

\vspace{-.5cm}
\begin{abstract}
\vspace{-.1cm}
State-of-the-art membership inference attacks (MIAs) typically require training many reference models, making it difficult to scale these attacks to large pre-trained language models (LLMs). 
As a result, prior research has either relied on weaker attacks that avoid training references (e.g., fine-tuning attacks), or on stronger attacks applied to small models and datasets.
However, weaker attacks have been shown to be brittle and insights from strong attacks in simplified settings do not translate to today's LLMs.
These challenges prompt an important question: 
are the limitations observed in prior work due to attack design choices, or are MIAs fundamentally ineffective on LLMs?
We address this question by scaling LiRA---one of the strongest MIAs---to GPT-2 architectures ranging from $10$M to $1$B parameters, training references on over $20$B tokens from the C4 dataset.
Our results advance the understanding of MIAs on LLMs in four key ways.
While (1) strong MIAs can succeed on pre-trained LLMs,
(2) their effectiveness, remains limited (e.g., $\auc{<}0.7$) in practical settings.
(3)~Even when strong MIAs achieve better-than-random $\auc$, aggregate metrics can conceal substantial per-sample MIA decision instability:
due to training randomness, many decisions are so unstable that they are statistically indistinguishable from a coin flip. 
Finally, (4) the relationship between MIA success and related LLM privacy metrics is not as straightforward as prior work has suggested.\looseness=-1
\end{abstract}

\vspace{-.4cm}
\section{Introduction}\label{sec:intro}
\vspace{-.1cm}

In a membership inference attack (MIA), an adversary aims to determine whether a specific data record was part of a model's training set~\citep{shokri2017membership,yeom2018privacy}.
MIAs pose a significant privacy risk to ML models, but state-of-the-art attacks are often too computationally expensive to run at the scale of pre-trained large language models (LLMs).
This is because strong MIAs require training multiple \shadow models to form membership predictions---and pre-training even one LLM is often prohibitively expensive in research settings. 
As a result, current work makes one of two compromises:
running weaker attacks that avoid training \shadow models (e.g., attacks that fine-tune an LLM), or running strong attacks that train small \shadow models on small datasets.  
However, both exhibit notable limitations (Section~\ref{sec:rw}). 
Weaker attacks are more practical, but they have been shown to be brittle---often performing no better than random guessing~\citep{fu2024membership,mireshghallah2022empirical,duanmembership}.
Stronger attacks, when run in simplified settings, fail to capture the complex dynamics of large-scale, pre-trained language models; as a result, their insights do not reliably generalize to modern LLMs~\citep{meeus2024did}.\looseness=-1

Results from both of these approaches leave key questions unanswered about the effectiveness of MIAs on LLMs. 
In particular, \emph{are the fidelity issues of weaker attacks due to omitting reference models, or do they point to a deeper, more fundamental challenge with applying membership inference to large language models?}
Current research has not offered an answer because, to date, there are no 
baselines of how the strongest MIAs perform on large-scale, pre-trained LLMs.\looseness=-1 

In this paper, we bridge this gap by running strong attacks at a scale significantly larger than previously explored.
We pre-train over $4{,}000$ GPT-2–like reference models, ranging from $10$ million to $1$ billion parameters~\citep{nanodo}, on subsets of the C4 dataset~\citep{raffel2020exploring} that are \emph{three orders of magnitude larger than those used in prior MIA studies}---over $50$ million samples, compared to fewer than $100{,}000$ in previous work~\citep{meeus2024sok}.
We use these models to conduct a detailed investigation of the Likelihood Ratio Attack (LiRA)~\citep{carlini2022membership}, one of the strongest MIAs in the literature.
This substantial effort proves worthwhile, as we uncover four key insights that advance the state of the art in understanding the potency and reliability of membership inference attacks on large language models:\looseness=-1 
\begin{itemize}[topsep=0cm, leftmargin=.65cm,itemsep=.05cm]
    \item \textbf{Strong membership inference attacks can succeed on pre-trained LLMs.}
    We are the first to execute strong attacks at this scale, and find that LiRA---in contrast to weaker fine-tuning attacks---can easily beat random $\rocauc$ baselines (Section~\ref{sec:exp1:realistic:warmup}). 
    Our results on Chinchilla-optimal models (trained for $1$ epoch) exhibit a non-monotonic relationship between model size and MIA vulnerability: 
    larger models are not necessarily more at risk (Section~\ref{sec:exp1:realistic:compute-optimal}).
    \item \textbf{The overall success of strong MIAs is limited on pre-trained LLMs.} 
    Even though we demonstrate that LiRA can succeed at LLM scale, we are only able to achieve impressive results (i.e., $\mathrm{AUC}{\geq}0.7$) when diverging from typical training conditions---specifically, by training for multiple epochs (Section~\ref{sec:exp2:limits:varyingtraining}) and varying training dataset sizes (Section~\ref{sec:exp2:limits:trainsize}).
    
    \item \textbf{Many correct  per‑sample MIA decisions for LLMs do not reflect reliable inference.}
    Even when an MIA achieves better-than-random $\auc$, the underlying per‑sample binary decisions are highly sensitive to training randomness.
    We quantify this by measuring per‑sample decision instability (Section~\ref{sec:instability:flip}) and find that, even at modest $\fpr$, many per‑sample decisions are statistically indistinguishable from a coin flip (Section~\ref{sec:instability:results}).\looseness=-1

    \item \textbf{The relationship between MIA success and related LLM privacy metrics is not straightforward.} 
    We show that samples seen later in training tend to be more at risk (Section~\ref{sec:exp3:per}); 
    however, this trend is complicated by sample length. 
    We also study if there is any relationship between training data extraction and MIA, and observe no correlation with MIA success.
    This suggests that the two privacy attacks may capture different signals related to memorization (Section~\ref{sec:exp3:extraction}).\looseness=-1
\end{itemize}

Our contributions serve as an extensive benchmark of strong MIAs, and also provide some initial answers to urgent open questions about the conditions under which MIAs exhibit a threat to privacy for LLMs. 
Our work also quantifies the performance gap between weaker (more feasible) and stronger attacks, establishing an upper bound for what weaker attacks could achieve in this setting.\looseness=-1 

\vspace{-.25cm}
\section{Background and related work}\label{sec:rw}
\vspace{-.15cm}

\textbf{Membership inference attacks (MIAs)} assess empirical privacy and information-leakage risk by asking whether an adversary can tell if \emph{a particular data point} $\vx$ was used to train a \textbf{target model} $h$.
Given knowledge of the target's architecture and training setup, the attacker trains multiple \textbf{reference models} $f \in \Phi$ on different subsets drawn from the same underlying distribution as the target's training data. 
For each $\vx$, references are partitioned into those trained with $\vx$ ($\Phi_\text{IN}$, where $\vx$ is a \textbf{member}) and those trained without $\vx$ ($\Phi_\text{OUT}$, where $\vx$ is a \textbf{non-member}).  
For a given $\vx$ and model $g$ (the target $h$ or a reference $f \in \Phi$), the attacker queries the model and computes an \textbf{observation statistic} $s(g, \vx)$ from the model's output on $\vx$ (e.g., loss, logit).  
MIAs transform these statistics into a \textbf{membership score} $\Lambda(\vx)$ that is used to infer whether $\vx$ was in the target's training data~\citep{carlini2022membership,sablayrolles2019white,watsonimportance,ye2022enhanced,zarifzadehlow}.\looseness=-1

Different attacks specify different ways of turning observation statistics into membership scores. 
For instance, for each query sample $\vx$, 
the \textbf{Likelihood Ratio Attack (LiRA)} collects two  sets of reference statistics,
$\{s(f,\vx):f\in\Phi_{\text{IN}}(\vx)\}$ and $\{s(f,\vx):f\in\Phi_{\text{OUT}}(\vx)\}$. 
These sets are treated as samples from two empirical distributions, to which density models ($p_\text{IN}$ and $p_\text{OUT}$) are fit. 
LiRA evaluates the target statistic $s(h,\vx)$ under the fitted densities to compute a likelihood ratio membership score $\Lambda(\vx)$ for $\vx$~\citep{carlini2022membership}. 
Given a score $\Lambda(\vx)$, the attacker outputs a binary membership decision via a threshold rule 
$b(\vx)=\1\{\Lambda(\vx)\ge \tau\}$. 
In practice, $\tau$ is typically calibrated on non-members to satisfy a fixed false positive rate ($\fpr$) $\eta$.  
Although membership inference is defined as a decision problem for a \emph{single} sample $\vx$, attack performance is evaluated as an \emph{average over many samples} (e.g., reporting $\tpr$ at fixed $\fpr$). 
Success is typically reported with threshold-agnostic metrics like $\rocauc$~\citep{yeom2018privacy, shokri2017membership} (Appendix~\ref{app:sec:background}).
To address this gap, we also run experiments that offer novel insights into sample-specific attack performance (Sections~\ref{sec:instability}~\&~\ref{sec:exp3:samples}, Appendices~\ref{app: instability} \&~\ref{app:moreperexamplemiaresults}). 

The number of reference models necessary for successful attacks varies across methods---from tens or hundreds for LiRA and Attack-R~\citep{ye2022enhanced}, to as few as $1$ or $2$ for RMIA~\citep{zarifzadehlow}.
While these attacks have been successfully applied to smaller settings, they are often considered impractical for contemporary language models due to the prohibitive computational cost of training even a single reference LLM.
As a result, prior work attempts to approximate stronger, reference-model-based attacks in various ways.\looseness=-1

\custompar{Small-scale, strong, reference-based attacks} 
The first work to evaluate MIAs in smaller language models (RNNs) trained $10$ references~\citep{song2019auditing}. 
However, insights from such settings do not translate to today's LLMs~\citep{meeus2024sok}, as the training dynamics differ significantly.
Other work has used a single reference model to attack a small, pre-trained masked language models~\citep{mireshghallah2022quantifying}, but this approach reduces precision, as effective  membership inference is difficult with fewer references.\looseness=-1

\custompar{Larger-scale, weak, reference-free attacks} 
To avoid the cost of training reference models, weaker attacks consider a range of observation statistics to infer membership, 
typically leveraging black-box access to the model. 
\citet{yeom2018privacy} use model loss, \citet{carlini2021extracting} use normalized model loss and \texttt{zlib} entropy, and \citet{mattern2023membership} compare the model loss to the loss achieved for neighboring samples.
More recent work experiments with token probabilities~\cite{shidetecting, zhang2025mink} and changes in loss based on prompting with different context~\citep{xie2024recall,wang2025recall}.  

Other work attempts to derive membership signal from changing the model.
For instance, prior work perturbs inputs or model parameters and observes resulting changes in target loss on the sample, or uses (parameter-efficient) fine-tuning on domain-specific datasets to detect privacy risks~\citep{fu2024membership,mireshghallah2022empirical,lukas2023analyzing,kandpal2024user,panda2025privacy, rossi2024auditing, meeus2025canary, chang2024context}. 
However, fine-tuning attacks introduce \emph{new} data to the problem setup, which may complicate the validity of using MIAs to detect benchmark contamination~\citep{oren2023proving,deng2024investigating,maini2024llm,maini2025peeking} and to draw reliable conclusions about other sensitive data issues~\citep{lee2023talkin,duarte2024cop,meeuscopyright,shidetecting,wei2024proving, cooper2024files, cooper2024unlearning, cooper2025books, zhang2025miacannot, cooper2023report}. 
A recent approach evaluates attacks on LLMs using post-hoc collected datasets. 
While prior work has reported high success rates on a variety of models and datasets ($\auc{\approx}0.8$)~\citep{shidetecting,meeus2024did,zhang2025mink,xie2024recall,wang2025recall},  
such evaluations rely on the model's training-date cutoff as a proxy for distinguishing between member and non-member data points~\citep{maini2024llm}.
These newer data introduce distribution shift, which can undermine the validity of the reported results~\citep{duanmembership,maini2024llm,das2024blind,meeus2024sok}.
Further, when current MIAs are evaluated in a controlled privacy game like this, they often barely outperform random guessing~\citep{duanmembership,meeus2024sok}.\looseness=-1 

\vspace{-.3cm}
\section{Examining strong MIAs in realistic settings for pre-trained LLMs}\label{sec:exp1:realistic}
\vspace{-.1cm}

Altogether, the limitations of prior work raise the key question that motivates our work: 
\emph{are the fidelity issues of weaker attacks due to omitting reference models, or do they point to a deeper, more fundamental challenge with applying membership inference to large language models?}
This is a big question, so we break it down into smaller ones that we can test with specific experiments that reveal different information about the effectiveness of strong MIAs on pre-trained LLMs. 
To start, we determine which strong MIA method to use across our experiments.
We evaluate two of the strongest attacks in the literature---LiRA~\citep{carlini2021extracting} and RMIA~\citep{zarifzadehlow}---in a variety of settings. 
For the experiments that follow, we use LiRA because we observed that it can achieve substantially higher $\rocauc$ when attacking pre-trained LLMs. 
We compare LiRA and RMIA in Appendix~\ref{app:sec:warmup}. 

In this section, we investigate the relationship between the number of reference models and attack success (Section~\ref{sec:exp1:realistic:warmup}). 
Based on these results, we decide to use $128$ reference models in all following experiments. 
Then, we test the effectiveness of strong attacks under realistic  settings---settings that reflect how LLMs are actually trained. 
To do so, we run LiRA on models of various sizes, which we train according to Chinchilla-scaling laws~\citep{hoffmann2022trainingcomputeoptimallargelanguage} (Section~\ref{sec:exp1:realistic:compute-optimal}).
Together, these experiments inform our first key result: with respect to overall $\rocauc$, 
\textbf{strong membership inference attacks can succeed on pre-trained LLMs}. 
In the following sections, we expand upon these results to other training and attack conditions; 
we will refine our first key result by investigating the limits of strong MIA success rates (Section~\ref{sec:exp2:limits}), and by digging beneath aggregate metrics like $\auc$ to better understand attack performance with respect to individual samples (Sections~\ref{sec:instability} \& \ref{sec:exp3:samples}).\looseness=-1

\custompar{General setup} 
For all experiments, we pre-train GPT-2 architectures of varying sizes---from 10M to 1B---on subsets of the C4 dataset~\citep{raffel2020exploring} using the open-source NanoDO library~\citep{nanodo}. 
The training datasets we use are \emph{$3$ orders of magnitude larger than those in prior MIA studies}:
over $50$M samples, compared to fewer than $100$K samples in previous work~\citep{meeus2024sok}. 
We explore datasets of this size because, while it is well established that MIA success depends on both model capacity and training dataset size~\citep{shokri2017membership,yeom2018privacy, ye2022enhanced}, the nature of this relationship remains unexplored pre-trained-LLM scale. 
For each attack, we start with a fixed dataset of size $2N$ (e.g., $20$M) drawn from C4, from which we randomly subsample (with different random seeds) reference training sets of size $N$ (e.g., $10$M).
So, for each reference $f$, half of the drawn samples are members and half are non-members. 
This yields the different member (IN) and non-member (OUT) distributions for each sample that we use to run LiRA. 
In our largest experimental setting, we use $2N{\approx}100$M. 
Specific experimental configurations vary, so we introduce additional setup as needed. 
(See \Cref{app:exp_details} for details.)\looseness=-1

\vspace{-.1cm}
\subsection{Warm-up: How many reference models should we use?}\label{sec:exp1:realistic:warmup}
\vspace{-.1cm}
\begin{wrapfigure}{r}{.48\linewidth}
\vspace{-.5cm}
\centering
    \includegraphics[width=\linewidth]{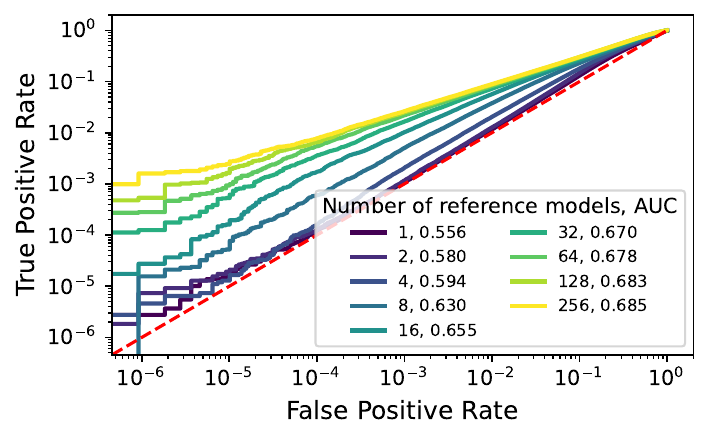}
\vspace{-.7cm}
\caption{\textbf{LiRA with different references.} 
We attack a $140$M model trained on ${\approx}7$M samples. 
As references increase, LiRA's performance improves (measured with $\rocauc$). 
However, there are diminishing returns: 
$\auc$ is effectively unchanged from $128$ to $256$ IN references.} 
\label{fig:lira_references}
\vspace{-.45cm}
\end{wrapfigure}
To determine the number of reference models to use for all of our experiments, we train $140$M-parameter models on ${\approx}7$M samples, which equates to approximately $2.8$B training tokens. (This is optimal for this model size, according to Chinchilla scaling laws with an over-training multiplier of $20$~\citep{hoffmann2022trainingcomputeoptimallargelanguage}.) 
As shown in Figure~\ref{fig:lira_references}, we test a range of reference models.
(We plot the number of IN references; the total number of IN/OUT references is $2\times$ this number.) 
The plot shows  multiple receiver operating characteristic ($\roc$) curves, indicating the observed true positive rate ($\tpr$) for the fixed false positive rate ($\fpr$) on a $\log$-$\log$ scale. 
Area under the curve ($\auc$) is provided for each $\roc$.
The dashed red line represents the baseline for which the attack is equivalent to random guessing (i.e., cannot distinguish between true and false positives so $\tpr{=}\fpr$; $\auc{=}0.5$). 
We report $\auc$ as our primary metric, as it is otherwise challenging to visualize $\tpr$ over a wide range of fixed $\fpr$. 
(For comparison, see Figure~\ref{fig:chinchilla:sizes}, which shows a limited range of $\fpr$, but does not surface  threshold-agnostic $\auc$.)
We also investigate the performance of different observation statistics (\Cref{app:sec:signal}), and choose to use model loss. 
Altogether, while LiRA clearly beats the random baseline, it is not remarkably successful in this setting: 
regardless of the number of references, it never achieves an $\auc$ of $0.7$.
Even though success increases with more references, there are diminishing returns. 
From $1$ to $8$ IN references ($2$ to $16$ references total), $\auc$ has a relative increase of $13.3\%$; for the next 8$\times$ increase (from $8$ to $64$), $\auc$ only increases $7.6$\%; and, doubling from $128$ to $256$ only yields a $0.2\%$ improvement.\looseness=-1
We opt to use $128$ total references ($64$ IN, $64$ OUT) in most experiments below.

\vspace{-.1cm}
\subsection{Training and attacking a compute-optimal model}\label{sec:exp1:realistic:compute-optimal}
\vspace{-.1cm}

In practice, models are typically trained based on observed scaling laws: 
for a given model size, the scaling law suggests the optimal number of tokens to use for training.
To assess strong MIA in realistic conditions for pre-trained LLMs, we attack models of various sizes, setting the number of training samples to be optimal according to Chinchilla scaling~\citep{hoffmann2022trainingcomputeoptimallargelanguage}.
Specifically, we set the number of training tokens to be $20\times$ larger than the number of model parameters and we \emph{only train for 1 epoch}---a common choice in large training runs~\citep{bai2023qwen, touvron2023llama}.
Specific training recipes and experimental details are in Appendices~\ref{app:sec:optimal} and \ref{app:exp_details}, including the number of samples used to train each model size.\looseness=-1 

\begin{figure}[t]
  \centering
\begin{subfigure}[t]{0.48\textwidth}
\centering
\vspace*{0cm}
    \includegraphics[width=.9\linewidth]{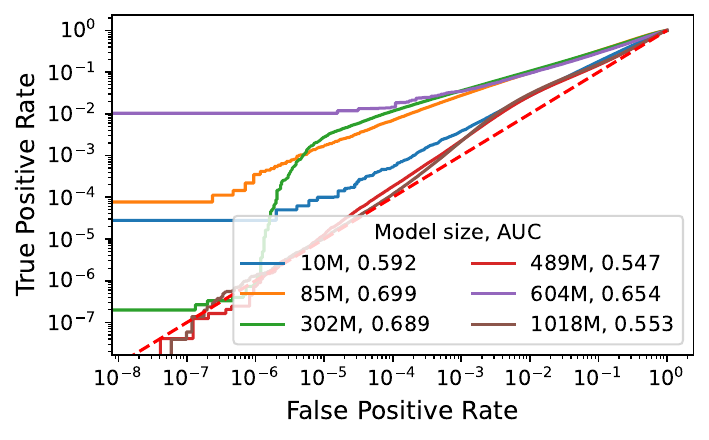}
    \vspace{-.1cm}
        \subcaption{$\rocauc$}
        \label{fig:chinchilla:roc}
\end{subfigure}\hfill
\hfill
\begin{subfigure}[t]{.48\textwidth}
\vspace*{0cm}
\centering 
    \includegraphics[width=.94\linewidth]{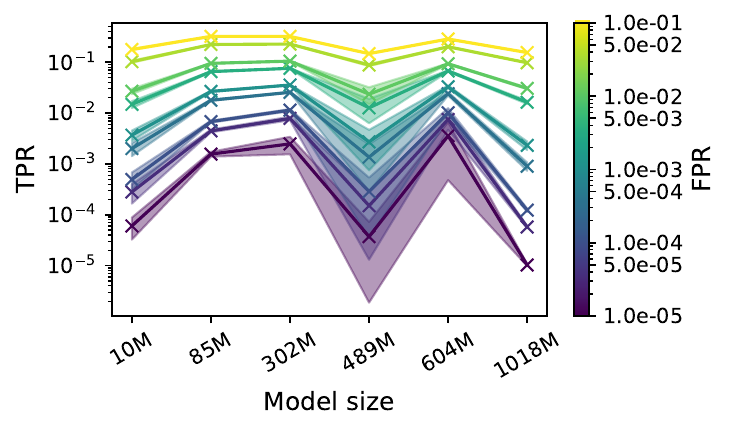}
        \vspace{-.1cm}
        \subcaption{$\tpr$ at fixed $\fpr$}
        \label{fig:chinchilla:sizes}
\end{subfigure}%
\vspace{-.1cm}
\caption{\textbf{MIA vulnerability for compute-optimally trained models.} 
We show attacks on $6$ models of different sizes under Chinchilla-optimal conditions for 1 epoch, using $128$ references.
(\textbf{a}) $\roc$ curves  demonstrate varying MIA susceptibility for $10$M ($\auc{=}0.592$), $85$M ($\auc{=}0.699$), $302$M ($\auc{=}0.689$), $489$M ($\auc{=}0.547$), $604$M ($\auc{=}0.654$) and $1018$M ($\auc{=}0.553$). 
The $85$M and $302$M models show the highest vulnerability, indicating that increasing model size does not uniformly decrease MIA risk in this setting.
(\textbf{b}) How $\tpr$ for each fixed $\fpr$ varies by model size. 
\looseness=-1}
\label{fig:compute-optimal-main}
\vspace{-.4cm}
\end{figure}

In Figure~\ref{fig:compute-optimal-main}, we show two views of the results of attacking $10$M-, $85$M-, $302$M-, $489$M-, $604$M- and $1018$M-parameter models. 
These model sizes come from the default configurations available in NanoDO~\citep{nanodo}.
For readability, we exclude the results for the $140$M model, as we investigate this architecture above.  
In Figure~\ref{fig:lira_references}, the attack on the $140$M model with $128$ IN/OUT references has  $\auc{=}0.678$, which puts its performance below the $85$M and $302$M models. 
Interestingly, we observe a non-monotonic relationship between model size and MIA vulnerability under these training conditions.
In Figure~\ref{fig:chinchilla:roc}, the $85$M model shows the highest $\auc{=}0.699$, followed by the $302$M ($\auc{=}0.689$). 
The $489$M model exhibits the lowest success  ($\auc{=}0.547$).\looseness=-1 

Figure~\ref{fig:chinchilla:sizes} provides a different view of the same results. 
By model size, each line compares the $\tpr$ for fixed settings of $\fpr$. 
Our expectation was that each line would look approximately horizontal, as the training set size is being scaled proportionally (and optimally, according to~\citet{hoffmann2022trainingcomputeoptimallargelanguage}) to model size. 
From $10$M to $302$M, there is a consistent pattern of the $\tpr$ increasing with model size;  
but then, at $489$M, there is a significant drop in $\tpr$. 
There are many reasons why this may have occurred.
First, the most pronounced differences in $\tpr$ are at extremely small values.  
Even subtle differences in training runs may flip samples from correct to incorrect member predictions (Section~\ref{sec:instability}), which, in the low $\tpr$ regime, can have a large effect on overall MIA success.
Second, Chinchilla scaling~\citep{hoffmann2022trainingcomputeoptimallargelanguage} is not the only such law.
\citet{sardana2023beyond}, \citet{hu2024minicpm}, and \citet{grattafiori2024llama} all introduce other ways to optimally select the number of training tokens for a given model. 
In future work, we will investigate if these other token-size-selection methods stabilize $\tpr$ as model size grows.\looseness=-1 

As we discuss next (Section~\ref{sec:exp2:limits:trainsize}), repeating this experiment by training these architectures on the same fixed dataset size exhibits vastly different results. 
We additionally test other training configurations. 
In Appendix~\ref{app:sec:limits}, we alter the learning rate schedule and observe that there is a modest effect on attack performance.  
(See Appendix~\ref{app:sec:optimal}, where, as a sanity check, we also confirm that larger models converge to lower loss values, reflecting their increased capacity to fit the training data.)

\vspace{-.2cm}
\section{Varying compute budget and training dataset size}\label{sec:exp2:limits}
\vspace{-.1cm}

Even in the most successful (i.e., highest $\auc$) 
case, overall attack performance is not particularly impressive when running LiRA with a large number of references on compute-optimal models trained for $1$  epoch.
Similar to our experiments with LiRA and varied numbers of references (Figure~\ref{fig:lira_references}), the maximum $\auc$ we observe remains under $0.7$ for all model sizes  (Figure~\ref{fig:compute-optimal-main}). 
This raises a natural follow-on question: 
if we free ourselves from the constraints of typical training settings, is it possible to improve success? 
\emph{Can we identify an upper bound on how strong MIAs could perform on pre-trained LLMs?}\looseness=-1 

To address this question, we run attacks on models trained on different-sized (not always Chinchilla-optimal) datasets (Section~\ref{sec:exp2:limits:trainsize}) for more than $1$  epoch (Section~\ref{sec:exp2:limits:varyingtraining}).
Our experiments show that diverging from typical settings can indeed improve attack success. 
However, while these experiments are a useful sanity check, they do not suggest conclusions about the effectiveness of strong MIAs in general. 
Instead, there appears to be an upper bound on how well strong MIAs can perform on LLMs under practical conditions.
In other words, these experiments inform our second main observation: \textbf{the success of strong MIAs is limited in typical LLM training settings.}

\vspace{-.1cm}
\subsection{Effects of scaling the compute budget (i.e., training for more epochs)}\label{sec:exp2:limits:varyingtraining}
\vspace{-.1cm}

\begin{figure}[t]
  \centering
\begin{subfigure}[t]{0.48\textwidth}
    \includegraphics[width=0.9\linewidth]{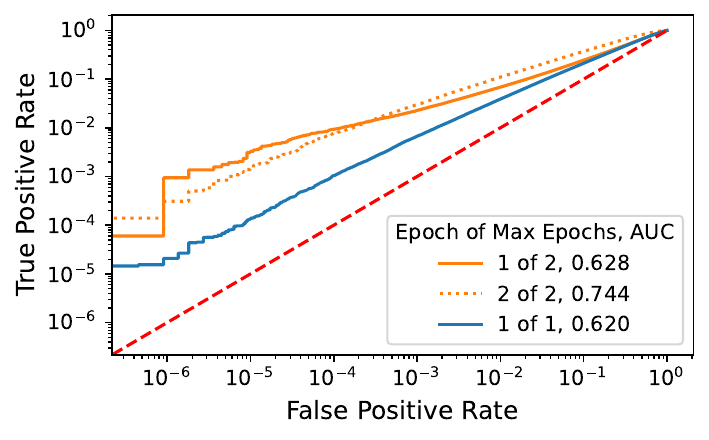}
        \vspace{-.1cm}
        \caption{$44$M model, split dataset in half and train for $2$ epochs, or train on the entire dataset for $1$ epoch}
        \label{fig:other-training:split}
\end{subfigure}\hfill
\begin{subfigure}[t]{.48\textwidth}
    \includegraphics[width=0.9\linewidth]{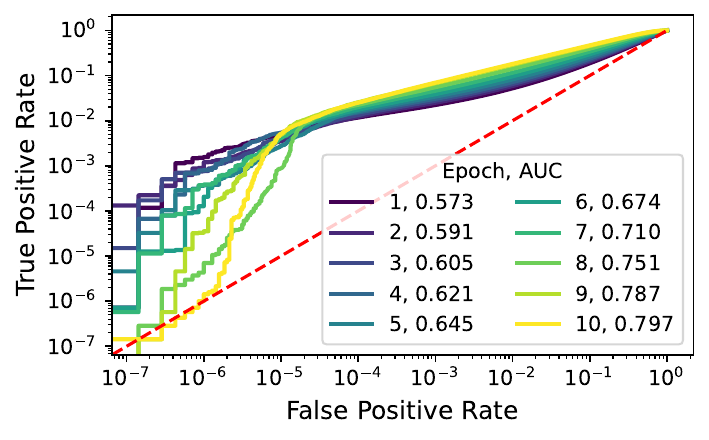}
        \vspace{-.1cm}
        \caption{$140$M model, training for $10$ epochs}
        \label{fig:other-training:epochs}
\end{subfigure}%
\vspace{-.1cm}
\caption{\textbf{Studying the effect of varying epochs.} (\textbf{a}) We compare attacking a $44$M model trained on the whole Chinchilla-optimal dataset in $1$ epoch ($\auc{=}0.620$ after $1$ of $1$ epoch) to training for $2$ epochs on only half of the dataset ($\auc{=}0.744$ after $2$ of $2$ epochs).
(\textbf{b}) We attack a $140$M model trained on the whole Chinchilla-optimal dataset for $10$ epochs. 
$\auc$ increases with more epochs.\looseness=-1 
}
\label{fig:other-training}
\vspace{-.4cm}
\end{figure}

In Figure~\ref{fig:other-training:split}, we compare MIA $\auc$ for the $44$M model under different training configurations.  
We keep the total number of tokens surfaced to the model during training Chinchilla-optimal, but we alter \emph{when} these tokens are surfaced. 
As a baseline, we train for $1$ epoch on the entire dataset, and achieve $\auc{=}0.620$ with LiRA. 
(See Figure~\ref{fig:other-training:split}, $1$ of $1$.)
We then take \emph{half} of the training dataset and train the same architecture for $2$ epochs.
In both settings the total number of training tokens is Chinchilla-optimal, however, in the latter, the model has processed each training sample twice rather than once.
For the $2$-epoch model, we observe a significant increase in MIA vulnerability:
$\auc{=}0.744$, which is higher than when this model has only completed $1$ epoch of training ($\auc{=}0.628$, $1$ of $2$) and than when the model is trained for $1$ epoch on the entire dataset ($\auc{=}0.620$, $1$ of $1$). 
Increasing training epochs---even on a smaller dataset to maintain Chinchilla optimality---amplifies vulnerability to MIA, compared to training for fewer epochs on a larger dataset. 
However, there is no significant uplift in $\tpr$ at small fixed $\fpr$ between epochs $1$ and $2$ for the $2$-epoch model.
The MIA at the second epoch is less successful than the one after $1$ epoch for small $\fpr$.
As above, this is perhaps due to subtle  differences in runs having an impact at these small values (Sections~\ref{sec:exp1:realistic:compute-optimal} \&~\ref{sec:instability}).\looseness=-1

To investigate this further, in Figure~\ref{fig:other-training:epochs}, we show how  
$\auc$ changes over the course of training the $140$M model for $10$ epochs. 
As expected, $\auc$ increases with more epochs, starting from $0.573$ and reaching $0.797$ at the end of the tenth epoch.\footnote{At epoch $1$, $\auc{=}0.573$, which differs from $\auc{=}0.678$ in Figure~\ref{fig:chinchilla:roc} (also $1$ epoch). 
    This is likely because of variance between runs (Section~\ref{sec:instability}) and substantially different learning rates between the two setups.\looseness=-1
}
As in \Cref{fig:other-training:split}, there is an $\fpr$ inflection point where $\tpr$ for later epochs is \emph{smaller} than earlier epochs.   
In Appendix~\ref{app:sec:limits}, we also train the $140$M model on fewer than the $\approx7$M Chinchilla-optimal samples, and (similar to Figure~\ref{fig:other-training:split}) we observe a more dramatic increase in MIA vulnerability.
Attacking a $140$M model trained on $2^{19}{\approx}500$K samples exhibits both greater absolute MIA success and a faster relative increase in success in the first few epochs.\looseness=-1 

\vspace{-.1cm}
\subsection{Effects of scaling the training dataset size}\label{sec:exp2:limits:trainsize}
\vspace{-.1cm}

\begin{figure}[t]
\vspace{-.1cm}
  \centering
\begin{subfigure}[t]{0.48\textwidth}
\vspace*{0cm}
    \includegraphics[width=0.9\linewidth]{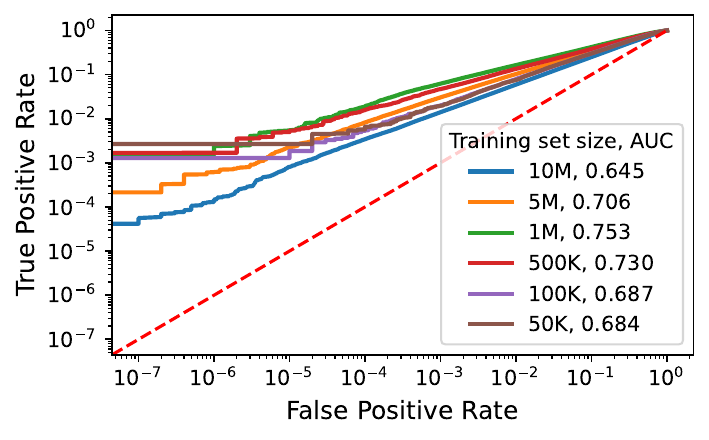}
    \vspace{-.1cm}
        \caption{$140$M model $\times$  various dataset sizes}
        \label{fig:other-training:sizes}
\end{subfigure}\hfill
\begin{subfigure}[t]{.48\textwidth}
\vspace*{0cm}
    \includegraphics[width=0.94\linewidth]{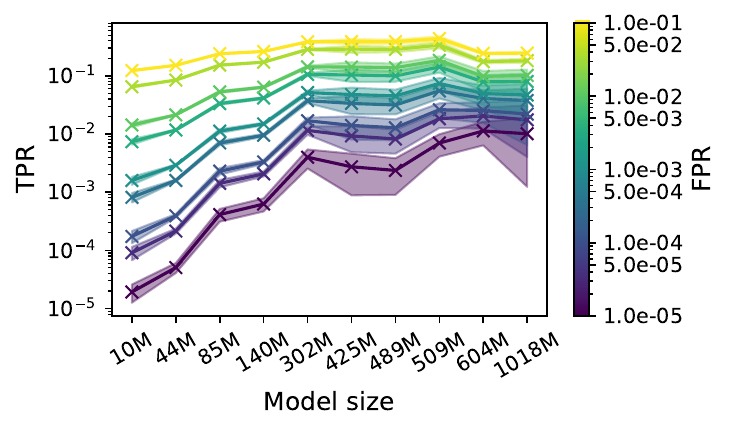}
        \vspace{-.1cm}
        \caption{Various model sizes $\times$ $2^{23}$sample training set}
        \label{fig:other-training:fixed-sized}
\end{subfigure}%
\vspace{-.1cm}
\caption{\textbf{Varying sizes of training dataset and model ($1$ epoch).}  
(\textbf{a}) We attack $140$M models trained on different-size datasets ($50$K to $10$M samples). 
MIA success does not monotonically increase with dataset size. 
(\textbf{b}) We attack different-size models trained on a fixed dataset size (${\approx}8.3$M samples), and plot how $\tpr$ varies at fixed $\fpr$. 
MIA success monotonically increases with model size.\looseness=-1} 
\label{fig:other-training-2}
\vspace{-.5cm}
\end{figure}

We next run two sets of experiments to study the role of training dataset size on MIAs---beyond training on the Chinchilla-optimal number of tokens.
We train $140$M models on datasets ranging from $50$K to $10$M samples (again for $1$ epoch) and attack these models with LiRA. 
In Figure~\ref{fig:other-training:sizes}, we show $\roc$ curves for the different models. 
As we train models on smaller datasets, for a given $\fpr$, $\tpr$ does not always increase.
This suggests that $\tpr$ at fixed $\fpr$ is not necessarily positively correlated with decreasing the 
training set size. 
Rather, $\auc$ is highest for moderately sized datasets (around $1$M samples, $\auc{=}0.753$), and decreases for both very small and very large datasets (under $\auc{=}0.7$ for both).
Indeed, the capacity of the model also has an effect on susceptibility to strong MIAs.\looseness=-1

In Figure~\ref{fig:other-training:fixed-sized}, we train different model sizes with a fixed training set size of $2^{23}{\approx}8.3$M samples---significantly more tokens than is Chinchilla-optimal for several models (e.g., $10$M, $44$M). 
We plot the mean and standard deviation of $\tpr$ at fixed $\fpr$, where we run the attack $16$ times using different random seeds, which has the effect of dictating the batch order. 
For each model size, we train $16$ sets of $128$ reference models, and we also vary the target model over each attack. 
We include the associated $\rocauc$ for each model size in Appendix~\ref{app:sec:limits}, which are consistent with the MIA prediction variability in Figure~\ref{fig:other-training:fixed-sized}. 
We observe a monotonic increase in $\tpr$ at different $\fpr$s as model size increases.
This is quite different from Figure~\ref{fig:chinchilla:sizes}, where we scale the training set size with model size.
As model capacity grows, vulnerability to MIA also grows if we keep the training set size constant. 
Further, there is significantly more variance in $\tpr$ for larger model sizes and at smaller fixed $\fpr$.\looseness=-1

\vspace{-.2cm}
\section{Uncovering per-sample MIA membership decision instability}\label{sec:instability}
\vspace{-.1cm}

The high degree of variance that we observe in the prior section raises a natural question:
how stable are the underlying per-sample membership decisions in strong MIAs? 
Next, we describe the setup we use to measure per-sample MIA decision instability (Section~\ref{sec:instability:flip}). 
In general, this is a reasonable thing to do: 
it is standard to report attack success with \emph{aggregate} metrics ($\auc$, $\tpr$ at fixed $\fpr$), but the MIA security game is defined with respect to determining if \emph{a particular sample} $\vx$ was used in training (Section~\ref{sec:rw}~\& Appendix~\ref{app:sec:background}). 
We show a selection of results for the $302$M model  (Section~\ref{sec:instability:results}), which reveal our third key takeaway: 
aggregate metrics may imply that a strong MIA on an LLM performs better than random guessing;
however, even at modest $\fpr$, due to training randomness \textbf{a large fraction of underlying, individual membership decisions are highly unstable, resembling coin flips}.
For these samples, strong MIAs are not producing reliable knowledge about membership.\looseness=-1

\vspace{-.1cm}
\subsection{Computing per-sample flip rate on calibrated membership decision rules}\label{sec:instability:flip}
\vspace{-.1cm}

We fix the IN/OUT references and target training set, varying only the random seed used to train targets. 
By isolating variability from equally plausible targets, we can probe the stability of using counterfactual references to derive membership signal for a given sample $\vx$. 
Substantial cross‑target disagreement on $\vx$'s membership would indicate the signal is not providing robust membership evidence for \(\vx\).\looseness=-1

Formally, let \(r\!\sim\!\mu\) denote a target trained on a \emph{fixed} dataset with randomness induced by the seed controlling batch order.
We train \emph{one set} of references for all attacks on different $r\!\sim\!\mu$.   
Let \(\Lambda_r(\vx)\in\R\) be the $r$-specific LiRA score for \(\vx\).
At a fixed $\fpr$ $\eta$, we calibrate a per-seed threshold $\tau_r(\eta)$ on non-members to form the binary decision rule $b_r^{(\eta)}(\vx){=} \1\{\Lambda_r(\vx)\ge \tau_r(\eta)\}$ (Section~\ref{sec:rw}). 
Per-seed calibration mirrors the standard threat model, in which an attacker runs the MIA on a \emph{single} target~\citep{carlini2022membership}. 
The (population) \textbf{flip rate}~\citep{cooper2024variance} at $(\eta$, $\vx)$ is the pairwise decision disagreement probability under \(\mu\):\looseness=-1

\vspace*{-.35cm}
\begin{equation*}
\textstyle
\mathrm{flip}_{\eta}(\vx)
\;\coloneqq\;
\Pr_{r,r'\stackrel{\mathrm{i.i.d.}}{\sim}\mu}\big[b_r^{(\eta)}(\vx)\neq b_{r'}^{(\eta)}(\vx)\big].
\vspace*{-.1cm}
\end{equation*}
In practice, with \(B\ge 2\) i.i.d.\ target replicas \(r_1,\ldots,r_B\sim\mu\), the canonical unbiased estimator is 

\vspace*{-.4cm}
\begin{equation}
\label{eq:flip:ustat}
\textstyle
\widehat{\mathrm{flip}}_{\eta,B}(\vx) \;=\;
\binom{B}{2}^{-1}\sum_{1\le i<j\le B}
\1\{b_{r_i}^{(\eta)}(\vx)\neq b_{r_j}^{(\eta)}(\vx)\} \;=\; \frac{2\,B_0(\vx)\,B_1(\vx)}{B\,(B-1)}, 
\end{equation}
where \(B_1(\vx){=}\sum_{i=1}^B b_{r_i}^{(\eta)}(\vx)\) and \(B_0(\vx){=}B{-}B_1(\vx)\) are the counts of member and non‑member decisions for $\vx$ at $\eta$ among the \(B\) target replicas. 
In practice, \(\widehat{\mathrm{flip}}_{\eta, B}(\vx) \in [0, {\approx}0.5]\); 
the finite-$B$ maximum exceeds $0.5$ and  converges to \(0.5\) as \(B{\to}\infty\) (Appendix~\ref{app:sec:instability:flip:overall}).
$\widehat{\mathrm{flip}}_{\eta,B}(\vx){\approx}0$  means the MIA decision for $\vx$ is stable across equally plausible targets;  
$\widehat{\mathrm{flip}}_{\eta,B}(\vx){\approx}0.5$ means the decision is statistically indistinguishable from a coin flip, with roughly half of decisions calling $\vx$ a member, and the other half calling $\vx$ a non-member. 
Figure~\ref{fig:unstable:member} provides an intuition for how this can occur. 
For a member $\vx$ at $\fpr$ $\eta{=}10^{-2}$, we plot the reference IN/OUT distributions and the median target statistic $s$ (across the $B{=}127$ targets). 
The IN/OUT distributions overlap substantially, making membership for this $\vx$ hard to disambiguate; 
accordingly, the $127$ decisions are split down the middle, so $\widehat{\mathrm{flip}}_{10^{-2},127}(\vx){\approx}0.5$.\looseness=-1

\vspace{-.1cm}
\subsection{Many membership decisions are statistically indistinguishable from a coin flip}\label{sec:instability:results}
\vspace{-.1cm}

\begin{figure}[t]
 \centering
\begin{subfigure}[t]{0.69\textwidth}
\includegraphics[trim={0cm 0cm 0cm 1cm},clip,width=\linewidth]{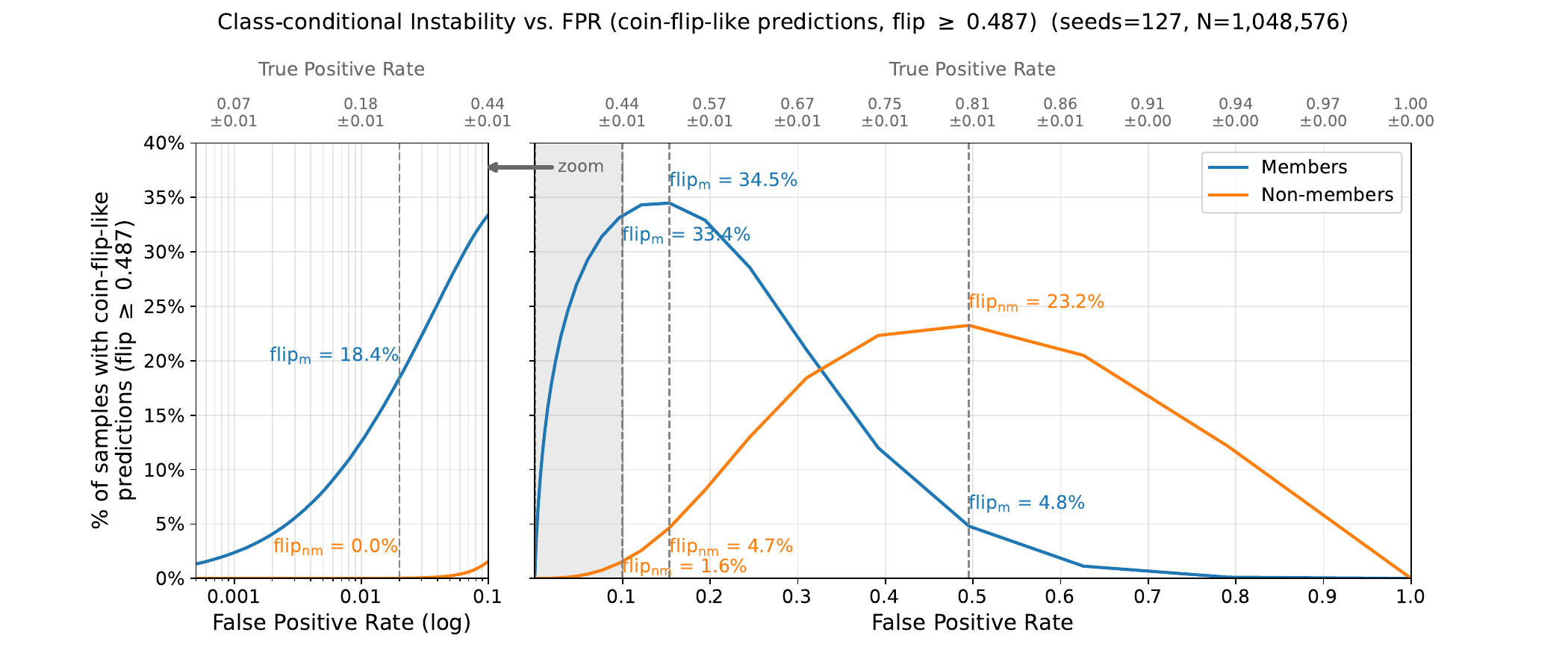}
\vspace{-.5cm}
\caption{Flip rate (Equation~\ref{eq:flip:ustat}) by membership at varied fixed $\fpr$, $B{=}127$}
\label{fig:unstable:curves}
\end{subfigure}\hfill
\begin{subfigure}[t]{0.3\textwidth}
\includegraphics[width=\linewidth]{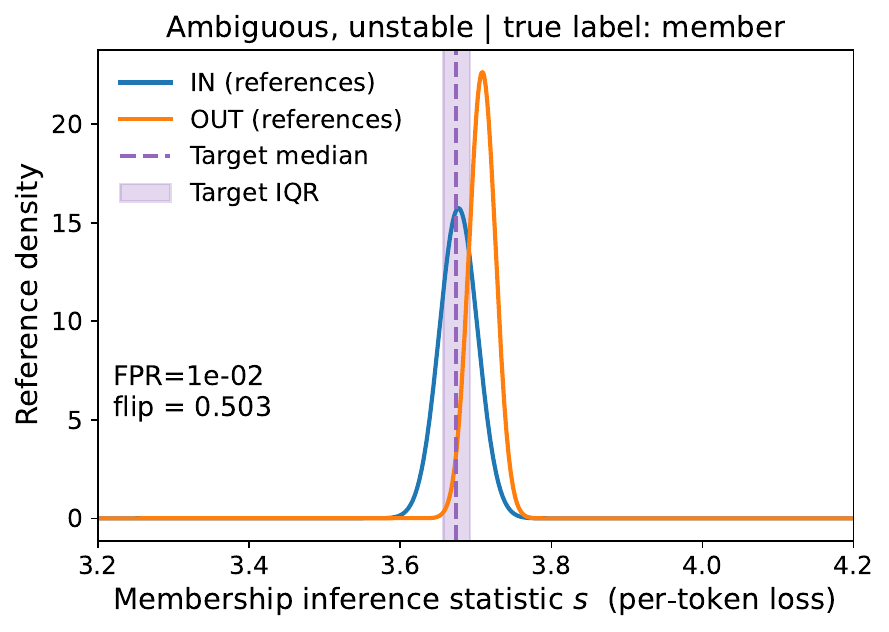}
\vspace{-.5cm}
\caption{Unstable member sample}
\label{fig:unstable:member}
\end{subfigure}
\vspace{-.1cm}
  \caption{\textbf{Visualizing decision instability.} 
  We train $B{=}127$ targets for the ${302}$M model  on $2^{19}$ samples, and one set of $128$ references to use for all attacks. 
  %We attack each target with these references. 
  LiRA achieves stable mean  $\auc{=}0.752\pm0.007$, yet many per-sample decisions behave like coin flips.
  (\textbf{left}) Share of coin-flip-like decisions across $\fpr$ ($\log$-scale for small $\fpr$; $\widehat{\mathrm{flip}}_{\eta, B}{\gtrsim}0.487$, the $\alpha{=}0.05$ cutoff, see Appendix~\ref{app:sec:instability:flip:arbitrary}). 
  Members exhibit more coin-flip-like decisions than non-members.
  (\textbf{right}) A representative unstable member:
  IN/OUT distributions overlap;
  the $B$ target decisions flip because $\vx$'s  seed-specific scores lie near the respective seed‑specific thresholds (Appendix~\ref{app:sec:instability:flip:results}).\looseness=-1 
  }
  \label{fig:instability:main}
\vspace{-.6cm}
\end{figure}

Flip rate (Equation~\ref{eq:flip:ustat}) lets us peer beneath average metrics to assess the reliability of the MIA procedure's decisions for individual samples (Appendix~\ref{app:sec:instability:flip:results}). 
%We provide extensive results in Appendix~\ref{app:sec:instability:flip:results}, and focus here on identifying $\vx$ for which predictions are statistically arbitrary. 
For the $302$M model, we train $128$ IN/OUT references to use for all attacks, and $127$ target replicas on the exact same ${\approx}500$K dataset with different seeds. 
While the population $\mathrm{flip}_\eta(\vx){=}0.5$ indicates coin-flip decisions for $\vx$, with finite replicas $B$, we need to determine a defensible cutoff above which  $\widehat{\mathrm{flip}}_{\eta, B}$ signifies this behavior.  
To do so, we set up a two-sided exact binomial test for $\theta{\coloneqq}\Pr_{r\sim\mu}[b_{r}^{(\eta)}(\vx){=}1]$, and use ``coin flip'' as shorthand for ``fail to reject the null $H_0: \theta{=}0.5$ at level $\alpha$.''
For $B{=}127$ target replicas and $\alpha{=}0.05$, this corresponds to the cutoff $\widehat{\mathrm{flip}}_{\eta, 127}{\gtrsim}0.487$ (Appendix~\ref{app:sec:instability:flip:arbitrary}).\looseness=-1

\custompar{Aggregate attack success is high and stable} 
A training set of ${\approx}500$K samples is significantly smaller than what is Chinchilla-optimal for the $302$M model (${\approx}15.1$M), so we expect higher overall MIA success (Section~\ref{sec:exp2:limits:trainsize}).
Indeed, aggregate attack success is stable, and substantially outperforms random guessing  ($\auc{=}0.752\pm0.007$). 
At fixed $\fpr$, $\tpr$ is also stable (Figure~\ref{fig:unstable:curves}, mean $\tpr\;\pm$ STD annotations). 
Nevertheless, models $r\!\sim\!\mu$ that yield similar accuracy can have very different decision rules, and therefore can disagree substantially on individual decisions~\citep{breiman2001multiplicity} (Appendix~\ref{app:sec:instability:flip:multiplicity}).\looseness=-1  

\custompar{At meaningful $\fpr$, flip rate rises with $\fpr$ and model size, and is higher for members} 
Figure~\ref{fig:unstable:curves} shows that, even at modest $\fpr$, large numbers of membership decisions behave like coin flips. 
Across fixed $\fpr$ $\eta$, we plot the proportion of samples with statistically coin-flip-like decisions, i.e., $\widehat{\mathrm{flip}}_{\eta, 127}{\gtrsim}0.487$ ($\alpha{=}0.05$);
the samples that satisfy this filter resemble the sample in Figure~\ref{fig:unstable:member}.
At $\eta{=}0.02$, ${\approx}18.4\%$ of members have coin-flip decisions (at $\alpha{=}0.05$); 
if we relax the flip threshold to also include highly unstable $\widehat{\mathrm{flip}}_{0.02, 127}{\geq}0.4$ decisions, this proportion becomes ${\approx}39.8\%$.
(By contrast, for non-members these proportions are ${\approx}0.03\%$ and ${\approx}0.2\%$, respectively.
This is unsurprising because decision thresholds are calibrated on non-members; see Appendix~\ref{app:sec:instability:flip:results}.)\looseness=-1

At lower $\fpr$, as $\fpr$ increases, the proportion of samples with coin-flip-like decisions increases. 
Each seed's calibrated threshold $\tau_r(\eta)$ moves into score regions where IN/OUT overlap is more extensive, expanding the overall set of samples (especially members) whose scores lie near the decision boundary. 
%In particular for members, this shift puts $\tau_r(\eta)$ in regions where many sample posteriors lie, and increases the proportion of samples whose seed-specific scores $\Lambda_r$ are near the decision boundary. 
As a result, small seed-induced score shifts (as well as across-seed variation in $\tau_r(\eta)$ itself, see Appendix~\ref{app:sec:instability:flip:results}) flip decisions more often. 
This effect is more pronounced for the $302$M model, compared to the $140$M model (Appendices~\ref{app:sec:instability:flip:multiplicity} \&~\ref{app:sec:instability:flip:results}).\looseness=-1 

\custompar{This is \emph{not} an attack, but a diagnostic of MIA signal robustness using counterfactuals} 
We are able to assess per-sample MIA decision instability by training \emph{many} different targets $r\!\sim\!\mu$, each of which is a plausible outcome of the same training process.
However, this is clearly not the same procedure that an attacker runs:
under the standard threat model (Section~\ref{sec:rw} \& Appendix~\ref{app:sec:background}), an attacker faces a \emph{single} target. 
%Importantly, for this diagnostic (Appendix~\ref{app:sec:instability:flip:alt}), the attacker does not have multiple targets to compute which predicted positives are unstable. 
This matters, even though a true positive is an MIA success: 
a coin-flip-like MIA \emph{decision} may be \emph{correct}, but it does \emph{not} show that the \emph{MIA procedure} is informing a \emph{reliable inferential claim} about membership for that sample. 
For such samples, flip rate shows that attack success is fragile:
the claim for $\vx$ is an artifact of seed-specific idiosyncrasies, rather than a reflection of stable signal obtained from running that specific MIA procedure (Appendix~\ref{app:sec:instability:flip:interp}).
An alternative diagnostic could fix the target, train $B$ independent IN/OUT reference sets on the same IN/OUT draws (varying only random seed), and measure disagreement with respect to those sets.
However, this is substantially more expensive. 
Our experiments varying the target similarly probe counterfactual robustness; 
we would expect to observe qualitatively similar instability under reference resampling (Appendix~\ref{app:sec:instability:flip:alt}).\looseness=-1 

Overall, flip rate diagnostics show that training randomness plays a significant role in per-sample MIA decisions. 
Even at $\fpr{=}10^{-3}$, we estimate for the $302$M model that roughly $15.4{\pm}0.6\%$ of true positives behave like coin flips---i.e.,  $\widehat{\mathrm{flip}}_{10^{-3}, 127}{\gtrsim}0.487$, at $\alpha{=}0.05$.
If we expand to include highly unstable decisions ($\widehat{\mathrm{flip}}_{10^{-3}, 127}{\geq}0.4$), these constitute $42.2\%{\pm}0.9\%$ of true positives (Appendix~\ref{app:sec:instability:flip:decompose}).\looseness=-1 

\vspace{-.25cm}
\section{Analyzing sample vulnerability to membership inference}\label{sec:exp3:samples}
\vspace{-.1cm}

The instability in membership predictions that we observe for individual samples  suggests a natural follow-on question: 
when does strong MIA succeed?
Which samples are actually vulnerable to MIA, and (how) does this vulnerability vary during training?
We approach these questions by digging deeper into our strong attacks on $140$M models---trained with a Chinchilla-optimal training set (${\approx}7$M samples) for $1$ epoch---with $128$ references. 
Samples seen later in training tend to be more vulnerable; 
however, this trend is complicated by sample length (Section~\ref{sec:exp3:per}).
While sample length has previously been linked to extraction risk~\citep{carlini2023quantifying, nasr2023scalable}, we observe no correlation between MIA and standard extraction methodology (Section~\ref{sec:exp3:extraction}). 
Together, this analysis informs our fourth key takeaway: \textbf{the relationship between MIA vulnerability and related privacy metrics is not straightforward for LLMs.}\looseness=-1 

\vspace{-.15cm}
\subsection{Identifying patterns in per-sample MIA vulnerability}\label{sec:exp3:per}
\vspace{-.1cm}

\begin{figure}[t]
\vspace{-.1cm}
  \centering
\begin{subfigure}[t]{0.45\textwidth}
    \includegraphics[width=0.95\linewidth]{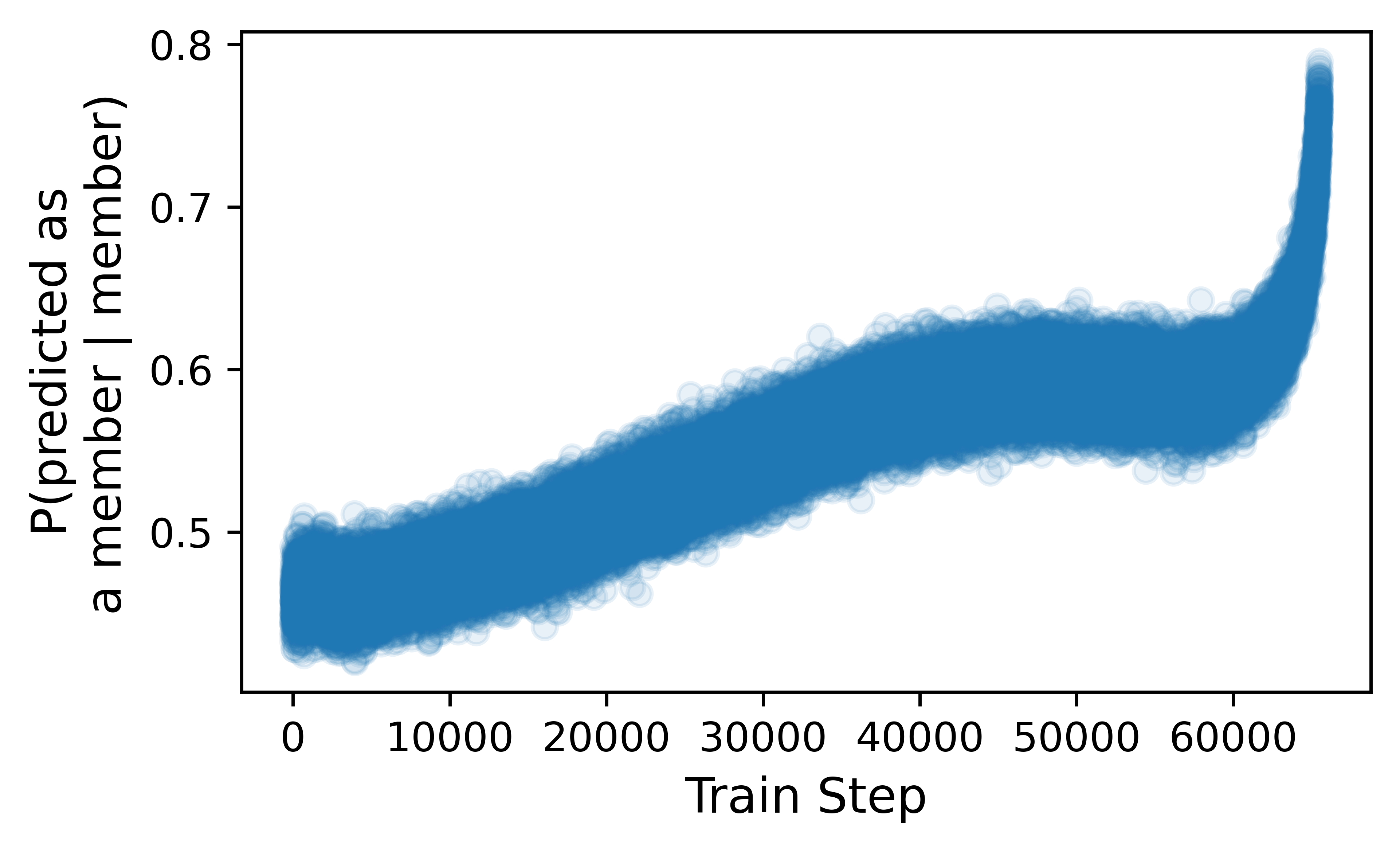}
    \vspace{-.2cm}
        \caption{Per-member success over training steps.}
        \label{fig:ex:steps}
\end{subfigure}\hfill
\begin{subfigure}[t]{.45\textwidth}
    \includegraphics[width=0.95\linewidth]{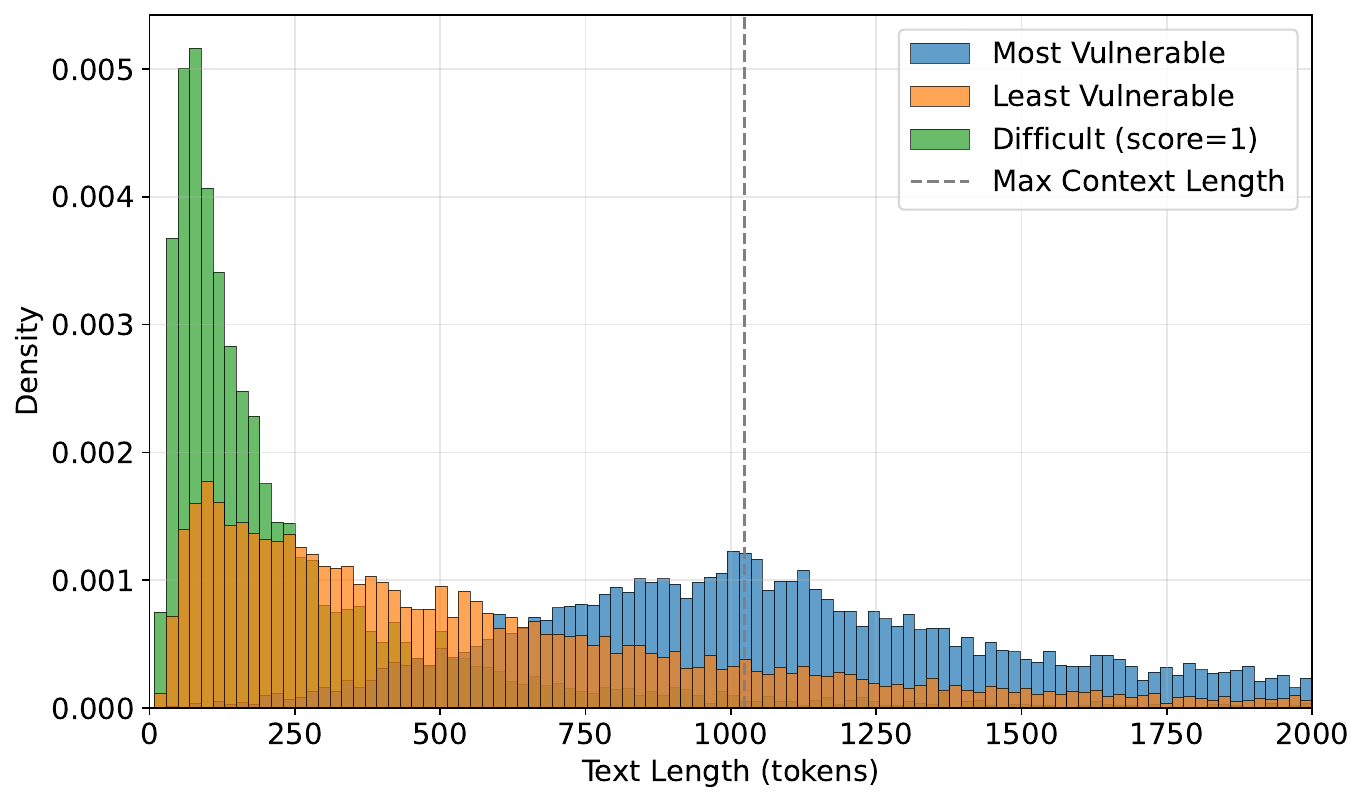}
    \vspace{-.2cm}
        \caption{Sample-length distributions by  vulnerability. 
        }
        \label{fig:ex:dist-vulnerable}
\end{subfigure}%
\vspace{-.1cm}
\caption{\textbf{Sample vulnerability to MIA.} 
For the $140$M model, 
(\textbf{a}) the evolution of sample vulnerability during  training, shown by sample true positive probabilities $\Pr(\text{predicted as a member}|\text{member})$ at each step.
(\textbf{b}) Distributions over sample lengths, according to MIA vulnerability for the $1{,}000$ samples that are least vulnerable, most vulnerable, and most difficult for MIA (i.e., with smallest, largest, and closest to $0.5$ 
$\Pr(\text{predicted as a member} | \text{member})$.) See Appendix~\ref{app:sec:instability:tp}.\looseness=-1} 
\label{fig:other-training-other}
\vspace{-.45cm}
\end{figure}

We first investigate how sample MIA vulnerability evolves over the course of training. 
In Figure~\ref{fig:ex:steps}, the scatter plot illustrates per-sample true positive probabilities by training step:  
we plot how the probability of a training sample being correctly predicted as a member changes as model training progresses, where the membership prediction for $\vx$
is computed using the reference distributions, i.e., $\tfrac{p_{\text{IN}}(\cdot|\vx)}{p_{\text{IN}}(\cdot|\vx)+p_{\text{OUT}}(\cdot|\vx)}{>}0.5$ (Section~\ref{sec:rw}~\&~Appendix~\ref{app:sec:background}). 
Across samples in the batch at each step, there is considerable variance in the underlying sample true positive probabilities $\Pr(\text{predicted as a member}|\text{member})$: 
it can vary by more than $15\%$, having an effect on overall attack success. 
For much of training, the mean $\Pr(\text{predicted as a member} | \text{member})$ is close to $0.5$, indicating many samples are challenging for MIA to distinguish as either members or non-members. 

The density of the points shifts upward toward the end of training (around step $60{,}000$).
Samples in batches that are processed in later epochs tend to be more vulnerable, as indicated by the higher probability of being correctly identified as members. 
This result highlights that the recency of exposure influences a sample's vulnerability to membership inference. 
Put differently, samples introduced earlier in training are more likely to be ``forgotten''~\citep{carlini2022privacy}: they are less vulnerable to MIA. 
This is perhaps a partial reason for LiRA decision instability for targets trained on the same dataset, but with different random seeds that control batch order (Section~\ref{sec:instability}). 
For some targets, a member $\vx$ may be seen late in training and exhibit a high true positive probability; 
for others, the same $\vx$ may appear early 
and be ``forgotten.'' (i.e., result in false negatives). 

While this appears to be the dominant trend, the details are more complicated. 
In Figure~\ref{fig:ex:dist-vulnerable}, we plot the distribution over members according to length, and partition this distribution according to vulnerability.
We consider members for which LiRA's predictions are confident but incorrect (i.e., predict non-member) 
to be least vulnerable, and members that LiRA correctly and confidently predicts as members to be most vulnerable. 
We also highlight members for which LiRA struggles to determine membership status (true positive probabilities ${\approx}0.5$). 
Figure~\ref{fig:ex:dist-vulnerable} suggests that vulnerable sequences tend to be longer. 
(See also Appendix~\ref{app:moreperexamplemiaresults}, which illustrates similar results for samples that have a higher proportion of \texttt{<unk>} tokens and higher average $\mathrm{TF}$-$\mathrm{IDF}$ scores.) 
This result is consistent with those in \citet{carlini2023quantifying}, which show that longer sequences tend to be more vulnerable to extraction attacks.\looseness=-1 

\vspace{-.15cm}
\subsection{Comparing MIA vulnerability and extraction}\label{sec:exp3:extraction}
\vspace{-.1cm}

Results such as those in Figure~\ref{fig:ex:dist-vulnerable} are consistent with prior work on memorization and extraction~\citep{carlini2021extracting}.
In general, it is assumed that a successful membership inference attack and successful extraction of training data imply that some degree of memorization has occurred for the attacked ML model. 
For MIA, this is assumed because the success of such attacks hinges on the model's tendency to behave differently for data it has seen during training (members) compared to unseen data (non-members) (Section~\ref{sec:rw}~\& Appendix~\ref{app:sec:background}). 
Prior work frequently ascribes this differential behavior to the model having memorized certain aspects of the training data.\looseness=-1

\begin{wrapfigure}{r}{.38\linewidth}
\vspace{-.45cm}
 \centering
\includegraphics[width=\linewidth]{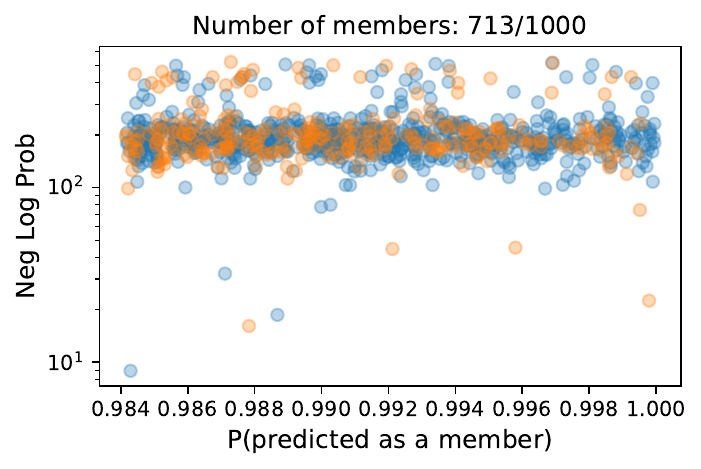}
  \caption{\textbf{Extraction for $\bf{140}$M.} 
  Neg\-ative $\log$-probability 
  of the $50$-token suffix (given the prior $50$ tokens as prefix) 
  for the $1{,}000$ samples predicted most strongly as members.
  }
  \label{fig:compare_memorization_largest}
\end{wrapfigure}

We therefore investigate whether samples that are vulnerable to  strong MIAs are also vulnerable to standard extraction attacks. 
In \Cref{fig:compare_memorization_largest}, for the $1{,}000$ samples identified as most vulnerable to strong MIA in the $140$M Chinchilla-optimal model (Figure~\ref{fig:lira_references}), we use the first $50$ tokens of each sample (prefix) to see if the next $50$ tokens (target suffix) is extractable.
We use a sample's negative $\log$-probability as a proxy for computing a probabilistic variant~\citep{hayes2025np} of \textbf{discoverable extraction}~\citep{carlini2023quantifying}---the standard extraction metric in research and model release reports~\citep{nasr2023scalable,reid2024gemini, biderman2023pythia, gemma2, llama3}.  
Discoverable extraction systematically underestimates extraction, relative to probabilistic extraction~\citep{hayes2025np, cooper2025books}.
We measure probabilistic extraction because we expect it to provide more reliable signal for memorization. 
A smaller negative $\log$-probability implies that a sample is easier to extract~\citep{cooper2025books}.\looseness=-1

After $1$ epoch, LiRA is able to identify members with better-than-random $\auc$ (Figure~\ref{fig:lira_references}). 
Out of the $1{,}000$ samples with the highest LiRA scores, 
$713$ are indeed members. 
The largest suffix extraction probability 
is ${\approx}0.0067$---for the member sample 
in \Cref{fig:compare_memorization_largest} 
that has the (smallest) negative $\log$ probability of ${\approx}5$. 
Most samples---members and non-members alike---have negative $\log$ probabilities $>100$, corresponding to probabilities on the order of $10^{-44}$ (measurements that do not register as successful extraction~\citep{hayes2025np, cooper2025books}). 
Altogether, while much prior work draws a direct connection between MIA and extraction vulnerability ~\citep[e.g.,][]{carlini2021extracting}, our results suggest a more nuanced story: 
the success of a strong MIA on a given member does not necessarily imply that the LLM is more likely to generate that sample than would be expected under the data distribution~\citep{hayes2025np,cooper2025books}.\looseness=-1

\vspace{-.3cm}
\section{Conclusion and future work}\label{sec:conclusion}
\vspace{-.2cm}

We perform dozens of experiments on thousands of GPT-2-like models (ranging from $10$M--$1$B parameters) on enormous training datasets sampled from C4 
(up to three orders of magnitude larger than those 
in prior work). 
In doing so, we address an urgent open question in ML privacy research: 
\emph{are the fidelity issues of weaker attacks due to omitting reference models, or do they point to a deeper, more fundamental challenge with applying membership inference to large language models?} 
We uncover four novel groups of findings.
While (1) strong MIAs can succeed on pre-trained LLMs (Section~\ref{sec:exp1:realistic}), (2) their success is limited (i.e., $\auc{<}0.7$) for LLMs trained using practical settings (Section~\ref{sec:exp2:limits}). 
Even when attacks achieve above‑chance $\auc$, (3) many per‑sample membership decisions are very unstable; under training randomness, they are statistically indistinguishable from a coin flip (Section~\ref{sec:instability}). 
Further, (4) the relationship between MIA vulnerability and related privacy metrics 
is not straightforward for LLMs  (Section~\ref{sec:exp3:samples}). 

As the first work to perform large-scale strong MIAs on pre-trained LLMs, we are also the first to clarify the extent of actual privacy risk MIAs pose in this setting.
By evaluating the effectiveness and limits of strong attacks, we are able to establish an upper bound on the accuracy that weaker, more feasible attacks can achieve. 
Together, our findings can guide others in more fruitful research directions to develop novel attacks and, hopefully, more effective defenses.\looseness=-1

\begin{ack}
Thank you to our anonymous reviewers, Nicholas Carlini,  Zachary Charles, and Christopher De Sa for feedback on earlier versions of this work.
A. Feder Cooper's contributions originated with a 2023--2024 student researcher position at Google Research.
Franziska Boenisch has received funding from the European Research Council (ERC) under the European Union's Horizon Europe research and innovation programme (grant agreement No.  101220235)

IS is currently employed at a startup;
CCC, MN, and KL are currently employed by OpenAI; 
MJ is currently employed by Anthropic;
NM is currently employed by Meta. 
\end{ack}

\bibliography{main}

\begin{thebibliography}{66}
\providecommand{\natexlab}[1]{#1}
\providecommand{\url}[1]{\texttt{#1}}
\expandafter\ifx\csname urlstyle\endcsname\relax
  \providecommand{\doi}[1]{doi: #1}\else
  \providecommand{\doi}{doi: \begingroup \urlstyle{rm}\Url}\fi

\bibitem[Bai et~al.(2023)Bai, Bai, Chu, Cui, Dang, Deng, Fan, Ge, Han, Huang, et~al.]{bai2023qwen}
Jinze Bai, Shuai Bai, Yunfei Chu, Zeyu Cui, Kai Dang, Xiaodong Deng, Yang Fan, Wenbin Ge, Yu~Han, Fei Huang, et~al.
\newblock Qwen technical report.
\newblock \emph{arXiv preprint arXiv:2309.16609}, 2023.

\bibitem[Biderman et~al.(2023)Biderman, Schoelkopf, Anthony, Bradley, O’Brien, Hallahan, Khan, Purohit, Prashanth, Raff, et~al.]{biderman2023pythia}
Stella Biderman, Hailey Schoelkopf, Quentin~Gregory Anthony, Herbie Bradley, Kyle O’Brien, Eric Hallahan, Mohammad~Aflah Khan, Shivanshu Purohit, USVSN~Sai Prashanth, Edward Raff, et~al.
\newblock {Pythia: A suite for analyzing large language models across training and scaling}.
\newblock In \emph{International Conference on Machine Learning}, pages 2397--2430. PMLR, 2023.

\bibitem[Breiman(2001)]{breiman2001multiplicity}
Leo Breiman.
\newblock {Statistical Modeling: The Two Cultures}.
\newblock \emph{Statistical Science}, 16\penalty0 (3):\penalty0 199--215, 2001.
\newblock ISSN 08834237.

\bibitem[Carlini et~al.(2021)Carlini, Tramer, Wallace, Jagielski, Herbert-Voss, Lee, Roberts, Brown, Song, Erlingsson, et~al.]{carlini2021extracting}
Nicholas Carlini, Florian Tramer, Eric Wallace, Matthew Jagielski, Ariel Herbert-Voss, Katherine Lee, Adam Roberts, Tom Brown, Dawn Song, Ulfar Erlingsson, et~al.
\newblock Extracting training data from large language models.
\newblock In \emph{30th USENIX security symposium (USENIX Security 21)}, pages 2633--2650, 2021.

\bibitem[Carlini et~al.(2022{\natexlab{a}})Carlini, Chien, Nasr, Song, Terzis, and Tramer]{carlini2022membership}
Nicholas Carlini, Steve Chien, Milad Nasr, Shuang Song, Andreas Terzis, and Florian Tramer.
\newblock Membership inference attacks from first principles.
\newblock In \emph{2022 IEEE Symposium on Security and Privacy (SP)}, pages 1897--1914. IEEE, 2022{\natexlab{a}}.

\bibitem[Carlini et~al.(2022{\natexlab{b}})Carlini, Jagielski, Zhang, Papernot, Terzis, and Tramer]{carlini2022privacy}
Nicholas Carlini, Matthew Jagielski, Chiyuan Zhang, Nicolas Papernot, Andreas Terzis, and Florian Tramer.
\newblock The privacy onion effect: Memorization is relative.
\newblock \emph{Advances in Neural Information Processing Systems}, 35:\penalty0 13263--13276, 2022{\natexlab{b}}.

\bibitem[Carlini et~al.(2023)Carlini, Ippolito, Jagielski, Lee, Tramèr, and Zhang]{carlini2023quantifying}
Nicholas Carlini, Daphne Ippolito, Matthew Jagielski, Katherine Lee, Florian Tramèr, and Chiyuan Zhang.
\newblock {Quantifying Memorization Across Neural Language Models}.
\newblock In \emph{International Conference on Learning Representations}, 2023.

\bibitem[Chang et~al.(2024)Chang, Shamsabadi, Katevas, Haddadi, and Shokri]{chang2024context}
Hongyan Chang, Ali~Shahin Shamsabadi, Kleomenis Katevas, Hamed Haddadi, and Reza Shokri.
\newblock Context-aware membership inference attacks against pre-trained large language models.
\newblock \emph{arXiv preprint arXiv:2409.13745}, 2024.

\bibitem[Chouldechova et~al.(2025)Chouldechova, Cooper, Barocas, Palia, Vann, and Wallach]{chouldechova2025comparison}
Alexandra Chouldechova, A.~Feder Cooper, Solon Barocas, Abhinav Palia, Dan Vann, and Hanna Wallach.
\newblock {Comparison requires valid measurement: Rethinking attack success rate comparisons in {AI} red teaming}.
\newblock In \emph{The Thirty-Ninth Annual Conference on Neural Information Processing Systems Position Paper Track}, 2025.
\newblock URL \url{https://openreview.net/forum?id=d7hqAhLvWG}.

\bibitem[Cooper and Grimmelmann(2024)]{cooper2024files}
A.~Feder Cooper and James Grimmelmann.
\newblock {The Files are in the Computer: Copyright, Memorization, and Generative AI}.
\newblock \emph{arXiv preprint arXiv:2404.12590}, 2024.

\bibitem[Cooper et~al.(2022)Cooper, Frankle, and De~Sa]{cooper2022lawless}
A.~Feder Cooper, Jonathan Frankle, and Christopher De~Sa.
\newblock {Non-Determinism and the Lawlessness of Machine Learning Code}.
\newblock In \emph{Proceedings of the 2022 Symposium on Computer Science and Law}, CSLAW '22, page 1–8, New York, NY, USA, 2022. Association for Computing Machinery.
\newblock ISBN 9781450392341.
\newblock \doi{10.1145/3511265.3550446}.

\bibitem[Cooper et~al.(2023)Cooper, Lee, Grimmelmann, Ippolito, Callison-Burch, Choquette-Choo, Mireshghallah, Brundage, Mimno, Choksi, Balkin, Carlini, Sa, Frankle, Ganguli, Gipson, Guadamuz, Harris, Jacobs, Joh, Kamath, Lemley, Matthews, McLeavey, McSherry, Nasr, Ohm, Roberts, Rubin, Samuelson, Schubert, Vaccaro, Villa, Wu, and Zeide]{cooper2023report}
A.~Feder Cooper, Katherine Lee, James Grimmelmann, Daphne Ippolito, Christopher Callison-Burch, Christopher~A. Choquette-Choo, Niloofar Mireshghallah, Miles Brundage, David Mimno, Madiha~Zahrah Choksi, Jack~M. Balkin, Nicholas Carlini, Christopher~De Sa, Jonathan Frankle, Deep Ganguli, Bryant Gipson, Andres Guadamuz, Swee~Leng Harris, Abigail~Z. Jacobs, Elizabeth Joh, Gautam Kamath, Mark Lemley, Cass Matthews, Christine McLeavey, Corynne McSherry, Milad Nasr, Paul Ohm, Adam Roberts, Tom Rubin, Pamela Samuelson, Ludwig Schubert, Kristen Vaccaro, Luis Villa, Felix Wu, and Elana Zeide.
\newblock {Report of the 1st Workshop on Generative AI and Law}.
\newblock \emph{arXiv preprint arXiv:2311.06477}, 2023.

\bibitem[Cooper et~al.(2024{\natexlab{a}})Cooper, Choquette-Choo, Bogen, Jagielski, Filippova, Liu, Chouldechova, Hayes, Huang, Mireshghallah, Shumailov, Triantafillou, Kairouz, Mitchell, Liang, Ho, Choi, Koyejo, Delgado, Grimmelmann, Shmatikov, Sa, Barocas, Cyphert, Lemley, danah boyd, Vaughan, Brundage, Bau, Neel, Jacobs, Terzis, Wallach, Papernot, and Lee]{cooper2024unlearning}
A.~Feder Cooper, Christopher~A. Choquette-Choo, Miranda Bogen, Matthew Jagielski, Katja Filippova, Ken~Ziyu Liu, Alexandra Chouldechova, Jamie Hayes, Yangsibo Huang, Niloofar Mireshghallah, Ilia Shumailov, Eleni Triantafillou, Peter Kairouz, Nicole Mitchell, Percy Liang, Daniel~E. Ho, Yejin Choi, Sanmi Koyejo, Fernando Delgado, James Grimmelmann, Vitaly Shmatikov, Christopher~De Sa, Solon Barocas, Amy Cyphert, Mark Lemley, danah boyd, Jennifer~Wortman Vaughan, Miles Brundage, David Bau, Seth Neel, Abigail~Z. Jacobs, Andreas Terzis, Hanna Wallach, Nicolas Papernot, and Katherine Lee.
\newblock {Machine Unlearning Doesn't Do What You Think: Lessons for Generative AI Policy, Research, and Practice}.
\newblock \emph{arXiv preprint arXiv:2412.06966}, 2024{\natexlab{a}}.

\bibitem[Cooper et~al.(2024{\natexlab{b}})Cooper, Lee, Choksi, Barocas, De~Sa, Grimmelmann, Kleinberg, Sen, and Zhang]{cooper2024variance}
A.~Feder Cooper, Katherine Lee, Madiha~Zahrah Choksi, Solon Barocas, Christopher De~Sa, James Grimmelmann, Jon Kleinberg, Siddhartha Sen, and Baobao Zhang.
\newblock {Arbitrariness and Social Prediction: The Confounding Role of Variance in Fair Classification}.
\newblock \emph{Proceedings of the AAAI Conference on Artificial Intelligence}, 38\penalty0 (20):\penalty0 22004--22012, March 2024{\natexlab{b}}.

\bibitem[Cooper et~al.(2025)Cooper, Gokaslan, Cyphert, Sa, Lemley, Ho, and Liang]{cooper2025books}
A.~Feder Cooper, Aaron Gokaslan, Amy~B. Cyphert, Christopher~De Sa, Mark~A. Lemley, Daniel~E. Ho, and Percy Liang.
\newblock Extracting memorized pieces of (copyrighted) books from open-weight language models.
\newblock \emph{arXiv preprint arXiv:2505.12546}, 2025.

\bibitem[Das et~al.(2024)Das, Zhang, and Tram{\`e}r]{das2024blind}
Debeshee Das, Jie Zhang, and Florian Tram{\`e}r.
\newblock Blind baselines beat membership inference attacks for foundation models.
\newblock \emph{arXiv preprint arXiv:2406.16201}, 2024.

\bibitem[Deng et~al.(2024)Deng, Zhao, Tang, Gerstein, and Cohan]{deng2024investigating}
Chunyuan Deng, Yilun Zhao, Xiangru Tang, Mark Gerstein, and Arman Cohan.
\newblock Investigating data contamination in modern benchmarks for large language models.
\newblock In \emph{Proceedings of the 2024 Conference of the North American Chapter of the Association for Computational Linguistics: Human Language Technologies (Volume 1: Long Papers)}, pages 8698--8711, 2024.

\bibitem[Duan et~al.(2024)Duan, Suri, Mireshghallah, Min, Shi, Zettlemoyer, Tsvetkov, Choi, Evans, and Hajishirzi]{duanmembership}
Michael Duan, Anshuman Suri, Niloofar Mireshghallah, Sewon Min, Weijia Shi, Luke Zettlemoyer, Yulia Tsvetkov, Yejin Choi, David Evans, and Hannaneh Hajishirzi.
\newblock Do membership inference attacks work on large language models?
\newblock In \emph{First Conference on Language Modeling}, 2024.

\bibitem[Duarte et~al.(2024)Duarte, Zhao, Oliveira, and Li]{duarte2024cop}
Andr{\'e}~V Duarte, Xuandong Zhao, Arlindo~L Oliveira, and Lei Li.
\newblock De-cop: detecting copyrighted content in language models training data.
\newblock In \emph{Proceedings of the 41st International Conference on Machine Learning}, pages 11940--11956, 2024.

\bibitem[Fisher et~al.(2019)Fisher, Rudin, and Dominici]{fisher2019all}
Aaron Fisher, Cynthia Rudin, and Francesca Dominici.
\newblock All models are wrong but many are useful: Variable importance for black-box, proprietary, or misspecified prediction models, using the {Rashomon} set.
\newblock In \emph{Proceedings of the AAAI/ACM Conference on AI, Ethics, and Society (AIES)}, pages 131--138, 2019.
\newblock \doi{10.1145/3306618.3314221}.

\bibitem[Fu et~al.(2024)Fu, Wang, Gao, Liu, Li, and Jiang]{fu2024membership}
Wenjie Fu, Huandong Wang, Chen Gao, Guanghua Liu, Yong Li, and Tao Jiang.
\newblock Membership inference attacks against fine-tuned large language models via self-prompt calibration.
\newblock In \emph{The Thirty-eighth Annual Conference on Neural Information Processing Systems}, 2024.

\bibitem[{Gemini Team} et~al.(2024){Gemini Team}, Savinov, Teplyashin, Lepikhin, Lillicrap, Alayrac, Soricut, Lazaridou, Firat, Schrittwieser, et~al.]{reid2024gemini}
{Gemini Team}, Nikolay Savinov, Denis Teplyashin, Dmitry Lepikhin, Timothy Lillicrap, Jean-baptiste Alayrac, Radu Soricut, Angeliki Lazaridou, Orhan Firat, Julian Schrittwieser, et~al.
\newblock Gemini 1.5: Unlocking multimodal understanding across millions of tokens of context.
\newblock \emph{arXiv preprint arXiv:2403.05530}, 2024.

\bibitem[Grattafiori et~al.(2024{\natexlab{a}})Grattafiori, Dubey, Jauhri, Pandey, Kadian, Al-Dahle, Letman, Mathur, Schelten, Vaughan, et~al.]{grattafiori2024llama}
Aaron Grattafiori, Abhimanyu Dubey, Abhinav Jauhri, Abhinav Pandey, Abhishek Kadian, Ahmad Al-Dahle, Aiesha Letman, Akhil Mathur, Alan Schelten, Alex Vaughan, et~al.
\newblock The llama 3 herd of models.
\newblock \emph{arXiv preprint arXiv:2407.21783}, 2024{\natexlab{a}}.

\bibitem[Grattafiori et~al.(2024{\natexlab{b}})]{llama3}
Aaron Grattafiori et~al.
\newblock {The Llama 3 Herd of Models}, 2024{\natexlab{b}}.
\newblock URL \url{https://arxiv.org/abs/2407.21783}.

\bibitem[Hayes et~al.(2025)Hayes, Swanberg, Chaudhari, Yona, Shumailov, Nasr, Choquette-Choo, Lee, and Cooper]{hayes2025np}
Jamie Hayes, Marika Swanberg, Harsh Chaudhari, Itay Yona, Ilia Shumailov, Milad Nasr, Christopher~A. Choquette-Choo, Katherine Lee, and A.~Feder Cooper.
\newblock Measuring memorization in language models via probabilistic extraction.
\newblock In Luis Chiruzzo, Alan Ritter, and Lu~Wang, editors, \emph{Proceedings of the 2025 Conference of the Nations of the Americas Chapter of the Association for Computational Linguistics: Human Language Technologies (Volume 1: Long Papers)}, pages 9266--9291, Albuquerque, New Mexico, April 2025. Association for Computational Linguistics.
\newblock ISBN 979-8-89176-189-6.
\newblock URL \url{https://aclanthology.org/2025.naacl-long.469/}.

\bibitem[Hoffmann et~al.(2022)Hoffmann, Borgeaud, Mensch, Buchatskaya, Cai, Rutherford, de~Las~Casas, Hendricks, Welbl, Clark, Hennigan, Noland, Millican, van~den Driessche, Damoc, Guy, Osindero, Simonyan, Elsen, Rae, Vinyals, and Sifre]{hoffmann2022trainingcomputeoptimallargelanguage}
Jordan Hoffmann, Sebastian Borgeaud, Arthur Mensch, Elena Buchatskaya, Trevor Cai, Eliza Rutherford, Diego de~Las~Casas, Lisa~Anne Hendricks, Johannes Welbl, Aidan Clark, Tom Hennigan, Eric Noland, Katie Millican, George van~den Driessche, Bogdan Damoc, Aurelia Guy, Simon Osindero, Karen Simonyan, Erich Elsen, Jack~W. Rae, Oriol Vinyals, and Laurent Sifre.
\newblock {Training Compute-Optimal Large Language Models}, 2022.
\newblock URL \url{https://arxiv.org/abs/2203.15556}.

\bibitem[Hu et~al.(2024)Hu, Tu, Han, He, Cui, Long, Zheng, Fang, Huang, Zhao, et~al.]{hu2024minicpm}
Shengding Hu, Yuge Tu, Xu~Han, Chaoqun He, Ganqu Cui, Xiang Long, Zhi Zheng, Yewei Fang, Yuxiang Huang, Weilin Zhao, et~al.
\newblock Minicpm: Unveiling the potential of small language models with scalable training strategies.
\newblock \emph{arXiv preprint arXiv:2404.06395}, 2024.

\bibitem[Kandpal et~al.(2024)Kandpal, Pillutla, Oprea, Kairouz, Choquette-Choo, and Xu]{kandpal2024user}
Nikhil Kandpal, Krishna Pillutla, Alina Oprea, Peter Kairouz, Christopher Choquette-Choo, and Zheng Xu.
\newblock User inference attacks on large language models.
\newblock In \emph{Proceedings of the 2024 Conference on Empirical Methods in Natural Language Processing}, pages 18238--18265, 2024.

\bibitem[Lee et~al.(2021)Lee, Ippolito, Nystrom, Zhang, Eck, Callison-Burch, and Carlini]{lee2021deduplicating}
Katherine Lee, Daphne Ippolito, Andrew Nystrom, Chiyuan Zhang, Douglas Eck, Chris Callison-Burch, and Nicholas Carlini.
\newblock Deduplicating training data makes language models better.
\newblock \emph{arXiv preprint arXiv:2107.06499}, 2021.

\bibitem[Lee et~al.(2023)Lee, Cooper, and Grimmelmann]{lee2023talkin}
Katherine Lee, A.~Feder Cooper, and James Grimmelmann.
\newblock {Talkin' 'Bout AI Generation: Copyright and the Generative-AI Supply Chain}.
\newblock \emph{arXiv preprint arXiv:2309.08133}, 2023.

\bibitem[Liu et~al.(2024)Liu, Novak, Lee, Wortsman, Xiao, Everett, Alemi, Kurzeja, Marcenac, Gur, Kornblith, Xu, Elsayed, Fischer, Pennington, Adlam, and Dickstein]{nanodo}
Peter~J. Liu, Roman Novak, Jaehoon Lee, Mitchell Wortsman, Lechao Xiao, Katie Everett, Alexander~A. Alemi, Mark Kurzeja, Pierre Marcenac, Izzeddin Gur, Simon Kornblith, Kelvin Xu, Gamaleldin Elsayed, Ian Fischer, Jeffrey Pennington, Ben Adlam, and Jascha-Sohl Dickstein.
\newblock {NanoDO: A minimal Transformer decoder-only language model implementation in {JAX}}, 2024.
\newblock URL \url{http://github.com/google-deepmind/nanodo}.
\newblock Version 0.1.0.

\bibitem[Loshchilov and Hutter(2017)]{loshchilov2017decoupled}
Ilya Loshchilov and Frank Hutter.
\newblock Decoupled weight decay regularization.
\newblock \emph{arXiv preprint arXiv:1711.05101}, 2017.

\bibitem[Lukas et~al.(2023)Lukas, Salem, Sim, Tople, Wutschitz, and Zanella-B{\'e}guelin]{lukas2023analyzing}
Nils Lukas, Ahmed Salem, Robert Sim, Shruti Tople, Lukas Wutschitz, and Santiago Zanella-B{\'e}guelin.
\newblock Analyzing leakage of personally identifiable information in language models.
\newblock In \emph{2023 IEEE Symposium on Security and Privacy (SP)}, pages 346--363. IEEE, 2023.

\bibitem[Maini and Bansal(2025)]{maini2025peeking}
Pratyush Maini and Hritik Bansal.
\newblock Peeking behind closed doors: Risks of llm evaluation by private data curators.
\newblock In \emph{The Fourth Blogpost Track at ICLR 2025}, 2025.

\bibitem[Maini et~al.(2024)Maini, Jia, Papernot, and Dziedzic]{maini2024llm}
Pratyush Maini, Hengrui Jia, Nicolas Papernot, and Adam Dziedzic.
\newblock Llm dataset inference: Did you train on my dataset?
\newblock \emph{Advances in Neural Information Processing Systems}, 37:\penalty0 124069--124092, 2024.

\bibitem[Marx et~al.(2020)Marx, Calmon, and Ustun]{marx2020predictive}
Charles Marx, Flavio Calmon, and Berk Ustun.
\newblock Predictive multiplicity in classification.
\newblock In \emph{Proceedings of the 37th International Conference on Machine Learning (ICML)}, pages 6765--6774. PMLR, 2020.
\newblock URL \url{https://proceedings.mlr.press/v119/marx20a.html}.

\bibitem[Mattern et~al.(2023)Mattern, Mireshghallah, Jin, Schoelkopf, Sachan, and Berg-Kirkpatrick]{mattern2023membership}
Justus Mattern, Fatemehsadat Mireshghallah, Zhijing Jin, Bernhard Schoelkopf, Mrinmaya Sachan, and Taylor Berg-Kirkpatrick.
\newblock Membership inference attacks against language models via neighbourhood comparison.
\newblock In \emph{Findings of the Association for Computational Linguistics: ACL 2023}, pages 11330--11343, 2023.

\bibitem[Meeus et~al.(2024{\natexlab{a}})Meeus, Jain, Rei, and de~Montjoye]{meeus2024did}
Matthieu Meeus, Shubham Jain, Marek Rei, and Yves-Alexandre de~Montjoye.
\newblock Did the neurons read your book? document-level membership inference for large language models.
\newblock In \emph{Proceedings of the 33rd USENIX Conference on Security Symposium}, pages 2369--2385, 2024{\natexlab{a}}.

\bibitem[Meeus et~al.(2024{\natexlab{b}})Meeus, Shilov, Faysse, and de~Montjoye]{meeuscopyright}
Matthieu Meeus, Igor Shilov, Manuel Faysse, and Yves-Alexandre de~Montjoye.
\newblock Copyright traps for large language models.
\newblock In \emph{Forty-first International Conference on Machine Learning}, 2024{\natexlab{b}}.

\bibitem[Meeus et~al.(2024{\natexlab{c}})Meeus, Shilov, Jain, Faysse, Rei, and de~Montjoye]{meeus2024sok}
Matthieu Meeus, Igor Shilov, Shubham Jain, Manuel Faysse, Marek Rei, and Yves-Alexandre de~Montjoye.
\newblock Sok: Membership inference attacks on llms are rushing nowhere (and how to fix it).
\newblock \emph{arXiv preprint arXiv:2406.17975}, 2024{\natexlab{c}}.

\bibitem[Meeus et~al.(2025)Meeus, Wutschitz, Zanella-B{\'e}guelin, Tople, and Shokri]{meeus2025canary}
Matthieu Meeus, Lukas Wutschitz, Santiago Zanella-B{\'e}guelin, Shruti Tople, and Reza Shokri.
\newblock The canary's echo: Auditing privacy risks of llm-generated synthetic text.
\newblock \emph{arXiv preprint arXiv:2502.14921}, 2025.

\bibitem[Mireshghallah et~al.(2022{\natexlab{a}})Mireshghallah, Goyal, Uniyal, Berg-Kirkpatrick, and Shokri]{mireshghallah2022quantifying}
Fatemehsadat Mireshghallah, Kartik Goyal, Archit Uniyal, Taylor Berg-Kirkpatrick, and Reza Shokri.
\newblock Quantifying privacy risks of masked language models using membership inference attacks.
\newblock In \emph{Proceedings of the 2022 Conference on Empirical Methods in Natural Language Processing}, pages 8332--8347, 2022{\natexlab{a}}.

\bibitem[Mireshghallah et~al.(2022{\natexlab{b}})Mireshghallah, Uniyal, Wang, Evans, and Berg-Kirkpatrick]{mireshghallah2022empirical}
Fatemehsadat Mireshghallah, Archit Uniyal, Tianhao Wang, David~K Evans, and Taylor Berg-Kirkpatrick.
\newblock An empirical analysis of memorization in fine-tuned autoregressive language models.
\newblock In \emph{Proceedings of the 2022 Conference on Empirical Methods in Natural Language Processing}, pages 1816--1826, 2022{\natexlab{b}}.

\bibitem[Nasr et~al.(2023)Nasr, Carlini, Hayase, Jagielski, Cooper, Ippolito, Choquette-Choo, Wallace, Tramèr, and Lee]{nasr2023scalable}
Milad Nasr, Nicholas Carlini, Jonathan Hayase, Matthew Jagielski, A.~Feder Cooper, Daphne Ippolito, Christopher~A. Choquette-Choo, Eric Wallace, Florian Tramèr, and Katherine Lee.
\newblock {Scalable Extraction of Training Data from (Production) Language Models}.
\newblock \emph{arXiv preprint arXiv:2311.17035}, 2023.

\bibitem[Oren et~al.(2023)Oren, Meister, Chatterji, Ladhak, and Hashimoto]{oren2023proving}
Yonatan Oren, Nicole Meister, Niladri~S Chatterji, Faisal Ladhak, and Tatsunori Hashimoto.
\newblock Proving test set contamination in black-box language models.
\newblock In \emph{The Twelfth International Conference on Learning Representations}, 2023.

\bibitem[Panda et~al.(2025)Panda, Tang, Choquette-Choo, Nasr, and Mittal]{panda2025privacy}
Ashwinee Panda, Xinyu Tang, Christopher~A. Choquette-Choo, Milad Nasr, and Prateek Mittal.
\newblock Privacy auditing of large language models.
\newblock In \emph{The Thirteenth International Conference on Learning Representations}, 2025.
\newblock URL \url{https://openreview.net/forum?id=60Vd7QOXlM}.

\bibitem[Raffel et~al.(2020)Raffel, Shazeer, Roberts, Lee, Narang, Matena, Zhou, Li, and Liu]{raffel2020exploring}
Colin Raffel, Noam Shazeer, Adam Roberts, Katherine Lee, Sharan Narang, Michael Matena, Yanqi Zhou, Wei Li, and Peter~J Liu.
\newblock Exploring the limits of transfer learning with a unified text-to-text transformer.
\newblock \emph{Journal of machine learning research}, 21\penalty0 (140):\penalty0 1--67, 2020.

\bibitem[Rossi et~al.(2024)Rossi, Marek, Hanke, Wang, Backes, Dziedzic, and Boenisch]{rossi2024auditing}
Lorenzo Rossi, Bart{\l}omiej Marek, Vincent Hanke, Xun Wang, Michael Backes, Adam Dziedzic, and Franziska Boenisch.
\newblock Auditing empirical privacy protection of private llm adaptations.
\newblock In \emph{Neurips Safe Generative AI Workshop 2024}, 2024.

\bibitem[Sablayrolles et~al.(2019)Sablayrolles, Douze, Schmid, Ollivier, and J{\'e}gou]{sablayrolles2019white}
Alexandre Sablayrolles, Matthijs Douze, Cordelia Schmid, Yann Ollivier, and Herv{\'e} J{\'e}gou.
\newblock White-box vs black-box: Bayes optimal strategies for membership inference.
\newblock In \emph{International Conference on Machine Learning}, pages 5558--5567. PMLR, 2019.

\bibitem[Sardana et~al.(2023)Sardana, Portes, Doubov, and Frankle]{sardana2023beyond}
Nikhil Sardana, Jacob Portes, Sasha Doubov, and Jonathan Frankle.
\newblock Beyond chinchilla-optimal: Accounting for inference in language model scaling laws.
\newblock \emph{arXiv preprint arXiv:2401.00448}, 2023.

\bibitem[Semenova et~al.(2022)Semenova, Rudin, and Parr]{semenova2022existence}
Lesia Semenova, Cynthia Rudin, and Ron Parr.
\newblock Existence, computation, and implications of {Rashomon} sets.
\newblock \emph{Machine Learning}, 111:\penalty0 3531--3569, 2022.
\newblock \doi{10.1007/s10994-022-06146-1}.

\bibitem[Shi et~al.(2024)Shi, Ajith, Xia, Huang, Liu, Blevins, Chen, and Zettlemoyer]{shidetecting}
Weijia Shi, Anirudh Ajith, Mengzhou Xia, Yangsibo Huang, Daogao Liu, Terra Blevins, Danqi Chen, and Luke Zettlemoyer.
\newblock Detecting pretraining data from large language models.
\newblock In \emph{The Twelfth International Conference on Learning Representations}, 2024.

\bibitem[Shokri et~al.(2017)Shokri, Stronati, Song, and Shmatikov]{shokri2017membership}
Reza Shokri, Marco Stronati, Congzheng Song, and Vitaly Shmatikov.
\newblock Membership inference attacks against machine learning models.
\newblock In \emph{2017 IEEE symposium on security and privacy (SP)}, pages 3--18. IEEE, 2017.

\bibitem[Song and Shmatikov(2019)]{song2019auditing}
Congzheng Song and Vitaly Shmatikov.
\newblock Auditing data provenance in text-generation models.
\newblock In \emph{Proceedings of the 25th ACM SIGKDD International Conference on Knowledge Discovery \& Data Mining}, pages 196--206, 2019.

\bibitem[Team et~al.(2024)]{gemma2}
Gemma Team et~al.
\newblock {Gemma 2: Improving Open Language Models at a Practical Size}, 2024.
\newblock URL \url{https://arxiv.org/abs/2408.00118}.

\bibitem[Touvron et~al.(2023)Touvron, Lavril, Izacard, Martinet, Lachaux, Lacroix, Rozi{\`e}re, Goyal, Hambro, Azhar, et~al.]{touvron2023llama}
Hugo Touvron, Thibaut Lavril, Gautier Izacard, Xavier Martinet, Marie-Anne Lachaux, Timoth{\'e}e Lacroix, Baptiste Rozi{\`e}re, Naman Goyal, Eric Hambro, Faisal Azhar, et~al.
\newblock Llama: Open and efficient foundation language models.
\newblock \emph{arXiv preprint arXiv:2302.13971}, 2023.

\bibitem[Wang et~al.(2025)Wang, Wang, Hooi, Cai, Peng, and Chang]{wang2025recall}
Cheng Wang, Yiwei Wang, Bryan Hooi, Yujun Cai, Nanyun Peng, and Kai-Wei Chang.
\newblock Con-recall: Detecting pre-training data in llms via contrastive decoding.
\newblock In \emph{Proceedings of the 31st International Conference on Computational Linguistics}, pages 1013--1026, 2025.

\bibitem[Watson et~al.(2022)Watson, Guo, Cormode, and Sablayrolles]{watsonimportance}
Lauren Watson, Chuan Guo, Graham Cormode, and Alexandre Sablayrolles.
\newblock On the importance of difficulty calibration in membership inference attacks.
\newblock In \emph{International Conference on Learning Representations}, 2022.

\bibitem[Watson-Daniels et~al.(2022)Watson-Daniels, Parkes, and Ustun]{watson2023multiplicity}
Jamelle Watson-Daniels, David~C. Parkes, and Berk Ustun.
\newblock {Predictive Multiplicity in Probabilistic Classification}, 2022.

\bibitem[Wei et~al.(2024)Wei, Wang, and Jia]{wei2024proving}
Johnny Wei, Ryan Wang, and Robin Jia.
\newblock Proving membership in llm pretraining data via data watermarks.
\newblock In \emph{Findings of the Association for Computational Linguistics ACL 2024}, pages 13306--13320, 2024.

\bibitem[Xie et~al.(2024)Xie, Wang, Huang, Zhang, Ge, Pei, Gong, and Dhingra]{xie2024recall}
Roy Xie, Junlin Wang, Ruomin Huang, Minxing Zhang, Rong Ge, Jian Pei, Neil Gong, and Bhuwan Dhingra.
\newblock Recall: Membership inference via relative conditional log-likelihoods.
\newblock In \emph{Proceedings of the 2024 Conference on Empirical Methods in Natural Language Processing}, pages 8671--8689, 2024.

\bibitem[Ye et~al.(2022)Ye, Maddi, Murakonda, Bindschaedler, and Shokri]{ye2022enhanced}
Jiayuan Ye, Aadyaa Maddi, Sasi~Kumar Murakonda, Vincent Bindschaedler, and Reza Shokri.
\newblock Enhanced membership inference attacks against machine learning models.
\newblock In \emph{Proceedings of the 2022 ACM SIGSAC Conference on Computer and Communications Security}, pages 3093--3106, 2022.

\bibitem[Yeom et~al.(2018)Yeom, Giacomelli, Fredrikson, and Jha]{yeom2018privacy}
Samuel Yeom, Irene Giacomelli, Matt Fredrikson, and Somesh Jha.
\newblock Privacy risk in machine learning: Analyzing the connection to overfitting.
\newblock In \emph{2018 IEEE 31st computer security foundations symposium (CSF)}, pages 268--282. IEEE, 2018.

\bibitem[Zarifzadeh et~al.(2024)Zarifzadeh, Liu, and Shokri]{zarifzadehlow}
Sajjad Zarifzadeh, Philippe Liu, and Reza Shokri.
\newblock Low-cost high-power membership inference attacks.
\newblock In \emph{Forty-first International Conference on Machine Learning}, 2024.

\bibitem[Zhang et~al.(2025{\natexlab{a}})Zhang, Das, Kamath, and Tramèr]{zhang2025miacannot}
Jie Zhang, Debeshee Das, Gautam Kamath, and Florian Tramèr.
\newblock {Membership Inference Attacks Cannot Prove that a Model Was Trained On Your Data}, 2025{\natexlab{a}}.
\newblock URL \url{https://arxiv.org/abs/2409.19798}.

\bibitem[Zhang et~al.(2025{\natexlab{b}})Zhang, Sun, Yeats, Ouyang, Kuo, Zhang, Yang, and Li]{zhang2025mink}
Jingyang Zhang, Jingwei Sun, Eric Yeats, Yang Ouyang, Martin Kuo, Jianyi Zhang, Hao~Frank Yang, and Hai Li.
\newblock Min-k\%++: Improved baseline for pre-training data detection from large language models.
\newblock In \emph{The Thirteenth International Conference on Learning Representations}, 2025{\natexlab{b}}.
\newblock URL \url{https://openreview.net/forum?id=ZGkfoufDaU}.

\end{thebibliography}
\bibliographystyle{plainnat}

%%%%%%%%%%%%%%%%%%%%%%%%%%%%%%%%%%%%%%%%%%%%%%%%%%%%%%%%%%%%

\newpage
\clearpage
\appendix

\section{Membership inference attacks}\label{app:sec:background}

\paragraph{Security game, threat model, and notation.}
Membership inference is formalized as a security game between a challenger and an attacker (i.e., adversary). 
Let \(\mathcal{D}\) denote the underlying data-generating distribution over samples (and labels, if applicable).
The challenger draws a finite training dataset \(\sD \sim \mathcal{D}^n\) and trains a target model \(h\) on \(\sD\).
A challenge record \(\vx\) is selected to be either a \textbf{member} (\(\vx\in\sD\)) or a \textbf{non-member} (\(\vx\notin\sD\)).
The attacker is given query access to \(h\) together with auxiliary resources and outputs a guess about \(\vx\)'s membership; success means accuracy exceeding random guessing.\looseness=-1

The strong attacks we study---LiRA and RMIA (Section~\ref{sec:exp1:realistic:warmup} and Appendix~\ref{app:sec:warmup})---assume the attacker can (i) query \(h\) on arbitrary inputs to obtain per-sample outputs (losses, logits, or confidence scores), and (ii) train \textbf{reference models} \(f \in \Phi\) by replicating the target's training recipe on datasets drawn from the same population \(\mathcal{D}\) that generated \(\sD\) (in practice, from a large proxy corpus approximating \(\mathcal{D}\)).
For a fixed query sample \(\vx\), each reference's training dataset either \emph{includes} \(\vx\) (IN) or \emph{excludes} \(\vx\) (OUT), yielding a per-\(\vx\) partition:
\[
\Phi_{\text{IN}}(\vx)\subseteq\Phi,\qquad
\Phi_{\text{OUT}}(\vx)\subseteq\Phi,\qquad
\Phi_{\text{IN}}(\vx)\cap\Phi_{\text{OUT}}(\vx)=\varnothing.
\]
This is the \textbf{online} setting; the \textbf{offline} setting assumes access only to \(\Phi_{\text{OUT}}(\vx)\).
Neither attack requires access to the target's parameters or \(\sD\); 
only queries to \(h\) and attacker-trained references are needed.
In research settings, one often controls both the  target and references, which allows evaluation across many \(\vx\) with known membership.
It is common (though not required) to choose \(\Phi\) so that \(|\Phi_{\text{IN}}(\vx)| \approx |\Phi_{\text{OUT}}(\vx)|\) for stability.
We do so in this work. 

\paragraph{Observation statistics and membership scores.}
For any model \(g\) and query sample \(\vx\), let
\[
s(g,\vx)\in\R
\]
denote a fixed scalar \textbf{observation statistic} from \(g\) on \(\vx\) (e.g., loss, negative $\log$-likelihood, or a monotone transform of confidence such as a logit).
A \textbf{membership inference attack (MIA)} maps the available statistics for \(\vx\) (from \(h\), and when used, from \(\Phi\)) to a real-valued \textbf{membership score} \(\Lambda(\vx)\in\R\), with larger values indicating stronger evidence that \(\vx\) is a member.

\paragraph{Baseline (reference-free) loss attack~\citep{yeom2018privacy}.}
Using only the target's statistic,
\[
\Lambda_{\text{Loss}}(\vx)\;=\; -\,s(h,\vx),
\]
so larger \(\Lambda_{\text{Loss}}(\vx)\) implies lower loss on \(\vx\).
Any strictly monotone transform preserves ranking and therefore \(\rocauc\).
No reference models are used in this baseline approach. 
Stronger attacks use reference models to yield improved membership signal.

\paragraph{Likelihood Ratio Attack (LiRA)~\citep{carlini2022membership}.}
LiRA uses references to model per-sample IN/OUT distributions over the chosen statistic \(s\).
For a fixed \(\vx\), the attacker forms
\[
\{s(f,\vx):f\in\Phi_{\text{IN}}(\vx)\}
\quad\text{and}\quad
\{s(f,\vx):f\in\Phi_{\text{OUT}}(\vx)\},
\]
fits univariate models (typically Gaussians) to obtain densities \(p_{\text{IN}}(\cdot\mid\vx)\) and \(p_{\text{OUT}}(\cdot\mid\vx)\), and evaluates the target's statistic $s(h,\vx)$ under these densities to form a likelihood ratio:
\begin{align}
\label{eq:lira}
\Lambda_{\text{LiRA}}(\vx)\;=\;
\frac{p_{\text{IN}}\big(s(h,\vx)\,\big|\,\vx\big)}{p_{\text{OUT}}\big(s(h,\vx)\,\big|\,\vx\big)}.
\end{align}
The online variant uses both IN and OUT; 
the offline variant performs a one-sided test using only OUT.
Working with \(\log\Lambda(\vx)\) is common for numerical stability; 
since this is monotone, \(\rocauc\) is unchanged.\looseness=-1

\paragraph{Robust Membership Inference Attack (RMIA)~\citep{zarifzadehlow}.}
RMIA also compares the target model's statistic on the sample $\vx$ to outputs for $\vx$ from a set of reference models $\Phi$, but uses a different construction based on a \emph{pairwise} likelihood ratio. 
This ratio is normalized by a reference population \(\sZ\) (e.g., a calibration set drawn from \(\mathcal{D}\) or a held-out proxy).
Define
\begin{align}
\label{eq:rmia-alpha-x}
\alpha(\vx) \;=\; \frac{s(h,\vx)}{\E_{f \in \Phi}\,s(f,\vx)}.
\end{align}

The expectation in the denominator is approximated empirically over the trained references. 
To improve robustness, RMIA contextualizes this ratio relative to population $\sZ$. 
For each $\vz \in \sZ$: 
\begin{align}
\label{eq:rmia-ratio}
\alpha(\vz) \;=\; \frac{s(h,\vz)}{\E_{f \in \Phi}\,s(f,\vz)},
\qquad
L(\vx,\vz) \;=\; \frac{\alpha(\vx)}{\alpha(\vz)}.
\end{align}

The computed membership score aggregates the pairwise tests at a threshold \(\gamma>0\):
\begin{align}
\label{eq:rmia}
\Lambda_{\text{RMIA}}(\vx) \;=\; \frac{1}{|\sZ|} \sum_{\vz \in \sZ} \1 \big[ L(\vx,\vz) \geq \gamma \big].
\end{align}
We focus on online (two-sided) variants of these attacks that use both IN and OUT references, as opposed to offline variants that only use OUT references.\looseness=-1 

\paragraph{Decision rules and calibration.}
Given a real-valued score \(\Lambda(\vx)\) (e.g., $\Lambda_{\text{Loss}}$, $\Lambda_{\text{LiRA}}$, or $\Lambda_{\text{RMIA}}$), the attacker outputs a binary decision about the membership of $\vx$ via
\[
b(\vx)\;=\;\1\{\Lambda(\vx)\ge\tau\}.
\]
To operate at a fixed false positive rate (\(\fpr\)) $\eta$, it is convenient to write
\[
b^{(\eta)}(\vx)\;=\;\1\{\Lambda(\vx)\ge\tau(\eta)\},
\]
where \(\tau(\eta)\) is calibrated for the target \(h\) using non-members (i.e., samples not in \(h\)'s training subset \(\sD\)).
(We will sometimes refer to the training set as \(\sD_\text{IN}\), when we want to refer to the set of non-members as \(\sD_\text{OUT}\).)\looseness=-1

\paragraph{Calibration to non‑members at fixed \(\fpr\).} 
Fix a target \(h\) and an $\fpr$ \(\eta\in[0,1]\), and assume larger scores are indicate stronger evidence that $\vx$ is a member. 
Let the non‑member (OUT) set be \(\sD_{\text{OUT}}\) with size \(N_\text{OUT} = |\sD_{\text{OUT}}|\).
(The attacker can draw i.i.d.\ samples from the population distribution \(\mathcal{D}\), or use auxiliary data from the same source, independently of the training set, to form \(\sD_{\text{OUT}}\).)
Write the scores as \(\{\Lambda(\vx):\vx\in\sD_{\text{OUT}}\}\). 
The empirical CDF of OUT scores is 
\[ \widehat{F}_{\text{OUT}}(t)\;=\;\frac{1}{N_{\text{OUT}}}\sum_{\vx\in\sD_{\text{OUT}}}\1\{\Lambda(\vx)\le t\}. 
\] 
We choose the right‑continuous empirical \((1-\eta)\)-quantile 
\[
\tau(\eta)\;=\;\inf\{\,t:\ \widehat{F}_{\text{OUT}}(t^{-})\;\ge\; 1-\eta\,\}.
\] 
Equivalently, if $\Lambda_{(1)}\le\cdots\le\Lambda_{(N_{\text{OUT}})}$ are the sorted OUT scores, let 
$k=\lceil(1-\eta)\,N_{\text{OUT}}\rceil,\ \bar k=\max\{j:\Lambda_{(j)}=\Lambda_{(k)}\}$, and set $\tau(\eta)=\Lambda_{(\bar k+1)} (\Lambda_{(N_{\text{OUT}}+1)}=+\infty)$.

We then apply the calibrated binary MIA decision rule \[ b^{(\eta)}(\vx)\;=\;\1\{\Lambda(\vx)\ge \tau(\eta)\}. 
\] 

By construction, this guarantees (finite-sample, with ties handled conservatively) that \[ \widehat{\fpr}(\eta)\;=\;\frac{1}{N_{\text{OUT}}}\sum_{\vx\in\sD_{\text{OUT}}}\1\{\Lambda(\vx)\ge \tau(\eta)\} \;=\;1-\widehat{F}_{\text{OUT}}\bigl(\tau(\eta)^{-}\bigr)\;\le\;\eta. \] This is because taking the right‑continuous quantile ensures that any mass tied at \(\tau(\eta)\) is counted on the \(\le\) side of the CDF. Therefore, the realized \(\fpr\) on OUT never exceeds \(\eta\) (and may be smaller in the presence of ties).

\paragraph{Common performance metrics.}
Because MIAs are typically compared across operating points, it is typical to report $\roc$ curves and $\auc$ (threshold-agnostic), and---when reporting $\tpr$ at a fixed $\fpr$---to set $\tau$ to achieve the target $\fpr$. 
For RMIA, the internal pairwise threshold $\gamma$ controls the per-comparison likelihood ratio test, while the final decision threshold $\tau$ controls the operating point.
Calibration may be global (single $\tau$) or conditional (e.g., per class/bucket).
All monotone transforms of the score \(\Lambda\) leave \(\rocauc\) invariant, while operating-point metrics (e.g., \(\tpr\) at fixed \(\fpr\)) depend on calibration.\looseness=-1

\paragraph{Practical note.}
Calibrating \(\fpr\) without knowledge of ground-truth membership can be challenging~\citep{zhang2025miacannot}.
In our experiments, we control training and evaluation, so membership labels are known; this enables exact calibration and measurement at desired operating points.

\section{Comparing membership inference attacks and observation statistics}\label{app:sec:warmup}

At the beginning of this project, we considered two candidates for strong membership inference attacks to use in our experiments: 
the Likelihood Ratio Attack (LiRA)~\citep{carlini2022membership} and the Robust Membership Inference Attack (RMIA)~\citep{zarifzadehlow}.
Both attacks involve training reference models (Section~\ref{sec:rw}) that enable the computation of likelihood ratios (which result in stronger attacks), though they differ in important ways.
LiRA~\citep{carlini2022membership} estimates membership by comparing the loss of a sample $\vx$ in a target model to empirical loss distributions from reference models trained with and without $\vx$. 
In contrast, RMIA~\citep{zarifzadehlow} performs and aggregates statistical pairwise likelihood ratio tests between $\vx$ and population samples $\vz$, using both reference models and $\vz$ to estimate how the inclusion of $\vx$ versus $\vz$ affects the probability of generating the observed model $\theta$ (Appendix~\ref{app:sec:background}).

By leveraging signal from both models and population samples, \citet{zarifzadehlow} observe that RMIA can outperform LiRA using fewer reference models. 
However, no prior work has compared these methods in the pre-trained LLM setting and with large numbers of reference models, leaving open the question of which attack fares better under these conditions.\looseness=-1

In this appendix, we investigate this question for the first time, and our results clearly indicate that LiRA outperforms RMIA for a large number of reference models in the online setting (Appendix~\ref{app:sec:background}).  
We observe limited cases where RMIA can outperform LiRA if the population dataset is large enough and the attack is performed for certain small numbers of reference models. 
However, we caution generalizing about comparative performance. 
LiRA seems to perform better with $1$ or $2$ IN references, while RMIA performs better with $4$--$16$, and then LiRA once again outperforms RMIA for ${>}16$ IN references.\looseness=-1

Overall, attacks with larger numbers of references perform better, as measured by  $\rocauc$. 
Since our aim is to test the strongest attacks possible---to investigate an upper bound on strong MIA  performance---this makes LiRA the best choice for our experiments. 
For those with smaller compute budgets that still wish to run strong attacks using ${\approx}16$ IN references, in some circumstances, RMIA may be a better choice. 

Following from our discussion of the threat model for membership inference, and how it is implemented with slight variations for LiRA and RMIA (Appendix~\ref{app:sec:background}),  
we next discuss our experiments comparing the performance of these two attacks. 
We first show how different choices for the  observation statistic impact attack performance (Appendix~\ref{app:sec:signal}).
This provides more detail about the choices we make in our overall experimental setup throughout the paper (introduced in Section~\ref{sec:exp1:realistic}). 
Then, we show our full results that compare the performance of LiRA and RMIA using different numbers of reference models (Appendix~\ref{app:sec:references}), which lead us to choose LiRA for the experiments that follow.\looseness=-1 

For all experiments comparing LiRA and RMIA, we train $140$M-parameter models on ${\approx}7$M samples, which equates to approximately $2.8$B training tokens (i.e., what is optimal for this model size, according to Chinchilla scaling laws~\citep{hoffmann2022trainingcomputeoptimallargelanguage} with an over-training multiplier of $20$).

\subsection{Different observation statistics}\label{app:sec:signal}

\begin{figure}[t]
 \centering
\includegraphics[width=0.6\textwidth]{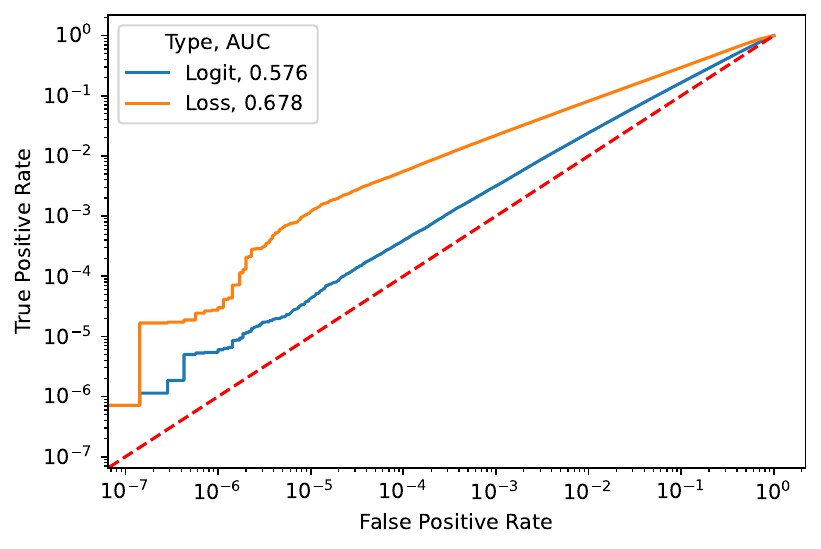}
  \caption{\textbf{Influence of observation statistic type on MIA Performance.} 
  For the $140$M model, we plot $\roc$ curves to compare the efficacy of using model logits ($\auc{=}0.576$) and model loss ($\auc{=}0.678$) as observation statistics for membership inference with LiRA. 
  In this setting, loss provides a stronger signal for distinguishing members from non-members.}
  \label{fig:compare_value_type}
\end{figure}

In our initial experiments in Section~\ref{sec:exp1:realistic},  we compare LiRA~\citep{carlini2022membership} and RMIA~\citep{zarifzadehlow} to decide which strong attack to use.
We also investigated the efficacy of different observation statistics for membership inference. 
We tested model loss and model logits (averaged over the entire sequence).
For example, in Figure~\ref{fig:compare_value_type}, we plot the $\roc$ curve for using LiRA to attack a
$140$M model trained on ${\approx}7$M samples with $128$ references. 
The plot shows the true positive rate ($\tpr$) against the false positive rate ($\fpr$) on a $\log$-$\log$ scale, with one $\roc$ curve each for logit and loss statistics. 
For the logit curve, $\rocauc{=}0.576$, while the loss curve has a higher $\rocauc{=}0.678$.
This indicates that, in this setup, using loss as the observation statistic results in a more effective attack compared to using logits.
Based on results like this, throughout this paper, we opt to use loss as observation statistic  $s$.\looseness=-1

\subsection{MIA attack performance for different numbers of reference models}\label{app:sec:references}

\begin{figure}[t]
  \centering
\begin{subfigure}[t]{0.47\textwidth}
    \includegraphics[width=0.95\linewidth]{figures/cooper/compare_ref_models_3.pdf}
        \caption{LiRA}
        \label{fig:compare:lira}
\end{subfigure}\hfill
\begin{subfigure}[t]{0.47\textwidth}
    \includegraphics[width=0.95\linewidth]{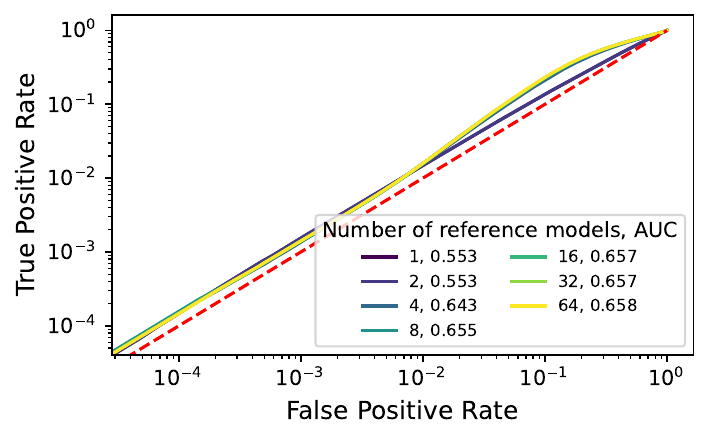}
        \caption{RMIA}
        \label{fig:compare:rmia}
\end{subfigure}%
\caption{\textbf{Comparing LiRA and RMIA.} 
We attack a $140$M-parameter model, with the target and references trained on ${\approx}7$M samples.
$\roc$ curves illustrate the effectiveness of (\textbf{a}) LiRA~\citep{carlini2022membership} and (\textbf{b}) RMIA~\citep{zarifzadehlow} for different numbers of reference models. 
As we increase the number of references, LiRA's performance surpasses RMIA's, measured by $\rocauc$.
These plots show the number of IN references.
(There are $2\times$ as many references in total, accounting for OUT.) 
}
\label{fig:compare_rmia_and_lira}

\end{figure}

\begin{figure}[t]
 \centering
\includegraphics[width=0.4\textwidth]{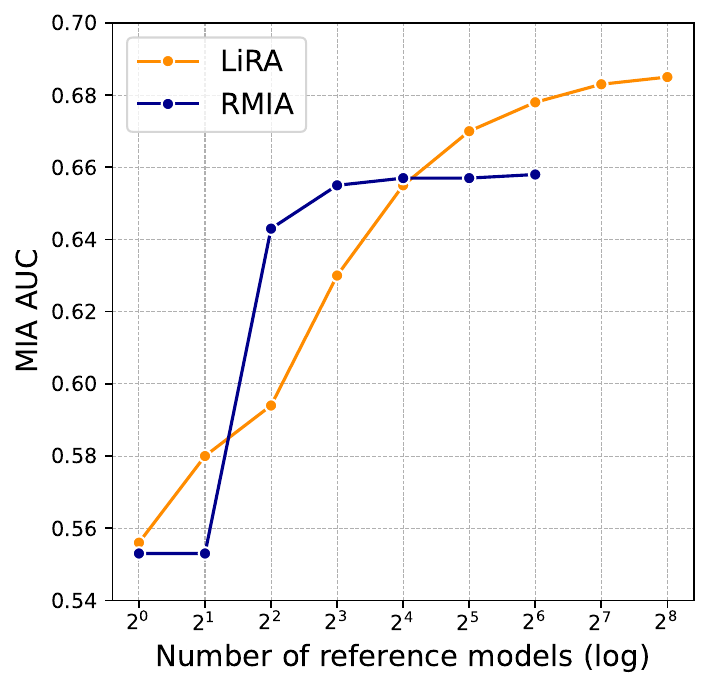}
  \caption{\textbf{Comparing LiRA and RMIA.} 
  As an alternative view of Figure~\ref{fig:compare_rmia_and_lira}, we plot the $\rocauc$ achieved by both attack methodologies for an increasing number of reference models.
  As the number of references increases, LiRA's performance continues to improve, while RMIA's gains saturate.
  Overall, LiRA is the stronger attack.
  This plot also only shows the number of IN references on the $x$-axis (there are the same number of OUT).\looseness=-1}
  \label{fig:compare_attacks_ref_models}
\end{figure}

Figure~\ref{fig:compare_rmia_and_lira} compares LiRA and RMIA, showing $\roc$ curves and $\rocauc$ for different numbers of reference models.
Figure~\ref{fig:compare_attacks_ref_models} provides an alternate view of the same results, plotting $\rocauc$ for both attacks as a function of reference models. 
LiRA's performance generally dominates RMIA's. 
LiRA continues to improve as we increase the number of reference models, while RMIA's effectiveness plateaus. 
However, with $4$-$16$ IN references, RMIA surpasses the performance of LiRA. 
It essentially matches LiRA using $16$ IN references.
That is, with $4$ references, LiRA exhibits $\rocauc{=}0.594$, which under-performs RMIA's corresponding $\rocauc{=}0.643$; 
but LiRA's $\rocauc$ increases to $0.678$ with $64$ IN references, which outperforms RMIA's $\rocauc{=}0.658$. 

Also note that RMIA exhibits a distinct diagonal pattern at low $\fpr$ (Figure~\ref{fig:compare_lira_rmia_simple}).
While RMIA aims to be a strong attack that 
is effective in low-compute settings, we find that a large population $\sZ$ is necessary to obtain meaningful $\fpr$ at very low $\fpr$ thresholds.
In particular, for a minimally acceptable $\fpr_{\text{min}}$, RMIA requires a population size $|\sZ|$ that is $\frac{1}{\fpr_{\text{min}}}$. 
In practice, this is quite expensive, as RMIA's membership score is computed via pairwise comparisons with these $|\sZ|$ reference points (i.e., there are $\mathcal{O}(|\sZ|)$ pairwise likelihood ratio tests for target record $\vx$, see Appendix~\ref{app:sec:background}). 
In these initial experiments we only used  $|\sZ|{=}10{,}000$ samples. 
We measure performance of RMIA on larger population sizes below in Appendix~\ref{app:sec:rmia}.\looseness=-1

Overall, as noted in Section~\ref{sec:exp1:realistic}, while both attacks clearly beat the random baseline of $\rocauc{=}0.5$, neither is remarkably successful in this setting: 
regardless of the number of reference models, neither attack achieves that meets or exceeds $\rocauc{=}0.7$.


\subsubsection{Further experiments on RMIA}\label{app:sec:rmia} 

We now further investigate RMIA, decoupling its different components. 
We investigate removing the dependence on the population $\sZ$, population sizes other than $|\sZ|{=}10{,}000$, and varying threshold $\gamma$.

\begin{figure}[t]
 \centering
\includegraphics[width=0.8\textwidth]{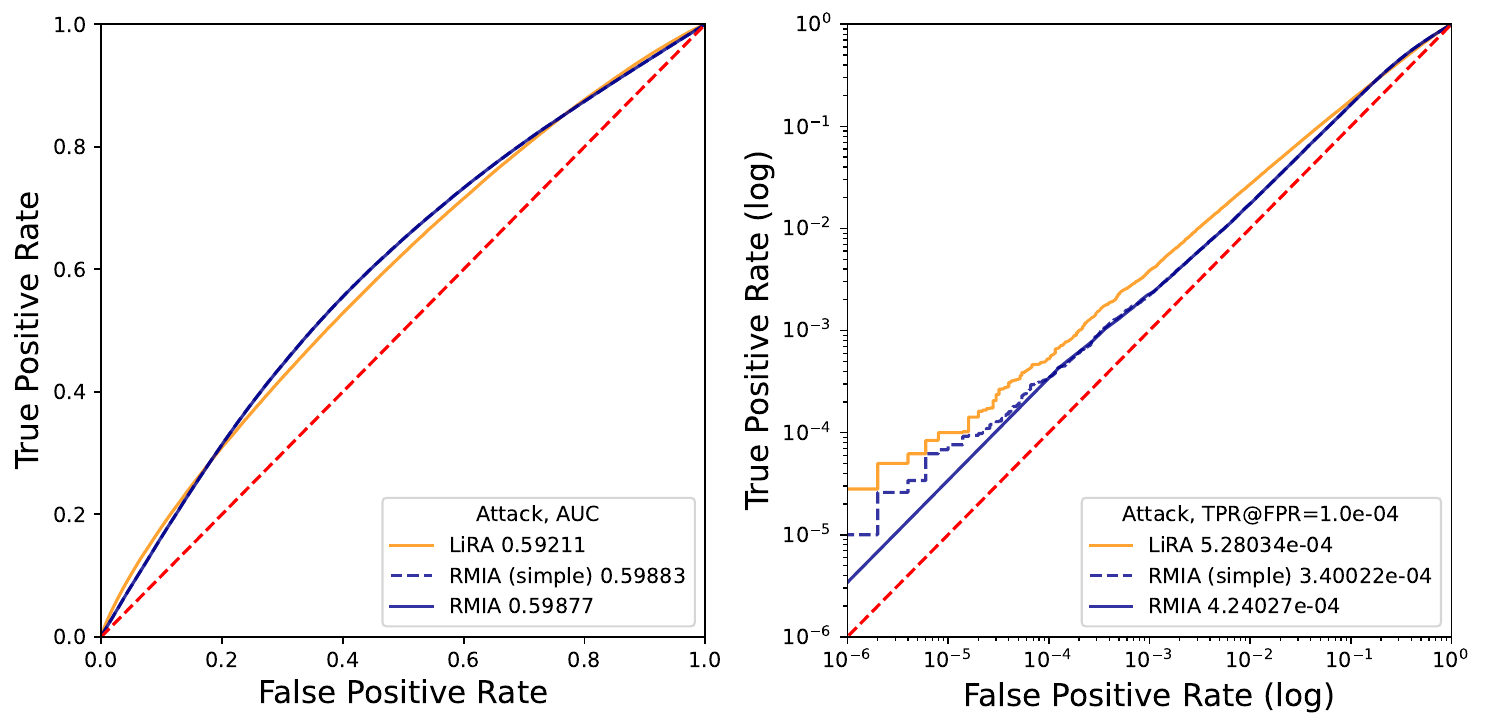}
  \caption{\textbf{Comparing LiRA, RMIA (simple) and RMIA.} Attacking a $10$M-parameter model trained for $1$ epoch with a training set size of $2^{19}$ samples.}
  \label{fig:compare_lira_rmia_simple}
\end{figure}

\paragraph{Eliminating dependence on population $\sZ$.} 
First, we consider the simplest form of RMIA (\emph{simple}), eliminating its dependence on a  population $\sZ$ and using $\alpha(\vx)$ directly (Equation~\ref{eq:rmia-alpha-x}). 
Figure~\ref{fig:compare_lira_rmia_simple} shows the $\roc$ curves for all three MIAs attacking one target model with $10$M parameters, trained for 1 epoch on a training set size of $2^{19}$ samples.
We use $128$ reference models and consider $2{\times} 2^{19}{=}2^{20}$ target records $\vx$ with (as elsewhere) balanced membership/non-membership to analyse MIA. 
We find all three attacks reach similar $\rocauc$ values. 

We also gauge MIA performance by evaluating the $\tpr$ at low, fixed $\fpr$. 
To understand the values RMIA reaches for $\tpr$ at low $\fpr$, an important subtlety arises from the entropy of the score distribution. 
Attacks that produce very coarse membership scores inherently limit achievable $\tpr$ at very low $\fpr$. 
For example, as RMIA compares $\alpha(\vx)$ to $\alpha(\vz)$ for all $\vz \in \sZ$ to compute its membership score $\Lambda_{\text{RMIA}}(\vx)$ (Equation~\ref{eq:rmia}), there are maximally $|\sZ|$ unique values $\Lambda_{\text{RMIA}}(\vx)$ can take for all $\vx$.
This limits the score's entropy and the possibility of achieving a meaningful $\tpr$ at very low $\fpr$. 
This explains the diagonal pattern for RMIA in Figure~\ref{fig:compare_lira_rmia_simple}, where $|\sZ|{=}10{,}000$. 
By contrast, both LiRA and RMIA (simple) provide a membership score that is not limited in entropy, leading to more meaningful values for $\tpr$ at lower $\fpr$. 

\begin{figure}[t]
 \centering
\includegraphics[width=0.8\textwidth]{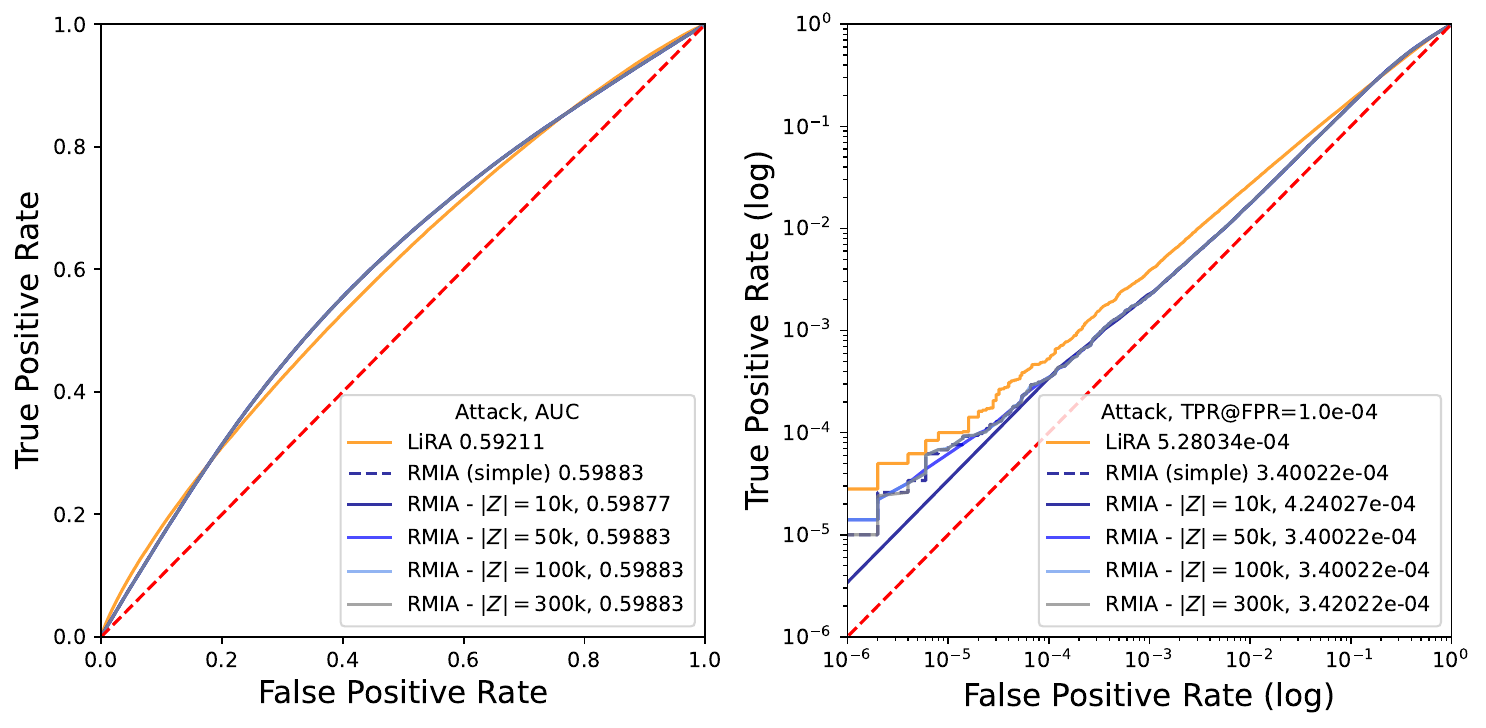}
  \caption{\textbf{Performance of RMIA for different population sizes $|\sZ|$.}
  We attack a $10$M-parameter model trained for $1$ epoch with a training set size of $2^{19}$ samples.}
  \label{fig:rmia_vary_z}
\end{figure}

\paragraph{Increasing the population size $|\sZ|$.}
We next test further increasing the size of the population $\sZ$ when computing RMIA. 
For the same setup as Figure~\ref{fig:compare_lira_rmia_simple}, Figure~\ref{fig:rmia_vary_z} shows how MIA performance varies with the size of $\sZ$. 
We observe very similar values for RMIA (simple) and RMIA $\rocauc$ for all population sizes that we test. 
When examining $\tpr$ at low $\fpr$, we find that increasing $|\sZ|$ improves the MIA performance. 
Indeed, the increased entropy in $\Lambda_{\text{RMIA}}(\vx)$ now allows the attack to reach meaningful values of $\tpr$ for $\fpr$ as low as $10^{-6}$. 
Notably, for all values of $|\sZ|$ we consider, LiRA still outperforms RMIA at low $\fpr$, while the  $|\sZ|$ likelihood comparisons in RMIA for every target record $\vx$ also incur additional computational cost.\looseness=-1 

\paragraph{Varying threshold $\gamma$.}
Finally, we evaluate RMIA under varying threshold $\gamma$. 
As $\gamma$ increases, in Equation~\ref{eq:rmia-ratio}, it becomes less likely that $\alpha(\vx)$ sufficiently exceeds $\alpha(\vz)$ for many $z \in \sZ$ to count toward the score---i.e., that $\alpha(\vx)/\alpha(\vz) \geq \gamma$ (Equation~\ref{eq:rmia}).

\begin{figure}[t]
  \centering
\begin{subfigure}[t]{0.8\textwidth}
    \includegraphics[width=0.95\linewidth]{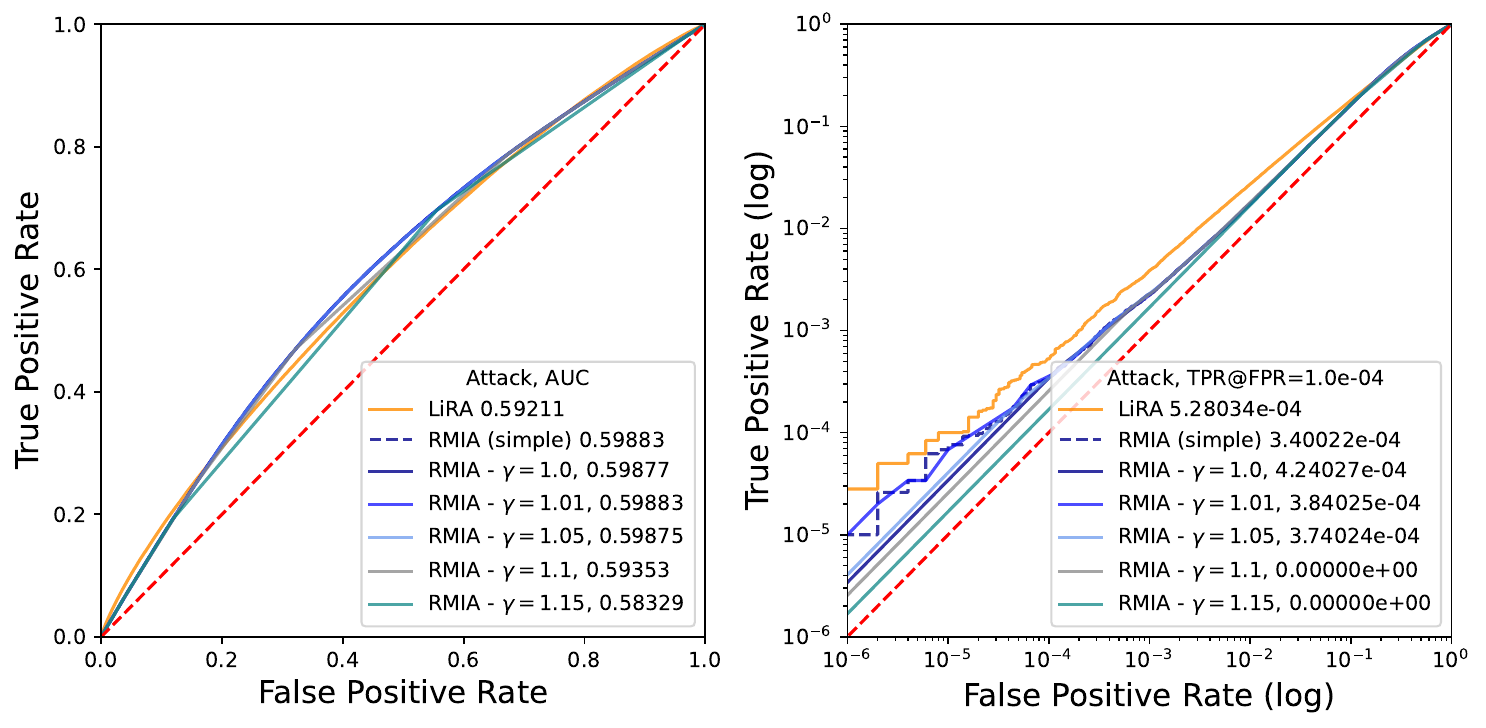}
        \caption{$|\sZ|=10{,}000$}
        \label{fig:rmia_vary_gamma_10k}
\end{subfigure}\hfill
\begin{subfigure}[t]{0.8\textwidth}
    \includegraphics[width=0.95\linewidth]{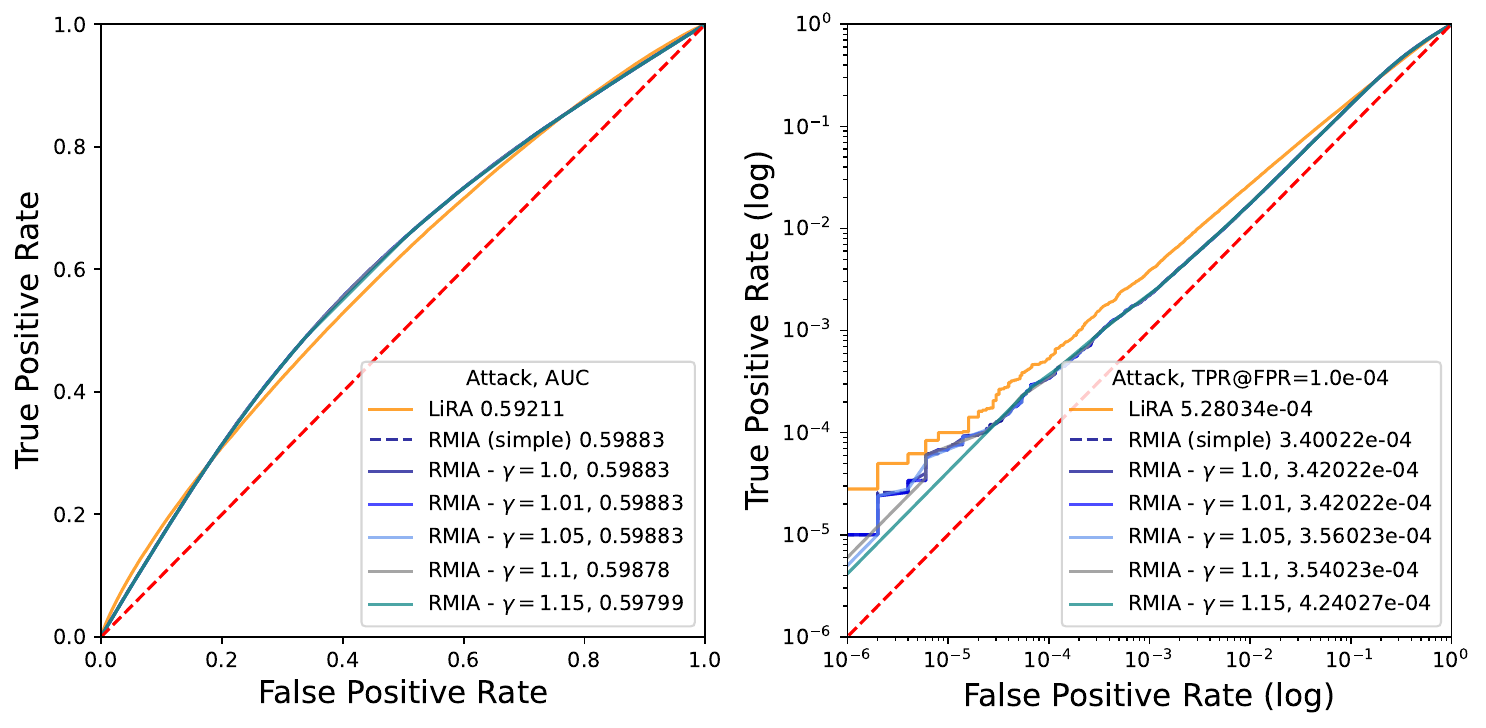}
        \caption{$|\sZ|=300{,}000$}
        \label{fig:rmia_vary_gamma_300k}
\end{subfigure}%
\caption{\textbf{Performance of RMIA for varied $\gamma$.}
We attack a $10$M-parameter model trained for $1$ epoch with a training set size of $2^{19}$ samples, varying the threshold $\gamma$ used to compute $\Lambda_{\text{RMIA}}$.} 
\vspace{-.3cm}
\label{fig:rmia_vary_gamma}
\end{figure}

Again for the same setup, Figure~\ref{fig:rmia_vary_gamma} shows how RMIA performs for varying values of $\gamma$, considering both $|\sZ|{=}10{,}000$ (Figure \ref{fig:rmia_vary_gamma_10k}) and $|\sZ|{=}300{,}000$ (Figure \ref{fig:rmia_vary_gamma_300k}). 
While MIA $\rocauc$ remains relatively stable as $\gamma$ increases, the $\tpr$ at low $\fpr$ varies. 
For $|\sZ|{=}10{,}000$, the $\tpr$ at $\fpr{=}10^{-4}$ decreases for increasing values of $\gamma$, reaching $0$ for $\gamma{\geq} 1.1$. 
This is due to the reduced granularity of RMIA's membership score: 
for larger $\gamma$, fewer $\vz$ satisfy $\alpha(\vx)/\alpha(\vz) \geq \gamma$;
this constrains the entropy of the RMIA score, making it harder to reach meaningful values of $\tpr$ at low $\fpr$. 
A larger reference population ($|\sZ|{=}300{,}000$) mitigates this issue, allowing meaningful $\tpr$ even at low $\fpr$ for similar $\gamma$ values.

Taking these three sets of results together, we find LiRA to outperform RMIA when a sufficiently large number of reference models is available, especially in the low-$\fpr$ regime. 
Since our aim is to study the strongest attacks, we adopt LiRA as the primary attack throughout our experiments.

\subsection{MIA performance in the offline setting}\label{app:sec:offline}

As stated in Section~\ref{sec:rw} and Appendix~\ref{app:sec:background}, the literature distinguishes between online and offline settings for reference-based MIAs~\citep{carlini2022membership,zarifzadehlow}. 
In the online setting, the attacker has access to reference models trained on data including the target sample $\vx$ ($\Phi_{\text{IN}}$) and excluding it($\Phi_{\text{OUT}}$).
In the offline setting, the attacker only has access to $\Phi_{\text{OUT}}$. 
Throughout this work, we consider the strongest attacker, and thus report all results in the online setting. 

For completeness, we instantiate MIAs in the offline setting in the same experimental setup as considered above for our additional RMIA tests (Appendix~\ref{app:sec:rmia}). 
We test the offline versions for both LiRA and RMIA, as originally proposed in ~\citet{carlini2022membership} and \citet{zarifzadehlow}, respectively.\looseness=-1 

For LiRA, without $\Phi_{\text{IN}}$, we are unable to approximate the probability $p_{\text{IN}}\big(s(h,\vx)\big)$ (Equation~\ref{eq:lira}), and so just consider the one-sided hypothesis test instead of the likelihood ratio:\looseness=-1
\[
\Lambda_{\text{LiRA}, \text{offline}}(\vx) = 1 - p_{\text{OUT}}\big(s(h,\vx)\big).
\]
For RMIA, we now compute the denominator in $\alpha(\vx)$ (Equation~\ref{eq:rmia-alpha-x}) by taking the expectation over the reference models that are available to the attacker, i.e.:
\[
\alpha_{\text{offline}}(\vx) \;=\; \frac{s(h,\vx)}{\E_{f \in \Phi_{\text{OUT}}}\,s(f,\vx)}.
\]
Note that~\citet{zarifzadehlow} propose to further adjust the denominator by using a variable $a$ (their Appendix B.2.2) to better approximate the $\E_{f \in \Phi}s(f, \vx)$, when only references $\Phi_{\text{OUT}}$ are available. 
We set $a{=}1$ and just compute the empirical mean across all reference models in $\Phi_{\text{OUT}}$ to approximate the expectation in the denominator. 
We then compute $\alpha_{\text{offline}}(\vz)$ and use membership inference score 
\[
\Lambda_{\text{RMIA}, \text{offline}}(\vx) \;=\; \frac{1}{|\sZ|} \sum_{\vz \in \sZ} \1 \left[ L_\text{offline}(\vx,\vz) \geq \gamma \right], \quad \text{where} \quad L_\text{offline}(\vx,\vz) = \frac{\alpha_{\text{offline}}(\vx)}{\alpha_{\text{offline}}(\vz)}.
\]

\begin{figure}[t]
  \centering
\begin{subfigure}[t]{0.8\textwidth}
    \includegraphics[width=0.95\linewidth]{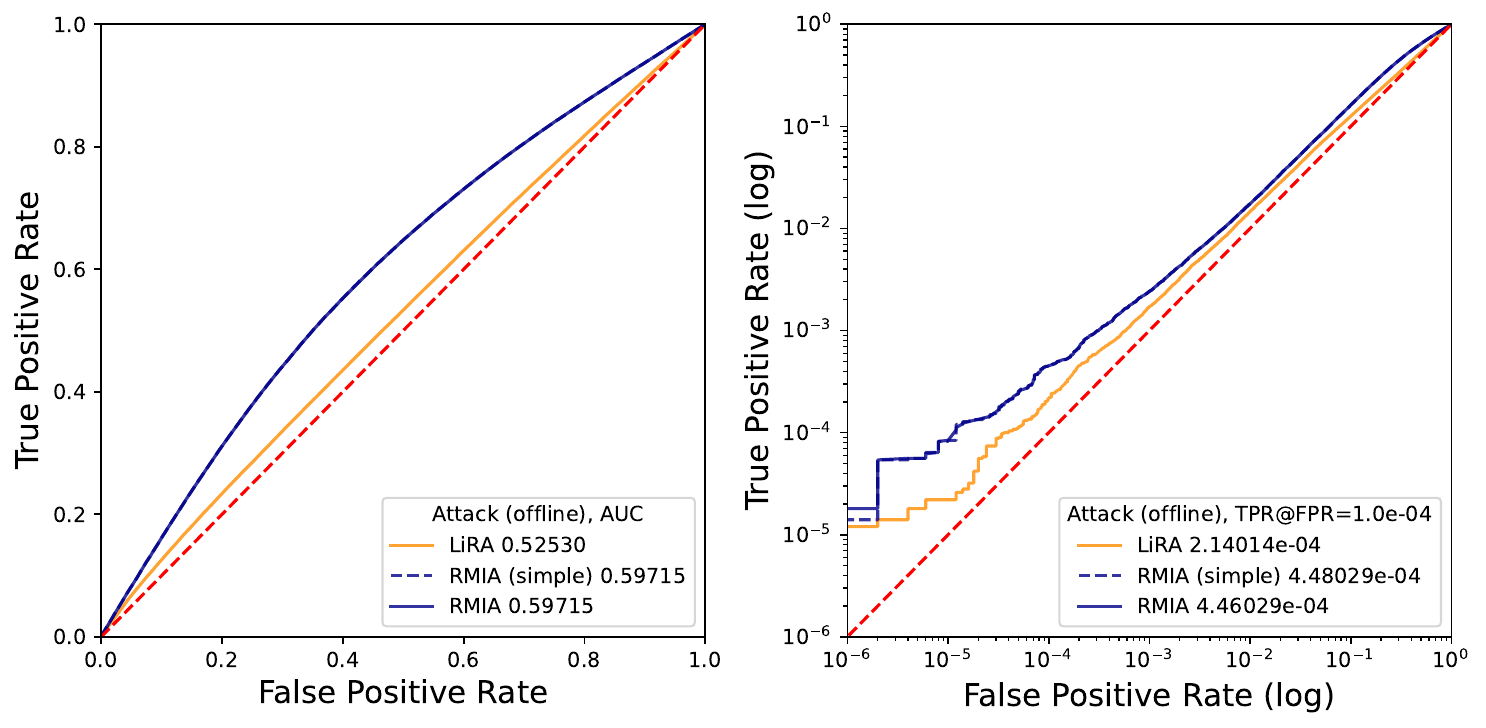}
        \vspace{-.1cm}
        \caption{Offline}
\end{subfigure}\hfill
\begin{subfigure}[t]{0.8\textwidth}
    \includegraphics[width=0.95\linewidth]{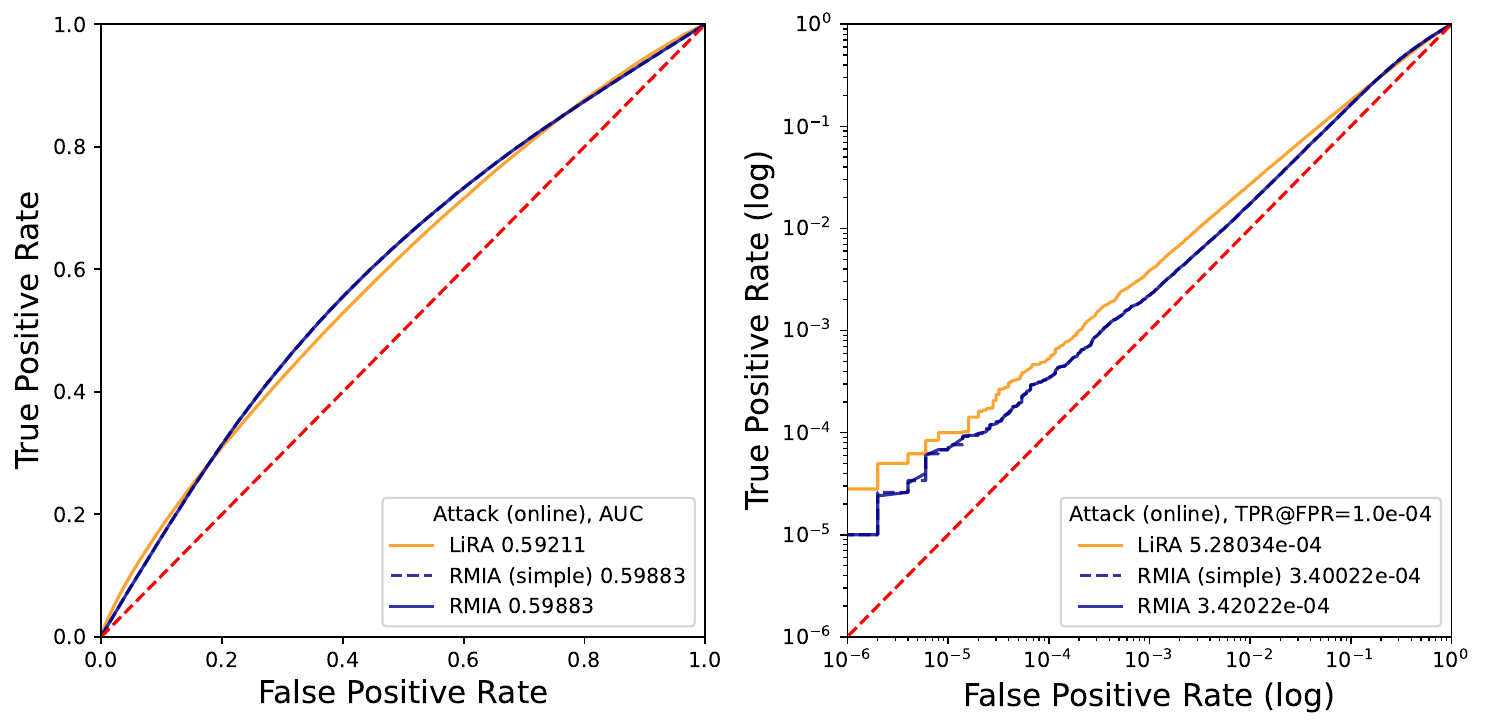}
        \vspace{-.1cm}
        \caption{Online}
\end{subfigure}%
\caption{\textbf{MIA performance in the offline and online setting.} We attack a $10$M-parameter model trained for $1$ epoch with a training set size of $2^{19}$ samples, considering $128$ references in the online setting and only the corresponding models $\Phi_{\text{OUT}}$ in the offline setting (on average $64$ references per $\vx$).\looseness=-1} 
\label{fig:rmia_offline_online}
\end{figure}

Figure~\ref{fig:rmia_offline_online} compares the MIA performance between the online and offline setting, for LiRA, RMIA (simple) (which does not use the reference population $\sZ$, Appendix~\ref{app:sec:rmia}), and RMIA;
we set $\gamma{=}1$ and $|\sZ|{=}300{,}000$. 
We again attack a $10$M-parameter model trained for $1$ epoch, using a training set size of $2^{19}$ samples.
We use $128$ reference models for the online setting and $64$ in the offline setting (on average, per target sample). 

We find that, in this configuration and with this number of reference models, offline RMIA outperforms offline LiRA, in terms of both $\rocauc$ and $\tpr$ at low fixed $\fpr$. 
This suggests that RMIA's offline variant more accurately captures membership signal compared to the one-sided hypothesis test used in offline LiRA. 
In contrast, in the online setting, LiRA and RMIA achieve similar $\rocauc$, with LiRA performing better than RMIA in the low-$\fpr$ regime. 

\vspace{-.2cm}
\section{More experiments on Chinchilla-optimal models}\label{app:sec:optimal}

In this appendix, we provide additional details on our experiments involving LiRA attacks on Chinchilla-optimal~\citep{hoffmann2022trainingcomputeoptimallargelanguage} models of different sizes in Section~\ref{sec:exp1:realistic:compute-optimal}: $10$M, $44$M, $85$M, $140$M, $489$M, and $1018$M. 
We summarize training hyperparameters in \Cref{app:exp_details}.

\begin{figure}[t]
  \centering
\begin{subfigure}[t]{0.49\textwidth}
    \includegraphics[width=0.95\linewidth]{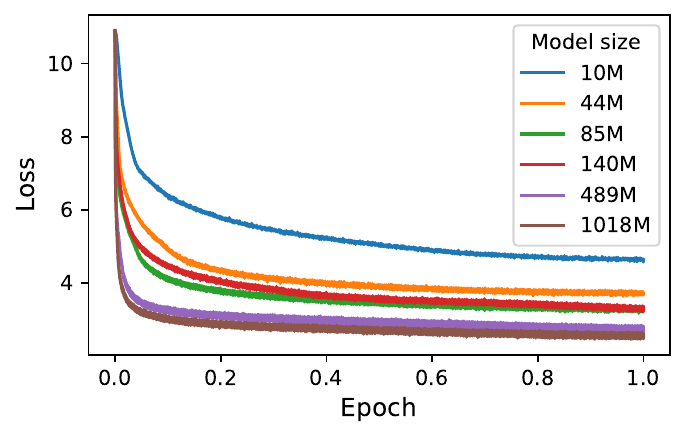}
        \vspace{-.1cm}
        \caption{Validation loss over time ($1$ epoch)}
        \label{fig:compare_model_sizes_loss}
\end{subfigure}\hfill
\begin{subfigure}[t]{0.49\textwidth}
    \includegraphics[width=0.95\linewidth]{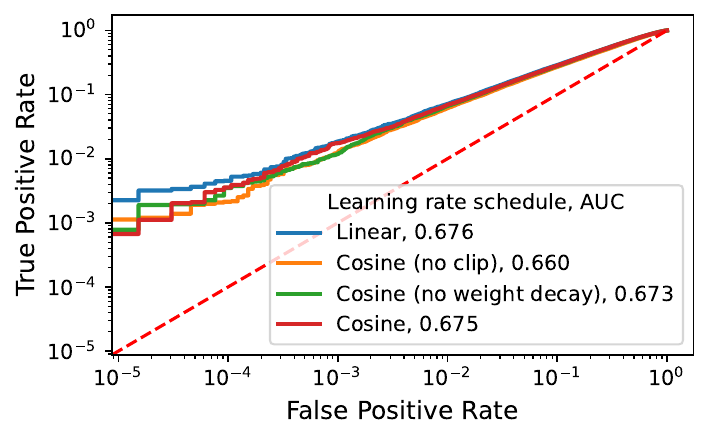}
        \vspace{-.1cm}
        \caption{LiRA $\mathrm{ROC}$ for different learning rate schedules}
        \label{fig:compare_lrs}
\end{subfigure}%
\caption{\textbf{Investigating training dynamics hyperparameters.}
    (\textbf{a}) Validation throughout the $1$ training epoch for our experiments involving Chinchilla-optimal trained models of various sizes. 
    (\textbf{b}) The effect of learning rate schedule on LiRA's attack success for $140$M models using $128$ references. 
    \vspace{-.2cm}
}
\label{fig:optimal-details}
\end{figure}

\paragraph{Observing changes in loss during training.}
In Figure~\ref{fig:compare_model_sizes_loss}, we show the decrease in validation loss over a single epoch.
The $x$-axis represents the fraction of the training epoch completed (from $0.0$ to $1.0$), and the $y$-axis shows the corresponding loss. 
As expected, all models exhibit a characteristic decrease in loss as training progresses. 
Larger models (namely, $489$M and $1018$M) demonstrate faster convergence to lower loss values, reflecting their increased capacity to fit the training data. 
They also maintain a lower loss throughout the epoch compared to smaller models ($10$M--$140$M).\looseness=-1

\paragraph{Investigating the role of learning rate schedule.}
In the Chinchilla-optimal setting, we also investigate the role of hyperparameters on MIA performance. 
In Figure~\ref{fig:compare_lrs}, we show $\roc$ curves that compare the MIA vulnerability (with LiRA) of $140$M-parameter models (trained on ${\approx}7$M records, with $128$ reference models), where we vary the learning rate schedule: 
Linear ($\auc{=}0.676$), Cosine (no global norm clipping, $\auc{=}0.660$), Cosine (no weight decay, $\auc{=}0.673$), and standard Cosine ($\auc{=}0.675$). 
As with all of our $\roc$ plots, the $\tpr$ is plotted against the $\fpr$ on a $\log$-$\log$ scale.
The $\rocauc$ values for each curve are relatively close.
This indicates that, while there are some minor differences in attack performance, the choice of learning rate schedule among those tested does not lead to drastically different MIA outcomes.

\section{Additional experiments exploring the limits of LiRA}\label{app:sec:limits}

In this appendix, we provide additional experiments that explore the limits of LiRA when there are duplicate samples in the training data, and (complementing results in Section~\ref{sec:exp2:limits}) when there are varying numbers of training epochs and varied dataset size. 

\begin{figure}[t]
 \centering
\includegraphics[width=0.7\textwidth]{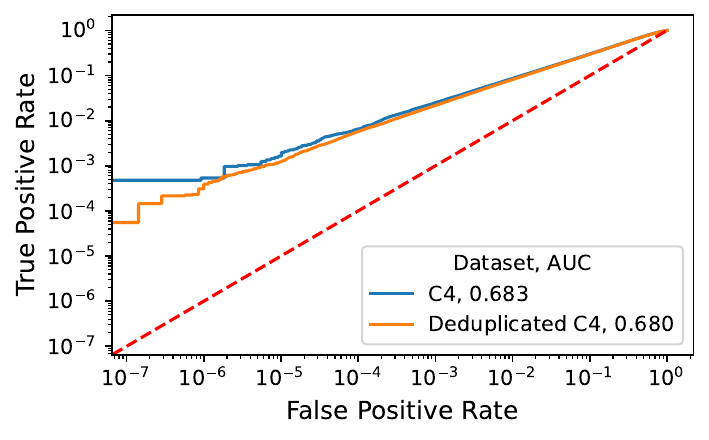}
  \caption{\textbf{The role of duplicates on MIA vulnerability.} We observe no significant differences (particularly as $\fpr$ increases) between models trained on C4 and de-duplicated C4.}
  \label{fig:duplicates}
\end{figure}

\paragraph{Investigating the role of duplicate training samples.} 
Given the relationship between MIA and memorization, and that prior work observes an important relationship between memorization and training-data duplication~\citep{lee2021deduplicating}, we test the relationship between MIA vulnerability and the presence of duplicate training samples.
In Figure~\ref{fig:duplicates}, we test the Chinchilla-optimally trained $140$M model on C4 and a de-duplicated version of C4. 
We de-duplicate C4 according to methodology described in \citet{lee2021deduplicating}, where we remove sequences that share
a common prefix of at least some threshold length.
This reduced the C4 dataset size from $364{,}613{,}570$ to $350{,}475{,}345$ samples.

We observe that the presence of duplicates has a negligible impact on $\auc$: 
it is $0.683$ for C4, and $0.680$ for de-duplicated C4. 
In other words, at least in terms of average attack success, the presence of duplicates does not seem to have a significant impact. 
However, further work is needed to assess how attack success changes with more stringent de-duplication, since our de-duplication procedure only removed $10$M samples from the dataset.\looseness=-1

\begin{figure}[t]
  \centering
\begin{subfigure}[t]{0.47\textwidth}
    \includegraphics[width=\textwidth]{figures/cooper/compare_num_epochs_10_3_Figure_5a.pdf}
  \caption{$140$M model, ${\approx}7$M samples, $10$ epochs.}
  \label{fig:compare_epochs_10_app}
\end{subfigure}
\hfill
\begin{subfigure}[t]{0.47\textwidth}
    \includegraphics[width=\textwidth]{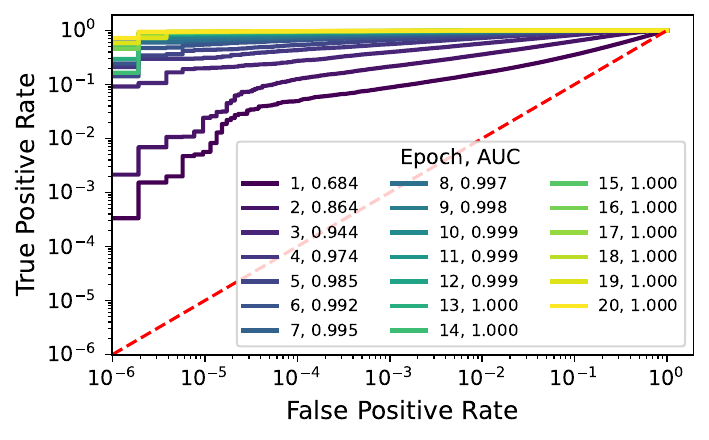}
  \caption{$140$M model, ${\approx}500K$, $20$ epochs.}
  \label{fig:compare_epochs_20_app}
\end{subfigure}%
\caption{\textbf{Over-training and MIA.}
$\roc$ curves demonstrate that MIA success significantly increases as models are trained for more epochs. 
(\textbf{a}) The $140$M model shows $\auc$ rising from $0.573$ ($1$ epoch) to $0.797$ ($10$ epochs). 
(\textbf{b}) Attacking a $140$M model trained on a smaller dataset shows a rapid escalation in $\auc$, from $0.604$ ($1$ epoch) to near-perfect membership inference ($\auc{=}1$) by $13$-$20$ epochs, highlighting that overfitting from prolonged training severely heightens privacy risks.}
\label{fig:compare_epochs}
\end{figure}

\paragraph{Varying training epochs and dataset size.} 
In \Cref{fig:compare_epochs}, we reduce the training set size from ${\approx}7$M (\Cref{fig:compare_epochs_10_app}) to  $2^{19}{\approx}500$K samples (\Cref{fig:compare_epochs_20_app}) on the $140$M model and train for $10$ (\Cref{fig:compare_epochs_10_app}) and $20$ epochs (\Cref{fig:compare_epochs_20_app}).
Both figures show $\roc$ curves that illustrate how MIA vulnerability changes with an increasing number of training epochs.
The goal of these experiments is to investigate if MIA becomes better with more training epochs, and if so, how attack performance improves over epochs as a function of training dataset size.

For the $140$M model trained on ${\approx}7$M samples for $10$ epochs, the $\auc$ increases with more epochs, starting from $0.573$ at $1$ epoch and reaching $0.797$ at $10$ epochs.
For the $140$M model trained on ${\approx}500$K samples for $20$ epochs, we observe a more dramatic increase in MIA vulnerability.
The $\auc$ starts at $0.604$ for $1$ epoch, rapidly increases to $0.864$ by 2 epochs, $0.944$ by 3 epochs, and approaches perfect MIA ($\auc$ close to $1.000$) after $13$ epochs. 
Of course, both of these experiments are effectively sanity checks.
We intentionally over-train in both, and use a relatively small training dataset size in the second. 

\begin{figure}[t]
  \centering
\begin{subfigure}[t]{0.33\textwidth}
    \includegraphics[width=\textwidth]{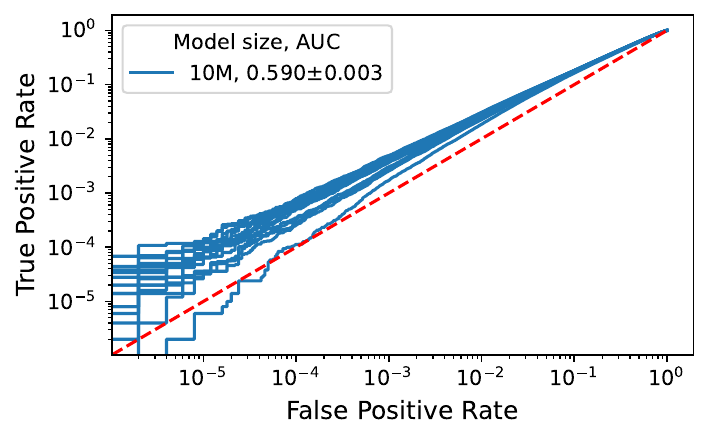}
  \caption{10M}
  \label{fig:compare_10m_roc_chinchilla}
\end{subfigure}\hfill
\begin{subfigure}[t]{0.33\textwidth}
    \includegraphics[width=\textwidth]{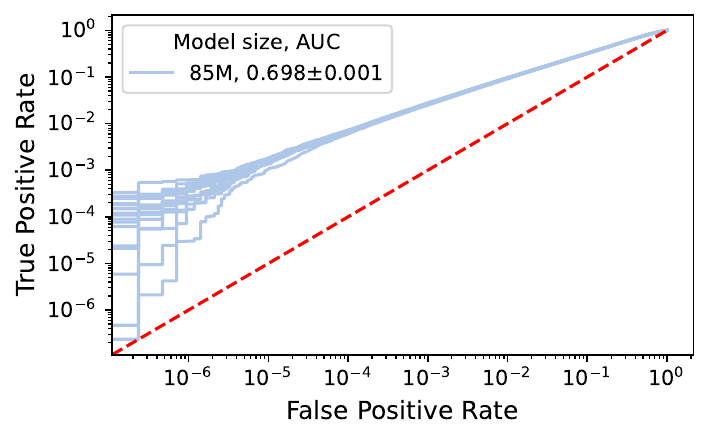}
  \caption{85M}
  \label{fig:compare_85m_roc_chinchilla}
\end{subfigure}\hfill
\begin{subfigure}[t]{0.33\textwidth}
    \includegraphics[width=\textwidth]{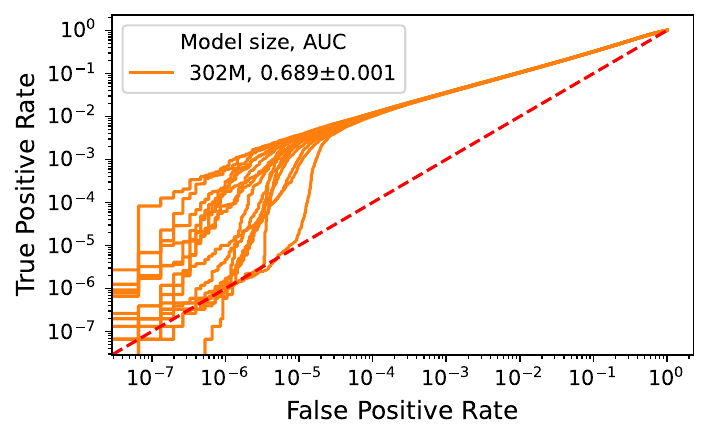}
  \caption{302M}
  \label{fig:compare_302m_roc_chinchilla}
\end{subfigure}\hfill
\begin{subfigure}[t]{0.33\textwidth}
    \includegraphics[width=\textwidth]{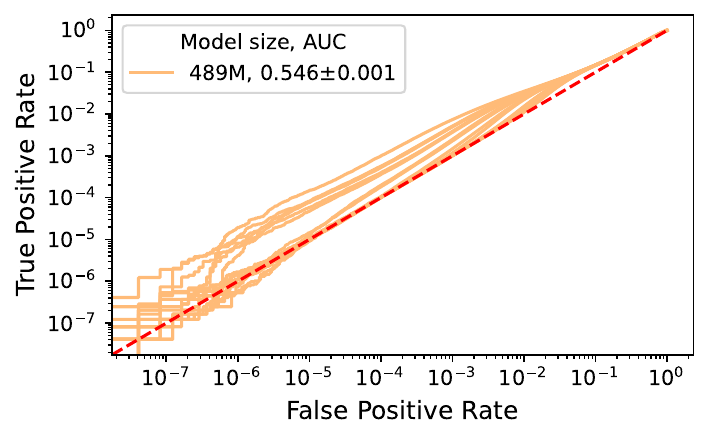}
  \caption{489M}
  \label{fig:compare_489m_roc_chinchilla}
\end{subfigure}\hfill
\begin{subfigure}[t]{0.33\textwidth}
    \includegraphics[width=\textwidth]{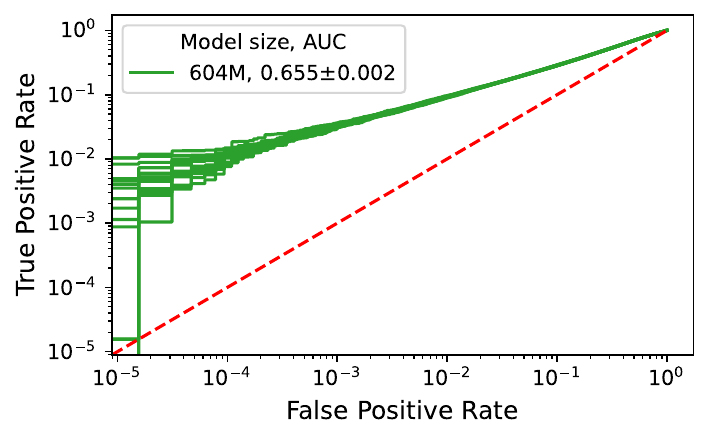}
  \caption{604M}
  \label{fig:compare_604m_roc_chinchilla}
\end{subfigure}\hfill
\begin{subfigure}[t]{0.33\textwidth}
    \includegraphics[width=\textwidth]{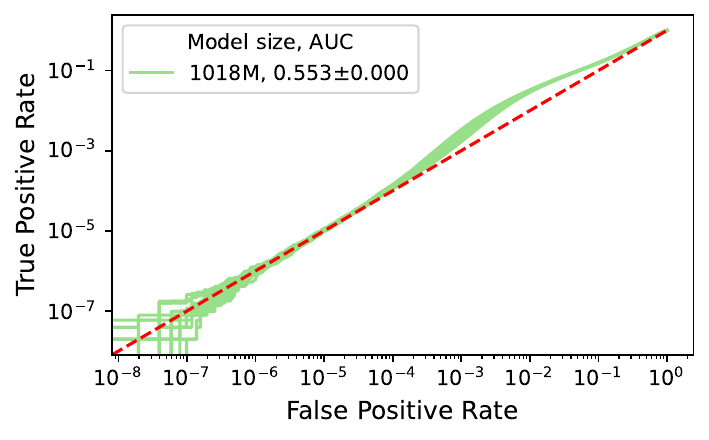}
  \caption{1018M}
  \label{fig:compare_1018m_roc_chinchilla}
\end{subfigure}
\caption{\textbf{$\mathrm{\bf{ROC}}$ curves and $\mathrm{\bf{AUC}}$ for Figure~\ref{fig:chinchilla:sizes}.} 
We attack different model sizes trained on the Chinchilla-optimal number of tokens. 
In each subplot, each line indicates a different attacked target.}
\label{fig:compare_model_sizes_chinchilla_dataset_rocs}
\vspace{-.05cm}
\end{figure}

\begin{figure}[h]
  \centering
\begin{subfigure}[t]{0.32\textwidth}
    \includegraphics[width=\textwidth]{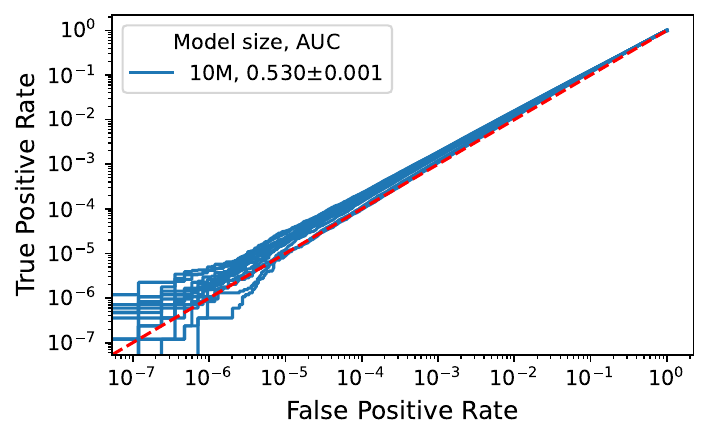}
  \caption{10M}
  \label{fig:compare_10m_roc}
\end{subfigure}\hfill
\begin{subfigure}[t]{0.32\textwidth}
    \includegraphics[width=\textwidth]{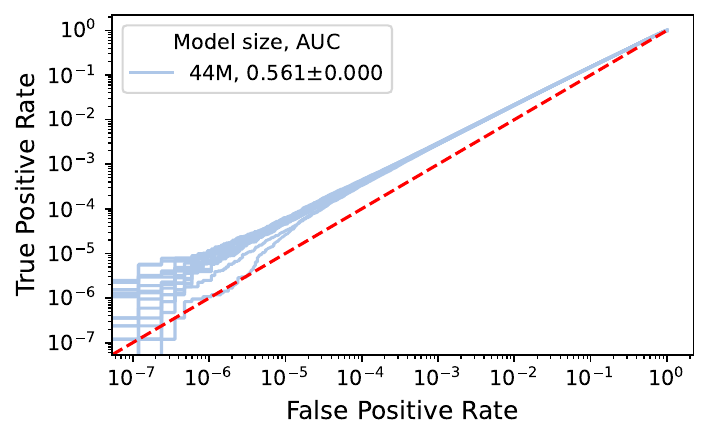}
  \caption{44M}
  \label{fig:compare_44m_roc}
\end{subfigure}\hfill
\begin{subfigure}[t]{0.32\textwidth}
    \includegraphics[width=\textwidth]{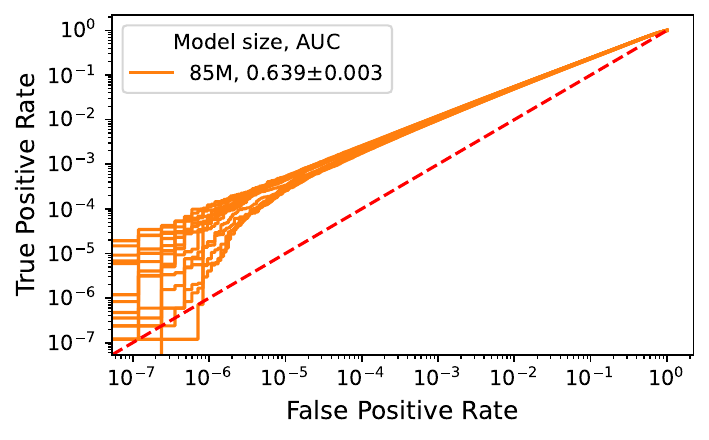}
  \caption{85M}
  \label{fig:compare_85m_roc}
\end{subfigure}\hfill
\begin{subfigure}[t]{0.32\textwidth}
    \includegraphics[width=\textwidth]{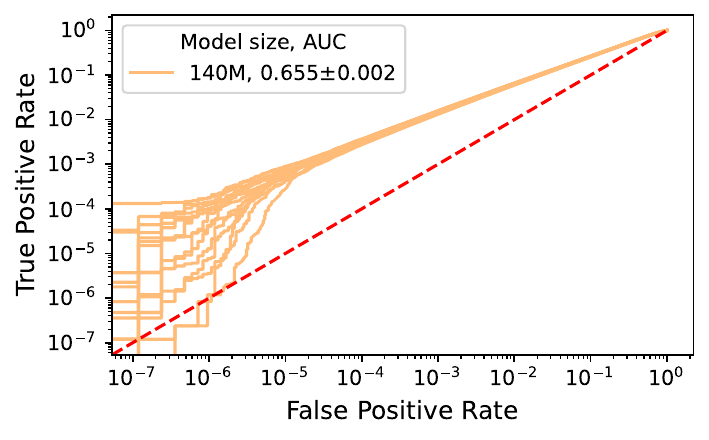}
  \caption{140M}
  \label{fig:compare_140m_roc}
\end{subfigure}\hfill
\begin{subfigure}[t]{0.32\textwidth}
    \includegraphics[width=\textwidth]{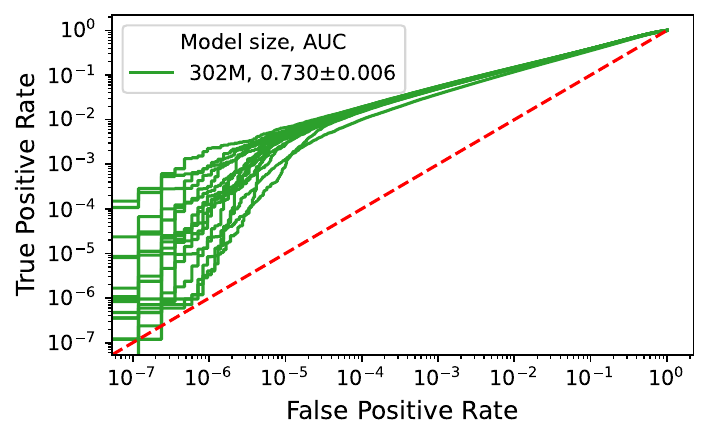}
  \caption{302M}
  \label{fig:compare_302m_roc}
\end{subfigure}\hfill
\begin{subfigure}[t]{0.32\textwidth}
    \includegraphics[width=\textwidth]{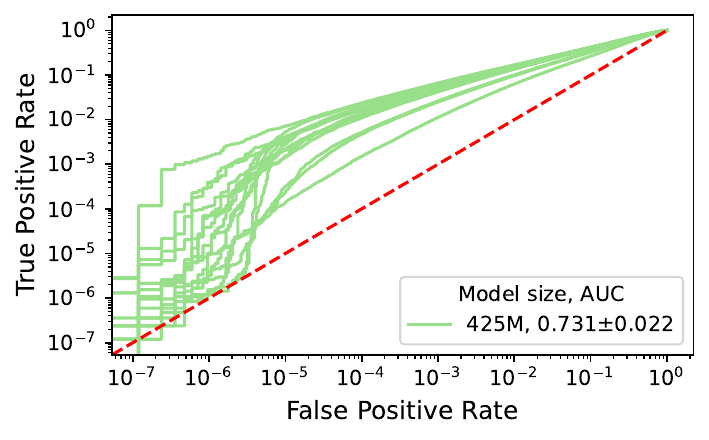}
  \caption{425M}
  \label{fig:compare_425m_roc}
\end{subfigure}\hfill
\begin{subfigure}[t]{0.32\textwidth}
    \includegraphics[width=\textwidth]{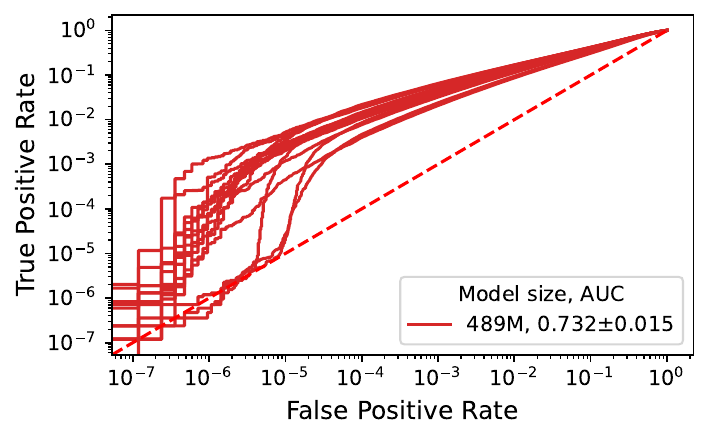}
  \caption{489M}
  \label{fig:compare_489m_roc}
\end{subfigure}\hfill
\begin{subfigure}[t]{0.32\textwidth}
    \includegraphics[width=\textwidth]{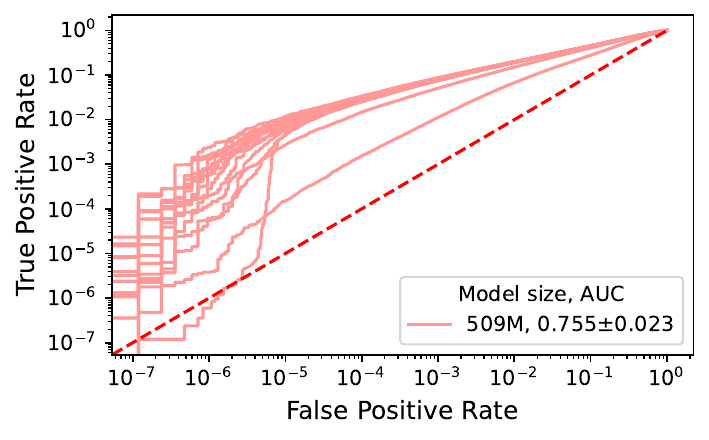}
  \caption{509M}
  \label{fig:compare_509m_roc}
\end{subfigure}\hfill
\begin{subfigure}[t]{0.32\textwidth}
    \includegraphics[width=\textwidth]{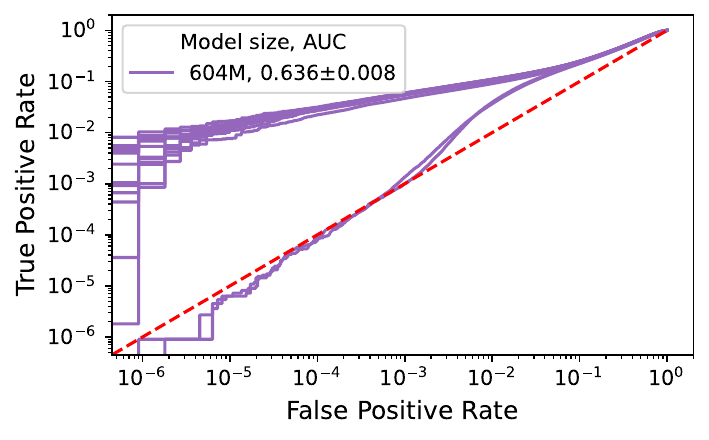}
  \caption{604M}
  \label{fig:compare_604m_roc}
\end{subfigure}\hfill
\begin{subfigure}[t]{0.32\textwidth}
    \includegraphics[width=\textwidth]{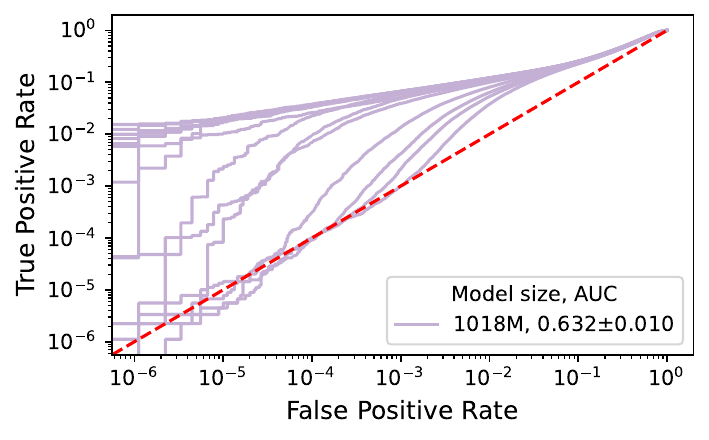}
  \caption{1018M}
  \label{fig:compare_1018m_roc}
\end{subfigure}
\caption{\textbf{$\mathrm{\bf{ROC}}$ curves and $\mathrm{\bf{AUC}}$ for Figure~\ref{fig:other-training:fixed-sized}.}
We attack different model sizes trained on the same number of samples (${\approx}8.3$M). 
In each subplot, each line indicates a different attacked target.}
\label{fig:compare_model_sizes_fixed_dataset_rocs}
\end{figure}

\paragraph{Full results for various-sized Chinchilla-trained models and fixed training set size.}
We provide full results for attacking Chinchilla-optimal models of various sizes for $1$ epoch (Figure~\ref{fig:chinchilla:sizes}), and attacking various model sizes trained on a fixed dataset of ${\approx}8.3$M samples for $1$ epoch (Figure~\ref{fig:other-training:fixed-sized}).
Both of these figures in the main paper show how $\tpr$ varies at fixed $\fpr$ in line plots.
Here, in Figures~\ref{fig:compare_model_sizes_chinchilla_dataset_rocs} and \ref{fig:compare_model_sizes_fixed_dataset_rocs}, we give individual $\roc$ curves for experimental results summarized in each of those figures, respectively.
For each subplot, each line indicates a different target model that we attack.
As discussed previously, some larger models appear to have more variance in their $\roc$ curves over different experimental runs. 
In \Cref{fig:compare_604m_roc}, we see that although $\auc$ is similar over different target models, there is catastrophic attack failure for one model at small $\fpr$s.\looseness=-1

\begin{figure}[h]
  \centering
\begin{subfigure}[t]{\textwidth}
    \includegraphics[width=\textwidth]{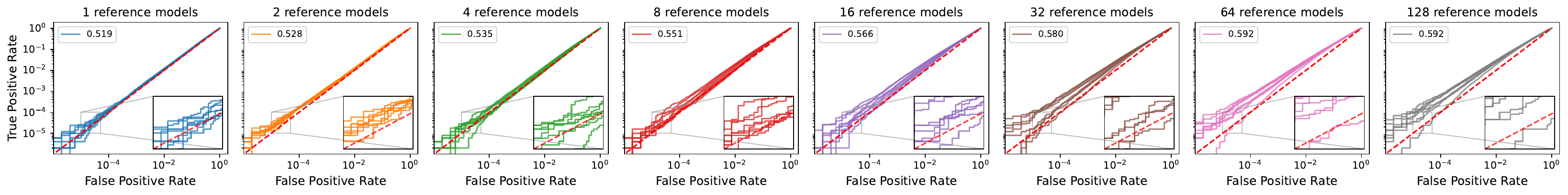}
  \caption{10M}
  \label{fig:compare_10m_roc_chinchilla_num_ref}
\end{subfigure}\hfill
\begin{subfigure}[t]{\textwidth}
    \includegraphics[width=\textwidth]{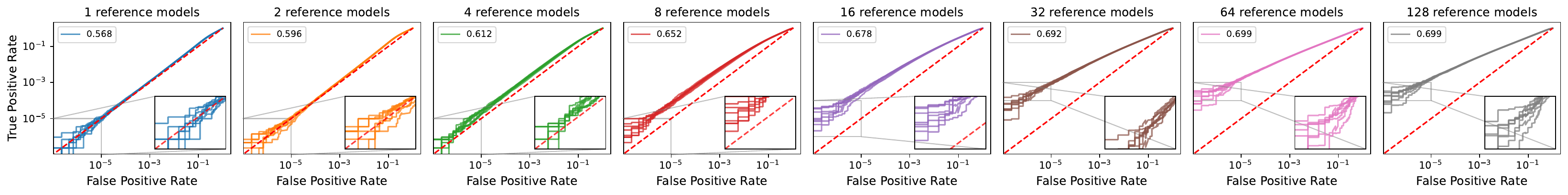}
  \caption{85M}
  \label{fig:compare_85m_roc_chinchilla_num_ref}
\end{subfigure}\hfill
\begin{subfigure}[t]{\textwidth}
    \includegraphics[width=\textwidth]{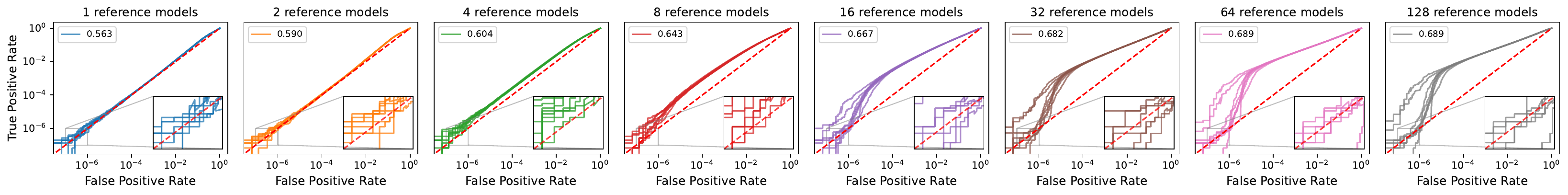}
  \caption{302M}
  \label{fig:compare_302m_roc_chinchilla_num_ref}
\end{subfigure}\hfill
\begin{subfigure}[t]{\textwidth}
    \includegraphics[width=\textwidth]{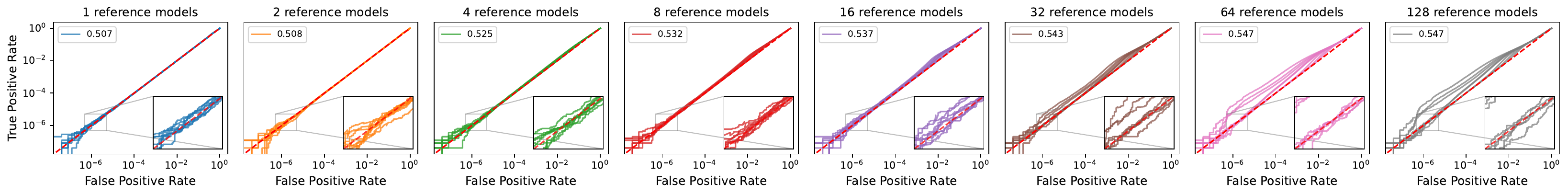}
  \caption{489M}
  \label{fig:compare_489m_roc_chinchilla_num_ref}
\end{subfigure}\hfill
\begin{subfigure}[t]{\textwidth}
    \includegraphics[width=\textwidth]{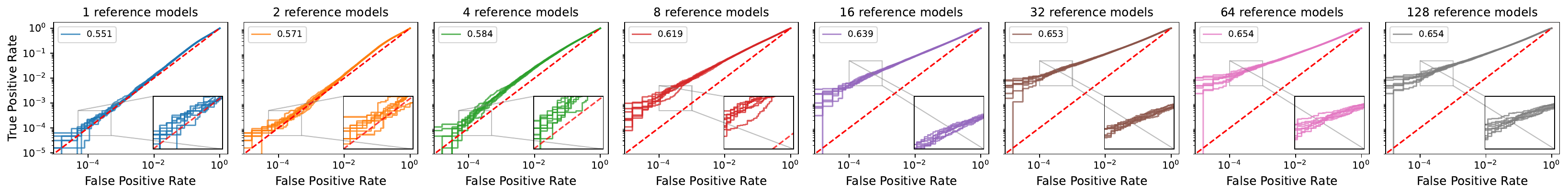}
  \caption{604M}
  \label{fig:compare_604m_roc_chinchilla_num_ref}
\end{subfigure}\hfill
\begin{subfigure}[t]{\textwidth}
    \includegraphics[width=\textwidth]{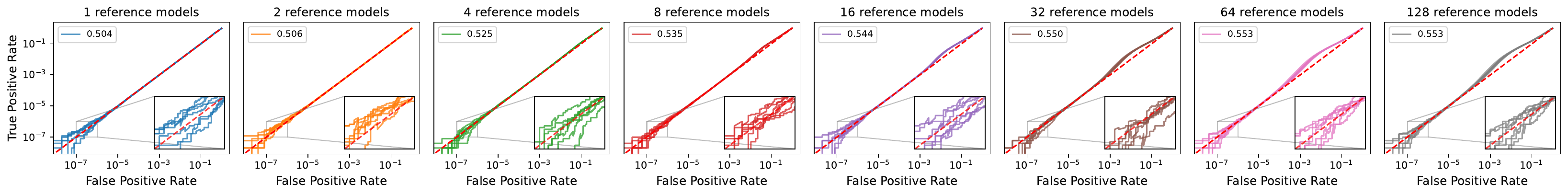}
  \caption{1018M}
  \label{fig:compare_1018m_roc_chinchilla_num_ref}
\end{subfigure}
\caption{\textbf{Extended $\mathrm{\bf{ROC}}$ curves and $\mathrm{\bf{AUC}}$ for Figure~\ref{fig:chinchilla:sizes}.} 
For each subplot, each line indicates a different target model that we attack.
Each row is a different model size.
Each column represents using LiRA with a different number of total reference models.  Each subplot also records the average $\auc$ across attacks on different targets.}
\label{fig:compare_model_sizes_chinchilla_dataset_rocs_num_refs}
\end{figure}

\clearpage
\paragraph{Varying reference models for all Chinchilla-optimally trained model sizes}
In \Cref{fig:compare_model_sizes_chinchilla_dataset_rocs_num_refs}, we replicate the experiments in \Cref{fig:compare_model_sizes_chinchilla_dataset_rocs}, but we vary the number of references.
Each row in the figure is for a different-sized model.
Each column uses a different number of total references (IN plotted) to perform the attack.
We attack $8$ targets trained on different training data subsamples in each plot. 

Unsurprisingly, MIA improves as we use more references. 
This mirrors our findings in Figure~\ref{fig:compare:lira}.
The key point of these figures is to show the general pattern of where the $\roc$ curve is relative to the reference line $y{=}x$.
We also show that there is variance (in the insets) across attack runs for the same model size.
These are not to be taken as detailed results that should be closely examined. 
(This is why the plots are not very large.) 
We investigate instability in Section~\ref{sec:instability} and Appendix~\ref{app: instability}.

\clearpage
\section{Investigating instability in per-sample membership decisions}\label{app: instability}

As noted in Section~\ref{sec:instability}, we observe substantial \emph{per-sample} instability in membership decisions. 
We also notice significant variability in $\rocauc$ across attacks in Figure~\ref{fig:compare_model_sizes_chinchilla_dataset_rocs_num_refs}.
However, because standard attack metrics such as $\rocauc$ are \emph{aggregates} over samples and decision thresholds;
they report metrics according to average $\fpr$/$\tpr$ over many samples.
As such, they can mask this instance-level variability.
We visualize and quantify individual-sample instability, and connect our analysis to prior work in other areas of statistics and machine learning. 

In Appendix~\ref{app:sec:instability:tp}, 
we provide extended results for Section~\ref{sec:exp3:per} on variation in per-sample true positive probabilities, and then in Appendix~\ref{app:sec:instability:flip:overall} we include more results and discussion on flip rate (Section~\ref{sec:instability}).\looseness=-1

\subsection{Variation in per-sample true positive probabilities}\label{app:sec:instability:tp} 

For the $140$M model, we plot the mean and standard deviation of the per-sample true positive probabilities, $\Pr(\text{predicted as member} | \text{member})$ for $2^{24}{=}16{,}777{,}216$ samples. 
For each sample, we compute variance across $64$ target models (for which the sample is a member); 
overall, this experiment trained $128$ models ($140$M size) on different random splits of the $2^{24}$ samples.
We compute $\Pr{\text{predicted as a member} | \text{member}}$, using $\frac{p_{\text{IN}}(\cdot|\vx)}{p_{\text{IN}}(\cdot|\vx)+p_{\text{OUT}}(\cdot|\vx)}>0.5$ to determine if the sample is predicted as a member (Section~\ref{sec:exp3:samples}).
We loop over each model, selecting it as the target model and the remainder as reference models used for LiRA. 
Since each sample had a probability of $0.5$ for inclusion in the training set, for each sample, we have on average $64$ target models where the sample was in training and $64$ for which it was not.\looseness=-1 

\begin{figure}[b]
  \centering
\begin{subfigure}[t]{0.32\textwidth}
    \includegraphics[width=\textwidth]{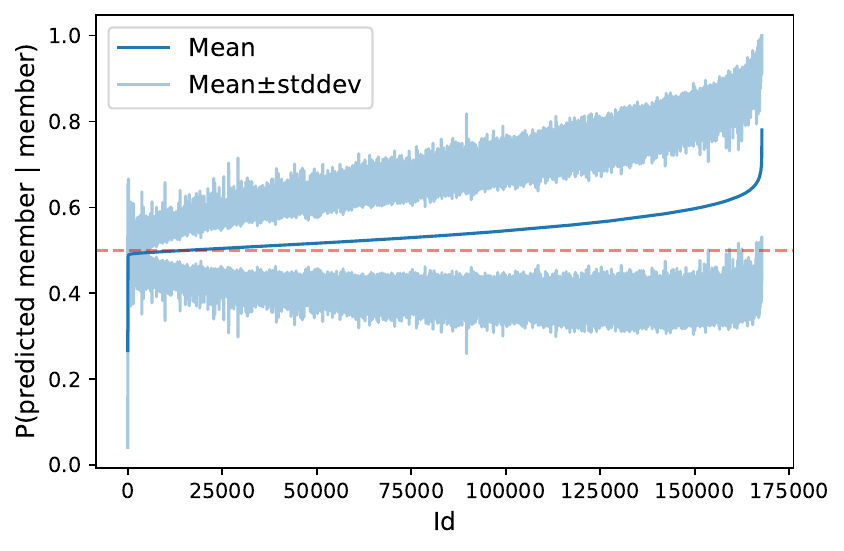}
  \caption{Per-sample true positive probabilities  (mean $\pm$ standard deviation),  ordered from smallest to largest.}
  \label{fig:variance-by-id}
\end{subfigure}\hfill
\begin{subfigure}[t]{0.32\textwidth}
    \includegraphics[width=\textwidth]{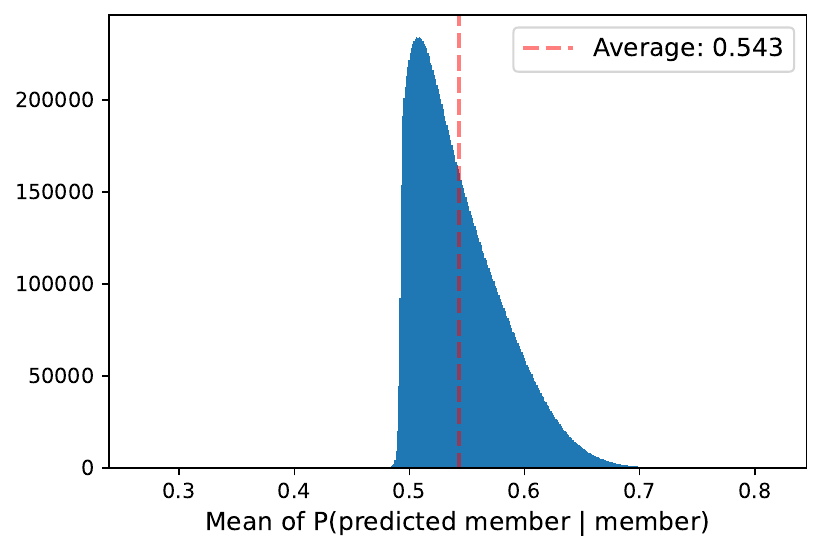}
  \caption{Histogram of average per-sample true positive probabilities from \Cref{fig:variance-by-id}.}
  \label{fig:example-means}
\end{subfigure}\hfill
\begin{subfigure}[t]{0.32\textwidth}
    \includegraphics[width=\textwidth]{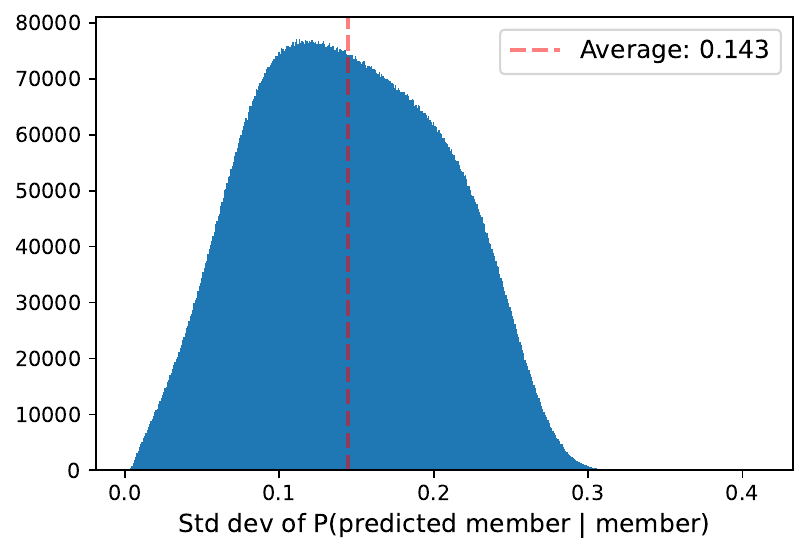}
  \caption{Histogram of standard deviation of per-sample true positive probabilities from \Cref{fig:example-means}.}
  \label{fig:example-stds}
\end{subfigure}\hfill
\caption{\textbf{Instability of per-sample true positive probabilities.}
For each of $2^{24}$ samples $\vx$, we compute the mean and standard deviation of $\Pr(\text{predicted as member}\mid \text{member})$ across $B{=}64$ target models. 
(\textbf{a}) After sorting samples by their mean, the mean and one standard deviation band.  
A histogram of (\textbf{b}) these per-sample means and (\textbf{c}) of the corresponding standard deviations.\looseness=-1}
\label{fig:instability}
\end{figure}

In Figure~\ref{fig:instability}, we provide three plots that give different views of the same data.
Figure~\ref{fig:variance-by-id} plots the true positive probability for each member. 
We sort members by the mean value of their true positive probability  (i.e., the mean of $\Pr(\text{predicted as member} | \text{member})$ over $64$ target models), so member ID corresponds to this ordering. 
We also show the variance over the $64$ target models by plotting the standard deviation. 

Together, Figures~\ref{fig:example-means} and~\ref{fig:example-stds} provide an alternate view of Figure~\ref{fig:variance-by-id}. 
Figure~\ref{fig:example-means} plots the histogram of the mean $\Pr(\text{predicted as member} | \text{member})$ for members across their respective $64$ target models. 
The average across these mean true positive probabilities for each member is $0.543$. 
However, note the distribution of per-sample means:
while the across-sample average of the per-sample means is $0.543$, a substantial mass of members exhibits mean $\Pr(\text{predicted as member} | \text{member}){>}0.6$. 
The spread is large: 
the average per-sample standard deviation is $0.143$, with many members exceeding a standard deviation of $0.2$. 

Overall, variance is significant.
The individual member true positive probabilities for each target are, when considered together, highly unstable. 
This variance can help explain why attack $\rocauc$ is perhaps lower than one might have hoped; 
there is considerable variance in the underlying sample binary decisions.
Altogether, this provides additional nuance concerning the extent of (alternatively, the limits of) attack robustness. 

\subsection{Analyzing per-sample membership decision instability}\label{app:sec:instability:flip:overall}

This appendix deepens our analysis of per‑sample MIA binary decision instability. 
We formalize \textbf{flip rate}~\citep{cooper2024variance}---the metric we use to measure instability at the per-sample decision level---and its unbiased empirical estimator (Appendix~\ref{app:sec:instability:flip}). 
We then explain how we measure flip rate in the MIA setting used in our experiments, and why the metric is informative for strong MIAs (Appendix~\ref{app:sec:instability:flip:mia}).
We connect our results to prior work on model/predictive multiplicity~\citep{breiman2001multiplicity} (Appendix~\ref{app:sec:instability:flip:multiplicity}).

Then, we derive an exact acceptance band (at level $\alpha$) for deciding when a sample's binary membership decisions are statistically indistinguishable from a coin flip;
for a finite number of targets $B$ and acceptance level $\alpha$, we obtain the resulting flip cutoff \(t_\alpha(B)\) that we deem the minimum required for $\vx$'s MIA decisions to be called ``statistically indistinguishable from a coin flip at level $\alpha$.'' (Appendix~\ref{app:sec:instability:flip:arbitrary}). 
Using these tools, we present extended empirical results for two model sizes, \(140\)M and \(302\)M (Appendix~\ref{app:sec:instability:flip:results}).
We estimate how much of standard attack performance (\(\rocauc\)) can be attributed to coin-flip-like decisions as opposed to reliable inference (Appendix~\ref{app:sec:instability:flip:decompose}). 
We also discuss additional ways to interpret the purpose and results of these experiments (Appendix~\ref{app:sec:instability:flip:interp}).  Finally, we discuss a more intuitive, but significantly more expensive, alternative approach for computing per-sample MIA decision instability, which, in principle, an attacker could compute (Appendix~\ref{app:sec:instability:flip:alt}). 
We expect this procedure to surface qualitatively similar instability that we observe with the setup we test.\looseness=-1 

For reference, the acceptance‑band cutoff values used in the figures/tables are \(t_{0.05}(125)\approx0.490\) and \(t_{0.05}(127)\approx0.487\) for the \(140\)M and \(302\)M models, respectively.

\paragraph{Key points.}
This is a long appendix, so we summarize key points here.
For a fixed sample \(\vx\) and $\fpr$ \(\eta\), we care whether the \emph{binary} membership decision produced by LiRA is \emph{reliable} (stable across equally plausible targets) or \emph{statistically indistinguishable from a coin flip}  with respect to target training randomness.
We compute flip rate with respect to the seed‑induced distribution \(\mu\). 
Different seeds reflect realistic training randomness (e.g., batch order); 
aggregate metrics are stable, indicating that \(\mu\) is not pathological/degenerate.

High flip rate (near \(0.5\)) means the decision for \(\vx\) is effectively a coin flip across plausible targets, so a true positive on a particular target is not evidence of \emph{reproducible} inference for \(\vx\); 
it is a lucky draw. 
Aggregate ranking performance (e.g., $\auc$) can still be \(>\!0.5\), but that is a different claim about \emph{averages}. 
We call the MIA decision for \(\vx\) ``indistinguishable from a coin flip at level \(\alpha\)'' if, under the exact two‑sided binomial test with \(B\) votes \(K\sim\mathrm{Binomial}(B,\theta)\), we fail to reject \(H_0:\theta=0.5\), where \(\theta \coloneqq \Pr_{r\sim\mu}\!\big[b_r^{(\eta)}(\vx)=1\big]\).
This yields a concrete cutoff \(t_\alpha(B)\) on \(\widehat{\mathrm{flip}}_{\eta,B}(\vx)\) via the equal‑tails acceptance region under \(H_0\); 
samples with \(\widehat{\mathrm{flip}}_{\eta,B}(\vx)\ge t_\alpha(B)\) are deemed indistinguishable from a coin flip.
(See Appendix~\ref{app:sec:instability:flip:arbitrary} for the derivation; the values we use in practice are noted above.) 
``Indistinguishable from a coin flip at level \(\alpha\)'' is a standard, finite‑sample exact test with a clear, observable cutoff \(t_\alpha(B)\).

Another way to understand these results is to see that, if  \(\theta=\Pr_{r\sim\mu}[b_r^{(\eta)}(\vx)=1]\approx 0.5\), then
\[
\Pr_{r,r'\sim\mu}\big[b_r^{(\eta)}(\vx)=b_{r'}^{(\eta)}(\vx)\big]\;=\;1-\mathrm{flip}_\eta(\vx)\;\approx\;0.5.
\]
Retraining the same pipeline on the same data would reproduce the \emph{same} decision for \(\vx\) only about half the time. 
This is the operational meaning of an ``indistinguishable from a coin flip'' per‑sample MIA decision.
Importantly, this claim concerning flip rate is about MIA decisions, not the underlying scores. 
Even if LiRA scores for \(\vx\) carry some signal, decision instability can be high when the calibrated threshold \(\tau_r(\eta)\) varies  across seeds; 
AUC may remain non‑trivial while flip rate is high. 
Our claim is specifically about the reliability of \emph{per‑sample decisions}.
(We further discuss interpretation details in Appendix~\ref{app:sec:instability:flip:interp}.) 
We apply this per‑sample test and report descriptive counts across many \(\vx\); 
the inferential claim is \emph{per sample} (``indistinguishable from a coin flip at level \(\alpha\)''), which is appropriate for the setup of the MIA security game.

The classifier threshold \(\tau_r(\eta)\) is calibrated on non‑members for each target (trained with seed $r$), anchoring non‑member decisions to their own distribution while leaving members more exposed to seed‑induced score and \(\tau\) variation, especially where IN/OUT overlap. 
This is a feature of the real attack protocol, not an evaluation artifact.
With finite \(B\), some truly non‑coin-flip-like \(\vx\) could be labeled coin-flip-like by chance. The exact binomial test controls Type‑I error at level \(\alpha\); 
the acceptance band and \(t_\alpha(B)\) make the rule explicit. 
In this decomposition (e.g., contributions to $\tpr$ and $\auc$), we filter only the ``coin-flip-like'' band (not all highly unstable cases like \([0.4,t_\alpha(B))\)), so reported  performance is a conservative \emph{upper bound} on reliable inference.

\subsubsection{Measuring instability of individual membership decisions with flip rate}\label{app:sec:instability:flip}

To complement our measurements of typical metrics from work on MIA, we adopt a metric from \citet{cooper2024variance} for measuring per-sample MIA decision instability.
We first review this metric, then specify our MIA-calibrated version and its unbiased estimator.

\paragraph{Self‑consistency across a distribution of models.}
Let \(g\sim\nu\equiv\nu_{\mathcal{A},\mathcal{D}}\) denote a model drawn from the distribution induced by training algorithm \(\mathcal{A}\) (a function of a random seed) with training data from distribution \(\mathcal{D}\). 
For a binary decision rule \(b_g(\vx)\in\{0,1\}\) (e.g., \(b_g(\vx)=\1\{g(\vx)\ge\tau\}\)), \citet{cooper2024variance} define the \textbf{self‑consistency} at \(\vx\) as the pairwise agreement probability under two i.i.d.\ draws:
\begin{align}
\label{eq:sc}
\mathrm{SC}(\vx)\;\coloneqq\; \Pr_{g,g'\stackrel{\mathrm{i.i.d.}}\sim\nu}\big[b_g(\vx)=b_{g'}(\vx)\big].
\end{align}
For such binary decisions, \(\mathrm{SC}(\vx)\in[0.5,1]\):  
values near \(1\) indicate stability among binary decisions for $\vx$ in spite of randomness in the training process; 
values near \(0.5\) indicate that the decision for $\vx$ using this training process is statistically indistinguishable from a coin flip~\citep{cooper2024variance}.
A standard U‑statistic yields an unbiased estimator: \(\E[\widehat{\mathrm{SC}}(\vx)]=\mathrm{SC}(\vx)\).
Note that $\mathrm{SC}$ is defined for any \(\vx\). \citet{cooper2024variance} estimate it for samples in a held‑out test.

\paragraph{Flip rate on calibrated MIA decision rules.}
In our setting, we fix the dataset $\sD\sim\mathcal{D}$ and vary only the training seed, which affects batch order during training. 
We adapt $\mathrm{SC}$ from $\mu_{\mathcal{A},\mathcal{D}}$ to the MIA decisions under $\mu_{\mathcal{A},\sD}$ calibrated at a fixed $\fpr$. 

Let \(r\!\sim\!\mu\equiv\mu_{\mathcal{A},\sD}\) denote a target model drawn from the seed‑induced distribution with the (fixed) training dataset \(\sD\). 
Let \(\Lambda_r(\vx)\in\R\) be the attack score (e.g., LiRA posterior, Equation~\ref{eq:lira}) for sample \(\vx\). 
For a desired false‑positive rate \(\eta\in[0,1]\), define the per‑seed calibrated threshold \(\tau_r(\eta)\) (e.g., the \((1-\eta)\)-quantile of \(\Lambda_r\) on non‑members for that seed), and the calibrated membership decision
\begin{align}
\label{app:eq:binaryrule}
b_r^{(\eta)}(\vx)\;=\;\1\{\Lambda_r(\vx)\ge\tau_r(\eta)\}
\end{align}
(as in Section~\ref{sec:rw} and Appendix~\ref{app:sec:background}). 
Unlike \citet{cooper2024variance}, we focus on \emph{dis}agreement between cross-seed MIA decisions for $\vx$ rather than agreement. 
The (population) \textbf{flip rate} at \(\vx\) under \(\mu\) and \(\fpr\) \(\eta\) is\looseness=-1
\begin{equation}
\label{app:eq:flip:pop}
\mathrm{flip}_{\eta}(\vx)
\;\coloneqq\;
\Pr_{r,r'\stackrel{\mathrm{i.i.d.}}{\sim}\mu}\big[b_r^{(\eta)}(\vx)\neq b_{r'}^{(\eta)}(\vx)\big]
\;=\;1-\mathrm{SC}_{\eta}(\vx),
\end{equation}
which lies in \([0,0.5]\) at the population level, with $0$ indicating that decision for $\vx$ does not flip/ is stable across target replicas and $0.5$ indicating that the decision for $\vx$ behaves like a coin flip.
(The operator point $\eta$ is left implicit in the use of $\mathrm{SC}$ in \citet{cooper2024variance}, as the authors always set $\tau{=}0.5$ in practice.)  

Note that we deliberately calibrate per seed, as this mirrors how MIAs are actually run in practice:
a single target is calibrated at its chosen $\fpr$.
Here, we vary the target (via seed) to expose instability across plausible targets $r\!\sim\!\mu$ using the same training recipe.

\paragraph{Unbiased estimator (order‑2 U‑statistic) and closed form.}
In practice, we estimate the population flip rate (Equation~\ref{app:eq:flip:pop}) for a concrete number of target replicas $B$ trained with different random seeds that control batch order. 
Given \(B\ge2\) i.i.d.\ target replicas \(r_1,\dots,r_B\!\sim\!\mu\) with calibrated rules \(b_{r_i}^{(\eta)}\), the canonical unbiased estimator of \(\mathrm{flip}_{\eta}(\vx)\) is
\begin{equation}
\label{app:eq:flip:ustat}
\widehat{\mathrm{flip}}_{\eta,B}(\vx)
\;=\;
\binom{B}{2}^{-1}\sum_{1\le i<j\le B}\1\{b_{r_i}^{(\eta)}(\vx)\neq b_{r_j}^{(\eta)}(\vx)\},
\qquad
\mathbb{E}\big[\widehat{\mathrm{flip}}_{\eta,B}(\vx)\big]=\mathrm{flip}_{\eta}(\vx).
\end{equation}
Let \(B_1(\vx)=\sum_{i=1}^B b_{r_i}^{(\eta)}(\vx)\) and \(B_0(\vx){=}B-B_1(\vx)\) be the numbers of ``member'' and ``non‑member'' binary decisions among the \(B\) replicas for \(\vx\). 
Then, Equation~\ref{app:eq:flip:ustat} has the closed form
\begin{equation}
\label{app:eq:flip:count}
\widehat{\mathrm{flip}}_{\eta,B}(\vx)\;=\;\frac{2\,B_0(\vx)\,B_1(\vx)}{B\,(B-1)}.
\end{equation}

Maximizing \(B_0(\vx)B_1(\vx)\) under \(B_0(\vx){+}B_1(\vx){=}B\) yields the finite‑\(B\) upper bound, since 
\[
\widehat{\mathrm{flip}}_{\eta,B}(\vx)\;\le\;
\frac{2\,\lfloor B/2\rfloor\,\lceil B/2\rceil}{B\,(B-1)}\;=\;
\begin{cases}
\dfrac{B}{2(B-1)}=\dfrac12+\dfrac{1}{2(B-1)}, & B \text{ even},\\[6pt]
\dfrac{B+1}{2B}=\dfrac12+\dfrac{1}{2B}, & B \text{ odd},
\end{cases}
\]
which exceeds \(0.5\) and converges to \(0.5\) as \(B\to\infty\) (e.g., \(B{=}125\Rightarrow 0.504\)).

To see why, note that \(B_0(\vx)B_1(\vx)\)  is maximized by the most balanced vote split (i.e., \(B_0(\vx)B_1(\vx) \leq \floor{B/2}\ceil{B/2}\)). 
If $B$ is even, i.e., $B{=}2k$, the maximum occurs at $B_0(\vx)=B_1(\vx)=\floor{B/2}=k$, so
\[
B_0(\vx)B_1(\vx) = k^2=\frac{B^2}{4}
\quad\Longrightarrow\quad
{\mathrm{flip}}_{\max}
= \frac{2\cdot (B^2/4)}{B(B-1)}
= \frac{B}{2(B-1)}
= \frac{1}{2}+\frac{1}{2(B-1)}.
\]
If $B$ is odd, $B=2k+1$, the maximum occurs at $B_0(\vx)B_1(\vx)=\floor{B/2}\ceil{B/2}=k(k+1)$,  so
\[
B_0(\vx)B_1(\vx) = k(k+1) = \frac{B^2-1}{4}
\quad\Longrightarrow\quad
{\mathrm{flip}}_{\max}
= \frac{2\cdot ((B^2-1)/4)}{B(B-1)}
= \frac{B^2-1}{2B(B-1)}
= \frac{1}{2}+\frac{1}{2B}.
\] 

Of course, this means that at low $B$, flip rate can have values that are quite far away from $0.5$. 
For example, when $B{=}2$, the $\mathrm{flip}_{\max}{=}1$. 
Nevertheless, this is the right choice of metric, as it is unbiased.
In our experiments, we ensure that the flip rate is easily interpretable by plotting results where the minimum $B{=}125$, such that $\mathrm{flip}_{\max}{\approx}0.504$.
We discuss this further in Appendix~\ref{app:sec:instability:flip:arbitrary}. 
\paragraph{Why the U-statistic is the right estimator (unbiasedness).}
Fix a sample \(\vx\) and an $\fpr$ \(\eta\).
Write \(b_r^{(\eta)}(\vx)\in\{0,1\}\) for the calibrated decision of target \(r\!\sim\!\mu_{\mathcal{A},\sD}\).
Let \(b\) denote a generic draw of \(b_r^{(\eta)}(\vx)\), and set
\[
\theta \;\coloneqq\; \Pr\big[b=1\big] \in [0,1].
\]
Draw \(B\ge2\) i.i.d.\ replicas \(b_1,\dots,b_B\overset{\text{i.i.d.}}{\sim}\mathrm{Bernoulli}(\theta)\).
The population flip rate at \((\vx,\eta)\) (the pairwise disagreement probability for two independent draws) is
\begin{align}
\label{eq:flipthetaform}
\mathrm{flip}_{\eta}(\vx)
\;&=\;\Pr[b\neq b']
\;=\;\Pr[b{=}1,b'{=}0]+\Pr[b{=}0,b'{=}1]\nonumber\\
\;&=\;\theta(1{-}\theta)+(1{-}\theta)\theta
\;=\;2\theta(1{-}\theta).
\end{align}
Because \((\theta-\tfrac12)^2\ge 0 \iff \theta(1-\theta)\le \tfrac14\), we have
\(\mathrm{flip}_{\eta}(\vx)=2\theta(1-\theta)\le \tfrac12\), i.e., the population flip rate never exceeds \(0.5\).

For a concrete \(B\), the empirical estimator (order‑2 U‑statistic) averages the pairwise indicator over all unordered pairs:
\[
\widehat{\mathrm{flip}}_{\eta,B}(\vx)
\;=\;\binom{B}{2}^{-1}\sum_{1\le i<j\le B}\1\{b_i\neq b_j\},
\]
as in Equation~\ref{app:eq:flip:ustat}.
By linearity of expectation and independence,
\[
\E\!\big[\widehat{\mathrm{flip}}_{\eta,B}(\vx)\big]
= \binom{B}{2}^{-1}\sum_{1\le i<j\le B}\E\!\big[\1\{b_i\neq b_j\}\big].
\]
For any fixed pair \((i,j)\) with \(i\ne j\),
\[
\E\!\big[\1\{b_i\neq b_j\}\big]
= \Pr[b_i\neq b_j]
= \Pr[b_i{=}1,b_j{=}0]+\Pr[b_i{=}0,b_j{=}1].
\]
Because \(b_i,b_j\) are independent \(\mathrm{Bernoulli}(\theta)\),
\[
\Pr[b_i{=}1,b_j{=}0]=\Pr[b_i{=}1]\Pr[b_j{=}0]=\theta(1-\theta),\quad
\Pr[b_i{=}0,b_j{=}1]=(1-\theta)\theta,
\]
so \(\Pr[b_i\neq b_j]=2\theta(1-\theta)\).
Therefore every term in the sum equals \(2\theta(1-\theta)\), so
\[
\E\!\big[\widehat{\mathrm{flip}}_{\eta,B}(\vx)\big]
= \binom{B}{2}^{-1}\sum_{1\le i<j\le B}2\theta(1-\theta)
= 2\theta(1-\theta)
= \mathrm{flip}_{\eta}(\vx),
\]
so \(\widehat{\mathrm{flip}}_{\eta,B}\) is exactly unbiased for all \(B\ge2\).

\paragraph{Showing unbiasedness via the vote fraction.}
For our discussion below and in Appendix~\ref{app:sec:instability:flip:arbitrary}, it is useful to see the same result via another argument.
As above in our discussion of flip rate (Equation~\ref{eq:flip:ustat}),
Let
\[
B_1(\vx)\;=\;\sum_{i=1}^B b_i^{(\eta)}(\vx),\qquad
B_0(\vx)\;=\;B-B_1(\vx).
\]
By construction, \(B_1(\vx)\) is the sum of \(B\) i.i.d.\ Bernoulli\((\theta)\) draws, so
\[
B_1(\vx)\sim \mathrm{Binomial}(B,\theta).
\]
Define the \textbf{vote fraction} \(v(\vx)\coloneqq B_1(\vx)/B\).
Therefore, we can write
\begin{align*}
    B_1(\vx) &= Bv(\vx)\\
    B_0(\vx) &= B(1 -v(\vx)).
\end{align*}
The number of disagreeing unordered pairs is \(B_1(\vx)\,B_0(\vx)\) (choose one ``member'' vote and one ``non‑member'' vote), so
\begin{align}
\label{app:eq:vote}
\widehat{\mathrm{flip}}_{\eta,B}(\vx)
\;&=\;\frac{B_1(\vx)\,B_0(\vx)}{\binom{B}{2}}
={B_1(\vx)\,B_0(\vx)}\frac{(B-2)!2!}{B!}
=\frac{2}{B(B-1)} B^2v(\vx)(1-v(\vx))\nonumber\\
\;&=\;\frac{2B}{B-1}\,v(\vx)\bigl(1-v(\vx)\bigr).
\end{align}
Since \(B_1(\vx)\sim\mathrm{Binomial}(B,\theta)\),
\begin{align}
    \label{eq:expectedvote}
    \E\big[v(\vx)\big]=\frac{\E[B_1(\vx)]}{B}=\frac{B\theta}{B}=\theta, \quad \text{and}
\end{align}
\begin{align}
    \label{eq:varvote}
    \Var\big[v(\vx)\big]=\frac{\theta(1-\theta)}{B},
\end{align}
because
\[
    \Var\big[v(\vx)\big]=\Var\Bigg[\frac{B_1(\vx)}{B}\Bigg]=\frac{\Var[B_1(\vx)]}{B^2},
\]
by the scaling law for variance:
\[
\Var[aX]=\E\!\big[(aX-\E[aX])^2\big]=\E\!\big[(a(X-\E[X]))^2\big]
= a^2\,\E\!\big[(X-\E[X])^2\big]=a^2\Var[X].
\]
Next,
\begin{align*}
    \Var[B_1(\vx)]
    &=\Var\Bigg[\sum_{r=1}^B b_r^{(\eta)}(\vx)\Bigg]\\
    &=\sum_{r=1}^B\Var\big[b_r^{(\eta)}(\vx)\big]
      + 2\sum_{r<j}\Cov\!\big[b_r^{(\eta)}(\vx), b_j^{(\eta)}(\vx)\big]\\
    &= \sum_{r=1}^B\Var\big[b_r^{(\eta)}(\vx)\big]
      \qquad\text{(independence: \(\Cov(\cdot,\cdot)=0\) for \(r\neq j\))}.
\end{align*}
Because $b_r^{(\eta)}(\vx)$ is a Bernoulli variable with success probability $\theta$, 
\(\Var[b_r^{(\eta)}(\vx)] = \theta(1-\theta)\).
Therefore
\[
\Var[B_1(\vx)] = \sum_{r=1}^B \theta(1-\theta) = B\theta(1-\theta),
\]
and so
\[
\Var\big[v(\vx)\big] = \frac{\Var[B_1(\vx)]}{B^2}=\frac{B\theta(1-\theta)}{B^2}
= \frac{\theta(1-\theta)}{B},
\]
as claimed in Equation~\ref{eq:varvote}. 
Finally, combining Equations~\ref{eq:expectedvote} and~\ref{eq:varvote} with the definition of variance,
\begin{align}
    \label{eq:expectvotesquared}
    \E\big[v(\vx)^2\big]\,=\,\Var[v(\vx)]+\E[v(\vx)]^2
\,=\,\frac{\theta(1{-}\theta)}{B}+\theta^2. 
\end{align}
Therefore, by Equations~\ref{eq:expectedvote} and~\ref{eq:expectvotesquared},
\begin{align*}
\E\big[v(\vx)(1-v(\vx))\big]
= \E[v(\vx)]-\E[v(\vx)^2]
&= \theta - \Big(\frac{\theta(1-\theta)}{B} + \theta^2\Big)\\
&= \big(\theta-\theta^2\big) - \frac{\theta(1-\theta)}{B}\\
&=\theta(1-\theta) - \frac{1}{B}\cdot\theta(1-\theta)\\
&= \theta(1-\theta)\Big(1-\frac{1}{B}\Big).
\end{align*}
Plugging into Equation~\ref{app:eq:vote} gives
\begin{align*}
\E\!\big[\widehat{\mathrm{flip}}_{\eta,B}(\vx)\big]
=\frac{2B}{B-1}\,\E\!\big[v(\vx)(1-v(\vx))\big] 
&=\frac{2B}{B-1}\,\theta(1-\theta)\Big(1-\frac{1}{B}\Big)\\
&=\frac{2B\theta(1-\theta)}{B-1}-\frac{2\theta(1-\theta)}{B-1}\\
&=\frac{2\theta(1-\theta)(B-1)}{B-1}\\
&=2\theta(1-\theta)
=\mathrm{flip}_{\eta}(\vx),
\end{align*}
by Equation~\ref{eq:flipthetaform}.
Therefore, the U‑statistic is unbiased for all \(B\ge2\).

\paragraph{Why a quadratic surrogate for ``lack of margin'' is biased.}
As we discuss in Appendix~\ref{app:sec:instability:flip:arbitrary}, interpreting empirical estimates of the flip rate can be a bit counter-intuitive.
Empirical estimates that are very close to \(0.5\) may actually reflect a vote split that seems a bit far from \(\lfloor B/2\rfloor/\lceil B/2\rceil\).
In other words, concrete splits for a given \(B\) might ``feel'' somewhat far from a $50/50$ split even if \(\widehat{\mathrm{flip}}_{\eta,B}\approx 0.5\).
As a result, it might seem natural to derive a metric that captures coin-flip behavior by showing how far the vote fraction (Equation~\ref{eq:varvote}) is from a completely split vote, rather than estimating the flip rate.\looseness=-1 

That is, consider that the raw margin from a completely split vote is \(v(\vx)-\tfrac{1}{2}\).
(Note that, if $v(\vx)=0.5$, then the raw margin is $0$; 
if $v(\vx)=1$, then the raw margin is $0.5$;
if $v(\vx)=0$, then the raw margin is $-0.5$;
and similarly, for any intermediate vote fraction.)
Scaling so the range becomes \([0,1]\) and taking absolute value so that there are no negative values gives
\[
m(\vx)\;\coloneqq\;\big|2v(\vx)-1\big|\in[0,1].
\]
Therefore, \(m(\vx)=0\) at a perfect split and \(m(\vx)=1\) at unanimity. 
But of course, $m(\vx)$ is neither smooth nor concave.
We show two convenient identities (by completing the square) that relate the margin and the quadratic in $v(\vx)$, so that we can have a smooth, concave alternative:
\begin{align}
\label{eq:margin-quad-forms}
v(\vx)\bigl(1-v(\vx)\bigr)
&= \tfrac{1}{4} - \bigl(v(\vx)-\tfrac{1}{2}\bigr)^2
= \tfrac{1}{4} - \tfrac{1}{4}\bigl(2v(\vx)-1\bigr)^2
= \tfrac{1}{4}\,\bigl(1 - m(\vx)^2\bigr),\\
2\,v(\vx)\bigl(1-v(\vx)\bigr)
&= \tfrac{1}{2} - 2\bigl(v(\vx)-\tfrac{1}{2}\bigr)^2
= \tfrac{1}{2} - \tfrac{1}{2}\,m(\vx)^2. \nonumber
\end{align}
So \(2v(\vx)(1-v(\vx))\) is a smooth, concave, symmetric surrogate for ``lack of margin'' (maximal at \(v(\vx)=\tfrac{1}{2}\), decreasing as the margin grows).

While this alternative seems to behave ``nicely'' in practice (i.e., is at most $\tfrac{1}{2}$, unlike $\widehat{\mathrm{flip}}_{\eta,B}$), it is biased (downward) for finite \(B\). 
That is,
\begin{align*}
    \E\!\left[2v(\vx)\big(1-v(\vx)\big)\right]
    &= 2\big(\E[v(\vx)] - \E[v(\vx)^2]\big)\\
    &= 2\Big(\E[v(\vx)] - \big(\Var[v(\vx)] + \E[v(\vx)]^2\big)\Big) &\text{(variance identity)}\\
    &= 2\Big(\theta - \big(\Var[v(\vx)] + \theta^2\big)\Big) &\text{(by Equation~\ref{eq:expectedvote})}\\
    &= 2\Big(\theta - \big(\tfrac{\theta(1-\theta)}{B} + \theta^2\big)\Big)
     &\text{(by Equation~\ref{eq:varvote})}\\
    &= 2\Big(\theta-\theta^2 - \tfrac{\theta(1-\theta)}{B}\Big)\\
    &= 2\,\theta(1-\theta)\Big(1-\tfrac{1}{B}\Big).
\end{align*}
The population flip rate is \(2\theta(1-\theta)\) (Equation~\ref{eq:flipthetaform}), so
\[
\E\left[2v(\vx)\big(1-v(\vx)\big)\right]
= \mathrm{flip_\eta}(\vx) \cdot \Big(1-\tfrac{1}{B}\Big),
\]
i.e., the quadratic surrogate metric for showing a ``lack of margin'' (i.e., coin-flip-like behavior of MIA decisions for $\vx$) is downward biased by $\tfrac{1}{B}$ (i.e., is $\tfrac{1}{B}$ below the population flip rate) for any finite \(B\ge2\), and becomes unbiased only as \(B\to\infty\).
By contrast, the U‑statistic \(\widehat{\mathrm{flip}}_{\eta,B}(\vx)\) (Equation~\ref{eq:flip:ustat}) is exactly unbiased at every \(B\ge2\).
This is why we report the U‑statistic, i.e., the pairwise decision disagreement probability (which we informally call the flip rate).
\subsubsection{Measuring flip rate for MIA}\label{app:sec:instability:flip:mia}

In \citet{cooper2024variance}, the authors train $B$ models using bootstrap replicates drawn from a dataset $\sD$.
They split $\sD$ into train and test sets, train $B$ models on bootstrap subsamples of the train set, and, for each held‑out test sample, compute an unbiased estimate of self‑consistency from the $B$ binary  decisions.\looseness=-1 

Here, we measure flip rate in a setup that mirrors strong MIA.
We fix a dataset $\sD$ of size $2N{=}2^{20}$ (so $N{=}2^{19}$) and train each target model on the \emph{same} $N$‑sized subset---i.e., the set of members (size $N$) and non‑members (size $N$) is identical across targets.
When training targets, we change \emph{only} the random seed that determines batch order during training.
Changing the batch order induces randomness in the training process.
Together with unavoidable hardware non‑determinism, this yields the variability we observe across target models~\citep{cooper2022lawless}.

For LiRA, we fix a reference set of $128$ independently trained models on different $N$‑sized subsamples, and we use these same references for \emph{every} target to compute per‑sample IN and OUT reference distributions, $p_{\text{IN}}(\cdot\mid\vx)$ and $p_{\text{OUT}}(\cdot\mid\vx)$.
At a chosen $\fpr$ $\eta\in[0,1]$, each target model $r$ calibrates its \emph{own} threshold $\tau_r(\eta)$ on that target’s non‑member scores (i.e., we perform per‑seed calibration), and then applies the calibrated decision rule in Equation~\ref{app:eq:binaryrule}.
We then compute the flip rate for a sample $\vx$ over the ensemble $\{r_i\}_{i=1}^B$ via Equation~\ref{app:eq:flip:ustat}, thereby isolating the effect of target‑training randomness while holding references fixed.
We run such experiments on two model sizes: $140$M and $302$M (Appendix~\ref{app:sec:instability:flip:results}).\looseness=-1 

For reference, we highlight some key points about calibration that will come up repeatedly in the rest of this appendix. 

\begin{tcolorbox}[title=Calibration asymmetry and its consequences,
                  colback=gray!3,colframe=black!55,left=6pt,right=6pt,top=6pt,bottom=6pt]
\textbf{What we calibrate.} For each target \(r\) and fixed $\fpr$ \(\eta\), the decision threshold is 
$\tau_r(\eta)\;=\;\inf\{\,t:\ \widehat{F}_{\text{OUT},r}(t^{-})\;\ge\; 1-\eta\,\}$, i.e., the empirical \((1{-}\eta)\)-quantile of that seed's \emph{non‑member} scores. 
This guarantees the \emph{non‑member} tail is controlled at level \(\eta\) for that seed (with the usual tie convention; see Appendix~\ref{app:sec:background}). 

\textbf{Why asymmetry arises.} Because \(\tau_r(\eta)\) is re‑estimated on non‑members for each seed, it ``tracks'' seed‑to‑seed shifts in \emph{non‑member} score distributions by construction. 
Members, however, are not used for calibration, so many member scores lie closer to (and straddle) the moving boundary across seeds (Figures~\ref{fig:error-v-tau},~\ref{fig:140-lowlevel-flip}, \&~\ref{fig:302-lowlevel-flip}).

\textbf{Empirical effect.} In regions where IN/OUT scores overlap (Figures~\ref{fig:140M-archetypes} \&~\ref{fig:302M-archetypes}), small seed‑induced shifts in either the score or the boundary can flip member decisions; 
consequently, members exhibit substantially higher flip rate than non‑members at the same \(\eta\), and the gap widens at larger \(\eta\) (up to a point) and with increased model size (Figures~\ref{fig:140M-instability-varied-fpr} 
\&~\ref{fig:302M-instability-varied-fpr}).

\textbf{Implication.} This calibration asymmetry explains why aggregate metrics (e.g., mean $\tpr$ at fixed $\fpr$, see Tables~\ref{tab:acc_metrics_by_fpr} \&~\ref{tab:acc_metrics_by_fpr_302M}) can look stable (Figures~\ref{fig:instability-roc} \&~\ref{fig:instability-roc-2}), while many \emph{member} decisions are individually unstable (Tables~\ref{tab:flip_by_range_140M}, \ref{tab:flip_by_range_302M}, \ref{tab:tp_by_flip_140M}, \& \ref{tab:tp_by_flip_302M}). 
It also motivates our hypothesis‑test cutoff \(t_\alpha(B)\) for flagging statistically coin-flip-like per‑sample decisions (Appendix~\ref{app:sec:instability:flip:arbitrary}). 
\end{tcolorbox}

\paragraph{Interpreting flip rate for MIA.} 
Each target model is a plausible outcome of this training process.
Any of them would be a reasonable choice for running LiRA, as they are i.i.d.\ draws from the same seed‑induced distribution.
Measuring flip rate across targets therefore quantifies how resilient LiRA's per‑sample decision is to randomness in target training.

If a sample's binary membership decisions are \emph{stable} (low flip), LiRA's decision for that sample is
\emph{robust} to target‑training randomness and more likely to reflect persistent signal, rather than seed-specific idiosyncrasies. 
Conversely, if binary decisions are \emph{unstable}  (flip near its population maximum $0.5$),
the per‑sample decision is effectively \emph{arbitrary} with respect to seed choice---even when
aggregate performance metrics (e.g., $\tpr$ at fixed $\fpr$, or $\mathrm{AUC}{>}0.5$) look stable and reasonably high-performance.
In this case, per-sample membership decisions are so influenced by randomness in the training process that we cannot draw a reliable conclusion about membership. 
Put differently, measuring per‑sample instability lets us peer beneath high-level, average metrics---e.g., for a fixed $\fpr$, mean $\tpr$ over all members across plausible targets $r\!\sim\!\mu$---to assess what strong MIAs can (and cannot) say reliably about individual samples. 
\subsubsection{Connections to prior work on model and predictive multiplicity}\label{app:sec:instability:flip:multiplicity}

This analysis connects to broader literature in statistics and machine learning outside membership inference.
Notably, Leo Breiman's seminal work on the \textbf{Rashomon effect} emphasized that, for a given dataset, there often exists a \emph{multiplicity} of distinct decision rules with essentially the same overall accuracy~\citep{breiman2001multiplicity}. 
The Rashomon set---the set of models within a small tolerance of the optimal risk---can be surprisingly large~\citep{fisher2019all, semenova2022existence}. 
More recent  work on  predictive multiplicity also shows that training processes can produce models with effectively indistinguishable overall test accuracy that nonetheless disagree widely at the per-sample level~\citep{marx2020predictive, cooper2024variance,watson2023multiplicity}.\looseness=-1

To the best of our knowledge, this connection has not been made in the MIA setting.
Our setup differs in that we fix $\sD$ and vary only algorithmic randomness (via seed controlling batch order for target replicas); 
we then observe targets with similar overall accuracy but substantial per-sample churn, quantified by flip rate (Appendix~\ref{app:sec:instability:flip:results}). 
(We make no claims about the optimality of the resulting MIA rules.) 
The key result of these experiments is that average attack performance can remain stable, while individual membership decisions vary across seeds---a phenomenon that bears directly on the reliability and validity of membership claims about specific samples (as the problem is set up in the membership inference security game). 
\subsubsection{Reasoning about the minimum empirical flip rate that reflects coin-flip MIA decisions}\label{app:sec:instability:flip:arbitrary}

As noted in Appendix~\ref{app:sec:instability:flip},  the population flip rate  $\mathrm{flip}_\eta(\vx) \in [0,0.5]$ (Equation~\ref{app:eq:flip:pop}):
$0$ reflects MIA decisions that are completely stable for $\vx$ (i.e., do not flip) and $0.5$ reflects coin-flip decisions for $\vx$.
In practice, we estimate the population flip rate with the U-statistic for flip rate using a concrete number of target replicas $B$, namely $\widehat{\mathrm{flip}}_{\eta,B}(\vx)$ (Equation~\ref{app:eq:flip:ustat}).
This empirical estimate also has a minimum of $0$, reflecting completely stable binary decisions, but its maximum (reflecting maximal disagreement) slightly exceeds $0.5$ and converges to $0.5$ as $B\to \infty$. 
This raises an important question: for concrete $B$ in practice, which measurements of $\widehat{\mathrm{flip}}_{\eta,B}(\vx)$ reflect that the MIA decisions for $\vx$ behave like a coin flip?
That is, we need to determine a reasonable cutoff for $\widehat{\mathrm{flip}}_{\eta,B}(\vx)$, indicating that the decisions for $\vx$ are statistically indistinguishable from a coin flip.\looseness=-1

A principled way to determine this cutoff is to set up a hypothesis test at level $\alpha$:
we call the MIA decision for a sample $\vx$ ``indistinguishable from a coin flip at level $\alpha$'' if a two‑sided exact binomial test fails to reject.
We do this for our experiments in Section~\ref{sec:instability} and Appendix~\ref{app:sec:instability:flip:results}.
For the experiment with the $140$M model ($B{=}125$), we call the MIA decision for $\vx$ indistinguishable from a coin flip at $\alpha{=}0.05$ if the MIA decisions for $\vx$ exhibit $\widehat{\mathrm{flip}}_{\eta,B}(\vx)\gtrsim0.490$;
for the $302$M model ($B{=}127$), we call the MIA decision for $\vx$ indistinguishable from a coin flip at $\alpha{=}0.05$ if the MIA decisions for $\vx$ exhibit $\widehat{\mathrm{flip}}_{\eta,B}(\vx)\gtrsim 0.487$ (Figure~\ref{fig:unstable:curves}). 
In the end, all this requires is finding the minimal number of member votes $k$ at which the CDF $F(k)$ of the binomial $\textrm{Binomial}(B, 0.5) \geq \tfrac{\alpha}{2}$, i.e., 
\begin{align}
\label{eq:minvote}
    k_{\mathrm{L}}=\min \{k : F(k) \geq \alpha/2\},
\end{align}
and computing the coin-flip cutoff as $\geq \widehat{\mathrm{flip}}_{\eta,B}$ with $B_1(\vx)=k_{\mathrm{L}}$ and $B_0(\vx)=B - k_{\mathrm{L}}$.\looseness=-1 

In this appendix, for the reader interested in a refresher, we walk through how we set up this exact test. 
We describe the hypothesis test at level $\alpha$, how this results in an acceptance region (in terms of the number of member votes), and how we convert that region into a minimum empirical  $\widehat{\mathrm{flip}}_{\eta,B}(\vx)$ that we can defensibly interpret as behaving like a coin flip.
(This depends on the the vote fraction $v(\vx)$ discussion from above.) 

\paragraph{Setting up a hypothesis test.}
We call the MIA decision for a sample $\vx$ indistinguishable from a coin flip if the probability of predicting ``member'' equals the probability of predicting ``non‑member''.
Let $r_1,\ldots,r_B \stackrel{\text{i.i.d.}}{\sim}\mu$ denote $B$ target replicas (varying only by seed) from the training pipeline.
As throughout, let
\[
B_1(\vx)\;=\;\sum_{i=1}^B b_{r_i}^{(\eta)}(\vx),
\]
and note that, for fixed $(\vx,\eta)$, the indicators $b_{r_i}^{(\eta)}(\vx)$ are i.i.d.\ Bernoulli with success probability
\[
\theta\;\coloneqq\;\Pr_{r\sim\mu}\!\big[b_{r}^{(\eta)}(\vx)=1\big].
\]
Therefore, 
\(
B_1(\vx)\;\sim\;\mathrm{Binomial}\big(B,\theta\big) 
\). 
Coin-flip behavior corresponds to \(\theta = 0.5\). 

We set up the null hypothesis
\begin{align}
    \label{eq:null}
    H_0:\;\theta=0.5 \quad \text{(two‑sided exact binomial test at level $\alpha$)}.
\end{align}

If we fail to reject $H_0$, then we do not have sufficient evidence to say that the MIA decision for $\vx$ is \emph{not} a coin flip, and so we deem the decision indistinguishable from a coin flip. 
The significance level $\alpha$ means that, if $H_0$ is true (i.e., the decision is indistinguishable from a coin flip), the probability that we incorrectly reject $H_0$ (i.e., say that the decision is not a coin flip) is at most $\alpha$. 
Smaller $\alpha$ imposes a stricter standard for rejecting $H_0$ (stronger evidence is required).
We will later show that, for $B$ replicas, “fail to reject” is equivalent to 
\[
\widehat{\mathrm{flip}}_{\eta,B}(\vx) \;\geq\; t_\alpha(B),
\]
with $t_\alpha(B)$ computed from the binomial acceptance region under $H_0$ (Equation~\ref{eq:talpha}).

\paragraph{Deriving the two‑sided exact $p$‑value at level $\alpha$.}
For $B$ replicas and operating point $\eta$, each replica $r$ outputs a binary membership decision $b_r^{(\eta)}(\vx)\in\{0,1\}$.
Going forward, we denote the member‑vote count
\[
K \;=\; B_1(\vx)\;=\;\sum_{i=1}^B b_{r_i}^{(\eta)}(\vx).
\]

Under the null hypothesis $H_0$ in Equation~\ref{eq:null}, each target replica's MIA decision behaves like a fair coin, so\looseness=-1
\[
K\;\sim\;\mathrm{Binomial}\big(B,0.5\big).
\]

Intuitively, the further $K$ is from the center $B/2$ (i.e., a split vote, indicating coin-flip behavior), the stronger the evidence against $H_0$. 

More formally, let the binomial PMF and CDF under $H_0$ be, respectively, 
\begin{align}
    \label{eq:pmf-and-cdf}
\Pr(K=i)\;=\;\binom{B}{i}2^{-B},\qquad
F(k)\;=\;\Pr(K\le k)\;=\;\sum_{i=0}^{k}\binom{B}{i}2^{-B}.
\end{align}
Because $\binom{B}{i}=\binom{B}{B-i}$,
\[
\Pr(K=i)\;=\;\Pr\big(K=B-i\big)\quad\text{for all $i$,}
\]
and so the distribution is symmetric about $B/2$.

We can reason about the tails of this distribution in terms of a concrete vote $k$ and the CDF, i.e., 
\begin{align}
\label{eq:tailsym}
F(k) &= \Pr(K\le k)
=\sum_{i=0}^{k}\Pr(K=i)
=\sum_{i=0}^{k}\Pr\big(K=B-i\big)
=\sum_{j=B-k}^{B}\Pr(K=j) \nonumber\\
&=\Pr\big(K\ge B-k\big).
\end{align}
Thus the left tail at $k$ equals the right tail at $B-k$.
This follows from a change of variable, setting $j=B-i$ (when $i=0$, $j=B$ and when $i=k$, $j=B-k$, so the index runs in reverse and $\sum_{i=0}^k \Pr(K=B-i) = \sum_{j=B-k}^B \Pr(K=j)$).

And so, 
\begin{align}
\label{eq:tailsym2}
\Pr(K\ge k)
&= 1 - \Pr(K \leq k - 1) 
= 1 - F(k - 1) \nonumber\\
&= 1 - \Pr\big(K \geq B - (k-1)\big) 
\quad\text{(by Equation~\ref{eq:tailsym} with $k\mapsto k-1$)}\nonumber\\
&= \Pr\big(K \leq B - k\big).
\end{align}

Intuitively, a two-sided $p$-value measures how surprising the actual observed count $K{=}k$
is under the null hypothesis (Equation~\ref{eq:null}): 
it sums the probabilities of outcomes \emph{at least as far from the center $B/2$ as $k$}, in both tails. 
Because the binomial for $0.5$ is symmetric and unimodal about $B/2$, ``equally or more extreme'' corresponds to the union of the left tail up to $k$ and the symmetric right tail from $B{-}k$ upward (or the mirror statement when $k$ is on the right).
So, to derive a $p$-value at $k$, there are two cases to consider.

\emph{Case 1}: $k \leq \floor{B/2}$

Here, the left tail is $K \leq k$ and the right tail is $K\geq B -k$. 
Therefore, 
\begin{align*}
    p\text{-value}(k) &= \Pr(K \leq k) + \Pr(K\geq B - k)\\
    &= 2 \cdot \Pr(K \leq k) \qquad \text{(by tail symmetry, Equation~\ref{eq:tailsym})}\\
    &= 2\cdot F(k) \qquad \text{(by definition of the CDF, Equation~\ref{eq:pmf-and-cdf})}.
\end{align*}

\emph{Case 2}: $k \geq \ceil{B/2}$
Here, the right tail is $K \geq k$ and the left tail is $K\leq B -k$. 
Therefore,
\begin{align*}
    p\text{-value}(k) &= \Pr(K \geq k) + \Pr(K\leq B - k)\\
    &= 2 \cdot \Pr(K \geq k) \qquad \text{(by tail symmetry, Equation~\ref{eq:tailsym2})}\\
    &= 2\cdot \big(1 - F(k - 1)\big). 
\end{align*}

From this, we derive the standard form of the two-sided exact $p$-value. 
That is, because the binomial is symmetric and unimodal about $B/2$, this can be written as
\begin{align*}
p\text{-value}(k)
&=\;2\min\Big\{\,\Pr(K\le k),\; \Pr(K\ge k)\,\Big\},
\end{align*}
which means we double the smaller tail in order to capture both-sided extremeness. 
Alternatively, using \(\Pr(K\ge k)=1-\Pr(K\le k-1)=1-F(k-1)\) (Equation~\ref{eq:tailsym2}), this becomes
\begin{align*}
p\text{-value}(k)
=\begin{cases}
2\,F(k), & k\le \lfloor B/2\rfloor,\\[4pt]
2\,\big(1-F(k-1)\big), & k\ge \lceil B/2\rceil.
\end{cases}
\end{align*}
Finally, to handle discreteness at the exact center (even $B$ and $k=B/2$, which counts the mass at $k$ twice), we cap $p$-values at $1$:
\begin{align}
\label{eq:pvalue-cases}
p\text{-value}(k)\;=\;\min\Bigl\{\,1,\; 2\,\min\bigl(F(k),\,1-F(k-1)\bigr)\Bigr\}.
\end{align}
These equal‑tail formulas handle discreteness conservatively: the acceptance region is defined so that \(p\text{-value}(k)\geq\alpha\) inside the region and \(p\text{-value}(k)< \alpha\) outside, with $\tfrac{\alpha}{2}$ in each tail.
Because the binomial is discrete, the equal‑tail construction is slightly conservative. We follow the convention ``reject if $p<\alpha$'' and fail to reject if $p\geq\alpha$,  so boundary points with $p\text{-value}=\alpha$ remain inside the acceptance region.

Using Equation~\ref{eq:pvalue-cases}, the acceptance region for member votes $k$ at level $\alpha$ is constructed by finding
\[
k_{\mathrm{L}}\;=\;\min\{k\in\{0,\dots,\lfloor B/2\rfloor\}:\; F(k)\ge \alpha/2\},
\]
and then setting
\[
\mathcal{A}_\alpha \;=\; \{k_{\mathrm{L}},\,k_{\mathrm{L}}{+}1,\,\dots,\,B{-}k_{\mathrm{L}}\},
\]
so that \(K\in\mathcal{A}_\alpha \iff p\text{-value}(K)\geq\alpha\) (fail to reject). 
By symmetry, the upper endpoint is \(B-k_{\mathrm{L}}\) and so the acceptance region is a symmetric band around $B/2$ (Equation~\ref{eq:tailsym}).

Equivalently, the critical (rejection) region is
\[
\{\,K \le k_{\mathrm L}-1\,\}\;\cup\;\{\,K \ge B-k_{\mathrm L}+1\,\},
\]
so $K\in\mathcal{A}_\alpha\iff p\text{-value}(K)\ge\alpha$ (fail to reject), and $K\notin\mathcal{A}_\alpha\iff p\text{-value}(K)<\alpha$ (reject).

And so, for a fixed $B$ and given $k$, we can check if $F(k) \geq \alpha/2$ simply by computing $F(k)=\sum_{i=0}^{k}\binom{B}{i}2^{-B} \geq \alpha/2$, as in Equation~\ref{eq:pmf-and-cdf}. 
For $\alpha{=}0.05$ and $B{=}127$, $k_{\mathrm{L}}{=}52$ (and $B-k_{\mathrm{L}}{=}75$) with $F(52){\approx}0.02524$; for $\alpha{=}0.05$ and $B=125$, it is also the case that  $k_{\mathrm{L}}{=}52$ with $F(52){\approx}0.03661$ (but $B-k_{\mathrm{L}}{=}73$).\looseness=-1 

\paragraph{From the acceptance band to a concrete flip cutoff.}
For fixed $B$ and $\fpr$ $\eta$, the empirical flip at $(\vx,\eta)$ as a function of the member‑vote count $K$ is
\[
\phi_{B}(K) \coloneqq \frac{2\,K\,(B-K)}{B(B-1)},
\]
where $\widehat{\mathrm{flip}}_{\eta,B}(\vx)\;=\;\phi_{B}(K)$ (to preserve notation/ defining $\widehat{\mathrm{flip}}_{\eta,B}$ on $\vx$ and continue using $K$, as in the rest of this section). 
Since this is just a rewrite of the empirical flip rate at $K$,
\[
\phi_B(K)=\phi_B(B-K)\quad\text{(symmetry about $B/2$).}
\]
This function is symmetric in $K$ about $B/2$ and unimodal. 
A one‑step discrete difference shows it is strictly increasing on the left half:
\[
\Delta(K)\;\coloneqq\;\phi_{B}(K{+}1)-\phi_{B}(K)
=\frac{2\,[\,B-2K-1\,]}{B(B-1)}\;>\;0\quad\text{for }K<\frac{B-1}{2}.
\]
Moreover, $\Delta(K)=0$ at $K=\frac{B-1}{2}$ and $\Delta(K)<0$ for $K>\frac{B-1}{2}$, so $\phi_B$ increases up to the center and then decreases (unimodal).

Therefore, on the symmetric acceptance band
\[
\mathcal{A}_\alpha=\{\,K: k_{\mathrm L}\le K\le B-k_{\mathrm L}\,\},
\]
the minimum flip occurs at the endpoints $K=k_{\mathrm L}$ or $K=B-k_{\mathrm L}$, and both give the same value by symmetry. 
And so, the \textbf{empirical flip cutoff} at level $\alpha$ is
\begin{equation}
\label{eq:talpha}
t_\alpha(B)\;\coloneqq\;\frac{2\,k_{\mathrm L}\,(B-k_{\mathrm L})}{B(B-1)}.
\end{equation}
Equivalently, similar to Equation~\ref{app:eq:vote}, in vote‑fraction form with $v_{\mathrm L}=k_{\mathrm L}/B$,
\[
t_\alpha(B)\;=\;\frac{2B}{B-1}\,v_{\mathrm L}\,(1-v_{\mathrm L}).
\]
We declare the MIA decision for sample $\vx$ \emph{indistinguishable from a coin flip at level $\alpha$} if
\[
\widehat{\mathrm{flip}}_{\eta,B}(\vx)\;\ge\;t_\alpha(B).
\]
Because the binomial is discrete, equal‑tail tests are slightly conservative. 
We follow the convention ``reject if $p<\alpha$'' and ``fail to reject if $p\ge\alpha$,'' so boundary points with $p\text{-value}=\alpha$ lie inside the acceptance region. 
This yields the monotone flip rule $\widehat{\mathrm{flip}}_{\eta,B}(\vx)\ge t_\alpha(B)$.

For the experiments in Section~\ref{sec:instability} and Appendix~\ref{app:sec:instability:flip:results}, we set $\alpha{=}0.05$.
For the $302$M model we have $B{=}127$ target replicas and for the $140$M model we have $B{=}125$ target replicas, respectively:
\begin{itemize}[leftmargin=0.75cm]
\item $B{=}127$: $k_{\mathrm L}{=}52$ (so the acceptance band is $K\in[52,\,75]$) and
\[
t_{0.05}(127)\;=\;\frac{2\cdot 52\cdot 75}{127\cdot 126}\;\approx\;0.487.
\]
\item $B{=}125$: $k_{\mathrm L}{=}52$ (acceptance band $K\in[52,\,73]$) and
\[
t_{0.05}(125)\;=\;\frac{2\cdot 52\cdot 73}{125\cdot 124}\;\approx\;0.490.
\]
\end{itemize}

(For $B{=}125$, our threshold is more conservative in part because the discrete CDF lands at $k_{\mathrm{L}}{=}52$, but a normal continuity-corrected approximation puts it closer to $51.6$.) 
\subsubsection{Extended results on flip rate}\label{app:sec:instability:flip:results}

We provide results for two model architectures: $140$M and $302$M. 
We use the same training dataset size for both: overall $2N{=}2^{20}$, so models are trained on $N{=}2^{19}{=}524{,}288{\approx}500$K samples each. 
Note that for both architectures, this training dataset size is significantly smaller than what is Chinchilla optimal (${\approx}7$M for the $140$M model and ${\approx}15.1$M for the $302$M model).
As a result, we expect attack success to be higher (as measured by $\rocauc$) compared to Chinchilla-optimal trained and attacked models (Sections~\ref{sec:exp1:realistic:compute-optimal} \&~\ref{sec:exp2:limits:trainsize}). 
For each model, we train one set of $128$ reference models (with $0.5$ probability that each sample is included as a member, so that member and non-member classes are balanced).
To measure flip rate, we then train many target models on the \emph{exact same} training dataset (i.e., the member and non-member samples are the same for all targets).
The only difference across models is the random seed, which controls the batch order in which samples are surfaced to the training algorithm. 

We intended to train $128$ target replicas for each architecture; however, some runs  crashed (and we ran out of time to re-run them), so in all we have $125$ targets for the $140$M model and $127$ for the $302$M model.
As noted in Appendix~\ref{app:sec:instability:flip:arbitrary}, the minimum values that we consider indistinguishable from a coin flip for $\widehat{\mathrm{flip}}_{\eta, B}$ are $t_{0.05}(125){\approx}0.490$ for the $140$M model and  $t_{0.05}(127){\approx}0.487$ for the $302$M model.\looseness=-1 

\paragraph{An intuition for per-sample flip rate.} 
Flip rate captures a sample $\vx$'s membership inference instability, computed across a set of target models where the only difference is the random seed that controls batch order. 
For a given $\vx$, it captures how much cross-decision disagreement there is---how much the MIA decisions for $\vx$ flip between both classes for equally plausible targets $r\!\sim\!\mu$. 

\begin{figure}[t]
  \centering
\begin{subfigure}[t]{0.4\textwidth}
\vspace*{0cm}
    \includegraphics[width=\textwidth]{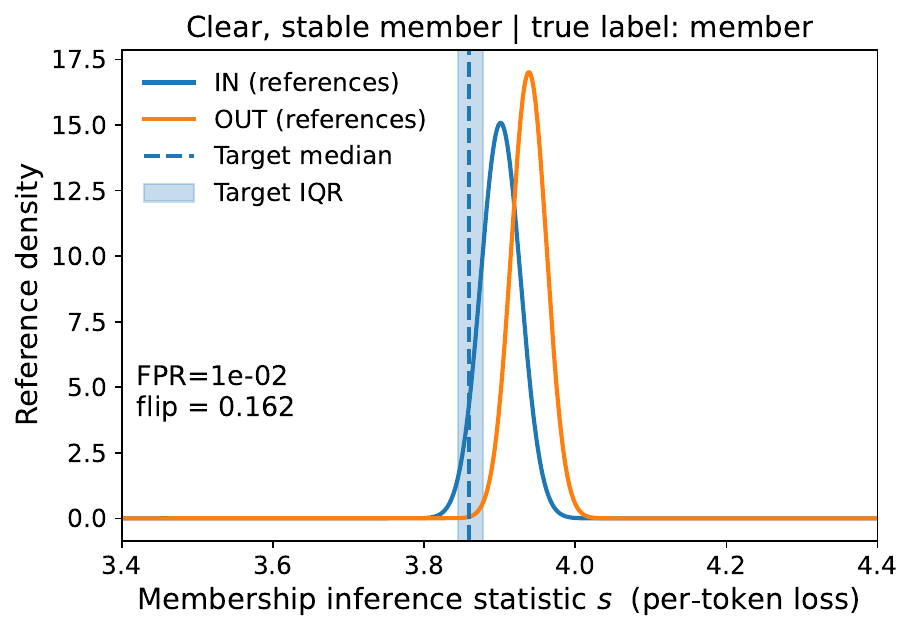}
  \caption{Clear, stable member}
  \label{fig:140M-clear-stable-member}
\end{subfigure}
\hspace{1cm}
\begin{subfigure}[t]{0.4\textwidth}
\vspace*{0cm}
    \includegraphics[width=\textwidth]{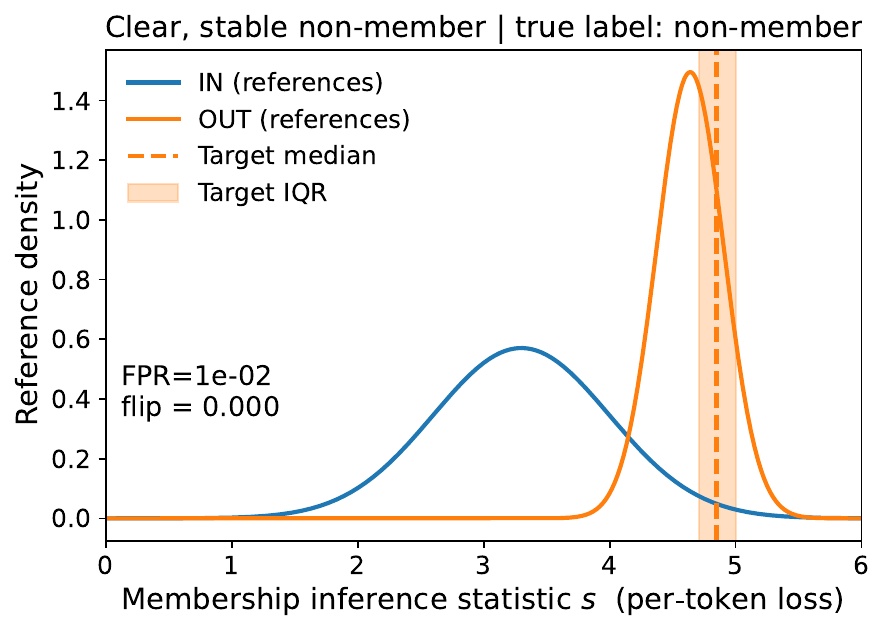}
  \caption{Clear, stable non-member}
  \label{fig:140M-clear-stable-nonmember}
\end{subfigure}%

\begin{subfigure}[t]{0.4\textwidth}
\vspace*{0cm}
    \includegraphics[width=\textwidth]{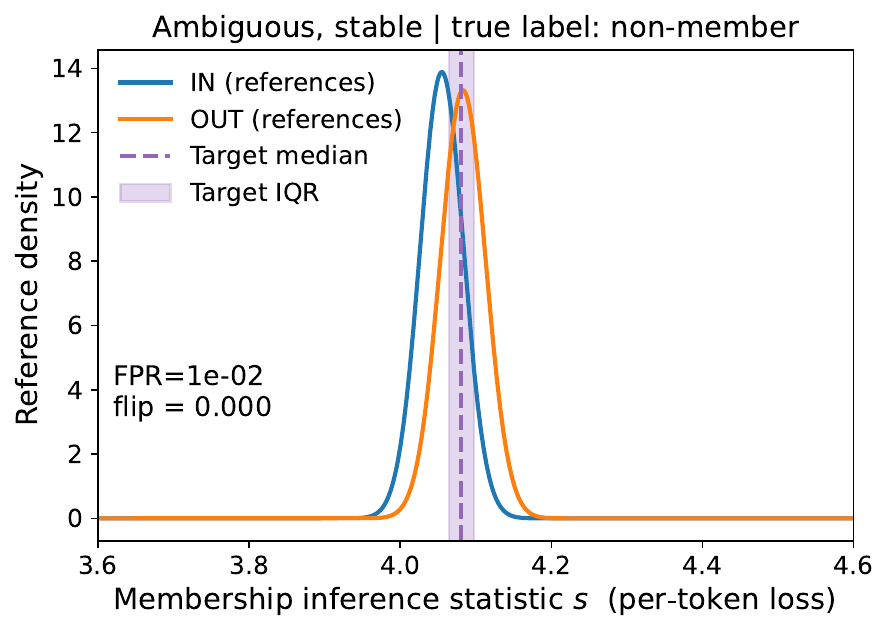}
  \caption{Ambiguous, stable sample}
  \label{fig:140M-ambig-stable-sample}
\end{subfigure}
\hspace{1cm}
\begin{subfigure}[t]{0.4\textwidth}
\vspace*{0cm}
    \includegraphics[width=\textwidth]{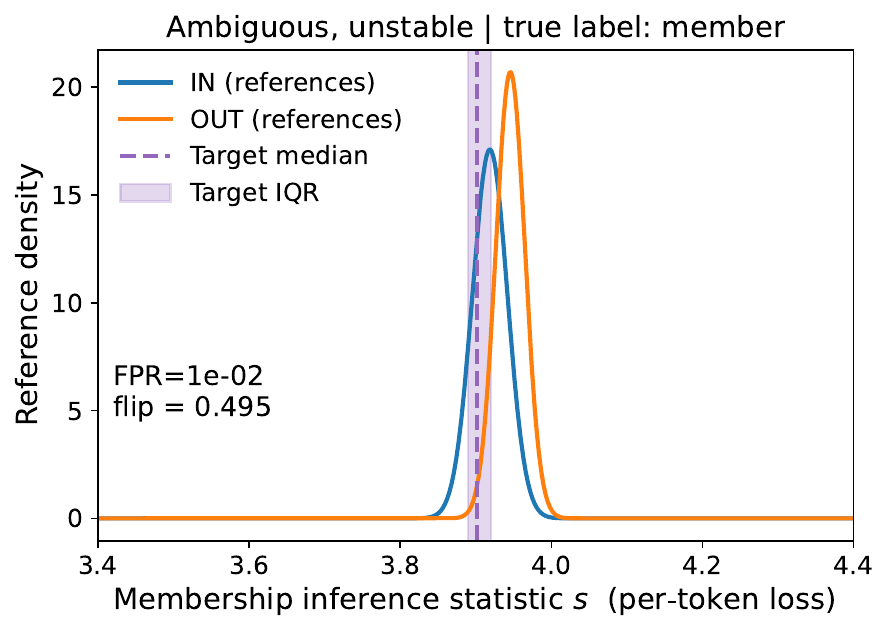}
  \caption{Ambiguous, unstable sample}
  \label{fig:140M-ambig-unstable-sample}
\end{subfigure}%
\caption{\textbf{Different sample ``archetypes'' for the $\bf{140}$M target models.} 
We plot the per-sample $\vx$'s reference distributions (IN and OUT), median target statistic $s$ (and IQR) for $\vx$ across the $125$ targets at $\fpr{=}10^{-2}$ for four different $\vx$: (\textbf{a}) clear, stable member; (\textbf{b}) clear, stable non-member; (\textbf{c}) ambiguous, stable sample; and, (\textbf{d}) ambiguous, unstable sample.
We annotate each plot with the sample's true label and empirical flip rate.  
For this architecture, we also provide snippets for the text of each sample in the main text. 
}
\vspace{-.5cm}
\label{fig:140M-archetypes}
\end{figure}

To give a sense of how this can happen, we provide plots at the sample-level that show where target membership observation statistics for a given $\vx$ fall in relation to $\vx$'s IN and OUT reference distributions, $p_\text{IN}(\cdot|\vx)$ and $p_\text{OUT}(\cdot|\vx)$---fitted from the statistics obtained for $\vx$ using the reference sets $\Phi_\text{IN}$ and $\Phi_\text{OUT}$, respectively. 
We plot four ``archetypes'' that capture different patterns in sample-specific MIA decision behavior, in relation to reference distributions:
(a) clear, stable member; (b) clear, stable non-member; (c) ambiguous, stable sample; and, (d) ambiguous, unstable sample.
We identify these archetypes at $\fpr{=}10^{-2}$. 
In Figure~\ref{fig:140M-archetypes}, we plot all four archetypes for the $140$M architecture.
In Figure~\ref{fig:302M-archetypes}, we plot archetypes (b)--(d), as we are unable to find clear, stable members at $\fpr{=}10^{-2}$.
Even for the $140$M model, we have to relax the flip rate in our search filter to allow for $\widehat{\mathrm{flip}}_{{10^{-2}},125}{\leq}0.2$ to identify a ``stable'' member (when arguably, such a flip rate is not particularly stable). 
We are unable to satisfy this relaxed filter for the $302$M architecture. 

Note that, for both model sizes, the IN and OUT reference distributions overlap considerably for member samples. 
This overlap is a reasonable explanation for MIA decision instability:
if LiRA has difficulty between establishing signal between members and non-members, then this will understandably impact the reliability of MIA decisions.
Across targets trained on different random seeds, this can also manifest as the MIA decision flipping from one class to the other. 
In contrast, we identify cases for non-member samples where there is clear separation of IN and OUT reference distributions (Figures~\ref{fig:140M-clear-stable-nonmember}~\&~\ref{fig:302M-clear-stable-nonmember}). 

For the $140$M archetypes, we include short snippets of the text for each sample:

\begin{figure}[t]
  \centering
\begin{subfigure}[t]{0.4\textwidth}
\vspace*{0cm}
    \includegraphics[width=\textwidth]{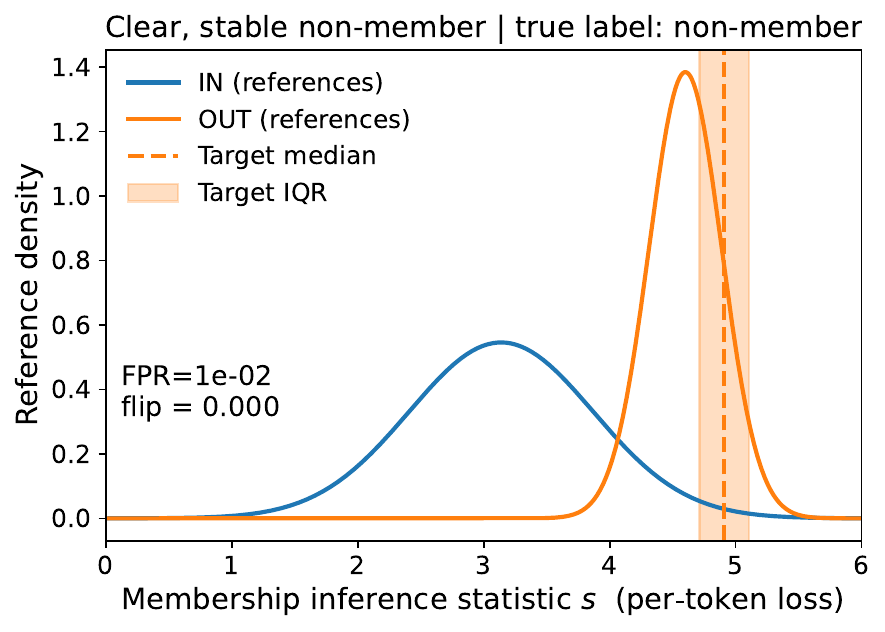}
  \caption{Clear, stable non-member}
  \label{fig:302M-clear-stable-nonmember}
\end{subfigure}

\begin{subfigure}[t]{0.4\textwidth}
\vspace*{0cm}
    \includegraphics[width=\textwidth]{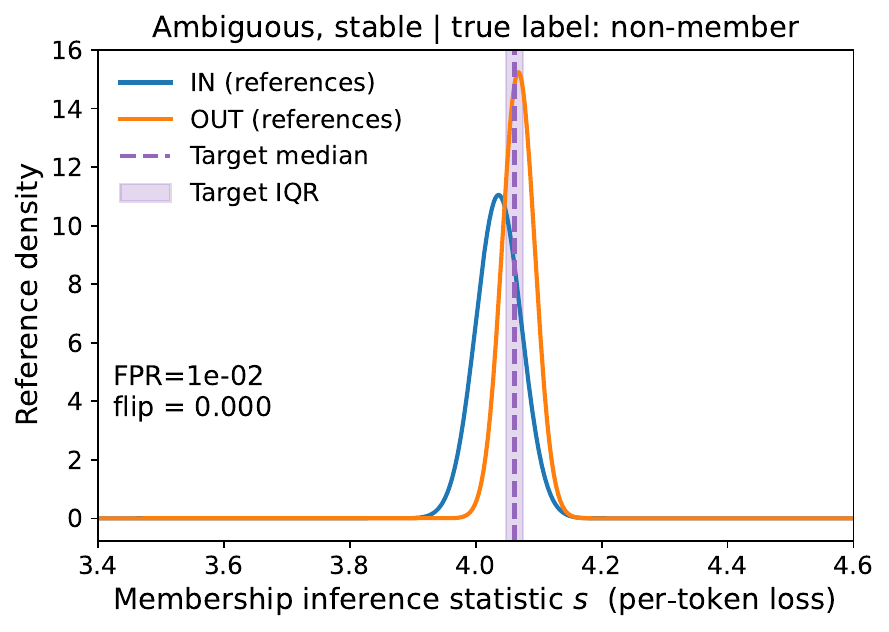}
  \caption{Ambiguous, stable sample}
  \label{fig:302M-ambig-stable-sample}
\end{subfigure}
\hspace{1cm}
\begin{subfigure}[t]{0.4\textwidth}
\vspace*{0cm}
    \includegraphics[width=\textwidth]{figures/cooper/stability/302M/archetype-302M-Ambiguous-unstable-831304.pdf}
  \caption{Ambiguous, unstable sample}
  \label{fig:302M-ambig-unstable-sample}
\end{subfigure}%
\caption{\textbf{Different sample ``archetypes'' for the $\bf{302}$M target models.} 
We plot the per-sample $\vx$'s reference distributions (IN and OUT), median target statistic $s$ (and IQR) for $\vx$ across the $127$ targets at $\fpr{=}10^{-2}$ for four different $\vx$: 
(\textbf{a}) clear, stable non-member;  (\textbf{b}) ambiguous, stable sample; and, (\textbf{c}) ambiguous, unstable sample.
We annotate each plot with the sample's true label and empirical flip rate.  
We are unable to identify a clear, moderately stable ($\widehat{\mathrm{flip}}_{{{10^{-2}}},127}\leq0.2$) member sample.\looseness=-1
}
\vspace{-.5cm}
\label{fig:302M-archetypes}
\end{figure}

\emph{$140$\emph{M}: clear, stable member.}
``Whether it’s your first time looking for a Personal Trainer and you are just starting out, or you are a veteran who has been around a long-time, SINA Fitness can help you reach your fitness goals.
Our Trainers are experienced, friendly and very energetic. We will help you set your fitness and lifestyle goals and most importantly help you achieve them. \ldots''

\emph{$140$\emph{M}: clear, stable non-member.}
``Ä Release Notes: AI War is an entirely unique large-scale RTS with aspects of TBS, tower defense, and grand strategy. It features single or cooperative play with as many as 8 humans against a pair of powerful, intelligent AIs. These AIs are driven by an AI Progress stat that players contribute to through aggressive actions such as taking control of planets and destroying key units, forcing tough decisions regarding which targets are worth capturing or destroying. \ldots''\looseness=-1

\emph{$140$\emph{M}: ambiguous, stable sample.}
``The Gingrich commentary came hours after The Wall Street Journal reported that Mueller empaneled a grand jury.

“The Mueller threat has probably been the most deadly, he has the power of the law, he has the ability to indict people, the ability to negotiate and let some people off if they’ll testify against other people,” said Gingrich, also a Fox News contributor. \ldots''

\emph{$140$\emph{M}: ambiguous, unstable sample.}
``Winner of the Junior Australian Open 2015 Tereza Mihalikova (20), who is going to participate at EMPIRE Women’s Indoor 2019 tournament, had spent the entire 2018 season under the guidance of tennis coach Martin Hromec. At the end of the year, the well-known fitness coach Jozef Ivanko, strengthened the team. Ivanko worked with Top 10 players in WTA ranking already. \ldots''

\begin{figure}[t]
  \centering
\begin{subfigure}[t]{0.4\textwidth}
\vspace*{0cm}
    \includegraphics[trim={0cm 0cm 0cm 0.65cm},clip,width=\textwidth]{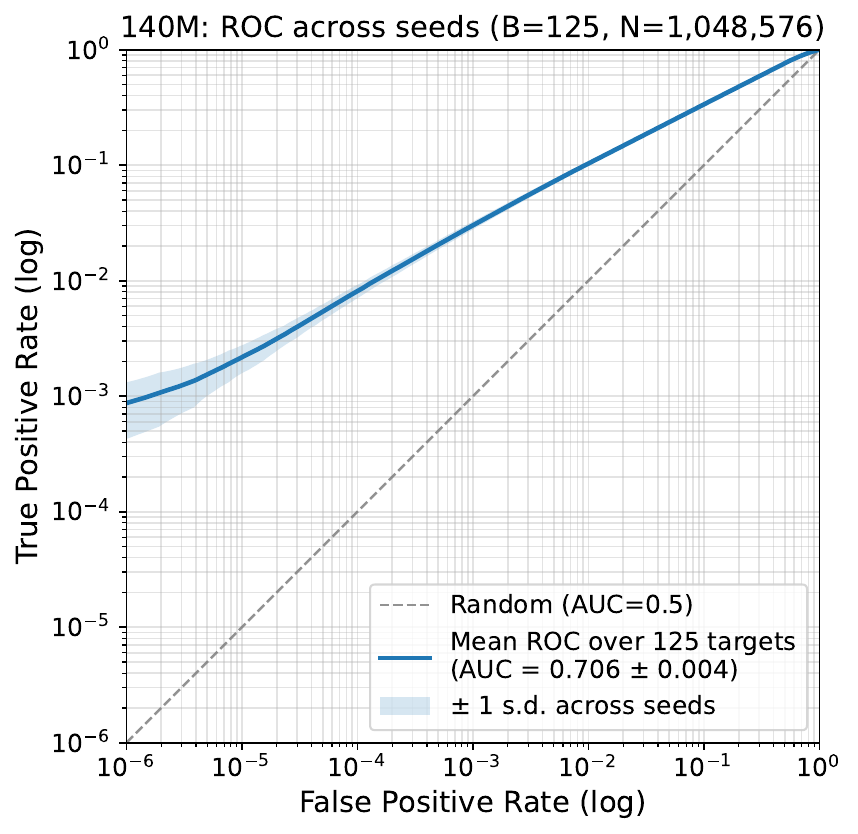}
  \caption{$140$M}
  \label{fig:140M-rocauc-instability}
\end{subfigure}
\hspace{1cm}
\begin{subfigure}[t]{0.4\textwidth}
\vspace*{0cm}
    \includegraphics[trim={0cm 0cm 0cm 0.65cm},clip,width=\textwidth]{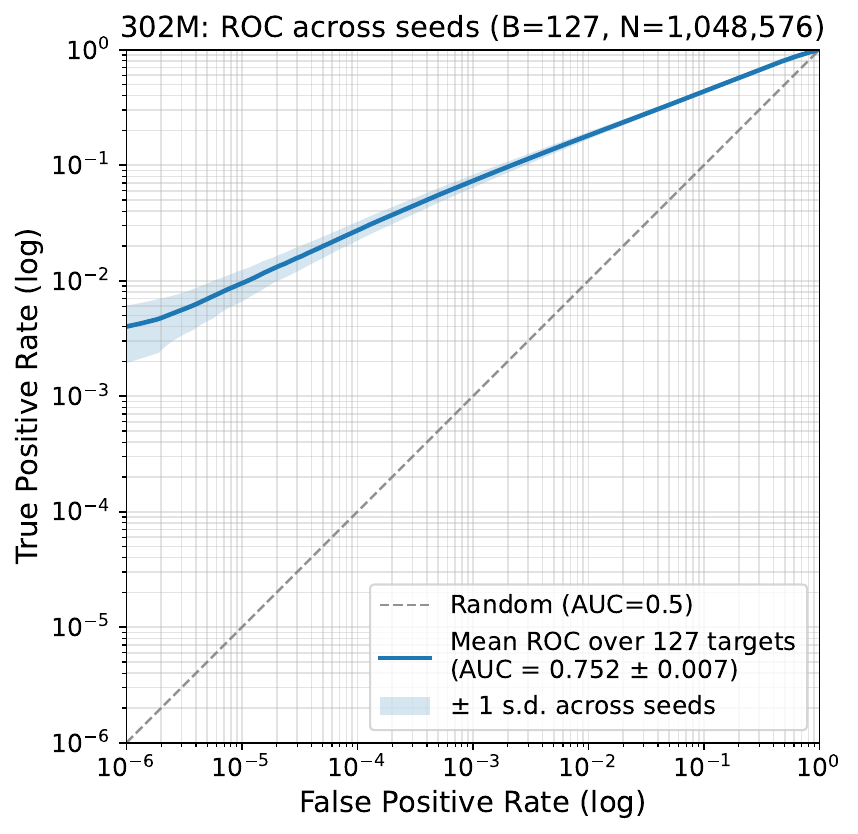}
  \caption{$302$M}
  \label{fig:302M-rocauc-instability}
\end{subfigure}
\caption{\textbf{Averaged $\mathrm{\bf{ROC}}$ curves and $\mathrm{\bf{AUC}}$ across targets.} 
We plot the mean $\roc$ across targets ($B{=}125$ for the $140$M architecture; $B{=}127$ for the $302$M architecture), and $\pm 1$ STD across seeds. 
Both models are trained on substantially fewer samples than is Chinchilla optimal (${\approx}500$K, compared to ${\approx}7$M and ${\approx}15.1$M, respectively). 
Mean $\rocauc$ is higher than for Chinchilla-optimal models (as in Section~\ref{sec:exp2:limits:trainsize}).
For the $140$M model (\textbf{left}), $\rocauc{=}0.706\pm0.004$; for the $302$M model (\textbf{right}), $\rocauc{=}0.752\pm0.007$. 
These are not typical attack $\roc$ curves, as they average over results for multiple targets.
In the standard MIA threat model, the attacker only has access to a single target.
These plots give a sense of the stability of overall attack performance, as computed over equally plausible targets $r\!\sim\!\mu$ where the only difference in targets is the seed controlling batch order.
}
\label{fig:instability-roc}
\end{figure}

\paragraph{Aggregate attack performance for MIA flip-rate experiments.}
While our main focus here is to measure per-sample instability, as a point of comparison, we also include measurements about attack averages. 
For both model sizes, we include mean (cross-seed) $\rocauc$ metrics and curves (Figure~\ref{fig:instability-roc}) and associated tables that show average  (cross-seed) accuracy and error rates by class at fixed $\fpr$ (Tables~\ref{tab:acc_metrics_by_fpr}~\&~\ref{tab:acc_metrics_by_fpr_302M}). 

We again emphasize that the models we attack in these experiments were \emph{not} trained on the Chinchilla-optimal number of tokens (${\approx}7$M and ${\approx}15.1$M samples, for $140$M and $302$M models, respectively; see Section~\ref{sec:exp1:realistic:compute-optimal} \& Appendix~\ref{app:sec:optimal}).
Both sets of experiments involved training models on only ${\approx}500$K samples.
As a result, we expect (and do observe) attack performance (in terms of $\rocauc$) to be higher than in the Chinchilla-optimal setting (Section~\ref{sec:exp2:limits:trainsize} \& Appendix~\ref{app:sec:limits}). 
For the $140$M architecture, we observe average $\auc{=}0.706\pm0.004$ across the $125$ targets (Figure~\ref{fig:140M-rocauc-instability}); 
for the $302$M architecture, we observe average $\auc{=}0.752\pm0.007$ across the $127$ targets (Figure~\ref{fig:302M-rocauc-instability}).
In both cases, $\auc$ is stable across targets (as indicated by the low standard deviation).
This same pattern of stability in overall attack metrics is also clear in Tables~\ref{tab:acc_metrics_by_fpr} and~\ref{tab:acc_metrics_by_fpr_302M}:
accuracy and error exhibit low standard deviation, with respect to these rates being aggregated across all samples (conditioned by class) and averaged across targets trained with different seeds. 

\begin{figure}[t]
  \centering
\begin{minipage}{0.045\textwidth}
    \textbf{\;}
    \vspace{.3cm}
\end{minipage}%
\hfill
\begin{minipage}{0.45\textwidth}
    \centering
    \textbf{\hspace{.2cm}$\mathrm{\bf{TPR}}$ at fixed $\mathrm{\bf{FPR}}$}
    \vspace{.3cm}
\end{minipage}%
\hfill
\begin{minipage}{0.45\textwidth}
    \centering
    \textbf{\hspace{.2cm}Threshold $\bf\tau$ variation}
    \vspace{.3cm}
\end{minipage}%

\begin{subfigure}[t]{0.045\textwidth}
\vspace{-1.65cm}
\hspace{-.3cm}\textbf{140M}
\vspace{.5cm}
\end{subfigure}%
\hfill
\begin{subfigure}[t]{0.45\textwidth}
    \includegraphics[width=\textwidth]{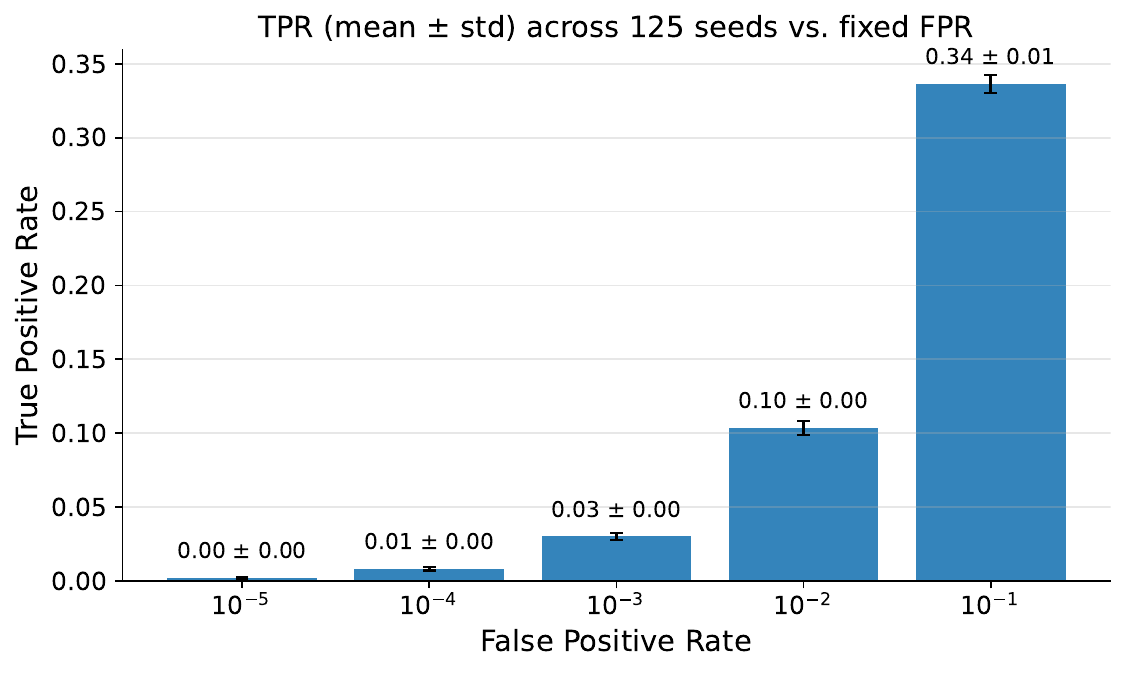}
\end{subfigure}
\hfill
\begin{subfigure}[t]{0.48\textwidth}
    \includegraphics[width=\textwidth]{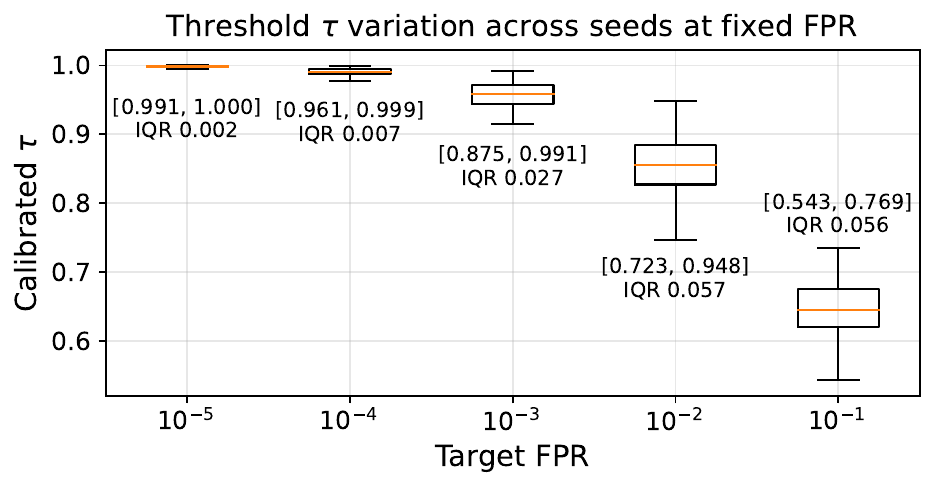}
\end{subfigure}
\begin{subfigure}[t]{0.045\textwidth}
\vspace{-1.65cm}
\hspace{-.3cm}\textbf{302M}
\vspace{.5cm}
\end{subfigure}%
\hfill
\begin{subfigure}[t]{0.45\textwidth}
    \includegraphics[width=\textwidth]{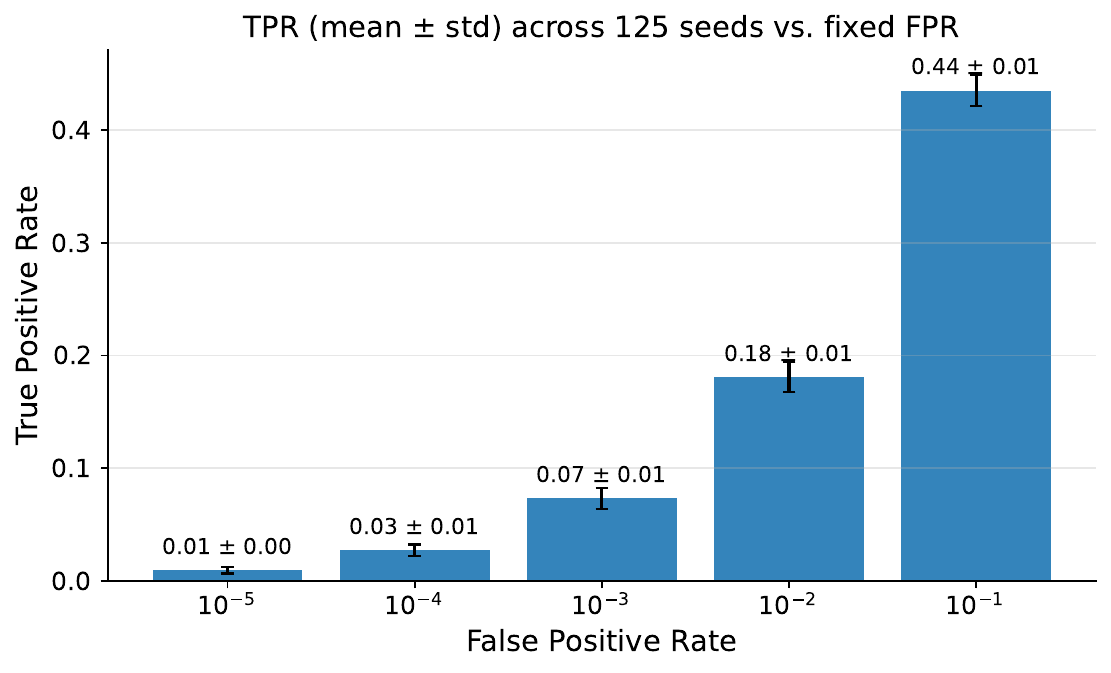}
\end{subfigure}
\hfill
\begin{subfigure}[t]{0.48\textwidth}
    \includegraphics[width=\textwidth]{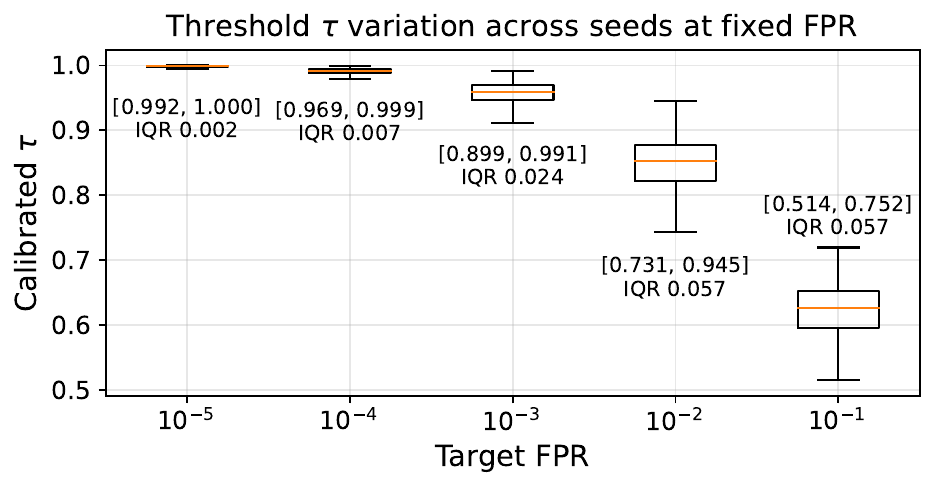}
\end{subfigure}
\caption{\textbf{Comparing attack performance and decision thresholds at fixed $\mathrm{\bf{FPR}}$.} 
Row shows results for model architectures: $140$M and $302$M
The left column shows mean $\pm$ standard deviation of the attack $\tpr$ at different fixed $\fpr$, computed across $B$ targets. 
The right column shows the decision threshold $\tau$ range (and IQR) at fixed $\fpr$ across $B$ targets, where the decision threshold for each target is calibrated with respect to non-member samples (Section~\ref{sec:rw}~\& Appendix~\ref{app:sec:background}). 
As is also surfaced by $\roc$ curves (Figure~\ref{fig:instability-roc}) at very low fixed $\fpr$, the $\tpr$ is also low. 
Here, we also show how this naturally results in a very high decision threshold $\tau$, which consequently exhibits low variability.
As we increase $\fpr$, $\tpr$ also increases and remains stable, with respect to low standard deviation.
However, the underlying decision rules for the targets start to vary considerably;
the underlying targets can have very different $\tau$, which (as we address in this appendix and in Section~\ref{sec:instability}) can result in very different/disagreeing membership decisions for the same sample $\vx$.
}
\label{fig:error-v-tau}
\end{figure}

For an alternate view of these results, we also include direct comparisons of attack performance (as measured by average $\tpr$ $\pm$ standard deviation at fixed $\fpr$) and variability in the underlying decision rule (with respect to threshold $\tau$) across targets.
In Figure~\ref{fig:error-v-tau}, we provide these comparisons for both the $140$M and $302$M model sizes.
Of course, as is also surfaced by $\roc$ curves (Figure~\ref{fig:instability-roc}) at very low fixed $\fpr$, the $\tpr$ is also low. 
Here, we also show how this naturally results in a very high decision threshold $\tau$, which also exhibits low variability.
As we increase $\fpr$, $\tpr$ also increases and remains stable, with respect to low standard deviation.
However, the underlying decision rules for the targets can vary considerably;
the underlying targets can have very different $\tau$.
This result is consistent with prior work on model and predictive multiplicity (Appendix~\ref{app:sec:instability:flip:multiplicity}): 
models with similar overall accuracy can have very different underlying decision rules.
As we address further below in this appendix and in Section~\ref{sec:instability}, even when overall accuracy is similar, the different decision rules can result in very different/disagreeing membership MIA decisions for the same samples.\looseness=-1

\begin{table}[b]
\centering
\caption{\textbf{$\mathbf{140}$M-parameter model error rate metrics.}
We report accuracy-related metrics as a function of fixed $\fpr$. 
Entries are rates (not percentages), as elsewhere in this paper.
We report mean $\pm$ STD where applicable.
Since we fix $\fpr$, there is no STD to report.
Since $1 - \fpr = \mathrm{TNR}$, similarly, there is no STD to report. 
$\mathrm{ACC} = \frac{\mathrm{TP} + \mathrm{TN}}{N}$, with  $2N{=}1{,}048{,}576$. 
Typically reported $\log$-scale $\fpr$ rows are highlighted in gray.}
\vspace{0.2cm}
\label{tab:acc_metrics_by_fpr}
\small
\begin{tabular}{l r r r r r}
\toprule
\textbf{$\mathrm{\mathbf{FPR}}$} &
\shortstack{$\mathrm{\mathbf{ACC}}$\\\textbf{All}} &
\shortstack{$\mathrm{\mathbf{FNR}}$\\\textbf{Members}} &
\shortstack{$\mathrm{\mathbf{FPR}}$\\\textbf{Non-members}} &
\shortstack{$\mathrm{\mathbf{TNR}}$\\\textbf{Non-members}} &
\shortstack{$\mathrm{\mathbf{TPR}}$\\\textbf{Members}} \\
\midrule
\rowcolor{gray!15}
$10^{-5}$ & $0.501 \pm 0.0$ & $0.998 \pm 0.001$ & $0.0$ & $1.0$ & $0.002 \pm 0.001$ \\
\rowcolor{gray!15}
$10^{-4}$ & $0.504 \pm 0.001$ & $0.992 \pm 0.001$ & $0.0 $ & $1.0 $ & $0.008 \pm 0.001$ \\
\rowcolor{gray!15}
$10^{-3}$ & $0.515 \pm 0.001$ & $0.97 \pm 0.002$ & $0.001 $ & $0.999 $ & $0.03 \pm 0.002$ \\
\rowcolor{gray!15}
$10^{-2}$ & $0.547 \pm 0.002$ & $0.896 \pm 0.005$ & $0.01 $ & $0.99 $ & $0.104 \pm 0.005$ \\
$0.02$ & $0.564 \pm 0.003$ & $0.852 \pm 0.005$ & $0.02 $ & $0.98 $ & $0.148 \pm 0.005$ \\
$0.05$ & $0.593 \pm 0.003$ & $0.764 \pm 0.006$ & $0.05 $ & $0.95 $ & $0.236 \pm 0.006$ \\
\rowcolor{gray!15}
$10^{-1}$ & $0.618 \pm 0.003$ & $0.664 \pm 0.006$ & $0.1 $ & $0.9 $ & $0.336 \pm 0.006$ \\
$0.2$ & $0.64 \pm 0.003$ & $0.52 \pm 0.006$ & $0.2 $ & $0.8 $ & $0.48 \pm 0.006$ \\
$0.5$ & $0.633 \pm 0.002$ & $0.234 \pm 0.004$ & $0.5 $ & $0.5 $ & $0.766 \pm 0.004$ \\
$0.75$ & $0.582 \pm 0.001$ & $0.085 \pm 0.002$ & $0.75 $ & $0.25 $ & $0.915 \pm 0.002$ \\
\rowcolor{gray!15}
$10^{0}$ & $0.5 \pm 0.0$ & $0.0 \pm 0.0$ & $1.0 $ & $0.0 $ & $1.0 \pm 0.0$ \\
\bottomrule
\end{tabular}
\end{table}

\begin{table}[t]
\centering
\caption{\textbf{$\mathbf{302}$M-parameter model error rate metrics.}
We report accuracy-related metrics as a function of fixed $\fpr$. 
Entries are rates (not percentages), as elsewhere in this paper.
We report mean $\pm$ STD where applicable.
Since we fix $\fpr$, there is no STD to report.
Since $1 - \fpr = \mathrm{TNR}$, similarly, there is no STD to report. 
$\mathrm{ACC} = \frac{\mathrm{TP} + \mathrm{TN}}{N}$, with  $2N{=}1{,}048{,}576$. 
Typically reported $\log$-scale $\fpr$ rows are highlighted in gray.}
\vspace{0.2cm}
\label{tab:acc_metrics_by_fpr_302M}
\small
\begin{tabular}{l r r r r r}
\toprule
\textbf{$\mathrm{\mathbf{FPR}}$} &
\shortstack{$\mathrm{\mathbf{ACC}}$\\\textbf{All}} &
\shortstack{$\mathrm{\mathbf{FNR}}$\\\textbf{Members}} &
\shortstack{$\mathrm{\mathbf{FPR}}$\\\textbf{Non-members}} &
\shortstack{$\mathrm{\mathbf{TNR}}$\\\textbf{Non-members}} &
\shortstack{$\mathrm{\mathbf{TPR}}$\\\textbf{Members}} \\
\midrule
\rowcolor{gray!15}
$10^{-5}$ & $0.505 \pm 0.001$ & $0.991 \pm 0.003$ & $0.0$ & $1.0$ & $0.009 \pm 0.003$ \\
\rowcolor{gray!15}
$10^{-4}$ & $0.514 \pm 0.003$ & $0.973 \pm 0.005$ & $0.0$ & $1.0$ & $0.027 \pm 0.005$ \\
\rowcolor{gray!15}
$10^{-3}$ & $0.536 \pm 0.005$ & $0.927 \pm 0.009$ & $0.001$ & $0.999$ & $0.073 \pm 0.009$ \\
\rowcolor{gray!15}
$10^{-2}$ & $0.585 \pm 0.007$ & $0.819 \pm 0.013$ & $0.01$ & $0.99$ & $0.181 \pm 0.013$ \\
0.02 & $0.608 \pm 0.007$ & $0.765 \pm 0.014$ & $0.02$ & $0.98$ & $0.235 \pm 0.014$ \\
0.05 & $0.642 \pm 0.007$ & $0.667 \pm 0.014$ & $0.05$ & $0.95$ & $0.333 \pm 0.014$ \\
\rowcolor{gray!15}
$10^{-1}$ & $0.668 \pm 0.007$ & $0.565 \pm 0.014$ & $0.1$ & $0.9$ & $0.435 \pm 0.014$ \\
0.2 & $0.685 \pm 0.006$ & $0.43 \pm 0.013$ & $0.2$ & $0.8$ & $0.57 \pm 0.013$ \\
0.5 & $0.655 \pm 0.004$ & $0.191 \pm 0.008$ & $0.5$ & $0.5$ & $0.809 \pm 0.008$ \\
0.75 & $0.588 \pm 0.002$ & $0.075 \pm 0.004$ & $0.75$ & $0.25$ & $0.925 \pm 0.004$ \\
\rowcolor{gray!15}
$10^{0}$ & $0.5 \pm 0.0$ & $0.0 \pm 0.0$ & $1.0$ & $0.0$ & $1.0 \pm 0.0$ \\
\bottomrule
\end{tabular}
\end{table}

\begin{figure*}[t!]
\begin{minipage}{0.045\textwidth}
    \textbf{\;}
    \vspace{.3cm}
\end{minipage}%
\hfill
\begin{minipage}{0.312\textwidth}
    \centering
    \textbf{\hspace{.6cm}Flip CCDF}
    \vspace{.3cm}
\end{minipage}%
\hfill
\begin{minipage}{0.312\textwidth}
    \centering
    \textbf{\hspace{.6cm}Flip vs. mean $|$distance$|$}
    \vspace{.3cm}
\end{minipage}%
\hfill
\begin{minipage}{0.312\textwidth}
    \centering
    \textbf{\hspace{.6cm}Flip vs. mean posterior}
    \vspace{.3cm}
\end{minipage}
\captionsetup[subfigure]{justification=centering}
  \centering
\begin{subfigure}[t]{0.045\textwidth}
\vspace{-1.65cm}
\hspace{-.3cm}$\bf{10^{-5}}$
\vspace{1cm}
\end{subfigure}%
\hfill
\begin{subfigure}[t]{0.312\textwidth}
\centering
    \includegraphics[width=1.\linewidth]{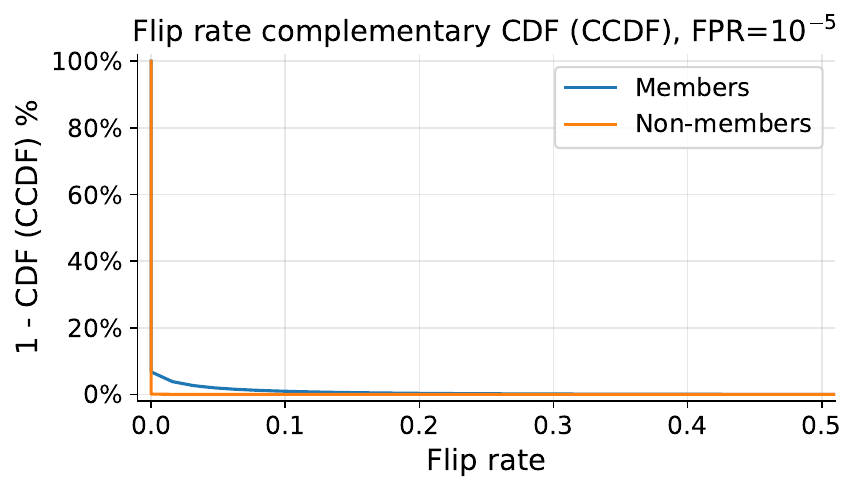}
\end{subfigure}%
\hfill
\begin{subfigure}[t]{0.312\textwidth}
\centering
    \includegraphics[width=1.\linewidth]{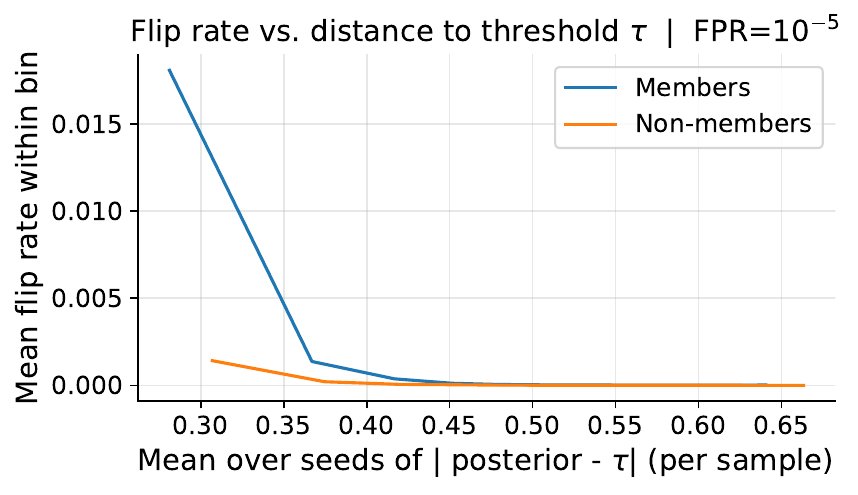}
\end{subfigure}%
\hfill
\begin{subfigure}[t]{0.312\textwidth}
\centering
     \includegraphics[width=1.\linewidth]{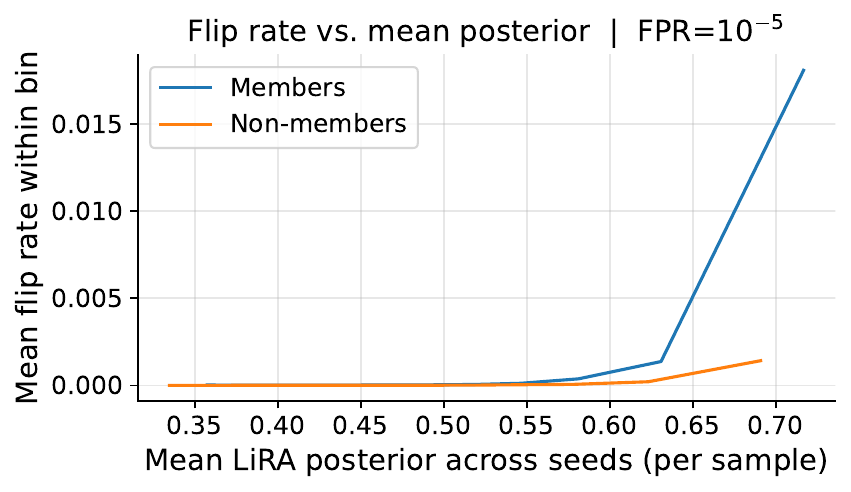}
\end{subfigure}\\%
\begin{subfigure}[t]{0.045\textwidth}
\vspace{-1.65cm}
\hspace{-.3cm}$\bf{10^{-4}}$
\vspace{1cm}
\end{subfigure}%
\hfill
\captionsetup[subfigure]{justification=centering}
  \centering
\begin{subfigure}[t]{0.312\textwidth}
\centering
    \includegraphics[width=1.\linewidth]{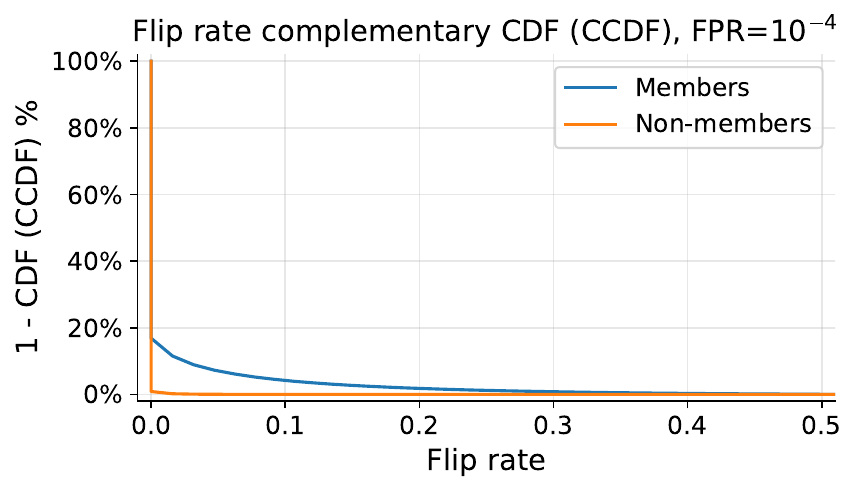}
\end{subfigure}%
\hfill
\begin{subfigure}[t]{0.312\textwidth}
\centering
    \includegraphics[width=1.\linewidth]{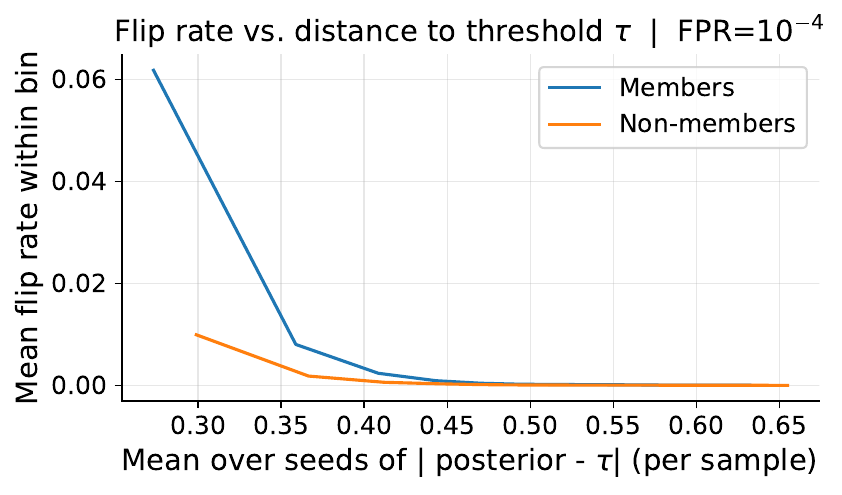}
\end{subfigure}
\hfill
\begin{subfigure}[t]{0.312\textwidth}
\centering
         \includegraphics[width=1.\linewidth]{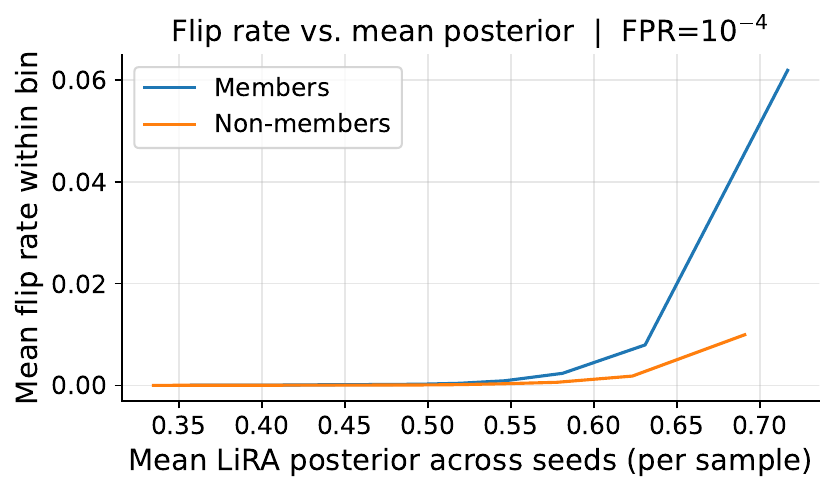}
\end{subfigure}\\
\begin{subfigure}[t]{0.045\textwidth}
\vspace{-1.65cm}
\hspace{-.3cm}$\bf{10^{-3}}$
\vspace{1cm}
\end{subfigure}%
\hfill
\begin{subfigure}[t]{0.312\textwidth}
\centering
    \includegraphics[width=1.\linewidth]{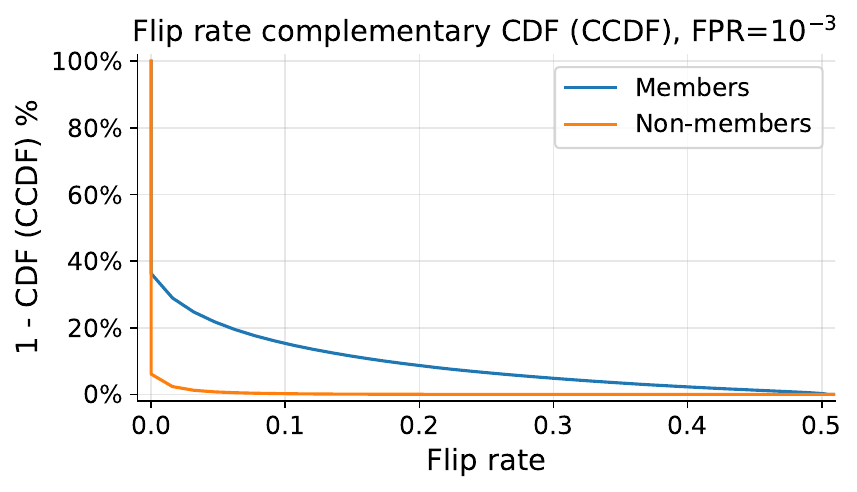}
\end{subfigure}%
\hfill
\begin{subfigure}[t]{0.312\textwidth}
\centering
    \includegraphics[width=1.\linewidth]{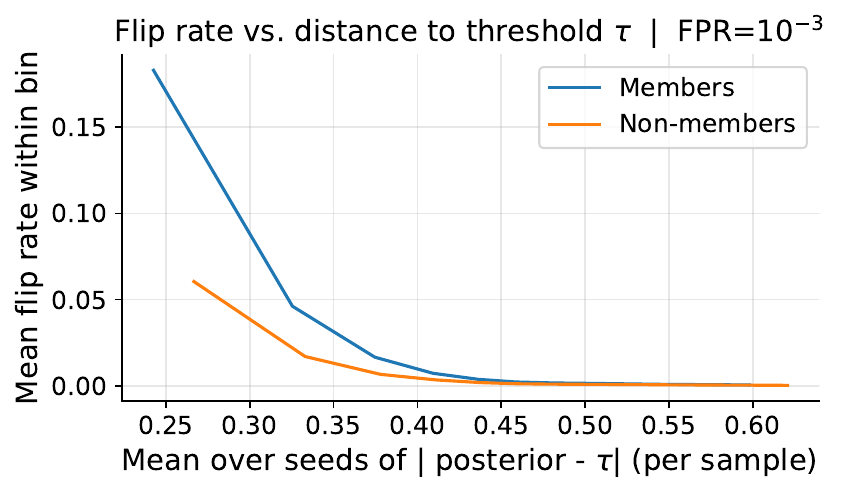}
\end{subfigure}
\hfill
\begin{subfigure}[t]{0.312\textwidth}
\centering
         \includegraphics[width=1.\linewidth]{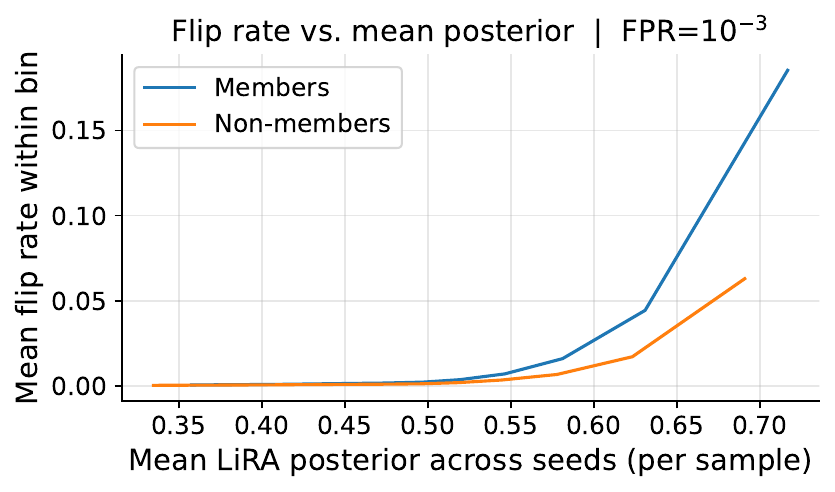}
\end{subfigure}\\%
\begin{subfigure}[t]{0.045\textwidth}
\vspace{-1.65cm}
\hspace{-.3cm}$\bf{10^{-2}}$
\vspace{1cm}
\end{subfigure}%
\hfill
\begin{subfigure}[t]{0.312\textwidth}
\centering
    \includegraphics[width=1.\linewidth]{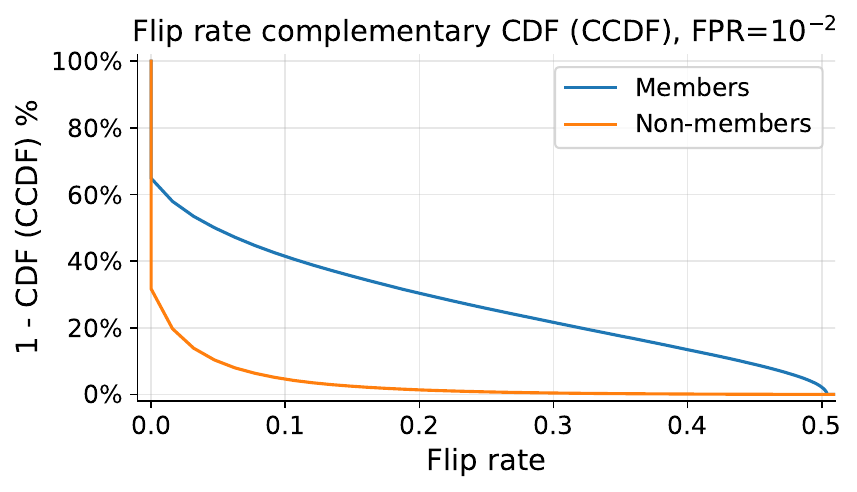}
\end{subfigure}%
\hfill
\begin{subfigure}[t]{0.312\textwidth}
\centering
    \includegraphics[width=1.\linewidth]{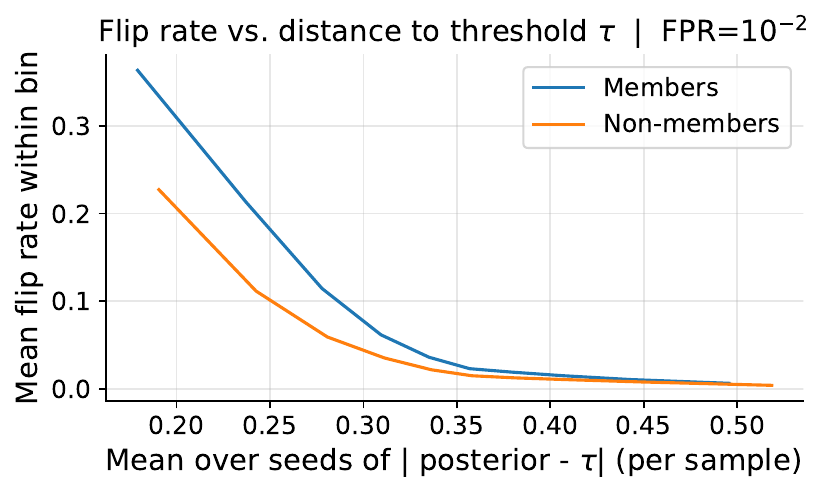}
\end{subfigure}
\hfill
\begin{subfigure}[t]{0.312\textwidth}
\centering
         \includegraphics[width=1.\linewidth]{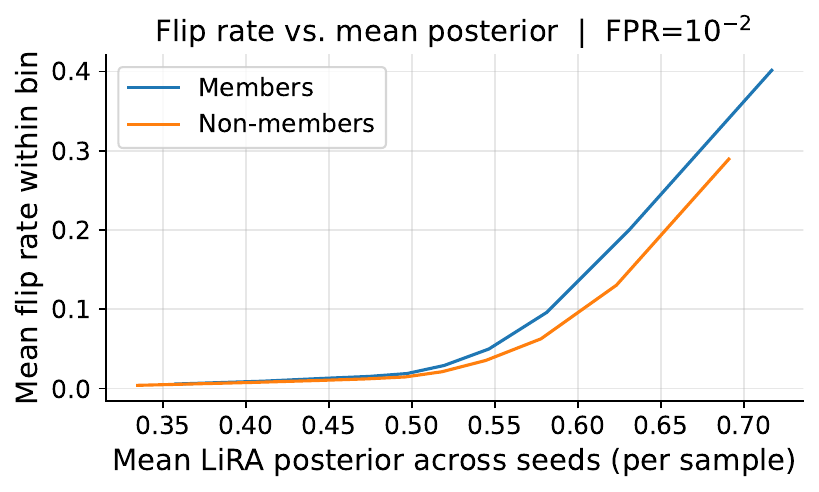}
\end{subfigure}\\%
\begin{subfigure}[t]{0.045\textwidth}
\vspace{-1.65cm}
\hspace{-.3cm}$\bf{10^{-1}}$
\vspace{1cm}
\end{subfigure}%
\hfill
\begin{subfigure}[t]{0.312\textwidth}
\centering
    \includegraphics[width=1.\linewidth]{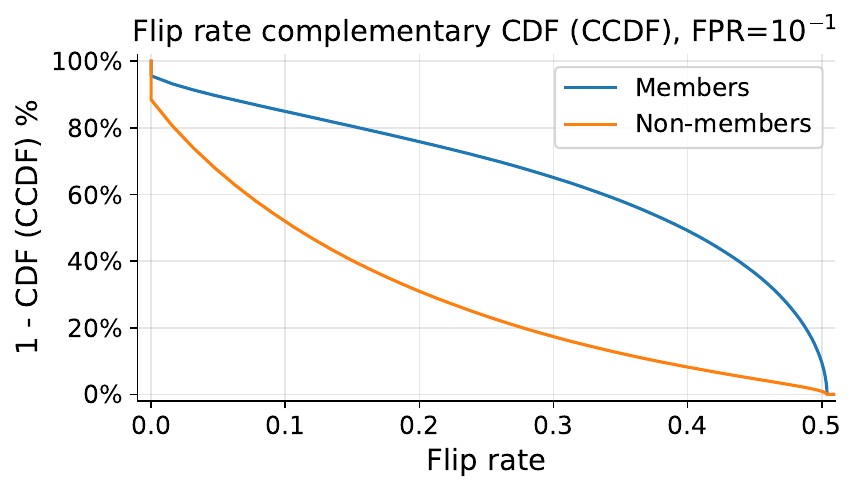}
\end{subfigure}%
\hfill
\begin{subfigure}[t]{0.312\textwidth}
\centering
    \includegraphics[width=1.\linewidth]{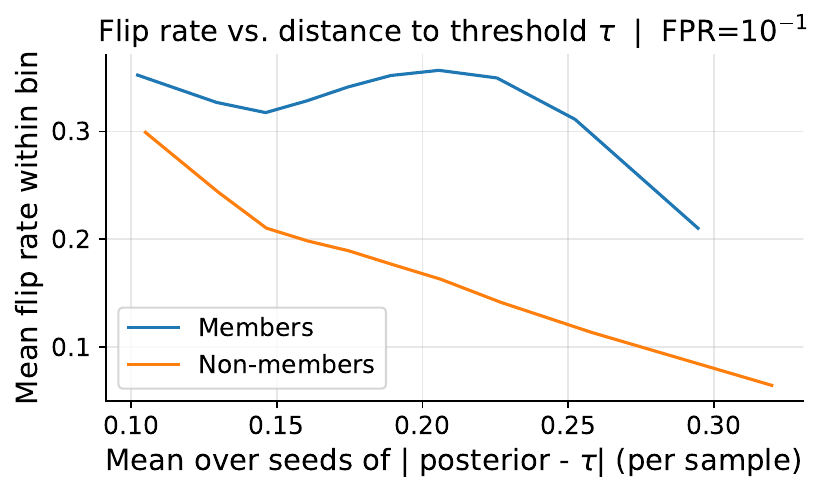}
\end{subfigure}
\hfill
\begin{subfigure}[t]{0.312\textwidth}
\centering
         \includegraphics[width=1.\linewidth]{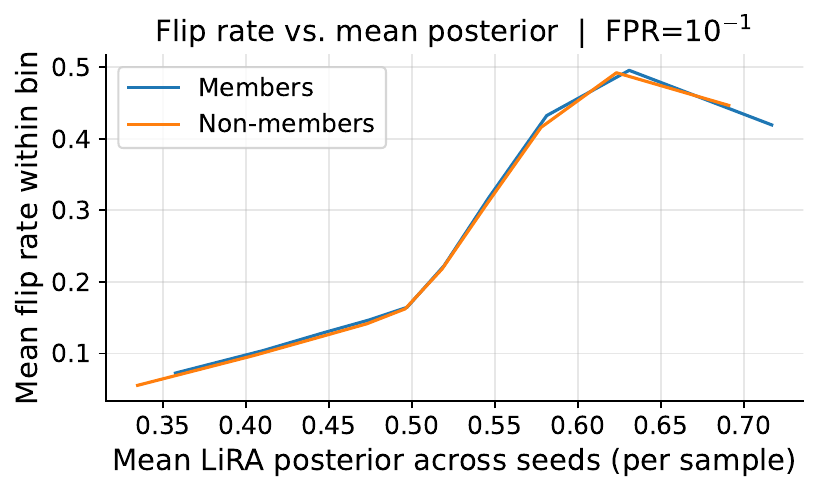}
\end{subfigure}\\
\caption{\textbf{Detailed flip rate results for the $\bf{140}$M model.} 
For the $140$M model, the number of target replicas $B{=}125$. 
(\textbf{rows}) For different fixed $\fpr \in \{10^{-5}, 10^{-4}, 10^{-3}, 10^{-2}, 10^{-1}\}$, we provide (\textbf{columns}) three different views on flip rate across member and non-member samples. 
(\textbf{left}) We plot the empirical complementary CDF (CCDF) for flip rate, conditioned on membership status.
Higher curves indicate more instability.
As $\fpr$ increases, the differences in flip rate across classes become more pronounced.
While flip rate is minimal at low $\fpr$, it is substantial---particularly for members---at higher $\fpr$.
For example, at $\fpr{=}10^{-1}$ approximately $50\%$ of member samples exhibit $\widehat{\mathrm{flip}}_{10^{-1},125}{\geq}0.4$, compared to approximately $10\%$ of non-members. 
(\textbf{middle}) We plot flip rate as a function of the mean of the magnitude of the distance (per sample) from the posterior to the decision threshold $\tau$.
Further to the left means closer to $\tau$.
This shows instability in terms of the distance to the threshold (regardless of direction of that distance). 
For higher $\fpr$, the flip rate for members is much higher than for non-members for the same mean absolute distance to $\tau$.
(\textbf{right}) We plot flip rate as a function of the mean posterior.
For higher $\fpr$, the member and non-member flip rates are more similar as a function of the mean LiRA posterior. 
For the last two columns, we use quantile (i.e., equal-count) bucketing on the $x$-axis so that each plotted point is based on (essentially) the same number of samples. 
That way, points on the curves are directly comparable. 
}
\label{fig:140-lowlevel-flip}
\end{figure*}

\begin{figure*}[t]
\begin{minipage}{0.045\textwidth}
    \textbf{\;}
    \vspace{.3cm}
\end{minipage}%
\hfill
\begin{minipage}{0.312\textwidth}
    \centering
    \textbf{\hspace{.6cm}Flip CCDF}
    \vspace{.3cm}
\end{minipage}%
\hfill
\begin{minipage}{0.312\textwidth}
    \centering
     \textbf{\hspace{.6cm}Flip vs. mean $|$distance$|$}
    \vspace{.3cm}
\end{minipage}%
\hfill
\begin{minipage}{0.312\textwidth}
    \centering
    \textbf{\hspace{.6cm}Flip vs. mean posterior}
    \vspace{.3cm}
\end{minipage}
\captionsetup[subfigure]{justification=centering}
  \centering
\begin{subfigure}[t]{0.045\textwidth}
\vspace{-1.65cm}
\hspace{-.3cm}$\bf{10^{-5}}$
\vspace{1cm}
\end{subfigure}%
\hfill
\begin{subfigure}[t]{0.312\textwidth}
\centering
    \includegraphics[width=1.\linewidth]{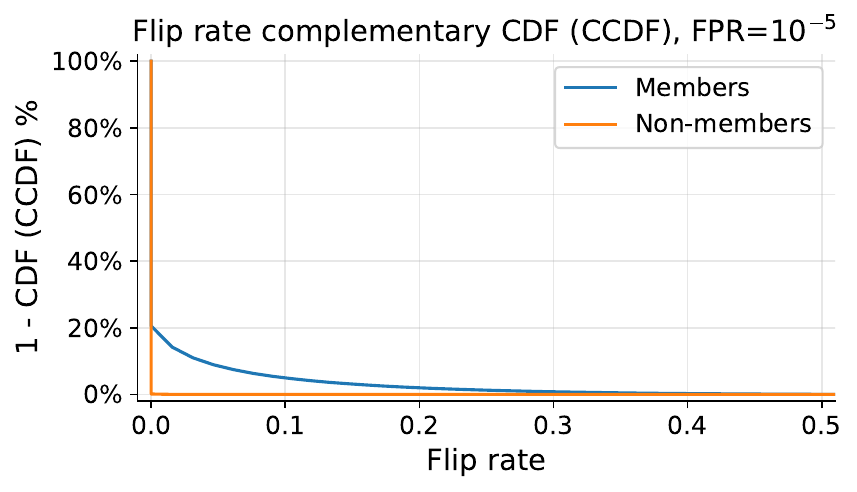}
\end{subfigure}%
\hfill
\begin{subfigure}[t]{0.312\textwidth}
\centering
    \includegraphics[width=1.\linewidth]{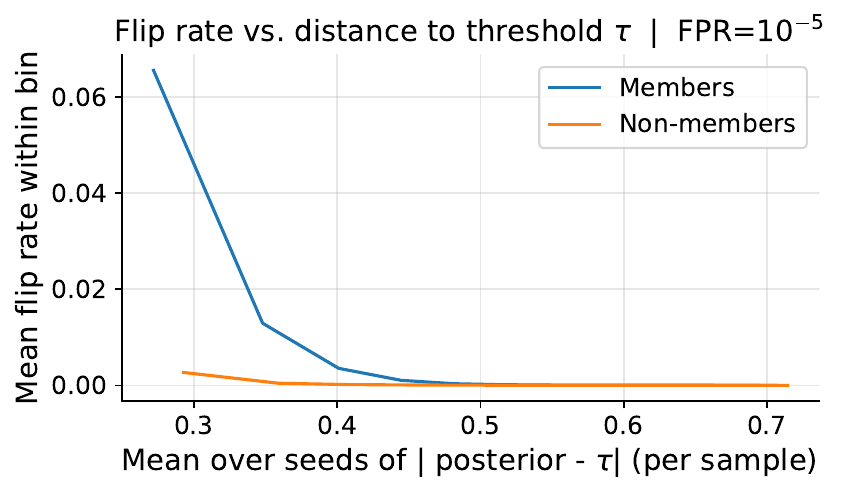}
\end{subfigure}%
\hfill
\begin{subfigure}[t]{0.312\textwidth}
\centering
     \includegraphics[width=1.\linewidth]{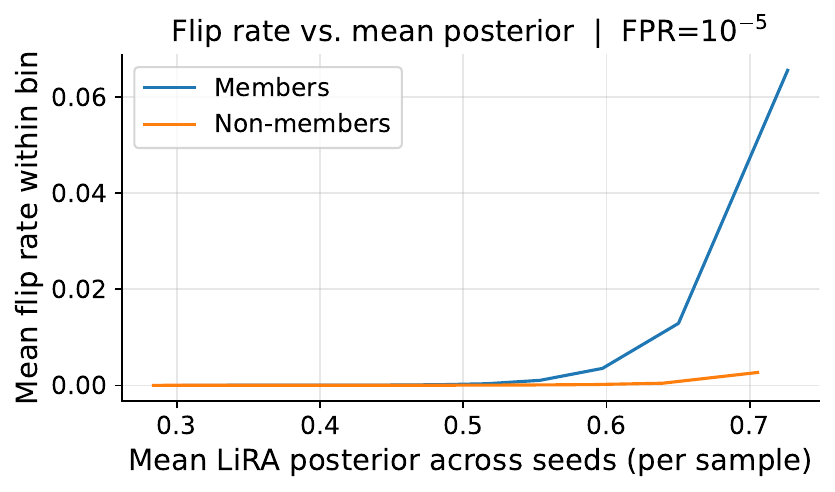}
\end{subfigure}\\%
\begin{subfigure}[t]{0.045\textwidth}
\vspace{-1.65cm}
\hspace{-.3cm}$\bf{10^{-4}}$
\vspace{1cm}
\end{subfigure}%
\hfill
\captionsetup[subfigure]{justification=centering}
  \centering
\begin{subfigure}[t]{0.312\textwidth}
\centering
    \includegraphics[width=1.\linewidth]{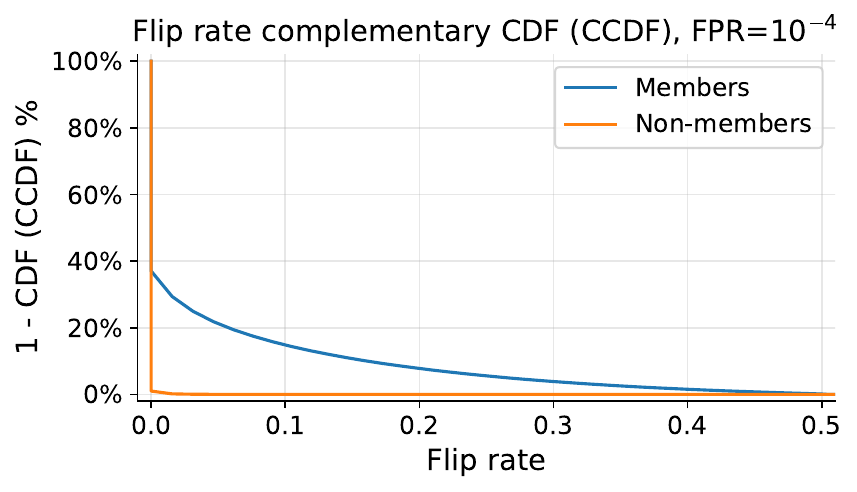}
\end{subfigure}%
\hfill
\begin{subfigure}[t]{0.312\textwidth}
\centering
    \includegraphics[width=1.\linewidth]{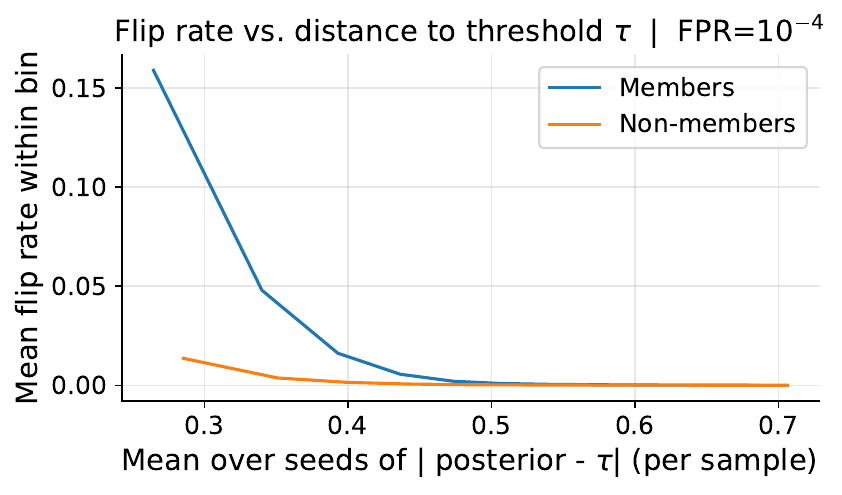}
\end{subfigure}
\hfill
\begin{subfigure}[t]{0.312\textwidth}
\centering
         \includegraphics[width=1.\linewidth]{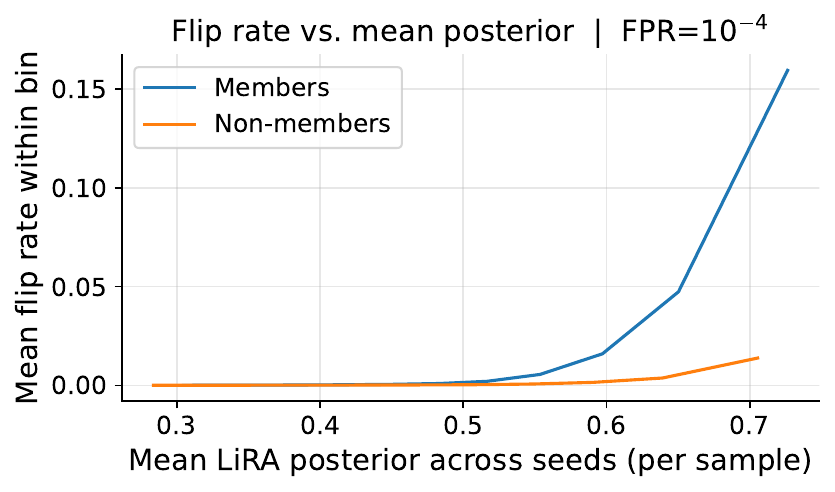}
\end{subfigure}\\
\begin{subfigure}[t]{0.045\textwidth}
\vspace{-1.65cm}
\hspace{-.3cm}$\bf{10^{-3}}$
\vspace{1cm}
\end{subfigure}%
\hfill
\begin{subfigure}[t]{0.312\textwidth}
\centering
    \includegraphics[width=1.\linewidth]{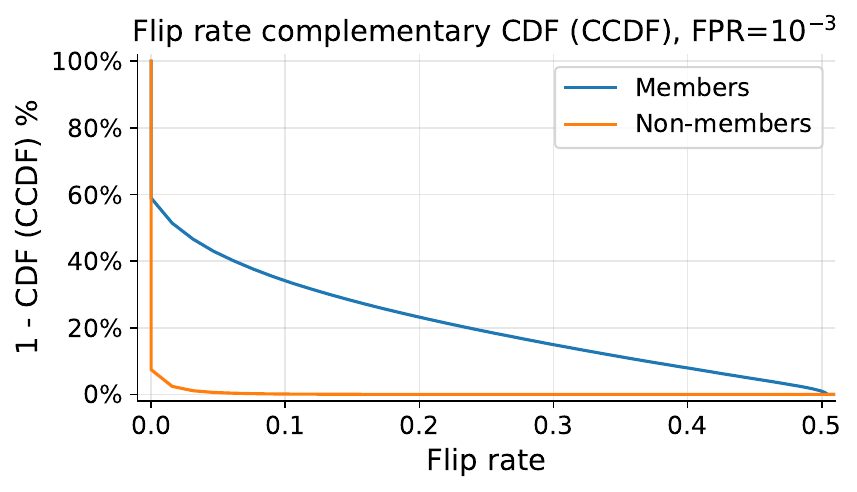}
\end{subfigure}%
\hfill
\begin{subfigure}[t]{0.312\textwidth}
\centering
    \includegraphics[width=1.\linewidth]{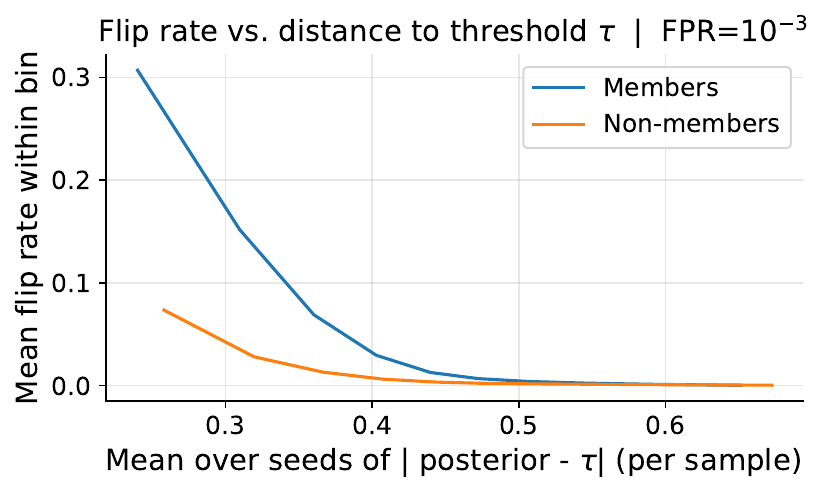}
\end{subfigure}
\hfill
\begin{subfigure}[t]{0.312\textwidth}
\centering
         \includegraphics[width=1.\linewidth]{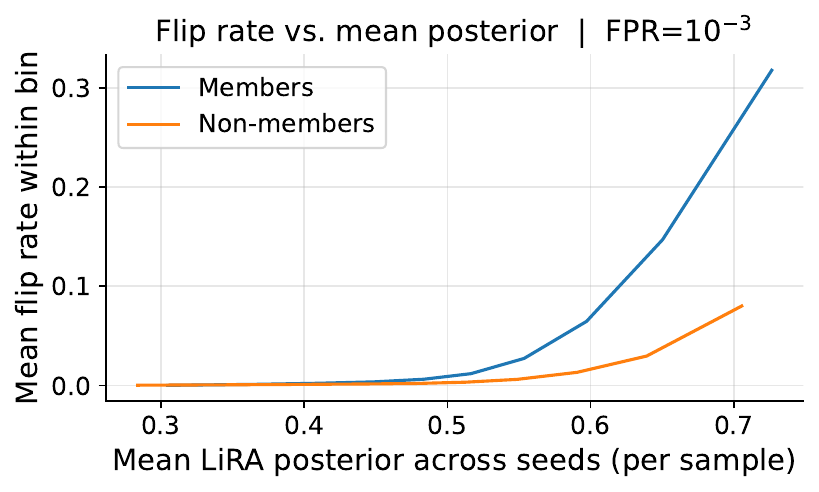}
\end{subfigure}\\%
\begin{subfigure}[t]{0.045\textwidth}
\vspace{-1.65cm}
\hspace{-.3cm}$\bf{10^{-2}}$
\vspace{1cm}
\end{subfigure}%
\hfill
\begin{subfigure}[t]{0.312\textwidth}
\centering
    \includegraphics[width=1.\linewidth]{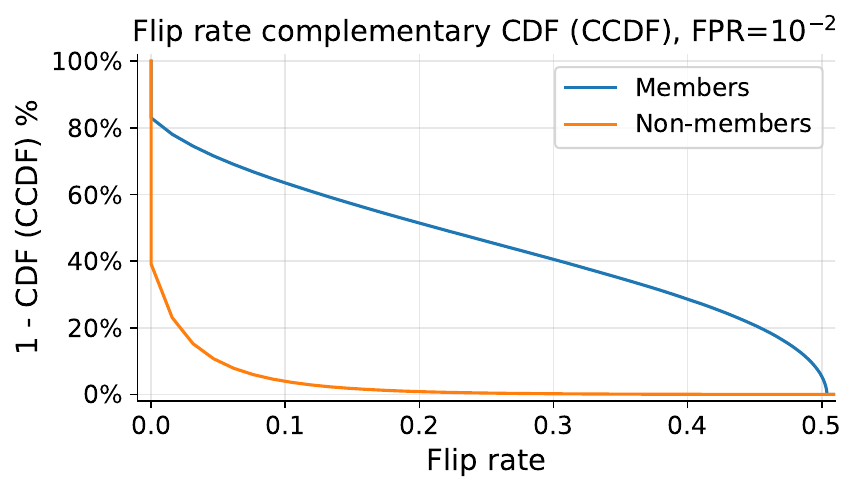}
\end{subfigure}%
\hfill
\begin{subfigure}[t]{0.312\textwidth}
\centering
    \includegraphics[width=1.\linewidth]{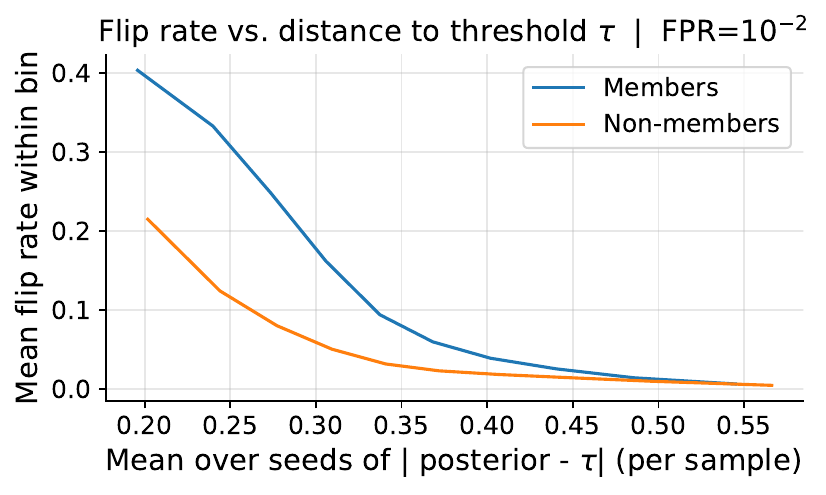}
\end{subfigure}
\hfill
\begin{subfigure}[t]{0.312\textwidth}
\centering
         \includegraphics[width=1.\linewidth]{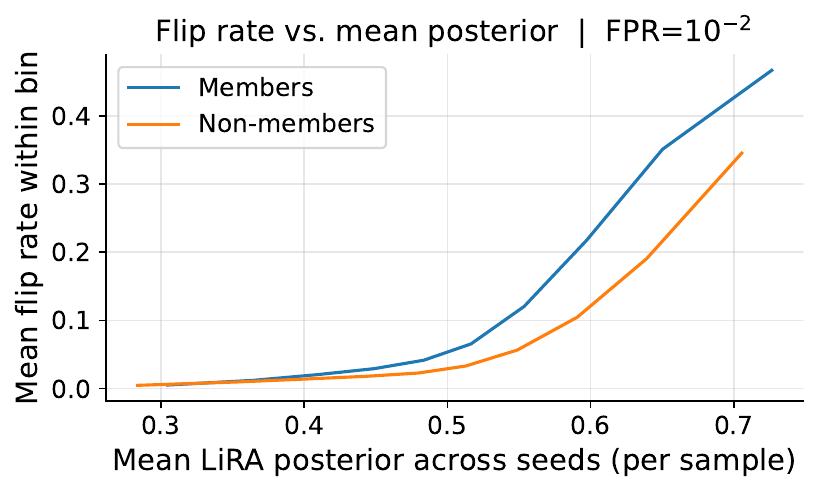}
\end{subfigure}\\%
\begin{subfigure}[t]{0.045\textwidth}
\vspace{-1.65cm}
\hspace{-.3cm}$\bf{10^{-1}}$
\vspace{1cm}
\end{subfigure}%
\hfill
\begin{subfigure}[t]{0.312\textwidth}
\centering
    \includegraphics[width=1.\linewidth]{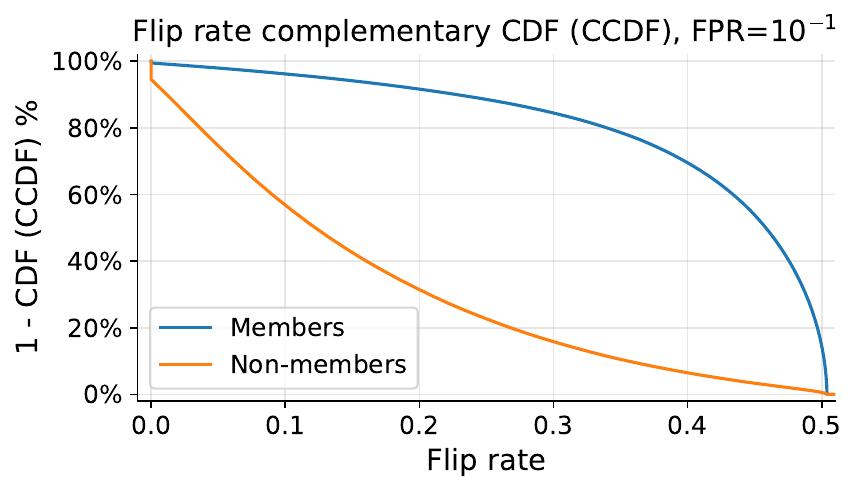}
\end{subfigure}%
\hfill
\begin{subfigure}[t]{0.312\textwidth}
\centering
    \includegraphics[width=1.\linewidth]{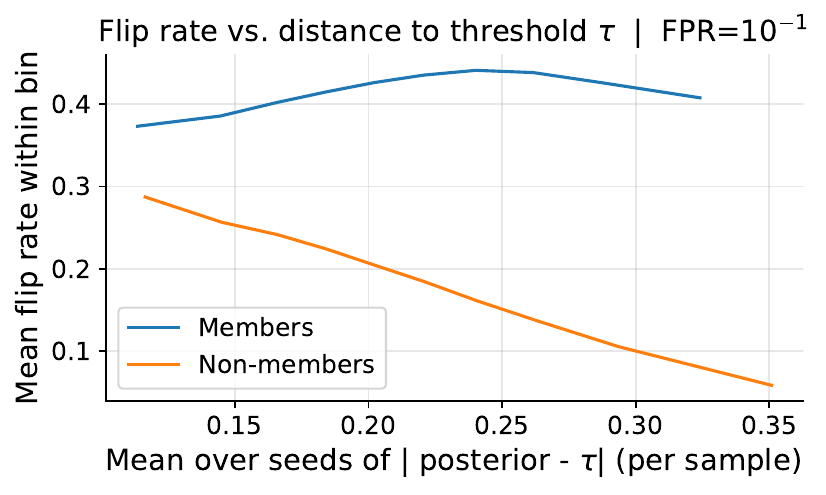}
\end{subfigure}
\hfill
\begin{subfigure}[t]{0.312\textwidth}
\centering
         \includegraphics[width=1.\linewidth]{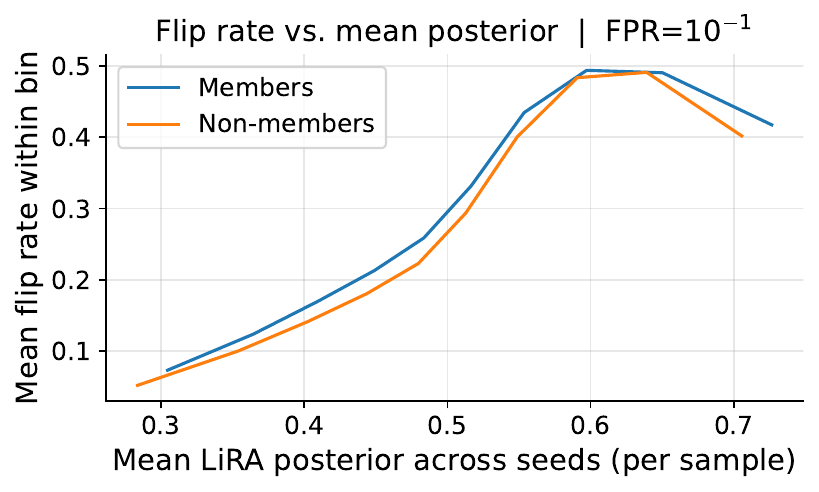}
\end{subfigure}\\
\caption{\textbf{Detailed flip rate results for the $\bf{302}$M model.} 
For the $302$M model, the number of target replicas $B{=}127$. 
(\textbf{rows}) For different fixed $\fpr \in \{10^{-5}, 10^{-4}, 10^{-3}, 10^{-2}, 10^{-1}\}$, we provide (\textbf{columns}) three different views on flip rate across member and non-member samples. 
(\textbf{left}) We plot the empirical complementary CDF (CCDF) for flip rate, conditioned on membership status.
Higher curves indicate more instability.
As $\fpr$ increases, the differences in flip rate across classes become more pronounced.
While flip rate is minimal at low $\fpr$, it is substantial---particularly for members---at higher $\fpr$.
For example, at $\fpr{=}10^{-1}$ approximately $70\%$ of member samples exhibit $\widehat{\mathrm{flip}}_{10^{-1},127}{\geq}0.4$, compared to approximately $10\%$ of non-members. 
(\textbf{middle}) We plot flip rate as a function of the mean of the magnitude of the distance (per sample) from the posterior to the decision threshold $\tau$.
Further to the left means closer to $\tau$.
This shows instability in terms of the distance to the threshold (regardless of direction of that distance). 
For higher $\fpr$, the flip rate for members is much higher than for non-members for the same mean absolute distance to $\tau$.
(\textbf{right}) We plot flip rate as a function of the mean posterior.
For higher $\fpr$, the member and non-member flip rates are more similar as a function of the mean LiRA posterior. 
For the last two columns, we use quantile (i.e., equal-count) bucketing on the $x$-axis so that each plotted point is based on (essentially) the same number of samples. 
That way, points on the curves are directly comparable. 
}
\label{fig:302-lowlevel-flip}
\end{figure*}

\clearpage
\paragraph{Sample flip rate variation at a single fixed $\fpr$.}
We provide three complementary views (by class) at fixed $\fpr$ to characterize per-sample instability: 
\textbf{(left)} complementary CDFs (CCDFs) of flip rate; \textbf{(middle)} flip rate vs.\ mean absolute distance to the calibrated decision boundary; and \textbf{(right)} flip rate vs.\ mean LiRA posterior.
We show these results for the $140$M and $302$M models in Figures~\ref{fig:140-lowlevel-flip} and~\ref{fig:302-lowlevel-flip}, respectively, each for $\fpr \in \{10^{-5},10^{-4},10^{-3},10^{-2},10^{-1}\}$.

For the middle and right columns, the $x$-axis uses equal-count (quantile) bins so that each plotted point aggregates (essentially) the same number of samples; points are therefore directly comparable across the curves. 
Together with Tables~\ref{tab:flip_by_range_140M} and \ref{tab:flip_by_range_302M}, these plots support our main takeaway in Section~\ref{sec:instability}: 
\emph{aggregate} attack metrics (e.g., mean $\tpr$ at fixed $\fpr$, $\rocauc$) can look stable while many \emph{individual} membership decisions are indistinguishable from a coin flip.

We organize observations by theme:

\begin{itemize}[leftmargin=0.75cm]
    \item \textbf{Flip rate rises with $\mathrm{\bf{FPR}}$.} 
    As is clear from the complementary CDFs for flip rate (left column), flip rate rises with $\fpr$.
    This is because, as $\fpr$ grows, the per-seed calibrated threshold $\tau_r(\eta)$ moves down 
    (right-tail quantile of the non‑member distribution), into regions where IN/OUT distribution overlap is more extensive.
    This boundary shift puts $\tau_r(\eta)$ in score regions where many member sample posteriors lie, and also increases the proportion of samples whose seed-specific scores lie near the boundary. 
    As a result, small seed-induced score shifts (as well as across-seed variation in $\tau_r(\eta)$ itself, see Figure~\ref{fig:error-v-tau}) flip the decision more often (i.e., increase per-sample MIA decision disagreement). 
    (More non-members will also be labeled as members, by construction; so, too will members.)
    
    The effect is modest at very low $\fpr$, where $\tau_r(\eta)$ sits deep in the extreme tail. 
    But it becomes more pronounced as we increase $\fpr$ (i.e., as the boundary moves toward denser parts of the score distribution). 
    The CCDFs (left column) and distance plots (middle column) both show this pattern: 
    at $\fpr{=}10^{-1}$, ${\approx}70\%$ of \emph{members} for the $302$M targets have $\widehat{\mathrm{flip}}_{10^{-1},127}{\ge}0.4$ vs.~${\approx}7\%$ of non‑members; 
    for the $140$M targets the corresponding figures for $\widehat{\mathrm{flip}}_{10^{-1},125}{\ge}0.4$ are ${\approx}49\%$ vs.~${\approx}8\%$ (see Table~\ref{tab:flip_by_range_302M} and Figure~\ref{fig:302-lowlevel-flip}, left; Table~\ref{tab:flip_by_range_140M} and Figure~\ref{fig:140-lowlevel-flip}, left).
    
    \item \textbf{Mean absolute distance to the boundary is a direct proxy for instability.} For each sample $\vx$, we define per-seed distance to the decision boundary and a cross-seed measure of closeness to the boundary, regardless of side: 
    \begin{align*}
    \textstyle
        d_r(\vx) = \Lambda_r(\vx) - \tau_r(\eta); \quad |\overline{d}|(\vx) = \tfrac{1}{B}\sum_r|d_r(\vx)|.
    \end{align*}
    For associated plots (middle column), quantile bins with small  $|\overline{d}|$ put the sample close to the decision boundary, resulting in high flip rate (i.e., many MIA decisions disagree). 
    Quantile bins with larger $|\overline{d}|$ are more reliably on one side of the decision boundary, which results in a lower flip rate (i.e., more decisions concentrate). 
    At $\fpr{=}10^{-1}$ for both model sizes, member flip rate is persistently high across a wide range of $|\overline{d}|$---evidence that IN/OUT overlap is substantial in the region where $\tau_r(\eta)$ lies for many seeds (middle column).
    In general, members exhibit markedly higher flip rate than non-members at the same $|\overline{d}|$, with the differences in the two becoming wider at higher $\fpr$.
    
    \item \textbf{Flip vs.\ mean posterior is non‑monotone at high $\fpr$.} 
    For the plots in the right column, we define the mean posterior across targets as 
    \begin{align*}
    \textstyle
        \overline{\Lambda}(\vx)=\tfrac{1}{B}\sum_r \Lambda_r(\vx).
    \end{align*}
    Flip rate increases as $\overline{\Lambda}(\vx)$ approaches $\tau_r(\eta)$ (more seeds straddle the boundary) and then \emph{declines} once $\overline{\Lambda}(\vx)$ is well above the boundary for most seeds (decisions re-concentrate on ``member'').
    This non‑monotonicity is most visible at $\fpr{=}10^{-1}$ (right column).
    (We similarly see this in the middle column, which directly plots distances to the boundary.) 
    
    \item \textbf{Members flip much more than non‑members, and the gap widens with model size.}
    Two forces seem to drive this.
    First, there is \emph{structural asymmetry across classes from calibration} (See box, Appendix~\ref{app:sec:instability:flip:mia}). 
    Thresholds are calibrated on non-members for each seed-specific target (Section~\ref{sec:rw}~\& Appendix~\ref{app:sec:background}), so $\tau_r(\eta)$ tracks seed‑to‑seed shifts in the non‑member distribution by design, and many non‑members remain far below $\tau_r(\eta)$ for modest $\fpr$.
    In contrast, the threshold is not anchored to the member score distribution. 
    Often, their scores straddle the moving decision boundary, so, small seed-induced shifts (either in the score or the decision boundary, see Figure~\ref{fig:error-v-tau}) can flip the decision.
    This effect is more pronounced for higher settings of $\fpr$, which push the threshold into a higher density region of the member score distribution. 
    Second, there is \emph{greater across‑seed score variability for members}. 
    Intuitively, training randomness primarily perturbs samples seen in training (as opposed to those that are not). 
    Empirically, IN/OUT reference distributions for many members overlap substantially, while some non‑members exhibit clearer separation (Figures~\ref{fig:140M-archetypes} \&~\ref{fig:302M-archetypes}). 
    Both of these effects are stronger for the larger model $302$M (compare Table~\ref{tab:flip_by_range_302M} vs.\ Table~\ref{tab:flip_by_range_140M}).
    
    \item \textbf{Effect of model size.} 
    In general, the observations above show that flip rate instability is worse for the larger ($302$M) model.
    Members flip much more than non‑members, and the gap widens with model size. 
    These results are also consistent with model-multiplicity-related results for higher capacity models:
    those with similar aggregate accuracy can exhibit more disagreement at the individual sample level~\citep{cooper2024variance}.
    
\end{itemize}

\begin{figure}[t]
  \centering
\begin{subfigure}[t]{\textwidth}
\centering
\vspace*{0cm}
    \includegraphics[trim={0cm 0cm 0cm 1cm},clip,width=.72\textwidth]{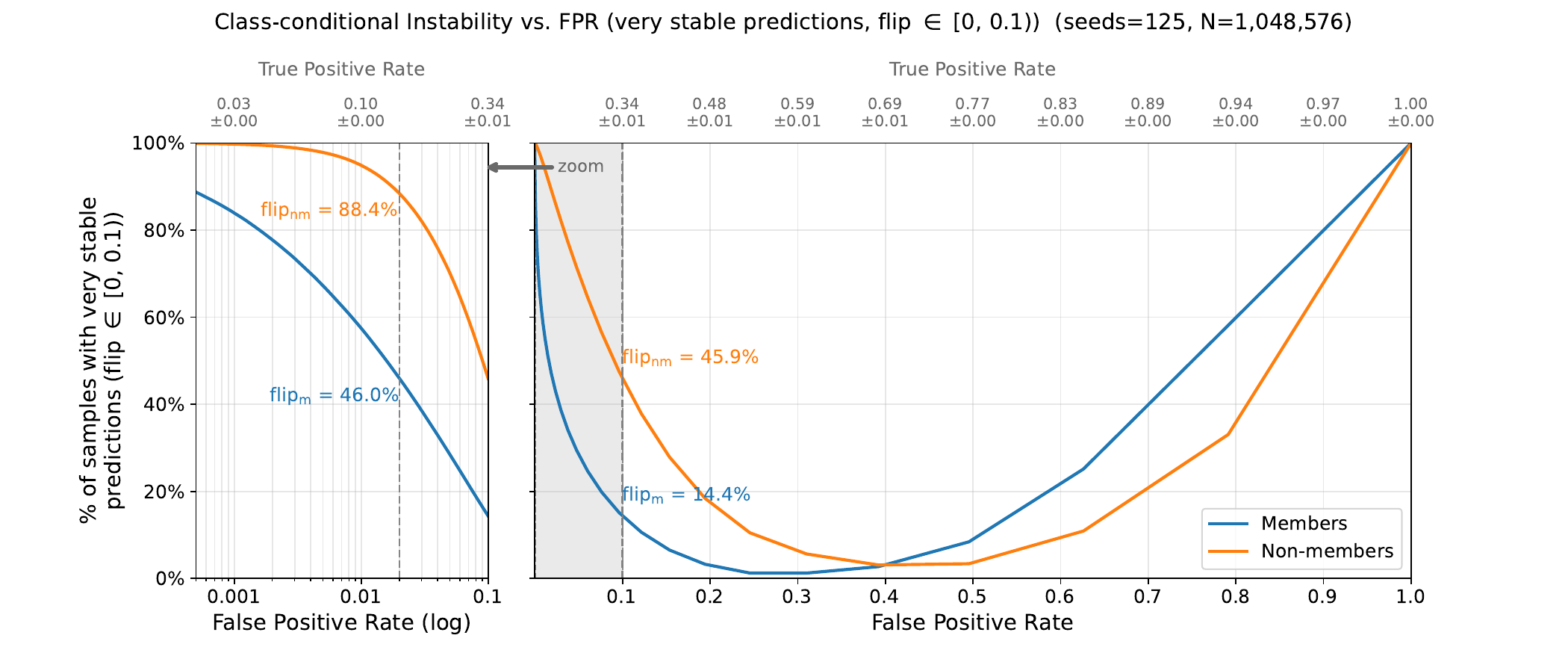}
    \vspace{-.12cm}
  \caption{Very stable MIA decisions: $\widehat{\mathrm{flip}}_{\eta,125} \in [0, 0.1)$}
  \label{fig:140M-stable}
\end{subfigure}

\begin{subfigure}[t]{\textwidth}
\vspace*{0cm}
\centering
    \includegraphics[trim={0cm 0cm 0cm 1cm},clip,width=.72\textwidth]{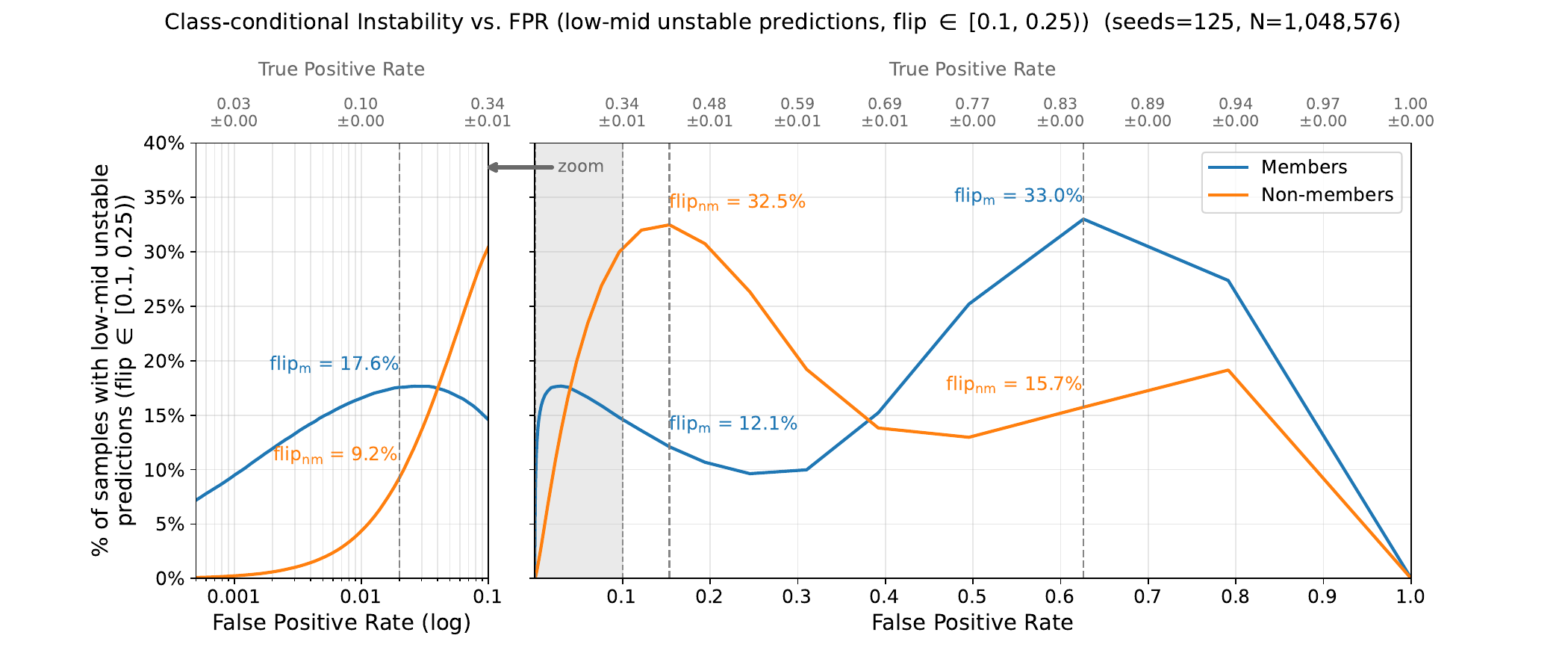}
    \vspace{-.12cm}
  \caption{Low/mid stable MIA decisions: $\widehat{\mathrm{flip}}_{\eta,125} \in [0.1, 0.25)$}
  \label{fig:140M-low-mid-stable}
\end{subfigure}

\begin{subfigure}[t]{\textwidth}
\vspace*{0cm}
\centering
    \includegraphics[trim={0cm 0cm 0cm 1cm},clip,width=.72\textwidth]{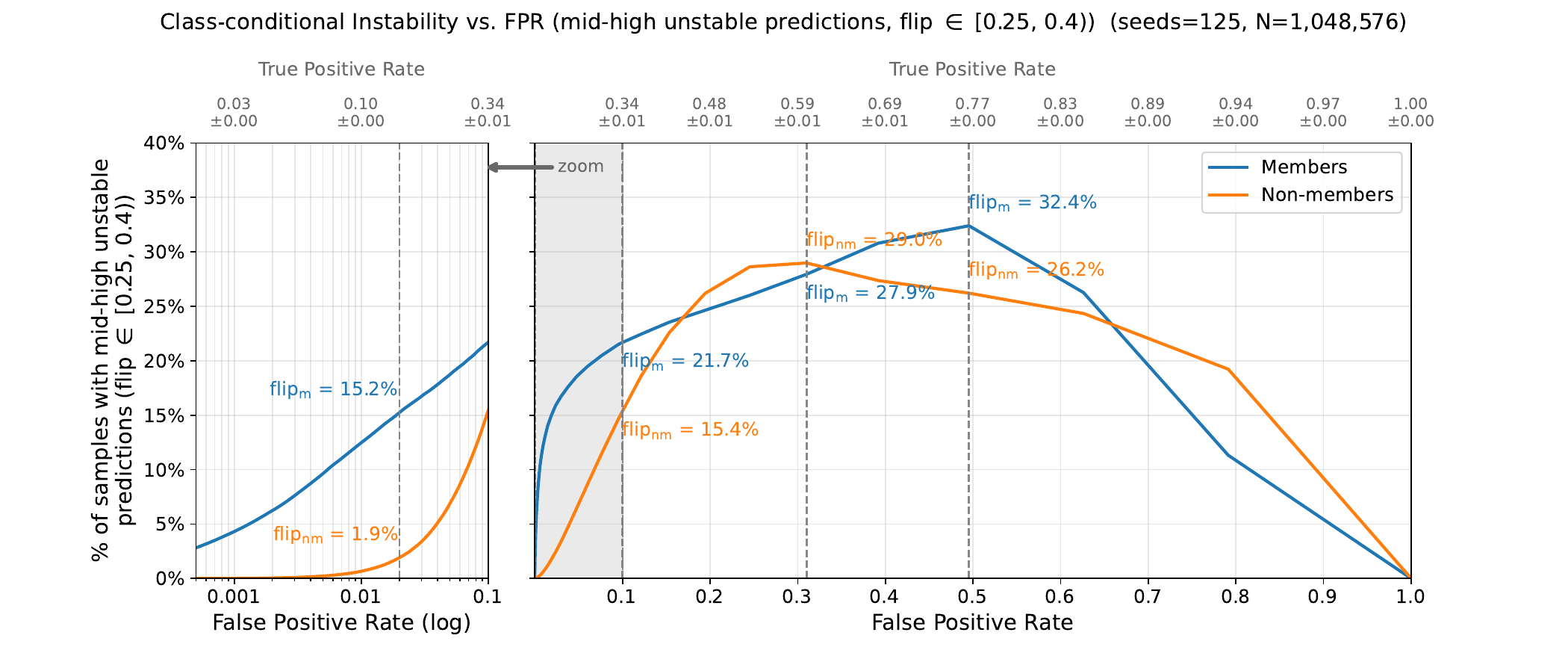}
    \vspace{-.12cm}
  \caption{Mid/high unstable MIA decisions: $\widehat{\mathrm{flip}}_{\eta,125} \in [0.25, 0.4)$}
  \label{fig:140M-mid-high-unstable}
\end{subfigure}

\begin{subfigure}[t]{\textwidth}
\vspace*{0cm}
\centering
  \includegraphics[trim={0cm 0cm 0cm 1cm},clip,width=.72\textwidth]{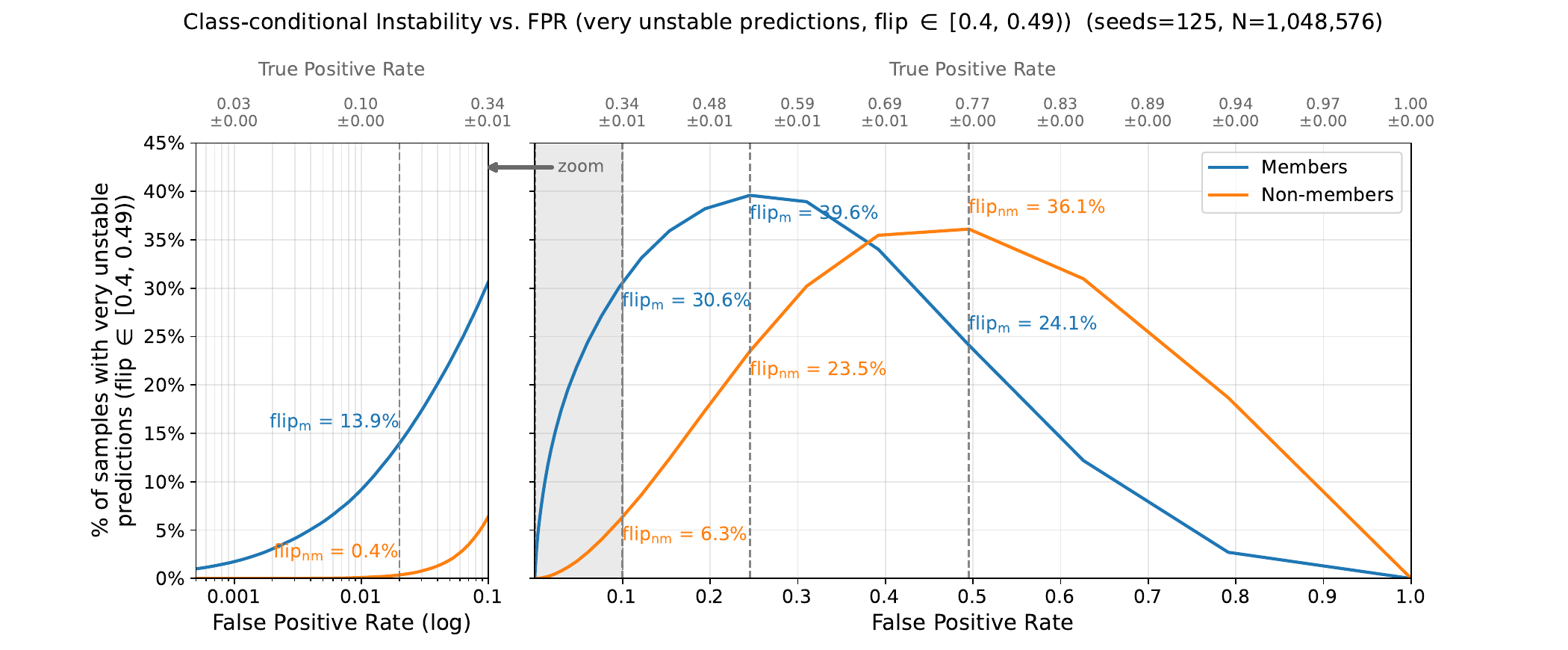}
  \vspace{-.12cm}
  \caption{Very unstable MIA decisions: $\widehat{\mathrm{flip}}_{\eta,125} \in [0.4, 0.490)$}
  \label{fig:140M-very-unstable}
\end{subfigure}

\begin{subfigure}[t]{\textwidth}
\vspace*{0cm}
\centering
    \includegraphics[trim={0cm 0cm 0cm 1cm},clip,width=.72\textwidth]{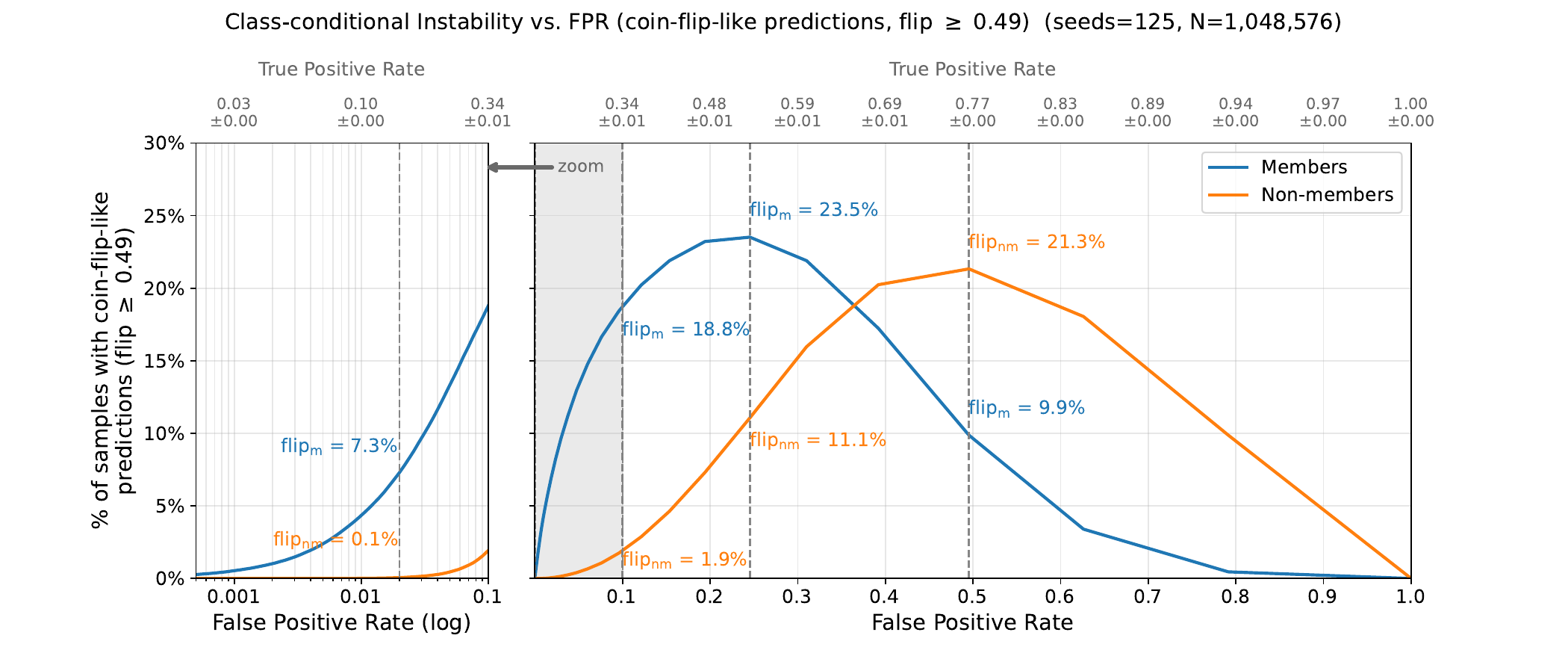}
    \vspace{-.12cm}
  \caption{Coin-flip-like MIA decisions: $\widehat{\mathrm{flip}}_{\eta,125} \in [0.490, \widehat{\mathrm{flip}}^{(\max)}_{\eta,125}] \quad$ ($\widehat{\mathrm{flip}}^{(\max)}_{\eta,125}{=}0.504$)}
  \label{fig:140M-arbitrary}
\end{subfigure}
\vspace{-.2cm}
\caption{\textbf{Flip rate variation by fixed $\mathrm{\bf{FPR}}$ for the $\bf140$M model}. For different ranges of $\widehat{\mathrm{flip}}_{\eta,125}$, we plot how class-conditional flip rate varies by $\fpr$. 
We annotate plots with corresponding mean $\pm$ standard deviation for the corresponding $\tpr$.
See main text for additional discussion. 
}
\label{fig:140M-instability-varied-fpr}
\end{figure}

\begin{figure}[t]
  \centering
\begin{subfigure}[t]{\textwidth}
\centering
\vspace*{0cm}
    \includegraphics[trim={0cm 0cm 0cm 1cm},clip,width=.72\textwidth]{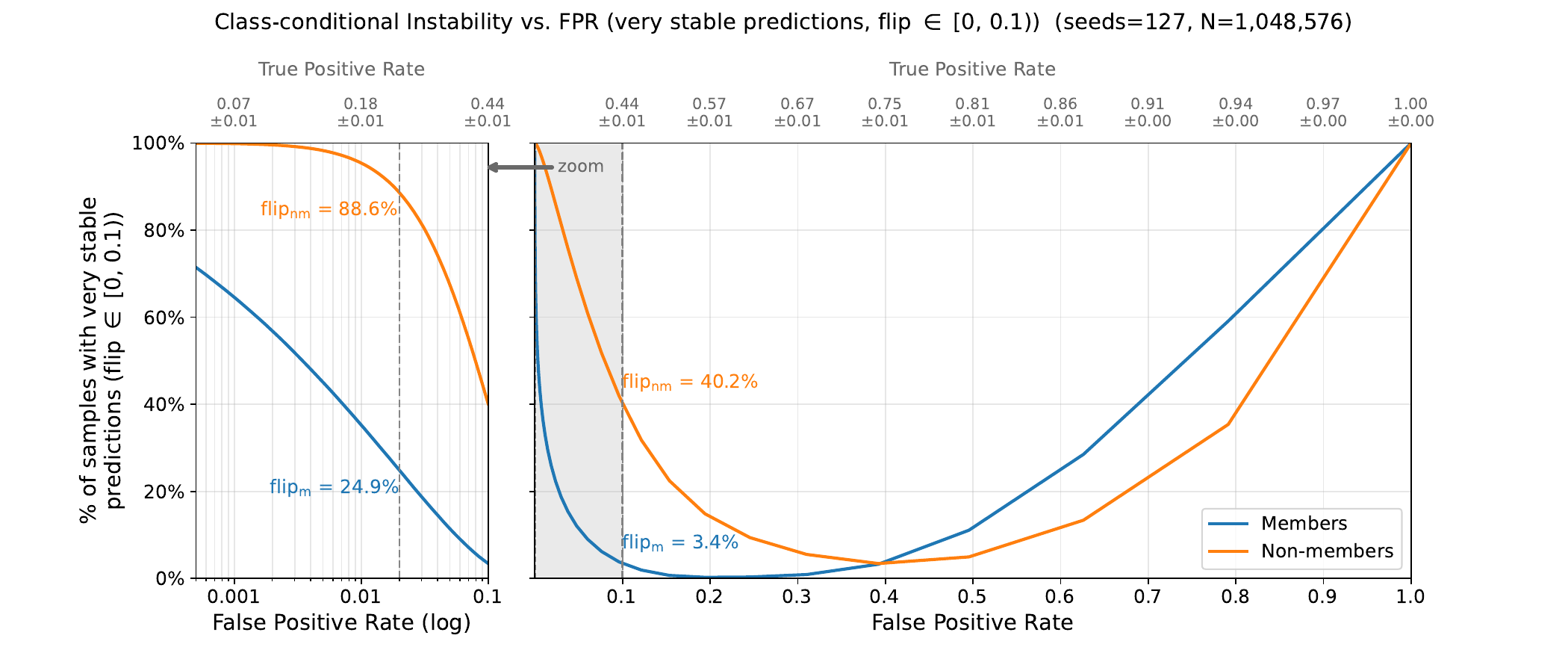}
    \vspace{-.12cm}
  \caption{Very stable MIA decisions: $\widehat{\mathrm{flip}}_{\eta,127} \in [0, 0.1)$}
  \label{fig:302M-stable}
\end{subfigure}

\begin{subfigure}[t]{\textwidth}
\vspace*{0cm}
\centering
    \includegraphics[trim={0cm 0cm 0cm 1cm},clip,width=.72\textwidth]{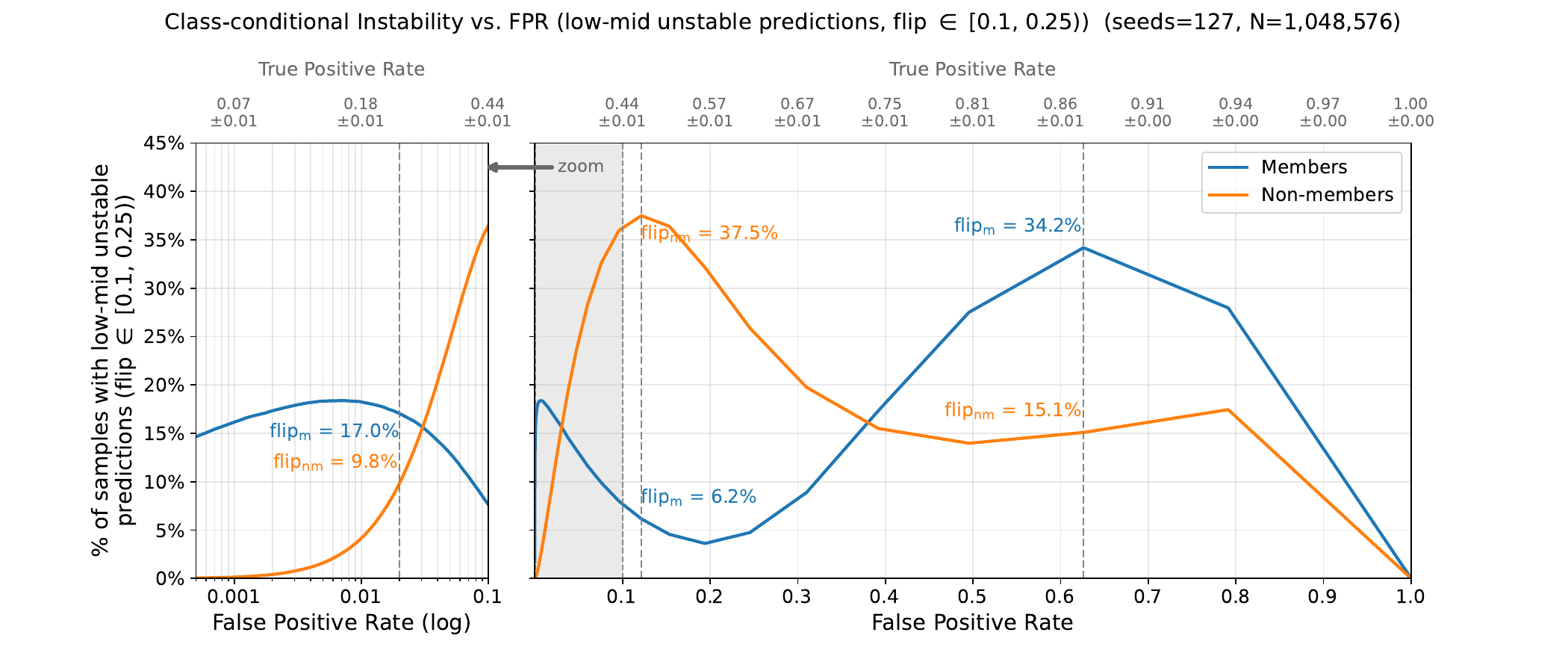}
    \vspace{-.12cm}
  \caption{Low/mid stable MIA decisions: $\widehat{\mathrm{flip}}_{\eta,127} \in [0.1, 0.25)$}
  \label{fig:302M-low-mid-stable}
\end{subfigure}

\begin{subfigure}[t]{\textwidth}
\vspace*{0cm}
\centering
    \includegraphics[trim={0cm 0cm 0cm 1cm},clip,width=.72\textwidth]{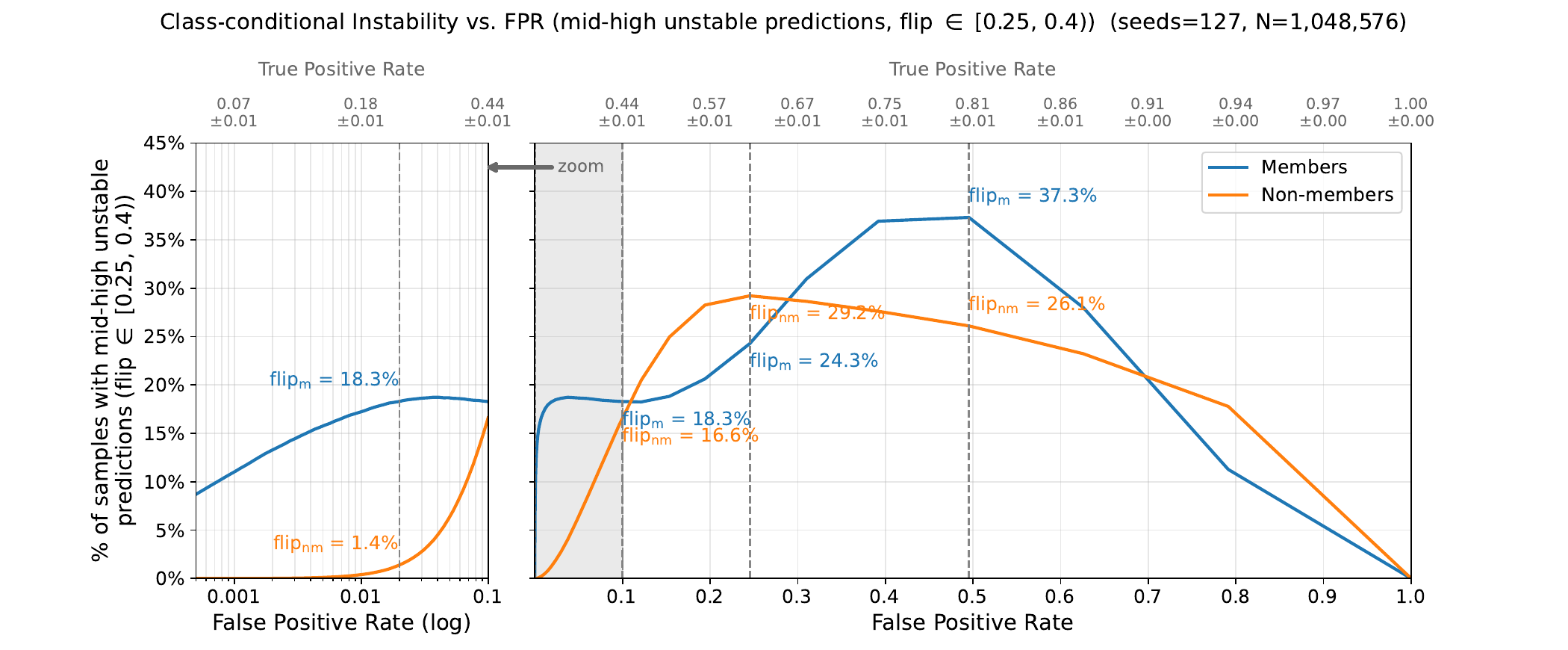}
    \vspace{-.12cm}
  \caption{Mid/high unstable MIA decisions: $\widehat{\mathrm{flip}}_{\eta,127} \in [0.25, 0.4)$}
  \label{fig:302M-mid-high-unstable}
\end{subfigure}

\begin{subfigure}[t]{\textwidth}
\vspace*{0cm}
\centering
  \includegraphics[trim={0cm 0cm 0cm 1cm},clip,width=.72\textwidth]{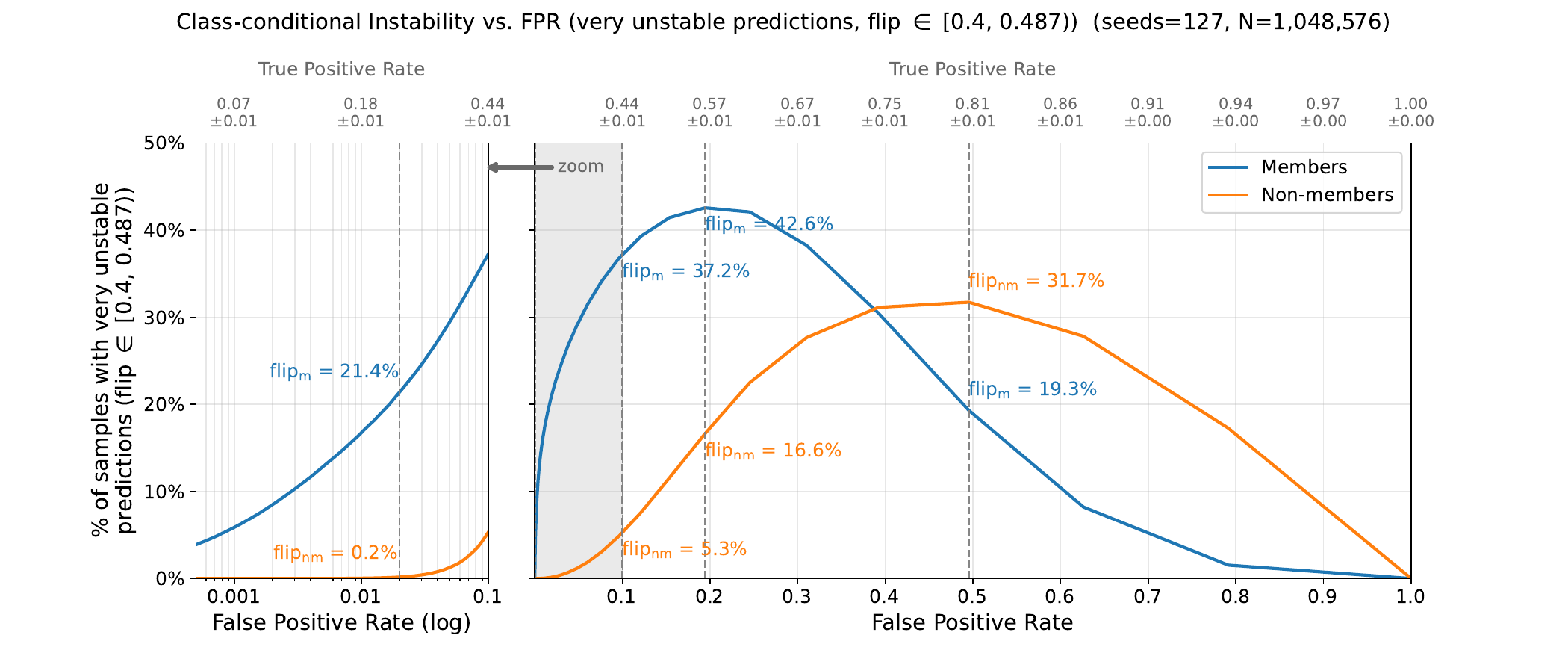}
  \vspace{-.12cm}
  \caption{Very unstable MIA decisions: $\widehat{\mathrm{flip}}_{\eta,127} \in [0.4, 0.487)$}
  \label{fig:302M-very-unstable}
\end{subfigure}

\begin{subfigure}[t]{\textwidth}
\vspace*{0cm}
\centering
    \includegraphics[trim={0cm 0cm 0cm 1cm},clip,width=.72\textwidth]{figures/cooper/stability/302M/instability-vs-fpr-fixed-flip-threshold-0.487-two-panel-302M.pdf}
    \vspace{-.12cm}
  \caption{Coin-flip-like MIA decisions: $\widehat{\mathrm{flip}}_{\eta,127} [0.487, \widehat{\mathrm{flip}}^{(\max)}_{\eta,127}]\quad$ ($ \widehat{\mathrm{flip}}^{(\max)}_{\eta,127}{=}0.50394$)}
  \label{fig:302M-arbitrary}
\end{subfigure}
\vspace{-.2cm}
\caption{\textbf{Flip rate variation by fixed $\mathrm{\bf{FPR}}$ for the $\bf302$M model}. For different ranges of $\widehat{\mathrm{flip}}_{\eta,B}$, we plot how class-conditional flip rate varies by $\fpr$. 
We annotate plots with corresponding mean $\pm$ standard deviation for the corresponding $\tpr$.
See main text for additional discussion. 
}
\label{fig:302M-instability-varied-fpr}
\end{figure}

\paragraph{Flip rate over varied fixed $\fpr$.}
We summarize across operating points in Figures~\ref{fig:140M-instability-varied-fpr} and~\ref{fig:302M-instability-varied-fpr}.
Each contains five sub-plots that show, for a given flip rate range, the class‑conditional proportion of samples in each range as a function of $\fpr$ (with the corresponding mean $\tpr$ $\pm$ STD annotated above the panels). 
We use the disjoint ranges
\([0,0.1)\) (very stable), \([0.1,0.25)\) (low/mid stable), \([0.25,0.4)\) (mid/high unstable), \( [0.4,t_\alpha(B))\) (very unstable), \([t_\alpha(B),\widehat{\mathrm{flip}}_{\eta,B}^{(\max)}]\),
where for $B{=}125$ we take $t_{0.05}(125){\approx}0.490$ and $\widehat{\mathrm{flip}}_{\eta,125}^{(\max)}{=}0.504$,  and for $B{=}127$ we take $t_{0.05}(127){\approx}0.487$ and $\widehat{\mathrm{flip}}_{\eta,127}^{(\max)}{\approx}0.50394$.

Our values for $t_{0.05}(125)$ and $t_{0.05}(127)$ are obtained from the exact two‑sided binomial acceptance region (Appendix~\ref{app:sec:instability:flip:arbitrary}). 
That is, we compute $t_\alpha(B)=\frac{2\,k_{\mathrm L}\,(B-k_{\mathrm L})}{B(B-1)}$ with $k_{\mathrm L}$  chosen as the smallest integer such that \ $F(k_{\mathrm L})\ge \alpha/2$ for $K\sim\mathrm{Binomial}(B,1/2)$. 
$\widehat{\mathrm{flip}}_{\eta,B}^{(\max)}{=}\frac{2\floor{B/2}\ceil{B/2}}{B(B-1)}$ (Appendix~\ref{app:sec:instability:flip}). 

These summary curves reinforce the fixed‑$\fpr$ views above: 
flip rate increases with $\fpr$, members flip far more than non‑members at all reasonable $\fpr$ (i.e., $\fpr{\lesssim}0.2$), and the $302$M model shows larger gaps between members and non-members as well as mass in the statistically indistinguishable-from-a-coin-flip range.

\FloatBarrier

\FloatBarrier

\FloatBarrier

\FloatBarrier

\begin{table}[t]
\centering
\caption{\textbf{$\mathbf{140}$M-parameter model $\mathrm{\bf{flip}}$ proportions for different $\mathrm{\bf{FPR}}$.}
For different settings of $\fpr$, we show the percentage of samples (by class) whose empirical flip lies in each range.
Lower flip corresponds to more stable MIA decisions; values near $0.5$ indicate coin‑flip-like behavior.
We split the high‑instability region into $[0.4,0.490)$ and $[0.490,0.504]$ to isolate cases that are statistically indistinguishable from a coin flip.
The $\max$ population flip is $0.5$; with $125$ seeds, the empirical $\max$ is ${\approx}0.504$.
($t_{0.05}(125){\approx}0.490$ is the minimum value at level $\alpha{=}0.05$ that our hypothesis test yields; see Appendix~\ref{app:sec:instability:flip:arbitrary}.) 
See Appendix~\ref{app:sec:instability:flip}.
Typically reported $\log$‑scale $\fpr$ rows are highlighted in gray.\looseness=-1}
\vspace{.2cm}
\label{tab:flip_by_range_140M}
\small
\begin{tabular}{l rr rr rr rr rr}
\toprule
\multicolumn{1}{c}{\textbf{$\mathrm{\mathbf{FPR}}$}} &
\multicolumn{2}{c}{\shortstack{\textbf{very stable}\\$\mathbf{\mathrm{flip}\in[0,0.1)}$}} &
\multicolumn{2}{c}{\shortstack{\textbf{low/mid unstable}\\$\mathbf{\mathrm{flip}\in[0.1,0.25)}$}} &
\multicolumn{2}{c}{\shortstack{\textbf{mid/high unstable}\\$\mathbf{\mathrm{flip}\in[0.25,0.4)}$}} &
\multicolumn{2}{c}{\shortstack{\textbf{very unstable}\\$\mathbf{\mathrm{flip}\in[0.4,0.49)}$}} &
\multicolumn{2}{c}{\shortstack{\textbf{coin flip}\\$\mathbf{\mathrm{flip}\geq 0.49}$}} \\
\cmidrule(lr){2-3} \cmidrule(lr){4-5} \cmidrule(lr){6-7} \cmidrule(lr){8-9} \cmidrule(lr){10-11}
 & \textbf{Mem.} & \shortstack{\textbf{Non‑}\\\textbf{Mem.}} & \textbf{Mem.} & \shortstack{\textbf{Non‑}\\\textbf{Mem.}} & \textbf{Mem.} & \shortstack{\textbf{Non‑}\\\textbf{Mem.}} & \textbf{Mem.} & \shortstack{\textbf{Non‑}\\\textbf{Mem.}} & \textbf{Mem.} & \shortstack{\textbf{Non‑}\\\textbf{Mem.}} \\
\midrule
\rowcolor{gray!15}
$10^{-5}$ & $98.93\%$ & $100.00\%$ & $0.86\%$ & $0.00\%$ & $0.17\%$ & $0.00\%$ & $0.03\%$ & $0.00\%$ & $0.01\%$ & $0.00\%$ \\
\rowcolor{gray!15}
$10^{-4}$ & $95.45\%$ & $99.99\%$ & $3.31\%$ & $0.01\%$ & $0.93\%$ & $0.00\%$ & $0.24\%$ & $0.00\%$ & $0.07\%$ & $0.00\%$ \\
\rowcolor{gray!15}
$10^{-3}$ & $83.94\%$ & $99.74\%$ & $9.47\%$ & $0.25\%$ & $4.31\%$ & $0.01\%$ & $1.63\%$ & $0.00\%$ & $0.65\%$ & $0.00\%$ \\
\rowcolor{gray!15}
$10^{-2}$ & $57.48\%$ & $94.88\%$ & $16.56\%$ & $4.34\%$ & $12.46\%$ & $0.66\%$ & $8.52\%$ & $0.12\%$ & $5.00\%$ & $0.00\%$ \\
$0.02$ & $45.97\%$ & $88.42\%$ & $17.55\%$ & $9.24\%$ & $15.23\%$ & $1.91\%$ & $12.95\%$ & $0.42\%$ & $8.29\%$ & $0.01\%$ \\
$0.05$ & $28.46\%$ & $70.12\%$ & $17.10\%$ & $20.63\%$ & $18.72\%$ & $6.88\%$ & $21.56\%$ & $2.37\%$ & $14.16\%$ & $0.01\%$ \\
\rowcolor{gray!15}
$10^{-1}$ & $14.37\%$ & $45.94\%$ & $14.60\%$ & $30.39\%$ & $21.69\%$ & $15.40\%$ & $30.34\%$ & $8.27\%$ & $19.00\%$ & $0.00\%$ \\
$0.2$ & $2.93\%$ & $17.29\%$ & $10.51\%$ & $30.34\%$ & $24.80\%$ & $26.59\%$ & $37.46\%$ & $17.83\%$ & $24.30\%$ & $8.95\%$ \\
$0.5$ & $8.85\%$ & $3.52\%$ & $25.61\%$ & $13.00\%$ & $32.29\%$ & $26.17\%$ & $24.92\%$ & $31.79\%$ & $8.34\%$ & $25.53\%$ \\
$0.75$ & $49.02\%$ & $26.08\%$ & $30.66\%$ & $18.59\%$ & $15.17\%$ & $20.91\%$ & $4.32\%$ & $22.21\%$ & $0.84\%$ & $12.21\%$ \\
\rowcolor{gray!15}
$10^{0}$ & $100.00\%$ & $100.00\%$ & $0.00\%$ & $0.00\%$ & $0.00\%$ & $0.00\%$ & $0.00\%$ & $0.00\%$ & $0.00\%$ & $0.00\%$ \\
\bottomrule
\end{tabular}
\end{table}

\begin{table}[t]
\centering
\caption{\textbf{$\mathbf{302}$M-parameter model $\mathrm{\bf{flip}}$ proportions for different $\mathrm{\bf{FPR}}$.}
For different settings of $\fpr$, we show the percentage of samples (by class) whose empirical flip lies in each range.
Lower flip corresponds to more stable MIA decisions; values near $0.5$ indicate coin‑flip-like behavior.
We split the high‑instability region into $[0.4,0.490)$ and $[0.490,0.50394]$ to isolate case that are statistically indistinguishable from a coin flip.
The $\max$ population flip is $0.5$; with $127$ seeds, the empirical $\max$ is ${\approx}0.50394$.
($t_{0.05}(127){\approx}0.487$ is the minimum value at level $\alpha{=}0.05$ that our hypothesis test yields; see Appendix~\ref{app:sec:instability:flip:arbitrary}.) 
See Appendix~\ref{app:sec:instability:flip}.
Typically reported $\log$‑scale $\fpr$ rows are highlighted in gray.\looseness=-1}
\vspace{.2cm}
\label{tab:flip_by_range_302M}
\small
\begin{tabular}{l rr rr rr rr rr}
\toprule
\multicolumn{1}{c}{\textbf{$\mathrm{\mathbf{FPR}}$}} &
\multicolumn{2}{c}{\shortstack{\textbf{very stable}\\$\mathbf{\mathrm{flip}\in[0,0.1)}$}} &
\multicolumn{2}{c}{\shortstack{\textbf{low/mid unstable}\\$\mathbf{\mathrm{flip}\in[0.1,0.25)}$}} &
\multicolumn{2}{c}{\shortstack{\textbf{mid/high unstable}\\$\mathbf{\mathrm{flip}\in[0.25,0.4)}$}} &
\multicolumn{2}{c}{\shortstack{\textbf{very unstable}\\$\mathbf{\mathrm{flip}\in[0.4,0.487)}$}} &
\multicolumn{2}{c}{\shortstack{\textbf{coin flip}\\$\mathbf{\mathrm{flip}\geq 0.487}$}} \\
\cmidrule(lr){2-3} \cmidrule(lr){4-5} \cmidrule(lr){6-7} \cmidrule(lr){8-9} \cmidrule(lr){10-11}
 & \textbf{Mem.} & \shortstack{\textbf{Non‑}\\\textbf{Mem.}} & \textbf{Mem.} & \shortstack{\textbf{Non‑}\\\textbf{Mem.}} & \textbf{Mem.} & \shortstack{\textbf{Non‑}\\\textbf{Mem.}} & \textbf{Mem.} & \shortstack{\textbf{Non‑}\\\textbf{Mem.}} & \textbf{Mem.} & \shortstack{\textbf{Non‑}\\\textbf{Mem.}} \\
\midrule
\rowcolor{gray!15}
$10^{-5}$ & $94.46\%$ & $100.00\%$ & $4.17\%$ & $0.00\%$ & $1.09\%$ & $0.00\%$ & $0.23\%$ & $0.00\%$ & $0.05\%$ & $0.00\%$ \\
\rowcolor{gray!15}
$10^{-4}$ & $84.15\%$ & $100.00\%$ & $10.06\%$ & $0.00\%$ & $4.15\%$ & $0.00\%$ & $1.31\%$ & $0.00\%$ & $0.33\%$ & $0.00\%$ \\
\rowcolor{gray!15}
$10^{-3}$ & $64.59\%$ & $99.85\%$ & $16.15\%$ & $0.14\%$ & $11.01\%$ & $0.00\%$ & $5.87\%$ & $0.00\%$ & $2.38\%$ & $0.00\%$ \\
\rowcolor{gray!15}
$10^{-2}$ & $35.27\%$ & $95.41\%$ & $18.24\%$ & $4.14\%$ & $17.25\%$ & $0.40\%$ & $16.69\%$ & $0.04\%$ & $12.55\%$ & $0.00\%$ \\
$0.02$ & $24.87\%$ & $88.61\%$ & $17.04\%$ & $9.81\%$ & $18.29\%$ & $1.39\%$ & $21.38\%$ & $0.17\%$ & $18.41\%$ & $0.03\%$ \\
$0.05$ & $11.42\%$ & $67.44\%$ & $12.96\%$ & $24.64\%$ & $18.67\%$ & $6.40\%$ & $29.49\%$ & $1.27\%$ & $27.46\%$ & $0.24\%$ \\
\rowcolor{gray!15}
$10^{-1}$ & $3.45\%$ & $40.16\%$ & $7.67\%$ & $36.41\%$ & $18.28\%$ & $16.59\%$ & $37.22\%$ & $5.28\%$ & $33.39\%$ & $1.56\%$ \\
$0.2$ & $0.26\%$ & $14.07\%$ & $3.62\%$ & $31.40\%$ & $20.97\%$ & $28.54\%$ & $42.65\%$ & $17.33\%$ & $32.49\%$ & $8.66\%$ \\
$0.5$ & $11.57\%$ & $5.13\%$ & $27.88\%$ & $13.98\%$ & $37.14\%$ & $26.02\%$ & $18.81\%$ & $31.66\%$ & $4.60\%$ & $23.21\%$ \\
$0.75$ & $50.77\%$ & $28.47\%$ & $31.22\%$ & $17.07\%$ & $15.27\%$ & $19.34\%$ & $2.53\%$ & $20.53\%$ & $0.22\%$ & $14.59\%$ \\
\rowcolor{gray!15}
$10^{0}$ & $100.00\%$ & $100.00\%$ & $0.00\%$ & $0.00\%$ & $0.00\%$ & $0.00\%$ & $0.00\%$ & $0.00\%$ & $0.00\%$ & $0.00\%$ \\
\bottomrule
\end{tabular}
\end{table}

\clearpage
\subsubsection{How many MIA true positives are statistically indistinguishable from a coin flip?}\label{app:sec:instability:flip:decompose}

From the above analysis, a natural follow-on is to attempt to estimate how many attack true positives are statistically indistinguishable from a coin flip.
That is, when we report $\rocauc$ at fixed $\fpr$, how much of the corresponding $\tpr$ is composed of positive MIA decisions that are essentially a coin flip, rather than reflecting reliable inference signal? 
In this appendix, we use the results from our experiments in Appendix~\ref{app:sec:instability:flip:results} to estimate an answer to this question. 
We split members into different bins that correspond to ranges for flip rate, and estimate the count (and rate) of true positives for each bin.\looseness=-1

This decomposition is a post‑hoc audit tool that uses target replicas. 
An attacker facing a single target cannot know which of its true positives are indistinguishable from a coin flip, when making per-sample membership claims.  
Therefore, the takeaway here is about reliability of per‑sample claims that an attacker makes, not about an attacker's observable signal.
Our aim is to surface the extent of unreliability that may affect the attacker's claims (regardless of the attacker's knowledge of reliability of those claims). 

We perform analysis that aggregates across attacks on multiple target models, so the numbers and figures we report are \emph{not} comparable with single-attack results in the typical MIA setup.
These results serve as diagnostics to assess attack reliability.
Individual attacks on specific targets of course vary in their performance, and are not directly comparable to what we present here that aggregates over many such attacks to try to better understand overall properties of attack behavior. 

\paragraph{Decomposing attack $\mathrm{\bf{TPR}}$ into contributions from flip rate bins.}
Fix a dataset $\sD_\text{IN}$ with $|\sD_\text{IN}|{=}M$ member examples and $B$ i.i.d.\ target replicas $r_1,\ldots,r_B\!\sim\!\mu$.
At a fixed $\fpr$~$\eta$, each target replica $r$ sets its threshold $\tau_r(\eta)$, calibrated on non-members (Section~\ref{sec:rw}~\&  Appendix~\ref{app:sec:background}).

For a member $\vx$, define the per-seed decision
\[
Y_r(\vx) \;=\; \1\!\{\Lambda_r(\vx)\ge \tau_r(\eta)\}\in\{0,1\},
\]
(This is just the binary membership decision rule  \(b_r^{(\eta)}(\vx)=\mathbf 1\{\Lambda_r(\vx)\ge \tau_r(\eta)\}\in\{0,1\}\), calibrated on non-members, but defined here specifically only on members for to reflect this analysis.)

So the define the \textbf{seed-wise true positive count} and \textbf{seed-wise true $\tpr$} are
\[
\mathrm{TP}_r \;=\; \sum_{\vx\in\sD_{\text{IN}}} Y_r(\vx),
\qquad
\tpr_r \;=\; \frac{\mathrm{TP}_r}{M}.
\]

\paragraph{Flip bins.}
Using the unbiased flip U-statistic $\widehat{\mathrm{flip}}_{\eta,B}(\vx)$ (Equation~\ref{app:eq:flip:ustat}, computed once from all $B$ seeds), we partition members into disjoint flip bins $\{\sB_j\}_j$,
e.g., $[0,0.1)$, $[0.1,0.25)$, $[0.25,0.4)$, $[0.4,t_\alpha(B))$, and $[t_\alpha(B), \widehat{\mathrm{flip}}^{(\max)}_{\eta,B}]$ with $t_\alpha(B)$ from Appendix~\ref{app:sec:instability:flip:arbitrary}.
For example, \(t_{0.05}(127){\approx}0.487\) and \(t_{0.05}(125){\approx}0.490\).

For bin $j$ and seed $r$ we define the \textbf{bin $\mathrm{\bf{TP}}$ count} and its \textbf{$\mathrm{\bf{TPR}}$ mass}:
\[
\mathrm{TP}_{r,j} \;=\; \sum_{\vx\in \sB_j} Y_r(\vx),
\qquad
\tpr_{r,j} \;=\; \frac{\mathrm{TP}_{r,j}}{M}.
\]
By construction, $\tpr_r=\sum_j \tpr_{r,j}$ for every seed $r$.

\paragraph{Across‐seed means and standard deviations.}
All means and STDs are computed across seeds. 
(We report STD across seeds, i.e., with denominator $B$, to match the rest of the paper.)
For totals,
\[
\overline{\tpr}=\frac{1}{B}\sum_{r=1}^B \tpr_r,
\quad
\operatorname{STD}(\tpr)=\sqrt{\frac{1}{B}\sum_{r=1}^B(\tpr_r-\overline{\tpr})^2},
\]
and the corresponding counts follow from the variance scaling law ($\Var[aX]=a^2\Var[X]$):
\[
\overline{\#\mathrm{TP}}=M\,\overline{\tpr},
\qquad
\operatorname{STD}(\#\mathrm{TP})=M\,\operatorname{STD}(\tpr).
\]

For each bin $j$,
\[
\overline{\tpr_j}=\frac{1}{B}\sum_r \tpr_{r,j},
\qquad
\operatorname{STD}(\tpr_j)=\sqrt{\frac{1}{B}\sum_r\!\bigl(\tpr_{r,j}-\overline{\tpr_j}\bigr)^2},
\]
and analogously for bin TP counts: $\overline{\#\mathrm{TP}_j}=M\,\overline{\tpr_j}$ and $\operatorname{STD}(\#\mathrm{TP}_j)=M\,\operatorname{STD}(\tpr_j)$.

\paragraph{Share of $\mathrm{\mathbf{TPR}}$ from a bin: mean-of-ratios.}
For seed $r$, define the \textbf{per-seed share} of $\tpr$ attributable to bin $j$:
\[
S_{r,j}\;=\;\frac{\mathrm{TP}_{r,j}}{\mathrm{TP}_r}
\quad(\text{set }S_{r,j}{=}0\text{ if }\mathrm{TP}_r{=}0).
\]
In our Tables~\ref{tab:tp_by_flip_140M} and~\ref{tab:tp_by_flip_302M}, we report the across-seed mean and STD of these shares:
\[
\overline{S}_j=\frac{1}{B}\sum_r S_{r,j},
\qquad
\operatorname{STD}(S_j)=\sqrt{\frac{1}{B}\sum_r \bigl(S_{r,j}-\overline{S}_j\bigr)^2}.
\]
This is a \emph{mean of ratios}, i.e., the average fraction of each seed's $\tpr$ that comes from bin $j$.

\paragraph{Alternative (ratio-of-means) and why it differs.}
A different but also reasonable summary is the \emph{ratio of means},
\[
R_j \;=\; \frac{\overline{\tpr_j}}{\overline{\tpr}}
\;=\; \frac{\frac{1}{B}\sum_r \mathrm{TP}_{r,j}/M}{\frac{1}{B}\sum_r \mathrm{TP}_r/M}
\;=\;\frac{\frac{1}{B}\sum_r \mathrm{TP}_{r,j}}{\frac{1}{B}\sum_r \mathrm{TP}_r}.
\]
$R_j$ answers ``what fraction of the \emph{expected} true positives lie in bin $j$?,'' whereas $\overline{S}_j$ answers ``for a \emph{typical seed}, what fraction of that seed's true positives lie in bin $j$?'' 
Because of seed‑to‑seed variability and the correlation between numerator and denominator, $R_j\neq \overline{S}_j$ in general. 
Our tables use $\overline{S}_j$ (mean of per‑seed shares) and its STD across seeds, as we are trying to estimate the fraction of the average/typical seed's true positives that lie in each bin (in particular, the statistically indistinguishable-from-a-coin-flip bin).\looseness=-1

\paragraph{Combining two  bins.}
Let the two high‑instability bins be $U{=}[0.4,t_\alpha(B))$ and $A{=}[t_\alpha(B),\widehat{\mathrm{flip}}^{(\max)}_{\eta,B}]$ and define the combined bin $C{=}U\cup A=[0.4, \widehat{\mathrm{flip}}^{(\max)}_{\eta,B}]$. 

For each seed $r$,
\[
S_{r,C}=S_{r,U}+S_{r,A},\quad
\mathrm{TP}_{r,C}=\mathrm{TP}_{r,U}+\mathrm{TP}_{r,A}.
\]
Therefore, means add by linearity of expectation: $\overline{S}_C=\overline{S}_U+\overline{S}_A$ and $\overline{\#\mathrm{TP}_C}=\overline{\#\mathrm{TP}_U}+\overline{\#\mathrm{TP}_A}$.  
However, for standard deviations, 
\[
\operatorname{STD}(S_C)=\sqrt{\operatorname{Var}(S_U)+\operatorname{Var}(S_A)+2\,\operatorname{Cov}(S_U,S_A)},
\]
Since we compute $C$ directly from per-seed $C$ values, we can easily combine bins in a way that is statistically correct. 
We report this combined bin in our tables.

\paragraph{What has no STD.}
Bin membership counts $|\sB_j|$ and their percentages $|\sB_j|/M$ have no across‑seed STD because bins are defined once from $\widehat{\mathrm{flip}}_{\eta,B}$ computed using all $B$ seeds.
True positives, though, do have across-seed STD, as these are estimated for each target. 

\paragraph{Assumptions and caveats for interpreting these numbers}
\begin{enumerate}[leftmargin=0.75cm]
\item \textbf{Single‑target vs.\ many‑seed view.}
These numbers diagnose instability that is hidden to an attacker with access to only a single target. 
A sample counted as a ``TP from coin-flip-like MIA decisions'' is not ``incorrect''---it is a member that this seed calls positive, but whose decision flips frequently across equally plausible seeds.
Even though it is not incorrect, it does \emph{not} reflect reliable knowledge about \emph{inferring} membership for that sample, since it is effectively a coin-flip decision.
We quantify how much of \(\overline{\tpr}\) is borne by such decisions.
However, any single target (that could reasonably be attacked by a real-world attacker) may deviate from this mean.\looseness=-1
\item \textbf{Calibration asymmetry.} 
Again, we note that thresholds \(\tau_r(\eta)\) are calibrated on non‑members for each seed, anchoring to non‑member behavior by construction. 
Member decisions are therefore more exposed to seed‑induced variation. 
This explains large member/non‑member flip gaps and is consistent with our seed‑to‑seed \(\tau\) dispersion at higher \(\eta\).
The true positive decomposition analysis we perform here is consistent with these other results. 
\item \textbf{Finite‑\(B\) effects and hypothesis testing.} 
The acceptance region cutoff \(t_\alpha(B)\) is derived from an exact two‑sided binomial test at level \(\alpha\) and is slightly conservative because of discreteness. 
Borderline cases (exactly at the tails) are included (fail‑to‑reject rule \(p\ge\alpha\)). 
This is important because all claims that we make about MIA decisions that are indistinguishable from a coin flip---including the decompositions here---hinge on the assumptions and results of this hypothesis test.
\item \textbf{Extremely low \(\mathrm{\bf{FPR}}\) \(\eta\).} When \(\overline{\tpr}\) is very small, the share $S_{r,j}$ can be numerically unstable; 
we suppress shares when \(\overline{\tpr}=0\) and note this where appropriate. 
\end{enumerate}

\paragraph{Results of the decomposition.} 
We show results for both model sizes in Tables~\ref{tab:tp_by_flip_140M} and~\ref{tab:tp_by_flip_302M}, respectively.
They indicate very large numbers of member decisions are indistinguishable from a coin flip or highly unstable  as $\fpr$ increases and moves the decision boundary into denser parts of the score distribution. 
Even at just $\fpr{=}10^{-3}$, $15.4\%{\pm}0.6\%$ of true positives for the $302$M model are indistinguishable from a coin flip---reflecting thousands of sample MIA decisions. 
If we consider both very unstable and coin-flip-like MIA decisions, they are responsible for $42.2\%{\pm}0.9\%$ of true positives at this $\fpr$.

\begin{table}[t]
\centering
\caption{\textbf{$\mathbf{140}$M model: Contribution of high-flip members to $\mathrm{\bf{TPR}}$ at fixed $\mathrm{\bf{FPR}}$.}
Shares are computed per seed ($\mathrm{TP}$s in bin divided by total $\mathrm{TP}$s for that seed) and then averaged; 
their STDs are across seeds. 
We additionally report the combined bin $\widehat{\mathrm{flip}}_{\eta,125}\in [0.4, 0.504]$ (``very unstable + coin flip'').
For the $140$M model, the coin-flip cutoff is $t_\alpha(125){\approx}0.490$, and $\widehat{\mathrm{flip}}^{\max}_{\eta,125}{=}0.504$. 
Typically reported $\log$-scale $\fpr$ rows are highlighted in gray.}
\vspace{.2cm}
\label{tab:tp_by_flip_140M}
\small
\hspace*{-2.5cm}
\setlength{\tabcolsep}{2.5pt}
\begin{tabular}{l r r rr rr rr}
\toprule
\textbf{$\mathrm{\mathbf{FPR}}$} &
\shortstack{\textbf{$\mathrm{\mathbf{TPR}}$}\\\textbf{(mean$\pm$STD)}} &
\shortstack{\textbf{$\mathrm{\mathbf{TP}}$}\\\textbf{(mean$\pm$STD)}} &
\multicolumn{2}{c}{\shortstack{\textbf{Very unstable}\\$\mathbf{[0.4, 0.490)}$}} &
\multicolumn{2}{c}{\shortstack{\textbf{Coin flip}\\$\mathbf{[0.490, 0.504]}$}} &
\multicolumn{2}{c}{\shortstack{\textbf{Very unstable + coin flip}\\$\mathbf{[0.4, 0.504]}$}} \\
\cmidrule(lr){4-5} \cmidrule(lr){6-7} \cmidrule(lr){8-9}
 &  &  &
\shortstack{$\mathrm{\mathbf{TP}}$s\\\textbf{(mean$\pm$STD)}} &
\shortstack{\textbf{Share of $\mathrm{\mathbf{TPR}}$}\\\textbf{(mean$\pm$STD)}} &
\shortstack{$\mathrm{\mathbf{TP}}$s\\\textbf{(mean$\pm$STD)}} &
\shortstack{\textbf{Share of $\mathrm{\mathbf{TPR}}$}\\\textbf{(mean$\pm$STD)}} &
\shortstack{$\mathrm{\mathbf{TP}}$s\\\textbf{(mean$\pm$STD)}} &
\shortstack{$\mathrm{\mathbf{TPR}}$\\\textbf{(mean$\pm$STD)}} \\
\midrule
\rowcolor{gray!15} $10^{-5}$ & \(0.002 \pm 0.001\) & \(1{,}114 \pm 305\) & \(61 \pm 9\) & \(5.8\% \pm 1.4\%\) & \(24 \pm 4\) & \(2.3\% \pm 0.7\%\) & \(84 \pm 11\) & \(8.0\% \pm 2.0\%\) \\
\rowcolor{gray!15} $10^{-4}$ & \(0.008 \pm 0.001\) & \(4{,}247 \pm 563\) & \(432 \pm 46\) & \(10.2\% \pm 0.7\%\) & \(160 \pm 13\) & \(3.8\% \pm 0.4\%\) & \(592 \pm 55\) & \(14.1\% \pm 1.0\%\) \\
\rowcolor{gray!15} $10^{-3}$ & \(0.030 \pm 0.002\) & \(15{,}826 \pm 1{,}239\) & \(3{,}048 \pm 304\) & \(13.3\% \pm 0.4\%\) & \(1{,}493 \pm 142\) & \(6.8\% \pm 0.3\%\) & \(4{,}541 \pm 413\) & \(19.9\% \pm 0.8\%\) \\
\rowcolor{gray!15} $10^{-2}$ & \(0.104 \pm 0.005\) & \(54{,}373 \pm 2{,}414\) & \(17{,}240 \pm 1{,}464\) & \(31.2\% \pm 1.0\%\) & \(12{,}164 \pm 1{,}070\) & \(22.8\% \pm 1.0\%\) & \(29{,}404 \pm 2{,}461\) & \(54.0\% \pm 2.5\%\) \\
\(0.02\) & \(0.148 \pm 0.005\) & \(77{,}667 \pm 2{,}718\) & \(27{,}455 \pm 2{,}031\) & \(34.7\% \pm 0.8\%\) & \(20{,}609 \pm 1{,}765\) & \(27.1\% \pm 1.1\%\) & \(48{,}064 \pm 3{,}698\) & \(61.8\% \pm 3.0\%\) \\
\(0.05\) & \(0.236 \pm 0.006\) & \(123{,}864 \pm 2{,}978\) & \(47{,}831 \pm 2{,}742\) & \(38.7\% \pm 0.7\%\) & \(37{,}910 \pm 3{,}223\) & \(30.5\% \pm 1.1\%\) & \(85{,}742 \pm 5{,}492\) & \(69.2\% \pm 3.2\%\) \\
\rowcolor{gray!15} \(0.1\) & \(0.336 \pm 0.006\) & \(176{,}346 \pm 3{,}180\) & \(70{,}670 \pm 3{,}050\) & \(39.2\% \pm 0.6\%\) & \(53{,}769 \pm 3{,}443\) & \(31.4\% \pm 0.9\%\) & \(124{,}439 \pm 6{,}231\) & \(70.5\% \pm 2.7\%\) \\
\(0.2\) & \(0.480 \pm 0.006\) & \(251{,}888 \pm 3{,}256\) & \(97{,}926 \pm 2{,}386\) & \(39.1\% \pm 0.3\%\) & \(67{,}384 \pm 1{,}688\) & \(26.4\% \pm 0.3\%\) & \(165{,}310 \pm 4{,}489\) & \(65.6\% \pm 1.2\%\) \\
\(0.5\) & \(0.766 \pm 0.004\) & \(401{,}354 \pm 2{,}242\) & \(74{,}851 \pm 610\) & \(18.7\% \pm 0.2\%\) & \(28{,}804 \pm 485\) & \(7.2\% \pm 0.1\%\) & \(103{,}655 \pm 807\) & \(25.8\% \pm 0.3\%\) \\
\(0.75\) & \(0.915 \pm 0.002\) & \(479{,}540 \pm 882\) & \(14{,}674 \pm 339\) & \(3.1\% \pm 0.1\%\) & \(2{,}644 \pm 75\) & \(0.6\% \pm 0.0\%\) & \(17{,}318 \pm 403\) & \(3.6\% \pm 0.1\%\) \\
\rowcolor{gray!15} $10^{0}$ & \(1.000 \pm 0.000\) & \(524{,}288 \pm 0\) & \(0 \pm 0\) & \(0.0\% \pm 0.0\%\) & \(0 \pm 0\) & \(0.0\% \pm 0.0\%\) & \(0 \pm 0\) & \(0.0\% \pm 0.0\%\) \\
\bottomrule
\end{tabular}
\end{table}

\begin{table}[t]
\centering
\caption{\textbf{$\mathbf{302}$M model: Contribution of high-flip members to $\mathrm{\bf{TPR}}$ at fixed $\mathrm{\bf{FPR}}$.}
Shares are computed per seed ($\mathrm{TP}$s in bin divided by total $\mathrm{TP}$s for that seed) and then averaged; 
their STDs are across seeds. 
We additionally report the combined bin $\widehat{\mathrm{flip}}_{\eta,127}\in [0.4, 0.50394]$ (``very unstable + coin flip'').
For the $302$M model, the coin-flip cutoff is $t_\alpha(127){\approx}0.487$, and $\widehat{\mathrm{flip}}^{\max}_{\eta,127}{=}0.50394$. 
Typically reported $\log$-scale $\fpr$ rows are highlighted in gray.}
\vspace{.2cm}
\label{tab:tp_by_flip_302M}
\small
\hspace*{-2.5cm}
\setlength{\tabcolsep}{2.5pt}
\begin{tabular}{l r r rr rr rr}
\toprule
\textbf{$\mathrm{\mathbf{FPR}}$} &
\shortstack{\textbf{$\mathrm{\mathbf{TPR}}$}\\\textbf{(mean$\pm$STD)}} &
\shortstack{\textbf{$\mathrm{\mathbf{TP}}$}\\\textbf{(mean$\pm$STD)}} &
\multicolumn{2}{c}{\shortstack{\textbf{Very unstable}\\$\mathbf{[0.4, 0.487)}$}} &
\multicolumn{2}{c}{\shortstack{\textbf{Coin flip}\\$\mathbf{[0.487, 0.50394]}$}} &
\multicolumn{2}{c}{\shortstack{\textbf{Very unstable + coin flip}\\$\mathbf{[0.4, 0.50394]}$}} \\
\cmidrule(lr){4-5} \cmidrule(lr){6-7} \cmidrule(lr){8-9}
 &  &  &
\shortstack{$\mathrm{\mathbf{TP}}$s\\\textbf{(mean$\pm$STD)}} &
\shortstack{\textbf{Share of $\mathrm{\mathbf{TPR}}$}\\\textbf{(mean$\pm$STD)}} &
\shortstack{$\mathrm{\mathbf{TP}}$s\\\textbf{(mean$\pm$STD)}} &
\shortstack{\textbf{Share of $\mathrm{\mathbf{TPR}}$}\\\textbf{(mean$\pm$STD)}} &
\shortstack{$\mathrm{\mathbf{TP}}$s\\\textbf{(mean$\pm$STD)}} &
\shortstack{$\mathrm{\mathbf{TPR}}$\\\textbf{(mean$\pm$STD)}} \\
\midrule
\rowcolor{gray!15} $10^{-5}$ & \(0.009 \pm 0.003\) & \(4{,}878 \pm 1{,}496\) & \(371 \pm 66\) & \(8.1\% \pm 2.1\%\) & \(122 \pm 16\) & \(2.8\% \pm 1.5\%\) & \(493 \pm 80\) & \(10.9\% \pm 3.6\%\) \\
\rowcolor{gray!15} $10^{-4}$ & \(0.027 \pm 0.005\) & \(14{,}254 \pm 2{,}631\) & \(2{,}233 \pm 349\) & \(15.8\% \pm 1.0\%\) & \(805 \pm 87\) & \(5.8\% \pm 0.9\%\) & \(3{,}038 \pm 432\) & \(21.6\% \pm 1.8\%\) \\
\rowcolor{gray!15} $10^{-3}$ & \(0.073 \pm 0.009\) & \(38{,}414 \pm 4{,}886\) & \(10{,}295 \pm 1{,}548\) & \(26.9\% \pm 0.9\%\) & \(5{,}883 \pm 651\) & \(15.4\% \pm 0.6\%\) & \(16{,}231 \pm 2{,}194\) & \(42.2\% \pm 0.9\%\) \\
\rowcolor{gray!15} $10^{-2}$ & \(0.181 \pm 0.013\) & \(94{,}823 \pm 6{,}939\) & \(32{,}549 \pm 3{,}549\) & \(34.3\% \pm 1.5\%\) & \(31{,}866 \pm 3{,}429\) & \(33.5\% \pm 1.5\%\) & \(64{,}477 \pm 6{,}973\) & \(67.8\% \pm 3.0\%\) \\
\(0.02\) & \(0.235 \pm 0.014\) & \(123{,}384 \pm 7{,}183\) & \(44{,}612 \pm 3{,}894\) & \(36.1\% \pm 1.3\%\) & \(47{,}301 \pm 4{,}805\) & \(38.2\% \pm 2.0\%\) & \(91{,}913 \pm 8{,}687\) & \(74.3\% \pm 3.3\%\) \\
\(0.05\) & \(0.333 \pm 0.014\) & \(174{,}687 \pm 7{,}312\) & \(69{,}431 \pm 4{,}149\) & \(39.7\% \pm 0.9\%\) & \(71{,}632 \pm 6{,}160\) & \(40.9\% \pm 2.2\%\) & \(141{,}062 \pm 10{,}270\) & \(80.6\% \pm 3.1\%\) \\
\rowcolor{gray!15} \(0.1\) & \(0.435 \pm 0.014\) & \(228{,}106 \pm 7{,}209\) & \(98{,}285 \pm 4{,}353\) & \(42.9\% \pm 0.6\%\) & \(88{,}061 \pm 5{,}522\) & \(38.6\% \pm 1.4\%\) & \(186{,}021 \pm 9{,}760\) & \(81.5\% \pm 2.0\%\) \\
\(0.2\) & \(0.570 \pm 0.013\) & \(299{,}000 \pm 6{,}649\) & \(128{,}904 \pm 3{,}649\) & \(43.1\% \pm 0.3\%\) & \(87{,}119 \pm 2{,}415\) & \(29.1\% \pm 0.2\%\) & \(216{,}023 \pm 5{,}982\) & \(72.2\% \pm 0.4\%\) \\
\(0.5\) & \(0.809 \pm 0.008\) & \(424{,}228 \pm 3{,}952\) & \(66{,}051 \pm 1{,}888\) & \(15.6\% \pm 0.4\%\) & \(12{,}913 \pm 241\) & \(3.0\% \pm 0.1\%\) & \(78{,}964 \pm 2{,}093\) & \(18.6\% \pm 0.4\%\) \\
\(0.75\) & \(0.925 \pm 0.004\) & \(484{,}974 \pm 1{,}908\) & \(9{,}094 \pm 372\) & \(1.9\% \pm 0.1\%\) & \(613 \pm 28\) & \(0.1\% \pm 0.0\%\) & \(9{,}708 \pm 394\) & \(2.0\% \pm 0.1\%\) \\
\rowcolor{gray!15} $10^{0}$ & \(1.000 \pm 0.000\) & \(524{,}288 \pm 0\) & \(0 \pm 0\) & \(0.0\% \pm 0.0\%\) & \(0 \pm 0\) & \(0.0\% \pm 0.0\%\) & \(0 \pm 0\) & \(0.0\% \pm 0.0\%\) \\
\bottomrule
\end{tabular}
\end{table}

\paragraph{Estimating the contribution of coin-flip-like MIA decisions to $\mathrm{\bf{ROC}}$-$\mathrm{\bf{AUC}}$.}
We similarly can use this type of analysis to estimate how much of $\rocauc$ can be attributed to MIA decisions that are indistinguishable from a coin flip.
Similarly, the point of this analysis is not to suggest that these MIA decisions are ``incorrect;'' 
the point is to attempt to distinguish the degree to which coin-flip-like MIA decisions are impacting overall claims about successful membership \emph{inference}. 
To do so, we distinguish between the whole (averaged) $\roc$ curve and associated mean $\auc$, and the mean $\roc$ curve (and mean $\auc$) computed for non-coin-flip-like decisions.

For a ROC curve written as $\tpr$ vs.\ $\fpr$, fix an $\fpr$ interval $[a,b]\subset(0,1)$ and write $w=b-a$.
For a given target $r$, denote $\tpr_r(\eta)$ as the true positive rate at operating point $\fpr{=}\eta$,
and let $\overline{\tpr}(\eta)$ be the mean across seeds.

We distinguish positives that come from coin-flip-like decisions as follows:   
at each $\fpr$, we remove the portion of the $\overline{\tpr}$ that is supported by decisions that are indistinguishable from a coin flip.
That is, fix $B$ targets and significance $\alpha$; let $t_\alpha(B)$ be the exact two–sided Binomial$(B,0.5)$ acceptance–band flip cutoff (Appendix~\ref{app:sec:instability:flip:arbitrary}).
Let $M=|\sD_\text{IN}|$ denote the number of members. 
For seed $r$ at FPR $\eta$, write $Y_r(\vx)\in\{0,1\}$ for the membership decision on member $\vx$, and let
\[
\mathrm{\tpr}_{\mathrm{raw},r}(\eta)\;=\;\frac{1}{M}\sum_{\vx\in\sD_{\mathrm{IN}}}\!\! Y_r(\vx),
\qquad
\tpr_{\mathrm{coin},r}(\eta)\;=\;\frac{1}{M}\sum_{\vx\in\sD_{\mathrm{IN}}}\! Y_r(\vx)\,\1\{\widehat{\mathrm{flip}}_{\eta,B}(\vx)\ge t_\alpha(B)\}.
\]
We define the non-coin-flip-like-induced $\tpr$ per seed by subtraction,
\[
\tpr_{\mathrm{noncoin},r}(\eta)\;=\;\tpr_{\mathrm{raw},r}(\eta)-\tpr_{\mathrm{coin},r}(\eta). 
\]
We compute the averages across $B$ seeds as
\[
\overline{\mathrm{\tpr}}_{\mathrm{raw}}(\eta)\;=\;\frac{1}{B}\sum_{r=1}^B \mathrm{\tpr}_{\mathrm{raw},r}(\eta),
\qquad
\overline{\mathrm{\tpr}}_{\mathrm{noncoin}}(\eta)\;=\;\frac{1}{B}\sum_{r=1}^B \mathrm{\tpr}_{\mathrm{noncoin},r}(\eta).
\]
By linearity of expectation, across seeds,
\begin{align}
\label{app:eq:decomp:tpr}
\overline{\tpr}_{\mathrm{noncoin}}(\eta)\;=\;\overline{\tpr}_{\mathrm{raw}}(\eta)-\overline{\tpr}_{\mathrm{coin}}(\eta). 
\end{align}

Note that $\overline{\tpr}_{\mathrm{noncoin}}$ is a diagnostic curve: 
it subtracts the statistically indistinguishable-from-a-coin-flip component and is \emph{not} itself the $\roc$ of a single threshold rule that could be produced by an attack.

We can then compare the overall $\roc$ to the non-coin-flip-like $\roc$.
It may be useful to do so for a specific range of $\fpr$, rather than for the entire $\roc$ curve. 
To do so, note that the unnormalized partial $\auc$ ($\mathrm{pAUC}$) over $[a,b]$ is
\[
\mathrm{pAUC}[a,b]
\;\coloneqq\;
\int_a^b \overline{\tpr}(\eta)\,d\eta.
\]
The normalized $\mathrm{pAUC}$ is the mean $\tpr$ averaged over $\fpr$ in $[a,b]$; it rescales to $[0,1]$ by dividing by the band width:
\[
\mathrm{pAUC}_{\mathrm{norm}}[a,b]
\;\coloneqq\;
\frac{1}{b-a}\int_a^b \overline{\tpr}(\eta)\,d\eta,
\quad\text{which equals the mean $\tpr$ over the band.}
\]
To quantify improvement over a random classifier, we subtract the area under the by-chance curve. 
For the random classifier $\overline{\tpr}(\eta)=\fpr(\eta)$, so
\[
\int_a^b \eta\,d\eta
\;=\;
\left.\frac{\eta^2}{2}\right|_{\eta=a}^{b}
\;=\;\frac{b^2-a^2}{2}.
\]
Dividing by $w=b-a$ gives the normalized random baseline \(\frac{1}{b-a} \cdot \frac{b^2-a^2}{2} = \frac{a+b}{2}\).
We can similarly report the \textbf{lift} above these random baselines: 
\[
\mathrm{Lift}[a,b]
\;\coloneqq\;
\int_a^b \big(\overline{\tpr}(\eta)-\eta\big)\,d\eta
\;=\;
\mathrm{pAUC}[a,b]-\frac{b^2-a^2}{2},
\]
\[
\mathrm{Lift}_{\mathrm{norm}}[a,b]
\;\coloneqq\;
\frac{1}{b-a}\int_a^b \big(\overline{\tpr}(\eta)-\eta\big)\,d\eta
\;=\;
\mathrm{pAUC}_{\mathrm{norm}}[a,b]-\frac{a+b}{2}.
\]

Note that $\mathrm{Lift}_{\mathrm{norm}}[a,b]=0$ for a random classifier and equals the average vertical gap from chance in $[a,b]$.

\begin{figure}[t]
  \centering
\begin{subfigure}[t]{0.4\textwidth}
\vspace*{0cm}
    \includegraphics[width=\textwidth]{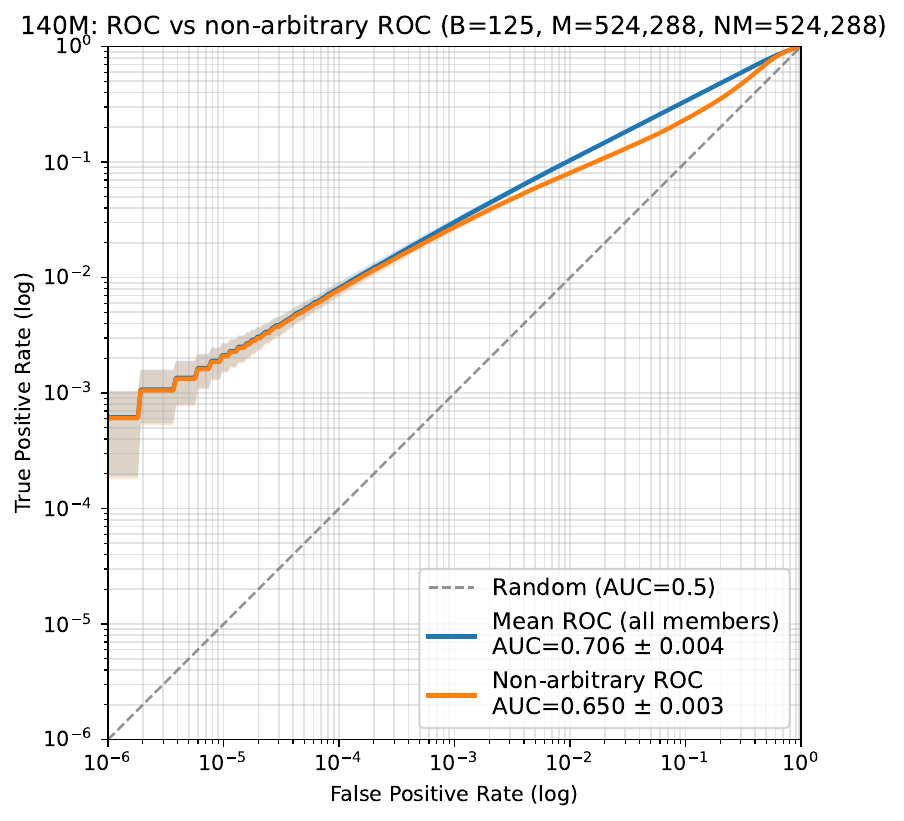}
  \caption{$140$M}
  \label{fig:140M-rocauc-instability-2}
\end{subfigure}
\hspace{1cm}
\begin{subfigure}[t]{0.4\textwidth}
\vspace*{0cm}
    \includegraphics[width=\textwidth]{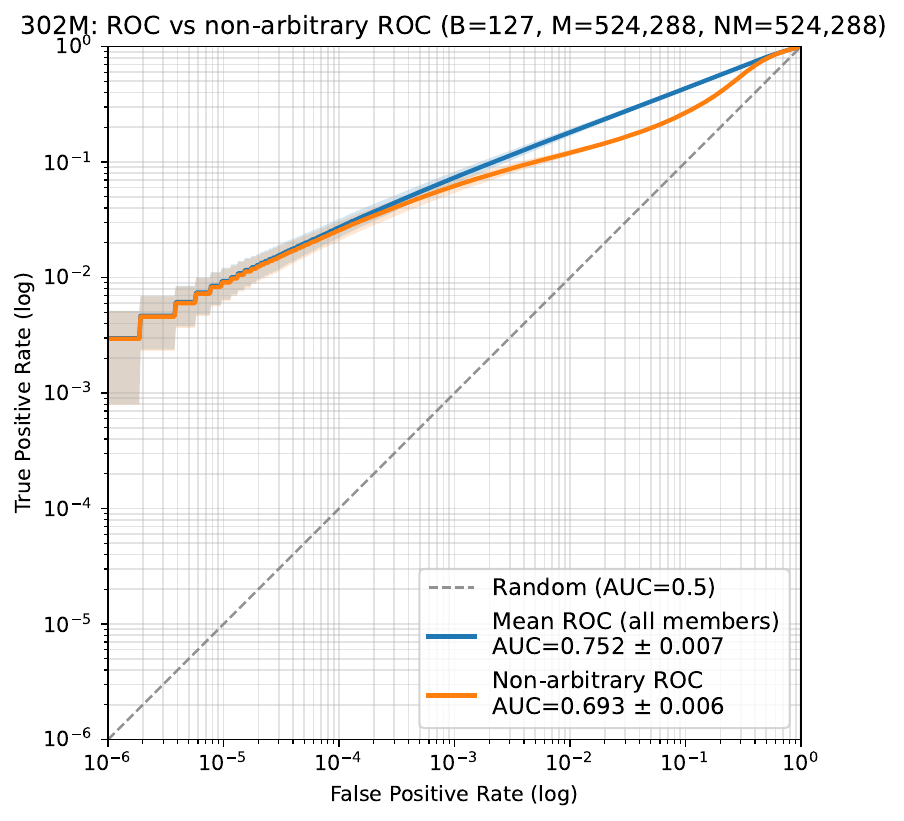}
  \caption{$302$M}
  \label{fig:302M-rocauc-instability-2}
\end{subfigure}
\caption{\textbf{Decoupling overall attack success from success based on coin-flip like sample decisions.} 
We produce the same mean $\roc$ curves and mean $\auc$ as in Figure~\ref{fig:instability-roc} for both the  (\textbf{a}) $140$M and (\textbf{b}) $302$M models ($B{=}125$ and $B{=}127$, respectively). 
At each $\tpr$ for fixed $\fpr$, we estimate how many true positives are attributable to MIA decisions that are statistically indistinguishable from a coin flip  ($t_\alpha(125){\approx}0.490$ and $t_\alpha(127){\approx}0.487$, respectively; see Appendix~\ref{app:sec:instability:flip:arbitrary}).
By Equation~\ref{app:eq:decomp:tpr}, at each $\eta$, we can then estimate how many true positives are from non-coin-flip-like decisions. 
Consistent with our other results, 
As both curves enter ranges for $\fpr$ with nontrivial $\tpr$, coin-flip-like MIA decisions make up a significant proportion of $\rocauc$, with a greater effect for the $302$M architecture. 
}
\vspace{-.3cm}
\label{fig:instability-roc-2}
\end{figure}

\paragraph{Worked example.}
We apply this procedure to the $\roc$ for the $302$M model.
We do so for the entire curve in Figure~\ref{fig:302M-rocauc-instability-2}, which shows that ${\approx}0.059$ of $\rocauc$ can be attributed to coin-flip-like MIA decisions. 
Of course, the region where this has the most impact is the range of $\eta$ that moves the decision boundary into the denser part of the score distribution where MIA calls more positives, but IN/OUT distribution overlap is more extensive.
$\tpr$ rises, but so does the share of coin-flip-like true positives.\looseness=-1

We can therefore also isolate the effect on this part of $\rocauc$ by setting $[a,b]=[10^{-4},10^{-1}]$.
Here $w=0.0999$, the normalized random baseline is $\tfrac{a+b}{2}=0.05005$,
and the unnormalized random area is $\tfrac{b^2-a^2}{2}=0.004999995$. 
With our measured areas,
\[
\mathrm{pAUC}_{\mathrm{norm}}^{\text{raw}}=0.314748,\quad
\mathrm{pAUC}_{\mathrm{norm}}^{\mathrm{noncoin}}=0.190810,
\]
so
\[
\mathrm{Lift}_{\mathrm{norm}}^{\text{raw}}=0.264698,\qquad
\mathrm{Lift}_{\mathrm{norm}}^{\mathrm{noncoin}}=0.140760.
\]

As a result, the non-coin-flip-like $\roc$ retains about $\frac{0.140760}{0.264698}\times100\approx53\%$ of the raw lift in this $\fpr$ band,
and the average $\tpr$ for the band drops by $1 - \frac{0.190810}{0.314748}\times100\approx39.4\%$ relative to raw.
In other words, filtering out coin-flip-like positives yields a substantially lower $\mathrm{pAUC}$ in this band. 
A sizable part of the apparent attack advantage in this $\fpr$ range comes from samples whose per‑seed decisions are statistically indistinguishable from coin flips (Appendix~\ref{app:sec:instability:flip:arbitrary}).
This weakens the value/reliability of per-sample positives in this $\fpr$ regime.

It is important to note that this is a conservative estimate of the parts of the attack that are reliable, as we only filter out positives that pass the coin-flip-like flip threshold, $t_\alpha(B)$. 
As a result, these numbers still include highly unstable cases (that are arguably also not reflective of meaningfully reliable membership inference, e.g., flip in $[0.4,t_\alpha(B))$).
In this respect, our non-coin-flip-like $\roc$ is a conservative upper bound on reliable membership inference. 
Given the extent of highly unstable MIA decisions in this band, we would expect larger decreases in partial $\auc$ and $\mathrm{Lift}$ if we filtered those decisions in our analysis.\looseness=-1

\paragraph{Caveats.}
Because bins are defined using $\widehat{\mathrm{flip}}_{\eta,B}$ pooled over $B$ seeds, per-seed variability in bin membership is not propagated.
This makes our estimates conservative for variability, but keeps the decomposition identity exact (i.e., the sum of the bin masses equals $\overline{\tpr}$). 
In our implementation, we also clip $\tpr_{\mathrm{noncoin},r}$ at $0$ to guard against finite-sample noise, i.e., $\tpr_{\mathrm{noncoin},r}(\eta)\leftarrow\max\{0, \tpr_{\mathrm{raw},r}(\eta) - \tpr_{\mathrm{coin},r}(\eta)\}$.

\subsubsection{Additional notes on interpreting flip rate measurements}\label{app:sec:instability:flip:interp}

Strong membership inference claims are \emph{conditional on a training pipeline}: 
given a fixed recipe (model class, optimizer, etc.), does the model's behavior on \(\vx\) indicate that \(\vx\) was in the training set? 
Strong MIAs such as LiRA estimate the sample $\vx$'s score given that $\vx\!\in\!\text{IN}$ vs.\ the score given that $\vx\!\in\!\text{OUT}$, computed from many reference models by varying the data, i.e., they probe \emph{data counterfactuals} under the same pipeline.

Our flip rate analysis probes the complementary axis: 
\emph{model counterfactuals}. 
We draw equally plausible targets \(r\!\sim\!\mu\) from the same pipeline using the same training dataset (changing only the seed that controls batch order), hold the set of references and calibration procedure fixed (i.e., thresholds calibrated per target on non‑members), and ask whether the binary decision for a sample \(\vx\) is stable across those targets.
References address \emph{data} variability; 
target flip rate quantifies \emph{model} variability. 
From a statistical reliability perspective, per-sample inference should be robust with respect to both.\looseness=-1

\paragraph{Descriptive vs.\ inferential claims.}
Put another way, both axes may matter depending on the type of claim we want to make. 
We can make a \emph{descriptive} claim~\citep{chouldechova2025comparison}: 
``On this specific model, we called \(\vx\) a `member' and were right.''
In contrast, we can also make an \emph{inferential} claim:
``If we retrained the same pipeline, we would still call \(\vx\) a `member' most of the time.''
Flip rate distinguishes between these. 
In the absence of flip rate analysis, we can make the first type of claim relatively easily, as in the prior work in the literature~\citep[e.g.,][]{carlini2022membership}. 
But without measuring something like flip rate, we cannot assess the second type of claim. 
When we do measure flip rate, observing high values (i.e., ${\approx}0.5$) means that the second claim fails: 
a positive call on \(\vx\) carries about the same evidential weight as a coin toss; 
it may be correct, but not because of stable signal.
This type of evidential weight is weak because this call is not reproducible---even if it is correct on the particular target that the attacker happens to be attacking.
(Conversely, think of a sample \(\vx\) that has low flip rate: 
the attack is able to produce reliable signal about membership for that \(\vx\).)

\paragraph{Why ML security should care about inferential (not just descriptive) claims.}
On a first read of our flip rate results, a security-focused reader may not necessarily be interested in making an inferential claim.
An attacker only gets one model, and so one might think, ``our positive call was correct: 
why care about reliability with respect to the training pipeline that produced that one model?''
From this perspective, the descriptive claim suffices.

However, in machine learning, focusing only on descriptive claims is arguably insufficient; 
for a statistical procedure (like a strong MIA), the standard for success is higher. 
That is, to develop \emph{scientific knowledge}, we require the \emph{entire attack procedure} to be reliable for making claims about \(\vx\).
From this perspective, a statistically coin-flip-like true positive is a lucky hit, not reliable evidence produced by this specific attack procedure.
(Again, conversely, a stable true positive is reliable evidence produced by this procedure.)\looseness=-1

This is why we perform flip rate analysis:
it is a useful diagnostic for making scientifically sound claims about the entire attack procedure, which is precisely what we are trying to address in this paper:
the extent to which strong MIAs (as an attack procedure) can succeed on LLMs.
For an attacker that attacks a single target, traditional MIA metrics ($\tpr$ at fixed $\fpr$, $\rocauc$) are still useful;
but our results surface what fraction of decisions are fragile. 
For reliable scientific knowledge, if that fraction is large, then claims about the MIA procedure's effectiveness should be couched as descriptive, not inferential.

\subsubsection{An alternative instability diagnostic: fixing the target, resampling reference sets}
\label{app:sec:instability:flip:alt}

The instability analysis in the main paper and the rest of this appendix (Section~\ref{sec:instability} \& Appendices~\ref{app:sec:instability:flip}--~\ref{app:sec:instability:flip:interp}) fixes the IN/OUT reference set and varies the target seed, probing \emph{model counterfactuals}.
A natural complementary diagnostic, closer in spirit to the typical threat model, fixes the target and varies the IN/OUT references, probing \emph{reference counterfactuals}.
Here we formalize that diagnostic and explain why we did not run it at scale.\looseness=-1

\paragraph{Setup and notation.}
Fix a trained target \(h\) on dataset \(\sD\).
Let \(\sS^{(1)},\ldots,\sS^{(B)}\) be \(B\) independent reference sets; each set
\(\sS^{(j)}\) contains \(|\Phi_\text{IN}|{=}R_\text{IN}\) IN references and \(|\Phi_\text{OUT}|{=}R_\text{OUT}\) OUT references (in our experiments, \(R_\text{IN}{+}R_\text{OUT}{=}128\)).
Given \(\sS^{(j)}\), define the LiRA score for a sample \(\vx\) as \(\Lambda_h^{(j)}(\vx)\in\mathbb{R}\), computed exactly as in the main setup (Appendix~\ref{app:sec:background}) but using only the models in \(\sS^{(j)}\) as the references.

\paragraph{Calibration to non-members at fixed \(\fpr\) (per reference set).}
Let the non-member dataset be \(\sD_{\text{OUT}}\) with size \(N_\text{OUT}\).
Form the empirical CDF of OUT scores using reference set \(j\):
\[
\widehat{F}_{\text{OUT}}^{(j)}(t)
=
\frac{1}{N_{\text{OUT}}}
\sum_{\vx\in\sD_{\text{OUT}}}\;\1\{\Lambda_h^{(j)}(\vx)\le t\}.
\]
For each reference set $j$ and $\fpr$ $\eta$, we calibrate a decision rule for the single target $h$, choosing the right-open (left-limit CDF) cutoff
\[
\tau_h^{(j)}(\eta)
\;=\;
\inf\bigl\{t:\ \widehat{F}_{\text{OUT}}^{(j)}(t^{-})\ge 1-\eta\bigr\},
\qquad
b_h^{(\eta,j)}(\vx)=\1\{\Lambda_h^{(j)}(\vx)\ge \tau_h^{(j)}(\eta)\}.
\]
Then the realized $\fpr$ on the non-member (OUT) dataset for the target $h$ using references \(j\) satisfies
\[
\widehat{\fpr}^{(j)}(\eta)
=
\frac{1}{N_{\text{OUT}}}
\sum_{\vx\in\sD_{\text{OUT}}}\1\{\Lambda_h^{(j)}(\vx)\ge \tau_h^{(j)}(\eta)\}
=
1-\widehat{F}_{\text{OUT}}^{(j)}\bigl(\tau_h^{(j)}(\eta)^{-}\bigr)
\;\le\;\eta,
\]
i.e., finite-sample, conservative with ties.
This mirrors how an attacker would calibrate a threshold for the target using the specific reference set they happened to train.
(In this case, we train $B$ such equally plausible reference sets.) 

\paragraph{Reference-resampling flip rate.}
Let \(J\) denote an index drawn uniformly from \(\{1,\ldots,B\}\), representing an i.i.d.\ draw of a reference set.
For this reference-resampling setup, define the population flip rate across possible reference sets for the fixed target \(h\) as
\[
\mathrm{flip}^{\text{ref}}_{\eta}(\vx)
\;\coloneqq\;
\Pr_{J,J'\stackrel{\text{i.i.d.}}{\sim}\{1,\ldots,B\}}
\big[b_{h}^{(\eta,J)}(\vx)\neq b_{h}^{(\eta,J')}(\vx)\big]
\;=\; 2\,\theta^{\text{ref}}\bigl(1-\theta^{\text{ref}}\bigr),
\]
where \(\theta^{\text{ref}} \coloneqq \Pr_{J}\big[b_h^{(\eta,J)}(\vx)=1\big]\).
Given \(B\ge 2\) reference sets, the unbiased estimator is
\[
\widehat{\mathrm{flip}}^{\text{ref}}_{\eta,B}(\vx)
=
\binom{B}{2}^{-1}
\sum_{1\le j<\ell\le B}
\1\big\{b_{h}^{(\eta,j)}(\vx)\neq b_{h}^{(\eta,\ell)}(\vx)\big\}
=
\frac{2\,B_0^{\text{ref}}(\vx)\,B_1^{\text{ref}}(\vx)}{B\,(B-1)},
\]
with \(B_1^{\text{ref}}(\vx)=\sum_{j=1}^B b_{h}^{(\eta,j)}(\vx)\) and \(B_0^{\text{ref}}(\vx)=B{-}B_1^{\text{ref}}(\vx)\).
Just as in our target-resampling analysis, we label a decision ``coin-flip-like at level \(\alpha\)'' by testing
\(H_0:\theta^{\text{ref}}=0.5\) with the exact two-sided binomial test on \(B\) votes; equivalently,
\(\widehat{\mathrm{flip}}^{\text{ref}}_{\eta,B}(\vx)\ge t_\alpha(B)\), where
\(t_\alpha(B)=\tfrac{2\,k_L(B-k_L)}{B(B-1)}\) and \(k_L\) is the equal-tail cutoff (Appendix~\ref{app:sec:instability:flip:arbitrary}).

\paragraph{Cost and feasibility.}
In the main paper target-resampling flip analyis, we trained one reference set of size \(R{=}128\) and \(B_\text{target}{=}127\) targets
(\(R{+}B_\text{target}=255\) total trained models for the \(302\)M setting).
For this reference-resampling diagnostic, to obtain  comparable statistical fidelity, one would fix the target and train \(B_\text{ref}\) independent reference sets, each of size \(R\), i.e., train \(R\times B_\text{ref}\) reference models.
Taking \(B_\text{ref}\approx 128\) yields \(128\times 128=16{,}384\) references (plus one target), requiring about \(4\times\) more models than all of the other experiments in this paper combined.
To approximate this diagnostic, we could perhaps subsample smaller reference sets from a single larger pool (e.g., subsample the $128$ references from the target flip rate experiments).
However, such subsampling would introduce dependence across reference sets, violating the i.i.d.\ assumption and complicating the exact binomial test we use to determine the coin-flip cutoff.

\paragraph{Why we expect qualitatively similar instability to our target-resampling experiments.}
Empirically, our cross-target flip rate analysis shows that many members have small margins relative to the calibrated threshold: 
for these \(\vx\), \(\Lambda_r(\vx)-\tau_r(\eta)\) is frequently near \(0\) (Appendix~\ref{app:sec:instability:flip:results}).

For the reference-resampling diagnostic, write the per-reference-set margin for the fixed target \(h\) as
\begin{align*}
\Delta^{(j)}(\vx)
\;&\coloneqq\;
\Lambda_h^{(j)}(\vx)-\tau_h^{(j)}(\eta)\\
\;&=\;
\underbrace{m_h(\vx)}_{\text{target + data fixed}}
\;+\;
\underbrace{\varepsilon_\Lambda^{(j)}(\vx)}_{\text{score variability from ref.\ sets}}
\;-\;
\underbrace{\varepsilon_\tau^{(j)}}_{\text{quantile variability from per-set calibration}},
\end{align*}
where \(\varepsilon_\Lambda^{(j)}\) captures sampling variability from fitting IN/OUT reference statistics and
\(\varepsilon_\tau^{(j)}\) captures quantile-estimation variability in \(\tau_h^{(j)}(\eta)\).
Our cross-target results show many members with margins near \(0\) (poor separability of reference distributions and proximity to the threshold).
For such samples \(\vx\), any nonzero variability in
\(\varepsilon_\Lambda^{(j)}(\vx)-\varepsilon_\tau^{(j)}\) will, with nontrivial probability, change the sign of \(\Delta^{(j)}(\vx)\) and thus flip the decision.
And so, given the sensitivity that we observe when training counterfactuals, we expect this reference-resampling diagnostic to surface qualitatively similar per-sample instability as the target-resampling experiments we run---especially for members---without running the much more expensive reference-resampling procedure.
%\clearpage
\section{Additional per-sample MIA vulnerability results}\label{app:moreperexamplemiaresults}

\begin{figure}[h]
\centering
\begin{subfigure}[t]{0.45\textwidth}
    \includegraphics[width=0.95\linewidth]{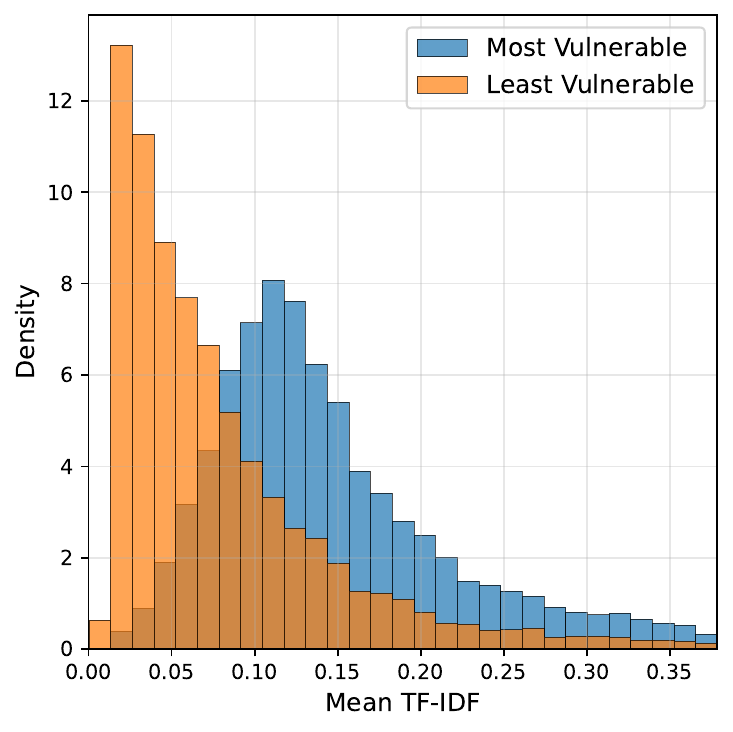}
        \caption{Mean $\mathrm{TF}${-}$\mathrm{IDF}$ scores by vulnerability}
        \label{fig:app:tfidf_distribution}
\end{subfigure}\hfill
\begin{subfigure}[t]{0.45\textwidth}
    \includegraphics[width=0.95\linewidth]{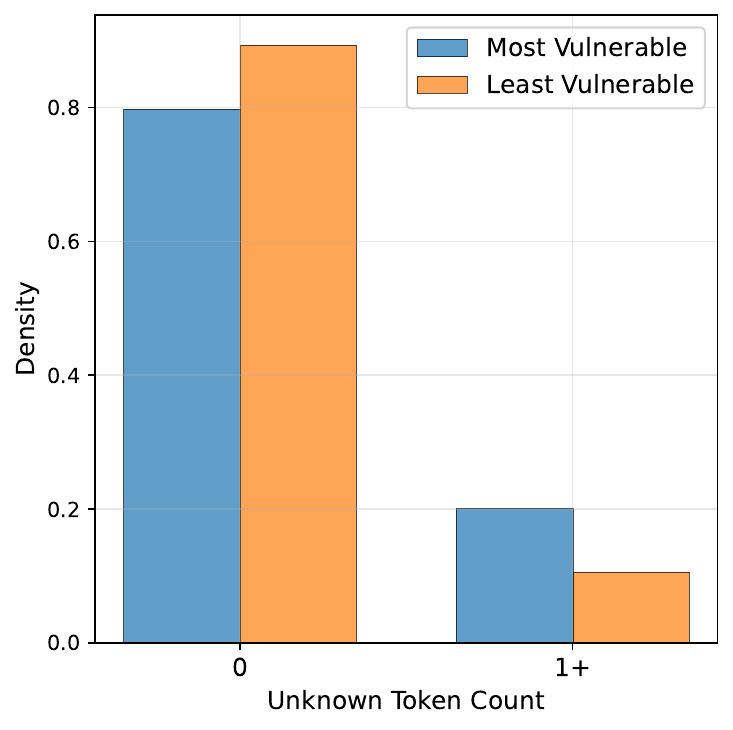}
        \caption{\texttt{<unk>} counts by vulnerability}
        \label{fig:app:unk_token_distribution}
\end{subfigure}%
\caption{\textbf{Text property distributions by MIA vulnerability.} The most vulnerable samples tend to (\textbf{a}) have higher $\mathrm{TF}${-}$\mathrm{IDF}$ scores compared to least vulnerable samples, and  (\textbf{b}) are more likely to contain at least one unknown token (\texttt{<unk>}).}
\label{fig:app:per_example_more_stats}
\end{figure}

Figure~\ref{fig:ex:dist-vulnerable} indicates that it is often the case that vulnerable sequences tend to be longer. 
Beyond sequence length, we observe that samples more vulnerable to MIA tend to have higher mean $\mathrm{TF}${-}$\mathrm{IDF}$ scores (\Cref{fig:app:tfidf_distribution}), suggesting that texts with distinctive, uncommon terms may exhibit stronger signals for membership inference. 
We compute these $\mathrm{TF}${-}$\mathrm{IDF}$ scores without normalization, collecting document frequency statistics over a random subsample of the original dataset and then taking the mean across all tokens in each sample. Similarly, samples containing unknown tokens (\texttt{<unk>}) appear more vulnerable to MIA (\Cref{fig:app:unk_token_distribution}). 

\subsection{Does memorization imply strong membership inference attacks?}

While memorization is a key factor that can make a model susceptible to membership inference attacks, it does not automatically guarantee that strong MIAs will always be successful. 
Memorization refers to a model learning specific details about its training data, rather than just general patterns.

When a model heavily memorizes training samples, it often exhibits distinct behaviours for these samples, which MIA attackers, in principle, can exploit. 
Indeed, studies have shown that the risk of membership inference is often highest for those samples that are highly memorized~\citep{carlini2021extracting}. 
However, our results show that the practical success and strength of a particular MIA can also depend on other factors, such as the model architecture, the type of data, the specifics of the attack method, and whether the memorization leads to clearly distinguishable outputs or behaviors for member versus non-member samples. 
Some models might memorize data in ways that are not easily exploitable by current MIA techniques, or the signals of memorization might be subtle for well-generalizing models, making strong attacks more challenging despite the presence of memorization.
%\clearpage
\subsection{Evolution of losses over different model sizes}\label{app:anecdotal_evidence}

\begin{figure*}[htbp!]
\begin{minipage}{0.045\textwidth}
    \textbf{\;}
    \vspace{.3cm}
\end{minipage}%
\hfill
\begin{minipage}{0.312\textwidth}
    \centering
    \textbf{\hspace{.6cm}Sample ID: 16777211, Member}
    \vspace{.3cm}
\end{minipage}%
\hfill
\begin{minipage}{0.312\textwidth}
    \centering
    \textbf{\hspace{.6cm}Sample ID: 16777212, Member}
    \vspace{.3cm}
\end{minipage}%
\hfill
\begin{minipage}{0.312\textwidth}
    \centering
    \textbf{\hspace{.6cm}Sample ID: 16777213, Non-member}
    \vspace{.3cm}
\end{minipage}
\captionsetup[subfigure]{justification=centering}
  \centering
\begin{subfigure}[t]{0.045\textwidth}
\vspace{-1.65cm}
\hspace{-.3cm}\textbf{10M}
\vspace{1cm}
\end{subfigure}%
\hfill
\begin{subfigure}[t]{0.312\textwidth}
\centering
    \includegraphics[width=1.\linewidth]{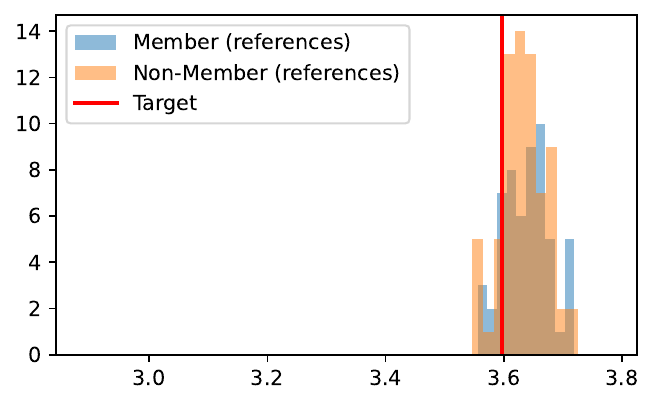}
\end{subfigure}%
\hfill
\begin{subfigure}[t]{0.312\textwidth}
\centering
    \includegraphics[width=1.\linewidth]{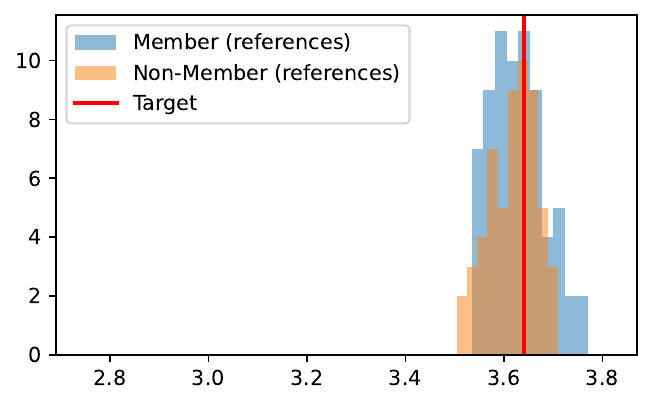}
\end{subfigure}%
\hfill
\begin{subfigure}[t]{0.312\textwidth}
\centering
     \includegraphics[width=1.\linewidth]{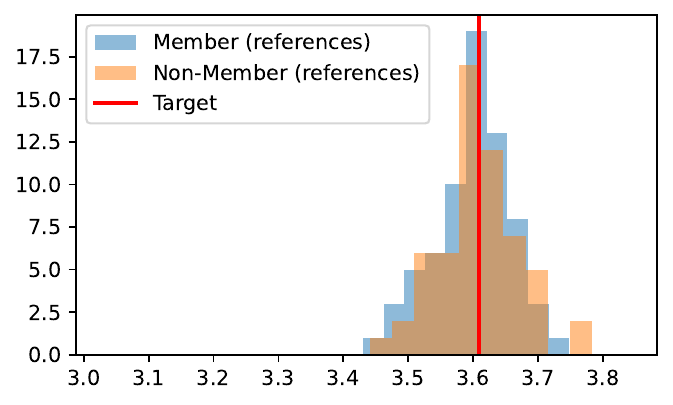}
\end{subfigure}\\%
\begin{subfigure}[t]{0.045\textwidth}
\vspace{-1.65cm}
\hspace{-.3cm}\textbf{44M}
\vspace{1cm}
\end{subfigure}%
\hfill
\captionsetup[subfigure]{justification=centering}
  \centering
\begin{subfigure}[t]{0.312\textwidth}
\centering
    \includegraphics[width=1.\linewidth]{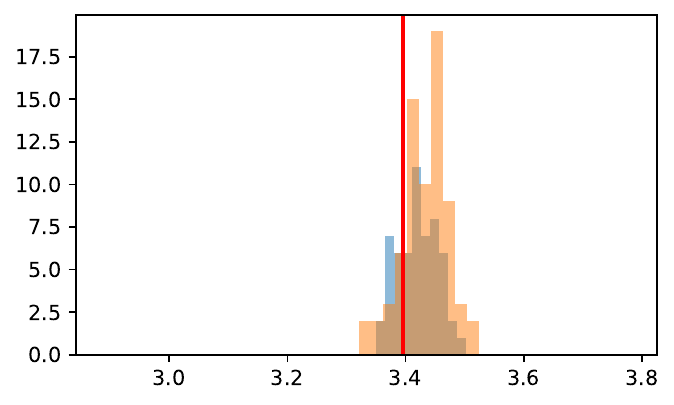}
\end{subfigure}%
\hfill
\begin{subfigure}[t]{0.312\textwidth}
\centering
    \includegraphics[width=1.\linewidth]{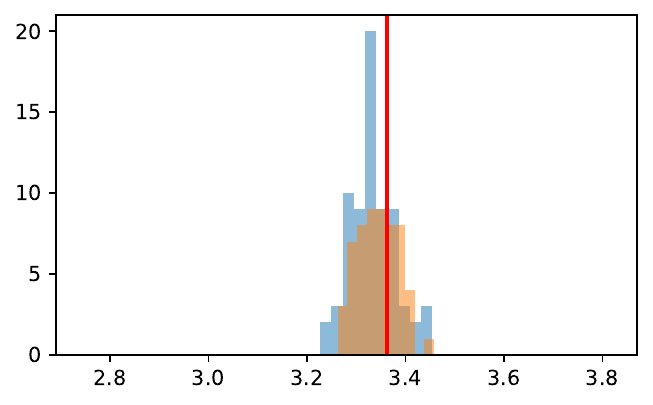}
\end{subfigure}
\hfill
\begin{subfigure}[t]{0.312\textwidth}
\centering
         \includegraphics[width=1.\linewidth]{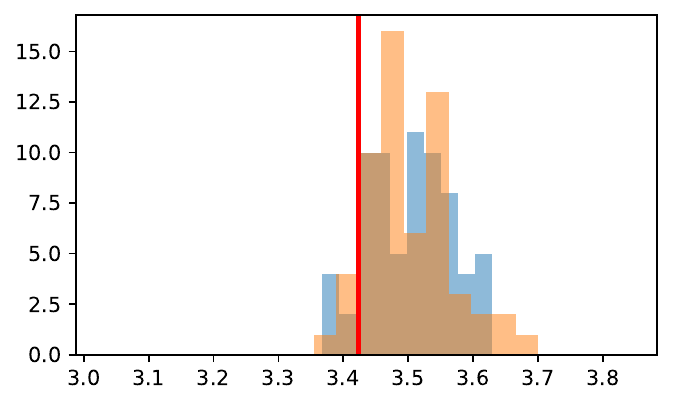}
\end{subfigure}\\
\begin{subfigure}[t]{0.045\textwidth}
\vspace{-1.65cm}
\hspace{-.3cm}\textbf{85M}
\vspace{1cm}
\end{subfigure}%
\hfill
\begin{subfigure}[t]{0.312\textwidth}
\centering
    \includegraphics[width=1.\linewidth]{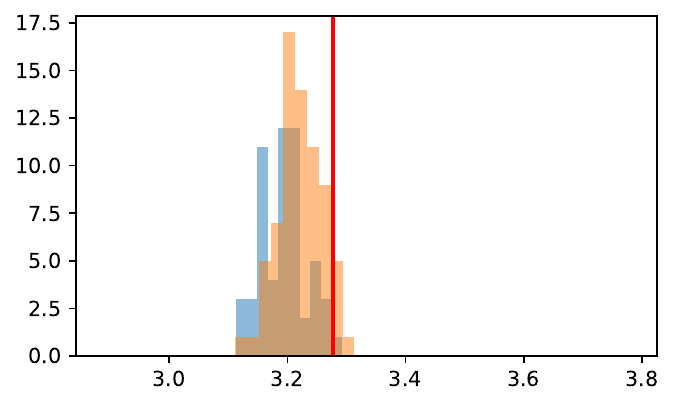}
\end{subfigure}%
\hfill
\begin{subfigure}[t]{0.312\textwidth}
\centering
    \includegraphics[width=1.\linewidth]{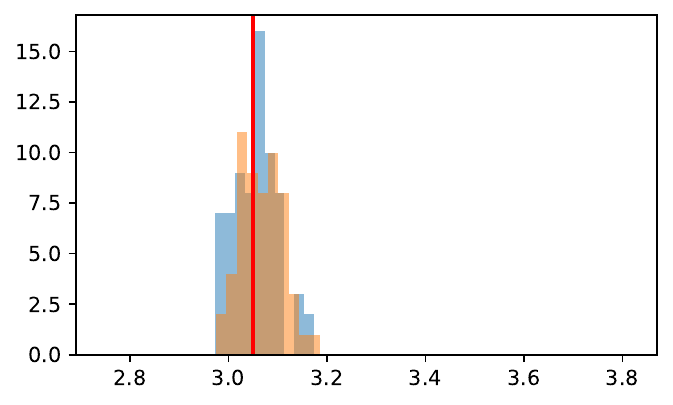}
\end{subfigure}
\hfill
\begin{subfigure}[t]{0.312\textwidth}
\centering
         \includegraphics[width=1.\linewidth]{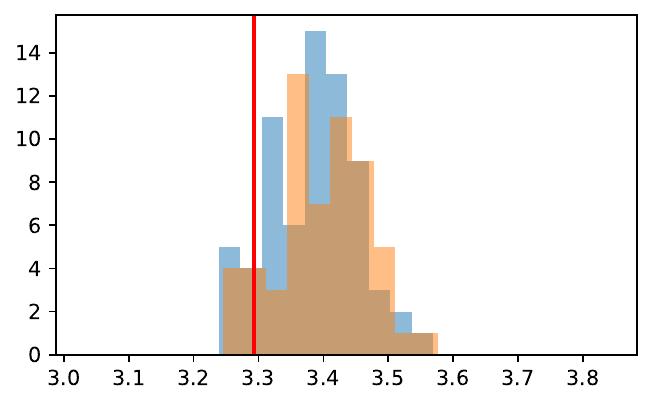}
\end{subfigure}\\%
\begin{subfigure}[t]{0.045\textwidth}
\vspace{-1.65cm}
\hspace{-.3cm}\textbf{140M}
\vspace{1cm}
\end{subfigure}%
\hfill
\begin{subfigure}[t]{0.312\textwidth}
\centering
    \includegraphics[width=1.\linewidth]{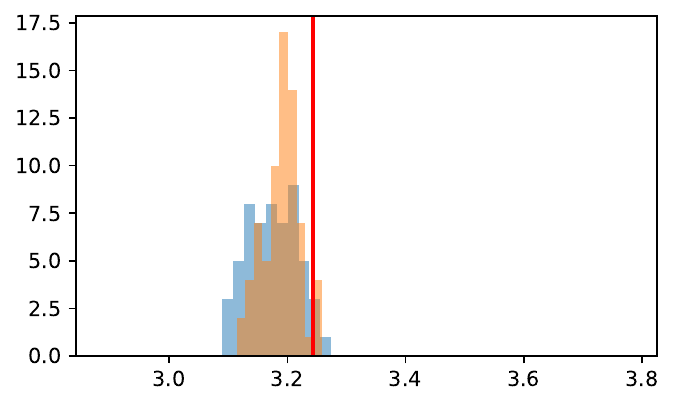}
\end{subfigure}%
\hfill
\begin{subfigure}[t]{0.312\textwidth}
\centering
    \includegraphics[width=1.\linewidth]{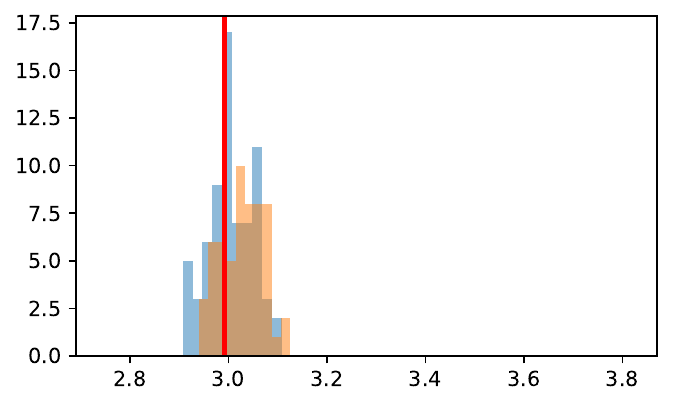}
\end{subfigure}
\hfill
\begin{subfigure}[t]{0.312\textwidth}
\centering
         \includegraphics[width=1.\linewidth]{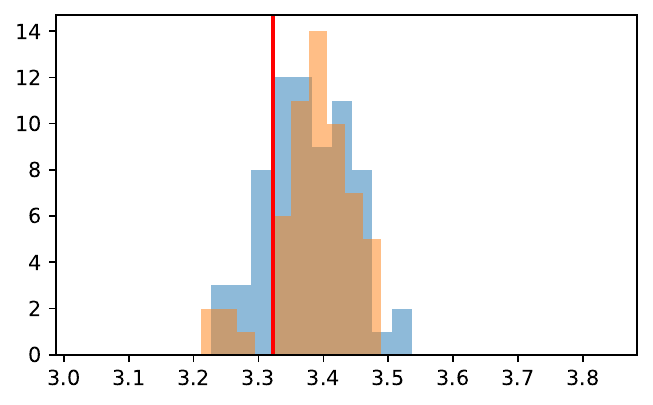}
\end{subfigure}\\%
\begin{subfigure}[t]{0.045\textwidth}
\vspace{-1.65cm}
\hspace{-.3cm}\textbf{302M}
\vspace{1cm}
\end{subfigure}%
\hfill
\begin{subfigure}[t]{0.312\textwidth}
\centering
    \includegraphics[width=1.\linewidth]{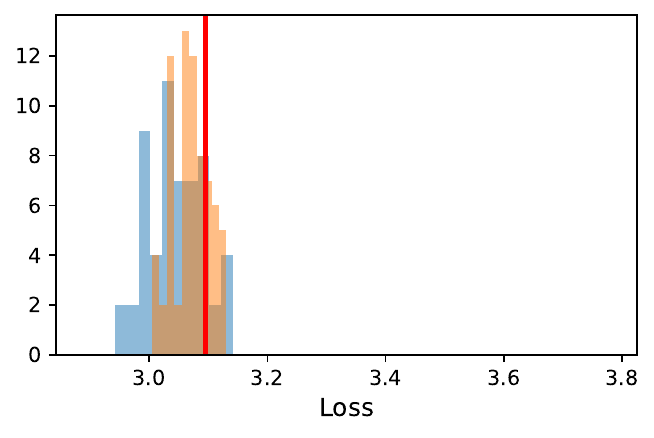}
\end{subfigure}%
\hfill
\begin{subfigure}[t]{0.312\textwidth}
\centering
    \includegraphics[width=1.\linewidth]{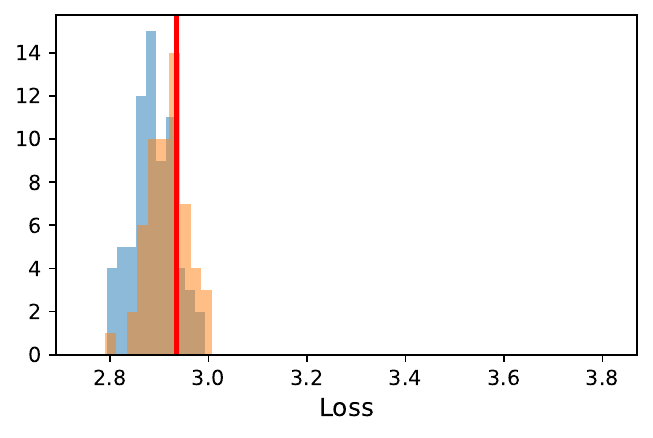}
\end{subfigure}
\hfill
\begin{subfigure}[t]{0.312\textwidth}
\centering
         \includegraphics[width=1.\linewidth]{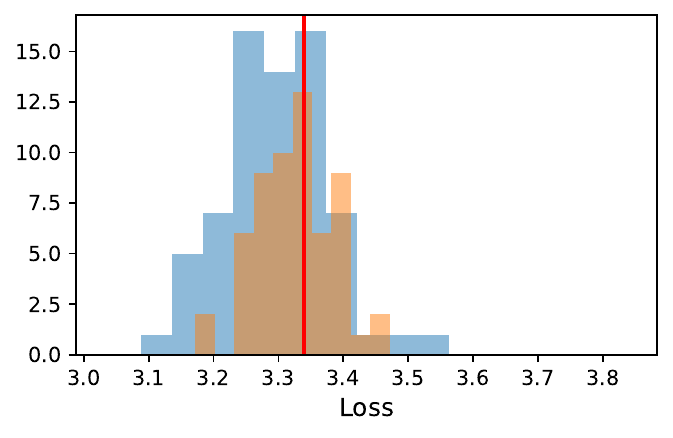}
\end{subfigure}\\
\caption{\textbf{Target loss and reference loss distributions for three samples.} 
For three different samples (referenced by their IDs in the C4 dataset), we plot the reference  distributions and the loss of the sample for the target model (as a vertical red line). 
Each row shows results for a different model size.\looseness=-1
}
\label{fig:in_out_dis_examples}
\end{figure*}

In \Cref{fig:in_out_dis_examples}, for three samples, we plot loss (target) and the per-sample reference distributions $p_\text{IN}(\vx)$ and $p_\text{OUT}(\vx)$ over different model sizes. 
Each of these models is trained for $1$ epoch on $2^{23}\approx8.3$M samples.
This is a sanity check that the losses decrease (on the same sample) as the model size increases.
Note that, for these samples, the distance between member and non-member reference distributions does not  significantly shift as the model size grows.

\FloatBarrier

\clearpage
\section{Experiment configuration details}\label{app:exp_details}

In \Cref{tab:exp_details}, we provide pecific experimental  settings.
Unless otherwise stated, we used the AdamW optimizer~\citep{loshchilov2017decoupled} with a cosine scheduler.
The initial learning rate is set to $10^{-7}$ and increases linearly over $750$ warm up steps to a peak learning rate of $3\cdot10^{-4}$, after which it decreases according to the cosine schedule to a final value of $3\cdot10^{-5}$. 
We typically use $128$ reference models and a single target model to measure MIA vulnerability, drawing a dataset of size $2N$ from C4 from which we subsample training datasets of size $N$.
For each reference and target model, the training set is subsampled from the same larger dataset of size $2N$.
This means each sample in this larger dataset is a member for ${\approx}64$ reference models.
The batch size is fixed to $128$ and sequence length to $1024$; 
if an sample has fewer tokens, we pad to $1024$. 
The weight decay is set to $0.1$, and a global clipping norm is set to $1.0$. 
Note that we can approximately convert the training set size to total number of training tokens by multiplying the training set size by $400$, as this the approximate average number of tokens within a C4 sample.
For example, this means the $1018$M model was trained on $20.4$B tokens in Figure~\ref{fig:compute-optimal-main}.\looseness=-1 

%\todo{Jamie: compute table could go here?}

\begin{table}[htbp]
\caption{\textbf{Experimental details.} Experiment (figure), training set size (approximate number of samples), model size, and specific details that diverge from default settings.}
\vspace{.2cm}
\label{tab:exp_details}
\resizebox{\textwidth}{!}{%
\begin{tabular}{llll}
\toprule
\multirow{2}{*}{Experiment} & \multirow{2}{*}{Training set size} & \multirow{2}{*}{Model size} & \multirow{2}{*}{Other information (which diverges from default experimental settings)}\\
&                 &            &  \\
\midrule
Figure~\ref{fig:lira_references} & $7$M & $140$M & Max. $512$ references\\
\cmidrule{2-4}
\multirow{8}{*}{Figures~\ref{fig:compute-optimal-main},~\ref{fig:compare_model_sizes_loss}} & $500$K & $10$M & \multirow{8}{*}{$128$ references}\\
& $2.2$M  & $44$M  & \\
& $4.25$M & $8$5M  & \\
& $7$M    & $140$M & \\
& $15.1$M & $302$M & \\
& $24.4$M & $489$M & \\
& $30.2$M & $604$M & \\
& $50.9$M & $1018$M & \\
\cmidrule{2-4}
\multirow{2}{*}{Figure~\ref{fig:other-training:split}} & $2.2$M & $44$M & \multirow{2}{*}{$2$ different variations; $1$ epoch and $2$ epochs (on the same $2.2$M samples, but split in different ways across epochs)}  \\
& $1.1$M & $44$M & \\
\cmidrule{2-4}
Figures~\ref{fig:other-training:epochs},~\ref{fig:compare_epochs_10_app}  & $7$M & $140$M & $10$ epochs \\
\cmidrule{2-4}
\multirow{6}{*}{Figure~\ref{fig:other-training:sizes}} & $50$K  & $140$M & \multirow{6}{*}{$80$ warm up steps} \\
& $100$K & $140$M & \\
& $500$K & $140$M & \\
& $1$M & $140$M & \\                                              
& $5$M & $140$M & \\
& $10$M & $140$M & \\
\cmidrule{2-4}
\multirow{10}{*}{Figure~\ref{fig:other-training:fixed-sized}} & \multirow{10}{*}{$2^{23}$} & $10$M & \multirow{10}{*}{} \\
& & $44$M & \\
& & $85$M & \\
& & $140$M & \\
& & $302$M & \\
& & $425$M & \\
& & $489$M & \\
& & $509$M & \\
& & $604$M & \\
& & $1018$M & \\
\cmidrule{2-4}
\shortstack{Figures~\ref{fig:instability:main},~\ref{fig:302M-archetypes},~\ref{fig:302M-rocauc-instability},\\~\ref{fig:error-v-tau},~\ref{fig:302-lowlevel-flip},~\ref{fig:302M-instability-varied-fpr},~\ref{fig:302M-rocauc-instability-2}} & $7$M & $302$M & $127$ target models (varying only in random seed; same training data); $128$ references\\
\cmidrule{2-4}
Figures~\ref{fig:other-training-other},~\ref{fig:compare_memorization_largest},~\ref{fig:compare_value_type} & $7$M & $140$M & $128$ references\\
\cmidrule{2-4}
Figure~\ref{fig:compare:lira} & $7$M & $140$M & Max. $256$ references \\
\cmidrule{2-4}
Figure~\ref{fig:compare:rmia} & $7$M & $140$M & Max. $64$ references, $10$K $\sZ$ population size \\
\cmidrule{2-4}
Figure~\ref{fig:compare_attacks_ref_models} & $7$M & $140$M & $256$ references \\
\cmidrule{2-4}
Figure~\ref{fig:compare_lira_rmia_simple} & $500$K & $10$M & $10$K $\sZ$ population size \\
\cmidrule{2-4}
Figures~\ref{fig:rmia_vary_z},~\ref{fig:rmia_vary_gamma} & $500$K & $10$M & $10$K-$300$K $\sZ$ population size; $64$ references \\
\cmidrule{2-4}
Figure~\ref{fig:rmia_offline_online} & $2^{19}$ & $10$M & up to $128$ references (testing online and offline variants)\\
\cmidrule{2-4}
Figure~\ref{fig:compare_lrs} & $50$K & $140$M & Learning rate schedules: cosine, cosine with $0$ weight decay, cosine with no clipping, linear. We use $50$ warm up steps. \\
\cmidrule{2-4}
Figure~\ref{fig:duplicates} & $7$M & $140$M & Comparing to de-duplicated training dataset\\
\cmidrule{2-4}
Figure~\ref{fig:compare_epochs_20_app} & $2^{19}$ & $140$M & $20$ epochs \\
\cmidrule{2-4}
Figure~\ref{fig:compare_model_sizes_chinchilla_dataset_rocs} & - & - &  Identical to Figure~\ref{fig:compute-optimal-main}, where we use $16$ different target models \\
\cmidrule{2-4}
Figure~\ref{fig:compare_model_sizes_fixed_dataset_rocs}& - & - & Identical to Figure~\ref{fig:other-training:fixed-sized}, where we use $16$ different target models \\ 
\cmidrule{2-4}
Figure~\ref{fig:compare_model_sizes_chinchilla_dataset_rocs_num_refs} & & & Identical to Figure~\ref{fig:compare_model_sizes_chinchilla_dataset_rocs}, except varying references (up to $128$).\\ 
\cmidrule{2-4}
Figure~\ref{fig:instability} & $2^{23}$ & $140$M       &  Up to $64$ targets, $64$ references \\ 
\cmidrule{2-4}
\shortstack{Figures~\ref{fig:140M-archetypes},~\ref{fig:140M-rocauc-instability},~\ref{fig:error-v-tau},\\~\ref{fig:140-lowlevel-flip},~\ref{fig:140M-instability-varied-fpr},~\ref{fig:140M-rocauc-instability-2}} & $7$M & $140$M & $125$ target models (varying only random seed; same training data); $128$ references\\
\cmidrule{2-4}
% Figure~\ref{fig:per_example_mia_influences}  
Figure~\ref{fig:app:per_example_more_stats} & - & - &  Identical to Figure~\ref{fig:ex:dist-vulnerable} \\ 
\cmidrule{2-4}
Figure~\ref{fig:in_out_dis_examples} & $2^{23}$ & - &  $10$M-$302$M model sizes \\ 
\bottomrule
\end{tabular}
}
\end{table}

\end{document}